\documentclass[british,a4paper,12pt]{phdthesis}
\usepackage[utf8]{inputenc}

\usepackage{graphicx}
\usepackage{array} % for \extrarowheight
\usepackage{algorithmic} % allow psuedocode

\usepackage[british]{babel}
\usepackage{amsmath}
\usepackage{amssymb}
\usepackage{setspace}
\usepackage{url}
\usepackage{fancyvrb}
\usepackage{listings}

\title{Internalising Interaction Protocols as First-Class Programming Elements in Multi Agent Systems}

\author{David J. Lillis, B.A., H.Dip., M.Sc.}

\authoraddress{
        School of Computer Science and Informatics \\
        University College Dublin \\
        Belfield, Dublin 4, Ireland
}

\date{\today}

\submission{The thesis is submitted to \\
  University College Dublin\\
  for the degree of Doctor of Philosophy\\
  in the\\
  College of Science}

\department{School of Computer Science and Informatics}
\hos{Mr. John Dunnion}
\supervision{Dr. Rem W. Collier \\ Mr. John Dunnion}
\submissionmonth{October}
\submissionyear{2012}

\begin{document}

% change default second-level bullet
\renewcommand{\labelitemii}{$\circ$}

\maketitle
\tableofcontents
\listoftables
\listoffigures

\begin{abstract}
\vspace{-1.5cm}
Since their inception, Multi Agent Systems (MASs) have been championed as a solution for the increasing problem of software complexity. Communities of distributed autonomous computing entities that are capable of collaborating, negotiating and acting to solve complex organisational and system management problems are an attractive proposition. Central to this is the requirement for agents to possess the capability of interacting with one another in a structured, consistent and organised manner.

This thesis presents the Agent Conversation Reasoning Engine (ACRE), which constitutes a holistic view of communication management for MASs. ACRE is intended to facilitate the practical development, debugging and deployment of communication-heavy MASs.

ACRE has been formally defined in terms of its operational semantics, and a generic architecture has been proposed to facilitate its integration with a wide variety of diverse agent development frameworks and Agent Oriented Programming (AOP) languages. A concrete implementation has also been developed that uses the Agent Factory AOP framework as its base. This allows ACRE to be used with a number of different AOP languages, while providing a reference implementation that other integrations can be modelled upon. A standard is also proposed for the modelling and sharing of agent-focused interaction protocols that is independent of the platform within which a concrete ACRE implementation is run.

Finally, a user evaluation illustrates the benefits of incorporating conversation management into agent programming.

\end{abstract}

\chapter*{STATEMENT OF ORIGINAL AUTHORSHIP}
  \vspace{-1.5cm}
I hereby certify that the submitted work is my own work, was completed while registered as a candidate for the degree stated on the Title Page, and I have not obtained a degree elsewhere on the basis of the research presented in this submitted work.

\begin{dedication}
  \vspace{-1.5cm}
  For Hailey. Making 2012 count.
 \end{dedication}

\begin{acknowledgements}
\vspace{-1.5cm}
This thesis would not have been possible but for the support and encouragement I have received over many years from a number of people. I owe a great debt of gratitude to my supervisor, Rem Collier. He has been a friend as well as a mentor, and has involved me in numerous research projects, which provides me with a strong platform for the rest of my career.

Fergus Toolan is one of the principal reasons why I now work in research. I am particularly thankful that he took the time to proofread this work in its entirety. He has been an important source of both guidance and sanity during my time in UCD.

Howell Jordan was also kind enough to act as a proofreader for this work, and his input on evaluation methods was invaluable. I am also very grateful that Howell was willing to present aspects of this work at ProMAS when I was unable to attend.

Of course, neither education nor living are free. I would like to thank all those who saw fit to employ me in some capacity during the time this work was ongoing. This includes Eamonn Nolan and Waseem Akhtar in Griffith College Dublin; Mark Hargaden in the Geary Institute UCD; Dorothy and Tom Meaney of Elite Accounting; and Greg O'Hare in the CLARITY centre, UCD.

In particular, I am extremely grateful to John Dunnion and Henry McLoughlin, who were instrumental in giving me the opportunity to greatly broaden my horizons by spending a semester lecturing in Fudan University, Shanghai. Both have also been great sources of advice and guidance throughout my academic career.

My parents deserve special mention. They have always been great believers in education and have greatly supported me, from forcing me to complete primary school homework through to giving me the opportunity to attend university. They have always encouraged me to achieve my potential and for that I am very grateful.

Finally, thank you Hailey. It is difficult to imagine having a more supportive, patient and loving person by my side throughout this journey and many journeys to come.

\end{acknowledgements}

\begin{listofpublications}

The following is a list of publications by the author in the area of Multi Agent Systems. Those publications that are directly relevant to the work in this thesis are marked with an asterisk.

\begin{itemize}

\item \emph{David Lillis}, Howell R. Jordan, and Rem. W. Collier. Evaluation of a Conversation Management Toolkit for Multi Agent Programming. In \emph{Proceedings of the 10th International Workshop on Programming Multi-Agent Systems (ProMAS 2012)}, Valencia, Spain, 2012.*

\item \emph{David Lillis} and Rem W. Collier. Augmenting Agent Platforms to Facilitate Conversation Reasoning. In \emph{M. Dastani, A. E. F. Seghrouchni, J. F. Hubner, and J. Leite, editors, Post-proceedings of the 3rd International Workshop on LAnguages, methodologies and Development tools for multi-agent systemS}, Lyon, France, 2011. Springer.*

\item Dinh Doan Van Bien, \emph{David Lillis}, and Rem W. Collier. Call Graph Profiling for Multi Agent Systems. In \emph{M. Dastani, A. El Fallah Segrouchni, J. a. Leite, and P. Torroni, editors, Languages, Methodologies, and Development Tools for Multi-Agent Systems, LADS '009 Post-Proceedings}, pages 153--167. Springer Berlin / Heidelberg, Sept. 2010.

\item Mauro Dragone, Howell R. Jordan, \emph{David Lillis}, and Rem W. Collier. Separation of Concerns in Hybrid Component and Agent Systems. \emph{International Journal of Communication Networks and Distributed Systems}, 2010.

\item Howell R. Jordan, Jennifer Treanor, \emph{David Lillis}, Mauro Dragone, Rem W. Collier, and G. M. P. O'Hare. AF-ABLE in the Multi Agent Contest 2009. \emph{Annals of Mathematics and Artificial Intelligence}, 2010.

\item \emph{David Lillis} and Rem W. Collier. ACRE: Agent Communication Reasoning Engine. In \textit{3rd International Workshop on LAnguages, Methodologies and Development Tools for Multi Agent SystemS (LADS'010)}, Lyon, 2010.*

\item Dinh Doan Van Bien, \emph{David Lillis}, and Rem W. Collier. Space-Time Diagram Generation for Profiling Multi Agent Systems. In \emph{Proceedings of the 7th International Workshop on PROgramming Multi-Agent Systems (PROMAS 2009), held at the 8th International Joint Conference on Autonomous Agents and Multi-Agent Systems (AAMAS 2009)}, Budapest, Hungary, May 2009.

\item Mauro Dragone, \emph{David Lillis}, Rem W. Collier, and G. M. P. O'Hare. Practical Development of Hybrid Intelligent Agent Systems with SoSAA. In \emph{Proceedings of the 20th Irish Conference on Artificial Intelligence and Cognitive Science}, Dublin, Ireland, Aug. 2009.

\item Mauro Dragone, \emph{David Lillis}, Rem W. Collier, and G. M. P. O'Hare. SoSAA: A Framework for Integrating Agents \& Components. In \emph{Proceedings of the 24th Annual Symposium on Applied Computing (ACM SAC 2009), Special Track on Agent-Oriented Programming, Systems, Languages, and Applications}, Honolulu, Hawaii, USA, Mar. 2009.

\item Mauro Dragone, G. M. P. O'Hare, \emph{David Lillis}, and Rem. W. Collier. Hybrid Agent \& Component-Based Management of Backchannels. In \emph{Proceedings of the 4th International Conference on Software and Data Technologoes (ICSOFT 2009)}, Sofia, Bulgaria, July 2009.

\item Howell R. Jordan, Jennifer Treanor, \emph{David Lillis}, Mauro Dragone, Rem W. Collier, and G. M. P. O'Hare. AF-ABLE: System Description. In \emph{Proceedings of the 10th International Workshop on Computational Logic in Multi-Agent Systems (CLIMA-X)}. Clausthal University of Technology, 2009.

\item \emph{David Lillis}, Rem W. Collier, Mauro Dragone, and G. M. P. O'Hare. An Agent-Based Approach to Component Management. In \emph{Proceedings of the 8th International Conference on Autonomous Agents and Multi-Agent Systems (AAMAS-09)}, Budapest, Hungary, May 2009.

\item Dinh Doan Van Bien, \emph{David Lillis}, and Rem W. Collier. Call Graph Profiling for Multi Agent Systems. In \emph{Proceedings of the Second Multi-Agent Logics, Languages, and Organisations Federated Workshops (MALLOW 2009)}, Turin, 2009.

\item Mauro Dragone, \emph{David Lillis}, Conor Muldoon, Richard Tynan, Rem Collier, and G. M. P. O'Hare. Dublin Bogtrotters: Agent Herders. In \emph{Postproceedings of the 6th International Workshop on Programming Multi-Agent Systems (PROMAS 2008)}, 2008.

\item \emph{David Lillis}, Rem W. Collier, Fergus Toolan, and John Dunnion. Evaluating Communication Strategies in a Multi Agent Information Retrieval System. In \emph{Proceedings of the 18th Irish Conference on Artificial Intelligence and Cognitive Science (AICS 2007)}, pages 81--90, Dublin, Ireland, 2007.

\item \emph{David Lillis}, Rem Collier, Fergus Toolan, and John Dunnion. Evaluating Communication Strategies in a Multi Agent Information Retrieval System. In \emph{Proceedings of the 5th European Workshop on Multi-Agent Systems (EUMAS'07)}, Hammamet, Tunisia, Dec. 2007.
\end{itemize}

\end{listofpublications}

\onehalfspacing

\chapter{Introduction} \label{chap:introduction}

In recent years, increased globalisation and reliance on technology has driven the need for complex software applications on an unprecedented scale. Vast improvements in networking capabilities have resulted in many such systems becoming increasingly distributed, with initiatives such as Grid Computing~\cite{Berman2003}, Cloud Computing~\cite{Armbrust2012} and Software As A Service~\cite{Turner2003} becoming commonly used.

Agent Oriented Programming (AOP) is a software development paradigm with intelligent agents at its core: autonomous computing entities that are capable of reacting to their environment, exhibiting proactive problem-solving behaviour and engaging in communication with one another to form Multi Agent Systems (MASs)~\cite{Wooldridge1995}. Such MASs are intended to collaboratively solve large-scale software engineering challenges.

AOP is originally rooted in the work carried out in the Artificial Intelligence community over a number of decades. With improving computer hardware resources, research began in the late 1980s to work towards practical deployments of intelligent agents by providing AOP languages, interpreters, toolkits and standards~\cite{Jennings2000}.

MASs are not solely an academic pursuit, however. Numerous spotlight deployments have been made, in a diverse range of application domains. This includes building agent-based approaches to supply chain management~\cite{Radjou2002}, simulations in games and films~\cite{MacNamee2009} and even space exploration~\cite{ESA2011}. Additionally, agents are widely considered to be a key enabling technology for the next generation of large-scale, self-managing software systems~\cite{Kephart2003,Tesauro2004,White2004}.

%\begin{itemize}
%   \item Deutsche Post developed an agent-based eProcurement system that matches their needs with the offerings of 6,000 freight carriers. This resulted in anticipated savings of \$17m per annum~\cite{Radjou2002}.
%   \item The use of agents to manage a DaimlerChrysler manufacturing process resulted in an estimated 10\% productivity increase~\cite{Radjou2002}.
%   \item An agent-based inventory management system deployed on the fulfilment network of Proctor \& Gamble cut inventory requirements by 30\% while substantially reducing the occurrences of stock shortages~\cite{Radjou2002}.
%   \item The MASSIVE agent system (\url{www.massivesoftware.com}) has been used in a number of Hollywood films, most notably in the battle scenes in the Lord of the Rings series of films~\cite{MacNamee2009}.
%   \item The European Space Agency incorporates agent technologies in its vision for next generation space exploration systems~\cite{ESA2011}.
%\end{itemize}

% others from Rem's notes: Whitestein Technologies (Living Systems Platform): telecoms, logistics/supply chain management, insurance/banking, engineering/manufacturing
% JACK agent platform: aerospace/air traffic control, oil trading/operations, defence, virtual actors/serious games
% Magenta Technology (maxoptra platform): retail distribution, oil/gas, asset rental, medical services, taxi/chauffeur operations, home delivery

Despite successes such as these, Object Oriented Programming (OOP) is still the dominant paradigm in industry, with AOP remaining a niche technology. The next Section discusses the current position of MASs in terms of its adoption thus far. Section~\ref{sec:introduction:conversation-management} then concentrates on the issue of the management of the communication between distributed agents, as this is the core focus of this thesis. Following this, Section~\ref{sec:introduction:motivations} outlines the motivations for the work presented in this thesis. Section~\ref{sec:introduction:core-contributions} then presents the principal contributions this thesis makes to the current state-of-the-art in interaction management for MASs. Finally, the structure of the remainder of the thesis is set out in Section~\ref{sec:introduction:outline}.

\section{The Position of Multi Agent Systems}

\begin{figure}[!hb]
   \includegraphics[width=\textwidth]{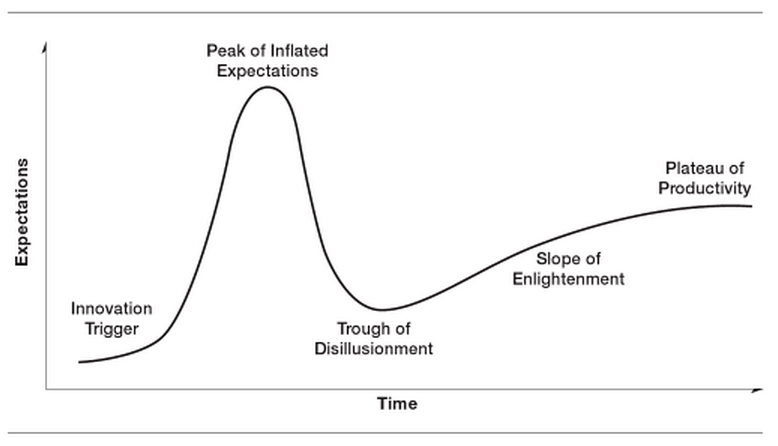}
   \caption{Gartner's hype cycle of innovation (taken from~\cite{Fenn2008}).}
   \label{fig:introduction:hype-cycle}
\end{figure}

Technology research firm Gartner Inc. have identified a pattern amongst many technological innovations over a long period of time. They describe this as the ``hype cycle of innovation''~\cite{Fenn2008}. This reflects the point from the initial optimism upon the launch of a new innovation, through a decline as companies fail to make the early gains they anticipated and then a slower rate of adoption as was originally expected as it reaches maturity. It is illustrated in Figure~\ref{fig:introduction:hype-cycle}. The stages of this hype cycle are as follows:

\begin{description}
   \item[Innovation Trigger:] A breakthrough piques interest in some innovation. As interest spreads, people become excited about its potential.
   \item[Peak of Inflated Expectations:] Early adopters have invested in the innovation and boast about its potential. Others begin to invest to avoid being left behind, causing a bandwagon effect to emerge.
   \item[Trough of Disillusionment:] Early over-hyped expectations are not satisfied. As impatience for results surpasses excitement, media coverage switches to challenges rather than potential.
   \item[Slope of Enlightenment:] Those early adopters that have persisted with the innovation begin to experience its benefits. The distinction between the realistic capabilities of the innovation and the unrealistic expectations that resulted in the early peak become clearer.
   \item[Plateau of Productivity:] With greater maturity and understanding comes stronger uptake. Risks are reduced and penetration accelerates.
\end{description}

In the area of MASs, the peak of inflated expectations had certainly been reached by the mid-to-late 1990s. Quotes describing agents as a ``new revolution in software''~\cite{Guilfoyle1994} or predicting that agents would become ``pervasive in every market by the year 2000''~\cite{Janca1995} serve to emphasise the degree to which they were seen as an essential tool for future software development. This led to Jennings et al. noting in 1998 that ``several observers feel that certain aspects of agents are being dangerously over-hyped''~\cite{Jennings1998b}.

However, since then MASs have failed to gain the traction that their claimed advantages would merit, and OOP remains the dominant paradigm in industrial software development. Recent research in the area indicates that the ``slope of enlightenment'' has been reached, however. Greater emphasis is being placed on bridging the gap between MAS development and more traditional software development practices. Much work has been done, for example, in integrating agent programs with external environments~\cite{Behrens2010,Ricci2007,Weyns2007} or distributed component-based systems~\cite{Lillis2009}.

Practitioners of OOP have long been accustomed to having tools for development, verification and debugging available to them, and it is only in recent years that an acknowledged lack of such tools in the MAS domain is being addressed~\cite{Bordini2006b}. This has resulted in the development of external debugging tools~\cite{DoanVanBien2010,DoanVanBien2009}, along with investigations into metrics for tasks such as informing programming guidelines and language improvements~\cite{Hindriks2012} or identifying bugs and predicting maintainability~\cite{Jordan2010b}. Other research has borrowed familiar concepts from the world of OOP to be adapted to AOP. Examples include the notion of inheritance to promote reuse~\cite{Jordan2011} and the introduction of strong typing to aid in debugging and error identification~\cite{Santi2012}.

This research demonstrates a conscious effort on the part of the MAS community to facilitate the development of practical MAS deployments.
 
\section{Conversation Management for Multi Agent Systems} \label{sec:introduction:conversation-management}

As MASs are a specialised form of distributed system, communication between agents is an essential capability. This is typically achieved by means of message passing (discussed in more detail in Chapter~\ref{chap:communication}). Traditional message-level processing treats each message as a separate event~\cite{Cost2000}. However, communication rarely occurs in isolation: a request for action is expected to result in a response, as is a query for information. In cases such as these, it is important to have some model of conversations to better reflect the realities of communication.

The advantages of allowing for the modelling of interactions in terms of conversations are outlined by Cost et al.~\cite{Cost1998}, as follows:
\begin{itemize}
   \item A conversation is a better fit than individual message passing for an intuitive model of agent communication.
   \item Existing theory and practice of network protocols can be applied to the domain of agent communication.
   \item If conversation definitions are kept separate from the actions the participating agents perform, the conversation structures can be shared amongst multiple agents.
   \item If such structures make use of existing Agent Communication Languages (ACLs) on the lower level, the advantages of these languages are preserved. ACLs are discussed in more detail in Chapter~\ref{chap:communication}.
\end{itemize}

Even where explicit support for conversation management is not a feature of an AOP language or framework, their importance is emphasised by the significant role they play in many development methodologies that have been created specifically for MASs (e.g Gaia~\cite{Wooldridge2000a}, MaSE~\cite{Wood2001}, MASSIVE~\cite{Lind2001}, PASSI~\cite{Cossentino2002}, SADAAM~\cite{Clynch2007}).

The core focus of this thesis is on the provision of conversation management capabilities for MASs. The motivations that inspire this work are presented in the following Section.

\section{Motivations} \label{sec:introduction:motivations}

The work presented in this thesis is motivated by the belief that greater standardisation in the area of communication between agents in a MAS is desirable and that insufficient progress has been made toward that end to date. The observations that have led to this work being carried out can be categorised under a number of headings, as follows.

\begin{itemize}
   \item Insufficient standardisation for agent communication.
   \begin{itemize}
   	   \item Some inroads have been made in the development of standard models of agent communication. This has occurred most prominently in the efforts that resulted in the Knowledge Query Manipulation Language (KQML) and the ACL developed by the Foundation for Intelligent Physical Agents (FIPA-ACL). These are discussed in Chapter~\ref{chap:communication}, with further focus on FIPA in Chapter~\ref{chap:fipa}. While these standards have seen widespread adoption for one-off messaging, less standardisation has occurred in the area of complex communication where multiple messages relate to the same conversation.
   	   \item KQML and FIPA-ACL both have particular semantics for different types of messages. However, support for these semantics is far from uniform. Additionally, these semantics are defined in terms of their effects on the mental models of the agents sending and receiving the messages. This is problematic for two reasons. Firstly, the mechanisms by which agents' mental models are implemented vary across languages. Secondly, and more importantly, it is impossible to verify whether the semantics of a message have been followed, as this would require the ability to examine the internal mental model of agents sending messages.
   	   \item Existing conversation standards require interpretation, causing incompatibilities. The closest the community has come to a standardised method of creating conversations has been FIPA's standard interaction protocols (discussed in Chapter~\ref{chap:fipa}). These standards are not unambiguously clear and as such require an element of interpretation in concrete implementations. This ambiguity can lead to differences in interpretation and consequently hinder interoperability.
   \end{itemize}
   \item Non-standard approaches are difficult to generalise.
   \begin{itemize}
   	  \item Numerous approaches to conversation management have been proposed in isolation (these are discussed in Chapter~\ref{chap:communication}). In some cases, this is tied to a particular AOP language and as such is not applicable more generally.
   	  \item Other approaches allow for dynamic conversation creation, where agents are capable of understanding the semantics of message exchange and decide on appropriate responses at run-time. As the understanding of message semantics is not a feature of all languages, this approach is also difficult to generalise.
   \end{itemize}
   	\item No single approach to conversation handling has been applied to multiple frameworks.
   	\begin{itemize}
   		\item Of the potential solutions to the problem of complex interactions (presented in Chapter~\ref{chap:communication}), none have been applied to multiple AOP languages or frameworks. Since communication is essential in promoting interoperability between diverse agent technologies, consistent standards for how this is to be approached should be applicable to a wide variety of languages and frameworks.
   \end{itemize}
   \item Conversations are difficult to debug.
   \begin{itemize}
   	   \item Due to the lack of conversation handling in many popular frameworks, software developers of agent-based systems are frequently required to implement conversations in an ad-hoc manner via individual messages with no formal link between them. This makes it more difficult to share descriptions with other developers of how conversations should be implemented in their own agents and also makes code less readable as all messages are handled in the same way. 

   \item In the absence of conversation handling capabilities in many languages, the burden falls on the developer to build these capabilities themselves, as mentioned in the previous point. However, it also falls on the developer to ensure that unusual communication situations do not result in unexpected or undesired behaviour. For example, creating an agent that can be manipulated into performing undesired actions by an adversary sending carefully crafted messages is easily done. This is discussed, along with other pitfalls, in Chapter~\ref{chap:evaluation}.

   \item Many AOP languages make use of rules that cause an agent to act when particular situations occur. For these languages, messages are typically handled by means of rules that catch anticipated incoming messages, which cause the agent to perform some reasoning and possibly reply if appropriate. For languages that follow this pattern, the receipt of messages that do not fit the expected format will generally cause no action whatsoever, as no rule will be fired. This leads to the silent loss of messages, which makes it unnecessarily difficult to debug interactions.
   \end{itemize}
\end{itemize}

\section{Core Contributions} \label{sec:introduction:core-contributions}

This thesis seeks to address the issues touched upon in Section~\ref{sec:introduction:motivations}. In doing so, it will make the following contributions:

\begin{itemize}
   \item To provide a holistic view of conversation management within a MAS, it will:
   \begin{itemize}
      \item Provide a formal specification of an approach to conversation management, so compatible implementations can be created in any language;
      \item Outline a generic architecture that specifies how ACRE should be integrated into any system;
      \item Discuss a reference implementation that integrates into an existing MAS framework so that it can be used by existing AOP languages;
      \item Provide tool support for developers to facilitate the creation and management of interaction protocols for MAS.
   \end{itemize}
   \item To demonstrate the effectiveness of the approach, a user evaluation will draw comparisons with more traditional methods of communication handling.
   \item In doing so, a testbed is required so that the evaluation of conversation management systems can be performed in a reliable manner.
   \item It will provide a standard for sharing protocol definitions via external repositories.
   \item In the achievement of these goals, a number of restrictions are required to aid in the achievement of greater compatibility:
   \begin{itemize}
      \item A centralised definition of protocols allows external monitoring to take place as part of debugging, system monitoring and verification.
      \item Minimal assumptions are made about the capabilities, features and attributes of agents that are candidates for ACRE integration. Programmers are free to use the features of their chosen language to implement deliberation regarding communication.
      \item Compatibility with agents fluent in FIPA-ACL that do not integrate directly with ACRE is maintained.
   \end{itemize}
\end{itemize}

\section{Thesis Outline} \label{sec:introduction:outline}

The organisation of this thesis is presented below. The Chapters are organised into two parts. The first part is concerned with the background to the work presented in the thesis. This covers Chapters~\ref{chap:aop}, \ref{chap:communication} and~\ref{chap:fipa}. Following this, the second part presents ACRE, beginning with a general introduction in Chapter~\ref{chap:acre-introduction} and becoming more specific before outlining an evaluation process in Chapter~\ref{chap:evaluation}. The chapters that constitute this thesis are as follows:

\begin{description}
\item[Chapter~\ref{chap:aop}: Agent Oriented Programming.] This Chapter provides an introduction to the notion of Agent Oriented Programming. It includes an exploration of what is meant by the term ``agent'', in addition to outlining a number of current or influential approaches to writing agent programs.
\item[Chapter~\ref{chap:communication}: Agent Communication] As the focus of this thesis is specifically on the communication between agents, this Chapter reviews past attempts to model and standardise agent communication. This includes standalone message-passing between agents and also attempts to model conversation-style communication.
\item[Chapter~\ref{chap:fipa}: FIPA Communication Standards] The Foundation for Intelligent Physical Agents (FIPA) has provided standards that agent developers are encouraged to adopt. The most influential of these are the standards relating to communication. These include specifications for how individual messages should be structured, in addition to outlining a number of common interaction patterns that can be modelled as conversations. The work in this thesis is based on these standards, which are outlined and analysed in this Chapter.
\item[Chapter~\ref{chap:acre-introduction}: Introduction to ACRE] This Chapter introduces the Agent Communication Reasoning Engine (ACRE), which is the core contribution of this thesis. The capabilities of ACRE are introduced informally, along with examples to illustrate how ACRE deals with a running conversation.
\item[Chapter~\ref{chap:semantics}: ACRE Formal Model] The operational semantics of ACRE's conversation handling are presented in this Chapter. For consistency, the examples from Chapter~\ref{chap:acre-introduction} are repeated, with the conversation being presented according to the mathematical model of the operational semantics.
\item[Chapter~\ref{chap:acre-architecture}: Generic Architecture] This Chapter presents a suggested generic architecture to guide ACRE's integration with existing AOP frameworks. This includes descriptions of the platform-independent aspects of ACRE and the capabilities a framework should have in order to be suitable for integration.
\item[Chapter~\ref{chap:agent-factory}: Integration with Agent Factory] Following the Generic Architecture, this Chapter outlines a specific implementation of ACRE within the Agent Factory AOP framework. It discusses the decisions made in conducting this integration and how the existing capabilities of the framework (and the AOP languages it supports) can be augmented by the adoption of ACRE.
\item[Chapter~\ref{chap:evaluation}: Evaluation] This Chapter presents the results of an evaluation that was conducted to compare the approach taken in ACRE with one based on individual standalone communication. The participants in the evaluation were two classes of students taking an Agent Oriented Software Engineering module. Participants were presented with a scenario for which to implement an agent-based solution.
\item[Chapter~\ref{chap:conclusions}: Conclusions and Further Work] The final chapter outlines some ideas for further extensions to the work presented this thesis and presents some concluding remarks.
\end{description}

\part{Background}
\chapter{Agent Oriented Programming} \label{chap:aop}

\section{Introduction}

Agent Oriented Programming (AOP) is a software development paradigm aimed at creating entities known as ``intelligent agents''~\cite{Jennings1998} or ``rational agents''~\cite{Wooldridge2000b}. This is typically aimed at the development of large-scale  distributed Multi Agent Systems (MASs), which are intended to provide higher level abstractions than traditional distributed programming~\cite{Singh1998}.

Because of their distributed nature, an essential aspect of a MAS is the communication between agents. The core aim of this thesis is to explore how agents within a MAS can be facilitated in performing this communication in reliable, predictable and verifiable ways. To facilitate this aim, this Chapter and the following two Chapters explore previous work that has been done in the area of AOP and agent communication. This Chapter begins this review by examining AOP in general.

It is first necessary to arrive at an appropriate definition of what exactly is meant by the term ``agent''. Several definitions have been proposed and full consensus has never been reached on what constitutes agency. Despite this, there is sufficient agreement on the core attributes of an agent to allow a relatively uncontroversial definition to be arrived at. This is done in Section~\ref{sec:aop:agent}. The Agent Oriented Programming is then introduced in Section~\ref{sec:aop:aop-intro}. This is followed in Section~\ref{sec:aop:aop} by a survey of the most commonly used AOP languages and frameworks. The work in this thesis is intended for use in as broad a set of agent technologies as possible. As such, it is important to identify the capabilities of frameworks that are already in use, so as to make consistent design decisions on interoperability.

\section{What is an ``agent''?} \label{sec:aop:agent}

In software development terms, the phrase ``software agent'' has attracted a myriad of competing definitions. This ranges from simple programs that perform tasks on behalf of users to sophisticated entities that leverage principles of Artificial Intelligence to perform complex planning, reasoning and actions. Researchers within the AOP community frequently make use of terms such as ``rational agent''~\cite{Wooldridge2000b} or ``intelligent agent''~\cite{Jennings1998} to describe their work. This indicates that the ``agent'' spoken about in this context is distinct from simpler programs.

The notion of \textit{intelligence} is generally taken to mean that the agent is capable of flexible autonomous action. Without the requirement for intelligence, Jennings and Wooldridge note that processes such as the UNIX \texttt{xbiff} utility, which perceives the environment in which it is situated (monitoring a user's incoming email) and takes actions when that environment changes (alerting a user to the presence of new mail), could be described as an agent~\cite{Jennings1998}. Similarly, many ``agents'' referred to by software developers are merely glorified search engines or user interfaces~\cite{Singh1998}.

Fisher considers it important that although autonomy is a central aspect to agency, rational agents should also exhibit \textit{reasonable} and \textit{explainable} behaviour~\cite{Fisher2006}. This requires that a number of aspects of an agent are represented, namely informational aspects (beliefs and knowledge), motivational aspects (goals and intentions) and deliberative aspects (why it makes the choices it does).

There has long been an acknowledgement in the agents community that total agreement has not been reached on what precisely an ``agent'' actually is~\cite{Jennings1998}. Wooldridge and Jennings propose a weak notion of agency that they consider to be uncontentious~\cite{Wooldridge1995}. This weak notion of agency includes \textit{autonomy}, \textit{social ability} and a combination of \textit{reactive} and \textit{proactive} behaviour. The precise meaning of each term is discussed in more detail in the Sections that follow. These four elements are sometimes accepted as the only key features of agency (e.g. in~\cite{Winikoff2005}). In other cases, the definition is expanded to include other aspects such as \textit{situatedness}, \textit{mobility} and \textit{benevolence}.

Beyond their weak notion of agency, Wooldridge and Jennings also propose a stronger notion of an agent that ``is either conceptualised or implemented using concepts that are more usually applied to humans''~\cite{Wooldridge1995}. This is in keeping with Shoham's definition of an agent as a computing entity ``whose state is viewed as consisting of mental components such as beliefs, capabilities, choices, and commitments''~\cite{Shoham1993}. This is an approach that many developers of AOP languages have followed, as can be seen in Section~\ref{sec:aop:aop}.

\subsection{Autonomy}
The notion of autonomy is considered to be a key element of agency in any discussion of intelligent agents (e.g. \cite{Castelfranchi1995,Collier2001,Jennings1998,Kakas2004,Luck2004,Winikoff2005,Singh1998,Shoham1993,Wooldridge1995}). It appears in the weak notion of agency proposed by Wooldridge and Jennings~\cite{Wooldridge1995}. Even where narrower definitions are used, autonomy still plays a central role. An example of this is Singh's definition, which considers only autonomy and interoperability as the two key attributes of agents~\cite{Singh1998}.

A number of studies have drawn a distinction between the autonomy of an agent and the concept of encapsulation as it applies to Object Oriented Programming~\cite{Jennings1998,Luck2004}. An object encapsulates a state and allows control over that state by means of the methods it provides. This is a feature of agents also. The object itself has no control over how or when these methods are invoked, however. For example, an object cannot prevent the execution of a method by another object. In contrast, agents are said to encapsulate not only state but also behaviour. They make their own decisions about the actions they take. Thus, agents can be thought of as requesting actions of each other, rather than invoking each other's methods.

\subsection{Social Ability/Interoperability}

Interoperability, or the ability of agents to communicate with one another, is the other key concept identified in Singh's narrow definition of an agent~\cite{Singh1998}. This social ability also appears in Wooldridge and Jennings' weak notion of agency~\cite{Wooldridge1995}.

Communication between agents is generally performed using an Agent Communication Language (ACL)~\cite{Genesereth1994,Labrou2001,Vieira2007}. An ACL is intended to allow an agent to inform another of some fact, request action, query the truth of a proposition and other related tasks. 

Several ACLs have been proposed, the most notable being the Knowledge Query Manipulation Language (KQML) and the ACL of the Foundation for Intelligent Physical Agents (FIPA). As communication between agents is the core focus of the work presented in this thesis, these are discussed in more detail in Chapter~\ref{chap:communication}, along with a survey of how more complex communication has been implemented in MASs.

\subsection{Perception and Environment}
Many agent definitions consider situatedness in an environment as being a central aspect to agency. Typically, this is in the context of the agent having the capability of perceiving its environment~\cite{Collier2003,Jennings1998,Goodwin1995,Maes1995,Singh1998}. Although not explicitly included in the weak notion of agency espoused by Wooldridge and Jennings, it is mentioned in the context of reactivity~\cite{Wooldridge1995}. In this definition, reactivity relates to the ability of an agent to respond in a timely manner to changes that occur in its environment.

Others have asserted that an agent is a mental entity and as such there is no requirement that it must interact with an environment~\cite{Hindriks1999}. Although many agents may be situated in an external environment (e.g. agents that are used to control a robot in a physical environment), it can be argued that it is not fundamental to the definition of agency.

\subsection{Proactivity and Reactivity}

Requirements of proactivity and reactivity are also contained in the weak notion of agency~\cite{Wooldridge1995} and have been accepted by many others (e.g.~\cite{Collier2003,Hindriks1999,Winikoff2005}). Here, reactivity refers to the ability of an agent to react to changes in its environment in a timely manner. An agent exhibits proactive behaviour when it is capable of taking the initiative towards achieving any goals that it may have. It is generally accepted that a successful agent will exhibit a combination of both types of behaviour.

\subsection{Benevolence}

Benevolence is the assumption that agents do not have conflicting goals. Rosenschein and Genesereth observe that almost all early work in distributed artificial intelligence (from which the idea of intelligent agents arose) operated on the assumption of benevolence~\cite{Rosenschein1985}. With this assumption, agents are assumed to be helping one another towards common or compatible goals.

Benevolence is not considered to be part of the weak notion of agency~\cite{Wooldridge1995} and it does not tend to be included in modern definitions of agency. Indeed, a significant body of research exists on handling conflict and trust issues amongst agents (surveys of the work conducted in this area can be found in~\cite{Pinyol2011,Ramchurn2004}).

\subsection{Mobility}

Mobile agents are those that are capable of moving from one environment to another, typically by migrating to a different host on a network~\cite{Wooldridge1995}. An advantage of mobility is that an agent may perceive some deficiencies in their present environment and cause a migration to occur to a more favourable host. An example of this occurs where two agents are required to communicate with one another frequently. If the communication load on the network is likely to be significant, it may be advantageous for the agents to migrate to the same platform and communicate locally~\cite{ElFallahSeghrouchni2003}.

Although this is considered an important facet for many researchers (examples include~\cite{ElFallahSeghrouchni2003,Muldoon2008}), it is not core to the definition of an agent. This is emphasised by the categorisation of agents as mobile or static agents, as done by Nwana~\cite{Nwana1996}.

\subsection{Other Properties}

The above is not an exhaustive list of all the properties that have, at some point, been included in agent definitions. Some others include the following:

\begin{description}
   \item[Rationality:] The agent works towards the achievement of its goals, and will not perform any action that is detrimental to their achievement~\cite{Maes1995,Wooldridge1995}.
   \item[Temporal Continuity:] The agent is a continuously running process~\cite{Franklin1996,Shoham1993}.
   \item[Learning:] The agent is capable of changing its behaviour based on past experience~\cite{Franklin1996}.
   \item[Flexible:] The actions of an agent are not scripted in advance~\cite{Franklin1996}.
   \item[Character:] An agent has a believable personality and an emotional state~\cite{Franklin1996}.
\end{description}

\subsection{Adopting a Definition} \label{sec:aop:definition}

% autonomy is fundamental
% so is social ability, given that we're working on comms
% proactivity: to engage in conversation based on information available to the agent
% reactivity: allow agents to become aware of messages sothey can react
% others not important: no assumption of benevolance, environment not relevant to our purposes so not something to worry about.

For the purposes of this work, it is sufficient to adopt the weak notion of agency set out by Wooldridge and Jennings~\cite{Wooldridge1995}, which considers agents to be autonomous, social, reactive and proactive.

As discussed above, autonomy is a fundamental feature of any agent definition. The consequence of accepting autonomy as a key agent feature is that it is important that the the work in this thesis should not limit the autonomy of the agents that make use of it.

As the work in this thesis focuses specifically on the communication between agents, the social ability of agents is also an acceptable key requirement. It is assumed that all agents are capable of sending and receiving communications to and from others. This work will not be applicable to any computational entity lacking these abilities.

The proactive and reactive nature of agents refers to how they respond to changes in their environment and how they plan the achievement of their goals. The fact that many agents will exhibit a combination of these behaviours means that the provision of a communication framework must allow for either approach when dealing with communication.

With regard to other potential elements of a definition of an agent, it can be left to the creators of MASs and MAS toolkits to decide whether they are relevant to their model of agency. The question of whether they are fundamental aspects of the definition of an agent is not one that requires an answer in the context of this work. For instance, the existence of an environment in which an agent is situated is not relevant to how it engages in communication. Similarly, the inclusion of mobility is not a requirement of this work. The location and possible relocation of agents is not core to modelling how agents communicate with one another, assuming some mechanism is in place that ensures that a message addressed to a particular agent will reach its intended destination.

Due to the volume of work conducted in the area of modelling and dealing with agent trust, benevolence is not assumed in this work. Agents are free to adopt any suitable model of trust.

%\section{Multi Agent System (MAS)}
%Per Singh~\cite{Singh1998}: ``Multiagent systems provide higher level abstractions than traditional distributed programming. These abstractions are closer to user expectations and allow the designer more flexibility in determining behaviour''.
%
%Luck et al.~\cite{Luck2004} state that ``Agents are often deployed in environments in which they interact, and maybe cooperate, with other agents (including both people and software) that have possibly conflicting aims. Such environments are known as \textit{multi-agent systems}''.

\section{Agent Oriented Programming} \label{sec:aop:aop-intro}

``Agent Oriented Programming'' (AOP) is a phrase coined by Shoham in his proposal for a framework of computation where the state of the entities (called ``agents'') in the system consists of mental concepts such as beliefs, decisions, capabilities and obligations~\cite{Shoham1993}. In this initial proposal, it is seen as a specialisation of Object Oriented Programming (OOP). The relationship between the two paradigms is shown in Table~\ref{tab:aop:oopaop}, which is taken from~\cite{Shoham1993}.

\begin{table}[!htb]
\caption{Comparison of Object Oriented Programming (OOP) concepts with Agent Oriented Programming (AOP) (taken from~\cite{Shoham1993}).}
\label{tab:aop:oopaop}

\begin{tabular}{lll}
\hline
 & OOP & AOP \\
\hline
Basic unit & object & agent \\
Parameters defining & unconstrained & beliefs, commitments,\\
state of basic unit  & & capabilities, choices, \ldots \\
Process of computation & message passing and & message passing and \\
 & response methods & response methods \\
Types of message & unconstrained & inform, request, offer, \\
 & & promise, decline, \ldots \\
Constraints on methods & none & honesty, consistency, \ldots \\
\hline
\end{tabular}
\end{table}

This relationship between agents and objects has been adopted by other researchers also, including Odell et al.~\cite{Odell2000}. Their basic definition of an agent, from a software development point-of-view, is that of ``an object that can say `go' (dynamic autonomy) and `no' (deterministic autonomy)''. That autonomy is key to this definition is in keeping with the definition of an agent adopted in Section~\ref{sec:aop:definition}. In this case, ``dynamic autonomy'' refers to an agent's ability to exhibit proactive behaviour whereas ``deterministic autonomy'' is the ability to refuse or modify an external request. This is in strong contrast to the behaviour of objects, which typically cannot act without external invocation, and whenever such an invocation occurs the associated action must occur.

Key to Shoham's proposal is the notion that agents exist within a societal construct that allows them to interact with one another. This interaction is proposed to follow the principles of speech act theory (discussed in Section~\ref{sec:communication:speech-act-theory}) to allow agents to inform, request, offer, accept, etc.

Much work had previously been done in the Artificial Intelligence community in an attempt to reason about mental attitudes using various forms of logic. Wooldridge and Jennings see AOP as developing practical programming languages that embody the various principles proposed by theorists~\cite{Wooldridge1995}. Indeed Shoham notes that it would be ``tempting to view AOP as a form of logic programming''~\cite{Shoham1993}.

One early and very influential model for intelligent agents is the \textit{Belief, Desire, Intention} (BDI) model~\cite{Bratman1987}. This models the internal workings of an agent in terms of three mental attitudes. Beliefs are intended to model information, desires to model motivation and intentions to model decisions~\cite{Rao1996}. Most practical implementations of BDI systems have their roots in the abstract interpreter proposed by Rao and Georgeff~\cite{Rao1995}. This operates in a cyclical fashion, with a number of stages required on each iteration. These include generating options (actions that the agent can undertake) from events that have been perceived, creating intentions to carry out some of those options, execution of adopted intentions, generation of new events and the dropping of successful or impossible intentions. This iterative approach to an agent interpreter is one that has been followed in many implementations since.

%According to Rao~\cite{Rao1996}, ``BDI agents are systems that are situated in a changing environment, receive continuous perceptual input, and take actions to affect their environment, all based on their internal mental state''. 

Approaches to tackling the issue of programming agents have varied from the purely declarative to the purely imperative. However many practical AOP languages adopt a hybrid approach. A number of languages and frameworks aimed at AOP are discussed in the next Section.

%FUNDAMENTAL LANGUAGES: AgentSpeak, GOAL, 2APL (all cited in Ancona2012 as targets for their work).
%Santi2012 mentions AF-APL, Jason, Goal, 2APL, Jack

%Quote directly from Muldoon's thesis:``Agent oriented programming is a paradigm for directly programming the behaviour of agents using languages whose semantics capture a theory of rational agency~\cite{Shoham1993}''

%From~\cite{Wooldridge1995}: ``Agent languages are programming languages that may embody the various principle proposed by theorists''.

%Survey~\cite{Bordini2006}

%A framework for programming agents is presented by Shoham in~\cite{Shoham1993}, including the initial presentation of the Agent0 AOP language. He references previous work by Cohen and Levesque, who present alternative mental aspects to be modelled by an AOP language.

% TODO: Cohen/Levesque papers are in Zim PhD Reading page. Wooldridge1995 references them too as developing a theory of intention for agents.

\section{AOP Frameworks and Languages} \label{sec:aop:aop}

This Section briefly introduces a number of programming languages and frameworks aimed at the facilitation of AOP development. It begins with a brief introduction to a variety of languages. This is followed by a more in-depth view of some current frameworks and languages that are candidates for integration with the work presented in this thesis. Those selected are languages cited as important in recent work (e.g.~\cite{Ancona2012, Santi2012}). The mechanisms by which these languages implement interaction between agents is omitted from this Section, as it is discussed in detail in the next Chapter (in Section~\ref{sec:communication:implementation}), which concentrates on communication.

The first practical agent programming language was AGENT0, which was developed by Shoham~\cite{Shoham1991,Shoham1993}. AGENT0 is a relatively simple language for creating purely reactive agents. It lacks the advanced planning capabilities of later AOP languages. The key entities included in AGENT0's mental model are beliefs, commitments and capabilities. Commitment rules operate on the belief base of the agent in order to determine when commitment to action should be made. The definition of capabilities allows an agent to determine whether it is capable of performing an action another agent requests of it.

Communication between agents is considered to be a key aspect of the language. This is based on speech act theory (see Section~\ref{sec:communication:speech-act-theory} for discussion of this). Three types of communication are permitted, which allow an agent to inform another of some fact, request another to perform some action and to retract unfulfilled requests. This approach has been adopted by many MAS languages and frameworks that are inspired by AGENT0.

Actions are standalone, in that advanced planning is not possible. Beliefs are generated either by the receipt of messages from other agents, or through private actions carried out by the agent itself. Here, no distinction is made between perception and action, as it is in many later languages. The capabilities of an agent are fixed, so agents may not learn new actions at any time. Also, the environment in which an agent sits is not expressly accounted for, although there is nothing to prevent the implementation of actions that interact with the environment.

AGENT0 was succeeded by PLACA (PLAnning Communicating Agents), which adds the ability to achieve goals~\cite{Thomas1995}. Additionally, plans can be defined for PLACA agents, which consist of sequences of actions and any sub-goals that are necessary.

An additional contribution of Shoham's work is in his vision of an interpreter for agents. This interpreter operates in a loop consisting of two key steps, executed at regular intervals. Firstly, the agent's current messages are read and its mental state is updated. This involves updating the belief base and adding any commitments that arise from the commitment rules. Next any active commitments are executed, which may also affect the belief base. This type of cyclical execution of agents has served as a model for many AOP frameworks since.

ConGolog~\cite{DeGiacomo2000,Lesperance1995} is a concurrent version of the earlier Golog~\cite{Levesque1994}. It is a logic programming language that reasons about actions. Agents have beliefs about an initial state of their environment, along with a set of beliefs about later states that are reachable from this initial state by means of executing some sequence of actions. The applicability of actions is known, so the agent can verify whether an action is appropriate to a given state. ConGolog offers a powerful language for the description of complex actions. Constructs such as sequencing, non-deterministic choice, iteration and prioritisation are supported. A variant called IndiGolog allows agents to gather new data from their environment during execution~\cite{DeGiacomo2009,Sardina2004}.

Concurrent M\textsc{etate}M is an agent programming language that relies on the direct execution of temporal logic statements~\cite{Fisher1995,Fisher2006,Fisher1997}. This temporal aspect distinguishes it from other logic-based approaches. A definition of the messages an agent can receive and send can be created, acting as a type of API for other agents. Extensions to the basic language allow for goals to be re-ordered (e.g. to begin by satisfying some of the more achievable goals) and supplementing beliefs with the concepts of \textit{ability} (knowledge that some agent is capable of performing some action) and \textit{confidence}, which expresses beliefs about future states and is derived from its confidence in other agents in the system.

MINERVA is an agent programming architecture based on the principles of Dynamic Logic Programming to achieve a balance between the reactive and rational behaviour of intelligent agents~\cite{Leite2001}. Each agent consists of a number of possibly concurrent sub-agents that each perform various tasks while utilising a common knowledge base. This knowledge base models information about the agent itself and its surrounding agent community and includes concepts such as capabilities, intentions, goals, plans, reactions, an object knowledge base and internal behaviour rules. This knowledge is represented by a combination of the Multi Dimensional Logic Programming language~\cite{Leite2000} (to represent the knowledge of an agent at each state) and LUPS~\cite{Alferes2002} (to represent transitions between states of the agent).

KGP is a model of agency that is so named because agents' mental state consists of \textbf{k}nowledge, \textbf{g}oals and \textbf{p}lans~\cite{Bracciali2005,Kakas2004}. Supplementing these is a set of reasoning capabilities that support planning, temporal reasoning, identifying the preconditions of actions, reactivity and decision-making with regard to goals. Transition rules define how the state of the agent can change. These rules include such actions as changing goals either reactively or through planning, sensing observations from the environment and executing actions. KGP draws on the principles of the classic BDI model, but uses Computational Logic so that formal analysis of the model and its computational feasibility can be facilitated.

\texttt{Go!} is a multi-paradigm programming language that uses both declarative and imperative programming to yield a BDI system~\cite{Clark2004}. It is based on previous work done on \texttt{April}~\cite{McCabe1995} and draws further influence from IC-Prolog II~\cite{Chu1994} and L\&O~\cite{McCabe1992}. Definitions of functions and relations are created declaratively, with an imperative style used to create action procedure definitions. The mental state of agents consists of beliefs, desires and intentions. These are maintained in tuple stores, which can be read and manipulated by the threads that constitute the agents. It is a strongly typed, multi-threaded language, with communication performed through asynchronous message passing.

%\subsection{JACK Intelligent Agents}

JACK is a Java-based language for BDI agents~\cite{Busetta1999,Winikoff2005}. It extends Java with a number of syntactic constructs to allow developers to define mental attitudes such as beliefs, plans and events. A number of statements are made available by default within the body of an agent plan. These include communication with other agents, waiting for a condition to become true and raising events. A JACK agent acts reactively in response to events. Those plans that are relevant to the event are evaluated to ensure that they are possible to execute given the agent's state. If such a plan is available, it is chosen for execution. This execution cycle is similar to that of earlier BDI systems such as the Distributed Multi-Agent Reasoning System (dMARS)~\cite{DInverno1997,DInverno2004} and the Procedural Reasoning System (PRS)~\cite{Georgeff1987}.

% Dastani 2008 argues that a lack of logical semantics for Jack beliefs and goals means that a Jack agent cannot generate plans that will help towards achieving a goal, and only plans that entirely satisfied a goal can be used.

% Active: Last release in November 2011. (Rem says it's dead)

%For communication, JACK's default mechanism is the sending of messages between agents that typically contain serialised Java objects. Its communications infrastructure does not prioritise inter-agent communication, rather concentrating on interaction with existing programs and software components, particularly those written in Java or C++. A third-party extension does add support for FIPA ACL~\cite{Winikoff2005} and in~\cite{Busetta1999} it is stated that integration with a high-level Agent Communication Language such as KQML would be possible.

Jazzyk is an agent programming language (and associated interpreter) developed by Peter Nov\'{a}k~\cite{Novak2009}. It is based on the idea of Behavioural State Machines (BSMs)~\cite{Novak2008}. BSMs are built as transition systems whereby the state of the system is the agent's mental state and transitions are brought about by atomic updates of this mental state. A BSM draws a clear distinction between two layers: the \textit{knowledge representation layer}, which may consist of multiple heterogeneous knowledge representation modules; and a \textit{behavioural layer}, which is responsible for controlling the agent.

In the early 2000s, the Foundation for Intelligent Physical Agents (FIPA) released a number of standards relating to MASs. This was an attempt to standardise agent platforms so they would function in a predictable and interoperable manner. These standards include the 
FIPA Abstract Architecture Specification~\cite{FIPA00001}, which outlines the architectural elements that should be present within an agent system, along with the relationship between them. Another such standard is the FIPA Agent Management Specification~\cite{FIPA00023} to define standard interfaces for accessing agent management services. This includes addressing such issues as how agents can be located, named and contacted, along with how the life cycle of agents should be handled and how they can register with an Agent Management Service running on an Agent Platform. Following this, a number of agent toolkits were implemented that followed these standards, including FIPA-OS~\cite{Poslad2000}, JADE~\cite{Bellifemine2007b} and Agent Factory~\cite{Collier2003}.

\subsection{AgentSpeak(L) and Jason} \label{sec:aop:jason}

AgentSpeak(L) is a programming language originally aimed at bridging the gap between the theoretical BDI model and the practicality of existing BDI-style programming languages that lacked a sound theoretical underpinning~\cite{Rao1996}. The original language has been extended through its use in the Jason MAS framework~\cite{Bordini2005,Bordini2007,Vieira2007}.

The language began as a formalisation of the operational semantics of the existing PRS~\cite{Georgeff1987} and dMARS~\cite{DInverno1997,DInverno2004} BDI systems. AgentSpeak(L) makes use of a restricted first-order language to define events and actions. Although beliefs, desires and intentions are not explicitly modelled in the language, these mental attitudes can easily be applied to the execution model of the agent.

The key mental components of the AgentSpeak(L) language are beliefs and goals. These model the agent's view of its environment, in addition to states the agent wishes to bring about. The addition or removal of beliefs or goals are described as triggering events that may cause the agent to execute a plan. A plan describes a sequence of actions that the agent should carry out in order to satisfy some goal.

Goals are divided into two categories: achievement goals and test goals. An achievement goal represents a goal whose aim is to bring about a particular state of affairs that is desired by the agent. The adoption of an achievement goal will result in a plan being selected that provides the means to result in the desired state. A plan that has been adopted to achieve a goal represents the ``intention'' of BDI parlance. Test goals are proactive checks on the belief base of the agent to test if the agent believes some proposition.

Plan selection is performed by means of rules. A rule consists of a triggering event (typically the adoption or removal of a goal or belief), a context (some precondition consisting of belief literals that specify when the rule is applicable) and a plan body (a sequence of actions or goals that the agent will perform or achieve when the plan is triggered).

Jason is an interpreter for AgentSpeak(L) that extended the original language with practical capabilities like integration with an environment (which can be implemented using Java), communication and the ability to distribute an AgentSpeak(L)-based MAS over a network. In doing so, it was necessary to extend the original operational semantics of the language~\cite{Bordini2005}.

The Jason interpreter operates iteratively. On each iteration, it firstly updates the list of events for each agent. Additions to this list will typically be generated from changes in the agent's environment or from its intentions. Next, an event is selected and a set of relevant plans is identified. A relevant plan is one whose trigger matches the selected event. This set is further narrowed by using the contexts of the relevant plans to find those that are applicable to the current situation. One of these applicable plans is selected to be a new intention. Finally, an active intention of the agent is selected for execution.

%Lots of references for AgentSpeak(L)/Jason in \cite{Vieira2007} (paper dealing specifically with communication between agents).

%Per \cite{Novak2008}, its underlying abstraction is a transition system. Novak also describes AgentSpeak(L) as a ``logic based state-of-the-art BDI agent programming language''.

\begin{figure}[!ht]
\begin{Verbatim}[numbers=left,frame=single,fontsize=\footnotesize]
skill(plasticBomb).
skill(bioBomb).
~skill(nuclearBomb).

safeArea(field1).

@p1
+bomb(Terminal, Gate, BombType) : skill(BombType)
  <- !go(Terminal, Gate);
     disarm(BombType).

@p2
+bomb(Terminal, Gate, BombType) : ~skill(BombType)
  <- !moveSafeArea(Terminal, Gate, BombType).

@p3
+bomb(Terminal, Gate, BombType) : not skill(BombType) &
                                  not ~skill(BombType)
  <- .broadcast(tell, alert).

@p4
+!moveSafeArea(T,G,Bomb) : true
  <- ?safeArea(Place);
     !discoverFreeCPH(FreeCPH);
     .send(FreeCPH, achieve, carryToSafePlace(T,B,Place,Bomb)).

...
\end{Verbatim}
\caption{Sample AgentSpeak(L) code in Jason (taken from~\cite{Bordini2005}).}
\label{fig:aop:jason-code}
\end{figure}

An example of AgentSpeak(L) code for use in Jason can be seen in Figure~\ref{fig:aop:jason-code}. This is taken from a sample disaster-recover scenario where agents must co-operate to disarm bombs at an airport. The early lines indicate initial beliefs that the agent has when it is started. Initially, it believes that it possesses the necessary skills to disarm plastic explosives (line 1) and biological weapons (line 2), but not nuclear bombs (line 3). The tilde operator negates beliefs. Additionally, the agent is aware that ``field1'' is a safe place to put bombs that it cannot disarm.

Following this are four plans (labelled \texttt{p1} through \texttt{p4}). For the first three plans, the triggering event is that a new event has been created that matches \texttt{+bomb(Terminal, Gate, BombType)}. Constant atoms begin with lowercase letters whereas those beginning with capital letters are variables (it is similar to Prolog in this sense). Hence any event that matches that pattern can trigger any of those rules. The plan to be executed depends on the rule's context. Plan \texttt{p1} applies where the agent believes it has the skill to disarm the relevant bomb type (the \texttt{BombType} variable will be bound to the particular type of bomb that was included in the event that triggered the rule). In this instance, the agent will adopt an achievement sub-goal (not shown) to travel to the terminal and gate that were included in the event (line 9). Achievement goals are preceded by the \texttt{!} character. It will then perform a basic action (not shown) to disarm the bomb (line 10).

In the event that the agent does not know about a bomb type (it neither believes it has the requisite skill nor that it lacks it), plan \texttt{p3} will fire, causing a broadcast message to be sent to other agents to alert them to the threat. The dot before the name of the \texttt{.broadcast} action on line 19 indicates that this is a built-in action that is available by default to all Jason agents.

If the agent believes it does not have the skill to disarm the bomb, plan \texttt{p2} will be triggered instead. An achievement goal is adopted to move the bomb to a safe area. In this agent, this will cause plan \texttt{p4} to be fired, which happens any time a \texttt{!moveSafeArea} goal is adopted. This is the case because the context of the rule (\texttt{true}) does not place any restrictions on when the plan can be adopted.

Once this occurs, the agent first queries its belief base using a test goal (beginning with a \texttt{?} character) to find a known safe place. This has the effect of binding the \texttt{Place} variable to a known safe area from the belief base (in this case, it will acquire the value ``field1''). It then adopts a further achievement sub-goal in line 24 that will cause the agent to attempt to find an agent capable of carrying a bomb to another place. On success, the identifier of this agent will be bound to the \texttt{FreeCPH} variable. Finally, a message is sent (using another internal action) to ask that agent to carry the bomb to the specified safe area.

\subsection{Agent Factory}

Agent Factory is an open source framework for the development and deployment of MASs~\cite{Collier2001,Collier2003,Muldoon2009}. It offers both a standard edition for regular computers and also a micro edition for use on resource-constrained devices (the micro edition is discussed in detail in~\cite{Muldoon2008,Muldoon2009}).

\begin{figure}[!hbt]
\centering
   \includegraphics[width=\textwidth]{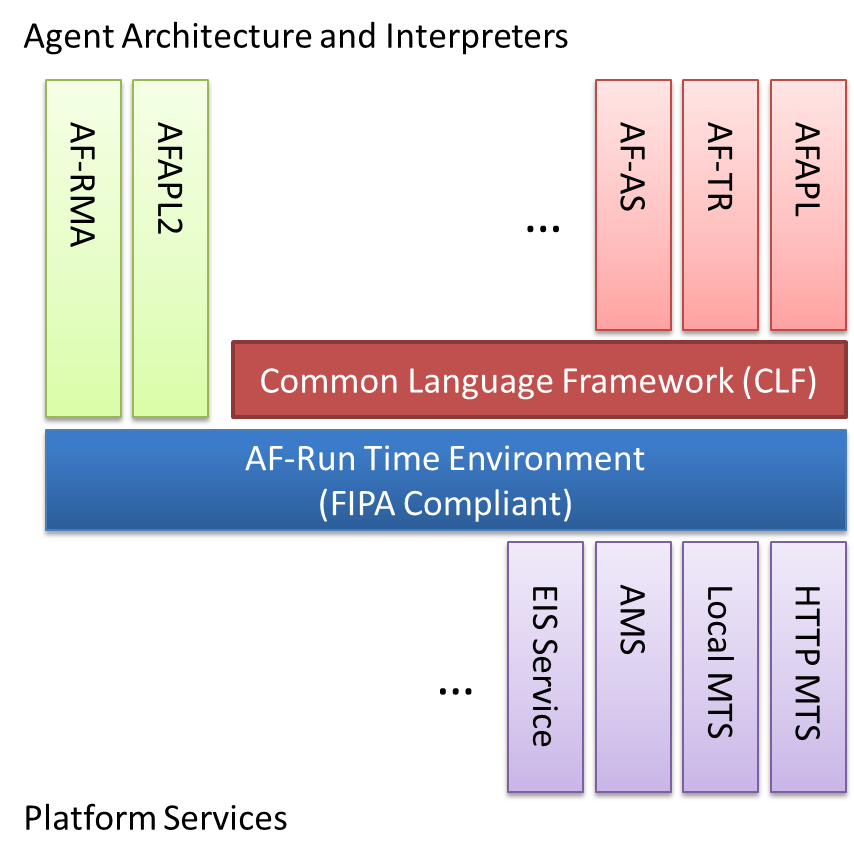}
   \caption{Agent Factory Architecture}
   \label{fig:aop:af-architecture}
\end{figure}

The standard edition offers a FIPA-compliant runtime environment at its core. The architecture is shown in Figure~\ref{fig:aop:af-architecture}.

Originally, the Agent Factory Agent Programming Language (AFAPL) and its successor AFAPL2 were the only AOP languages supported by the framework. This has since been joined by numerous other AOP languages through the Common Language Framework (CLF)~\cite{Russell2011}. These include AF-AgentSpeak (an adaptation of AgentSpeak(L)) and AF-TR, (based on Nilsson's teleoreactive model~\cite{Nilsson1994}). A hybrid of these approaches, called AS-TR, combines elements of AgentSpeak(L) and teleoreactive behaviour, reflecting similar work done by Coffey and Clark~\cite{Coffey2006}.

The Platform Service model allows agents to access shared services that provide capabilities such as Message Transport Services (MTSs), an Agent Management Service (AMS) and access to the environment. Figure~\ref{fig:aop:af-architecture} shows environment access through an Environment Interface Standard (EIS) interface~\cite{Behrens2010}, but the framework is extensible with platform services offering access to other environment models.

The AFAPL language is primarily based on beliefs, commitments and commitment rules~\cite{Collier2009}. It has been formally specified using a multi-modal branching time first-order logic of commitment~\cite{Collier2001}. A commitment models a decision that an agent has made to perform some action or series of actions. Commitment rules cause commitments to be adopted when particular belief states arise. Basic actions and sensors are available to effect change to and sense changes in an agent's environment. Actions can be used to compose plans using a number of plan operators that allow for sequential or parallel execution, looping, waiting and querying of the belief base.

Later additions to AFAPL include support for goals~\cite{Muldoon2009} and roles~\cite{Collier2005}. Roles add a level of abstraction that promotes software reuse. Where agents can play numerous roles within a MAS, it is useful to package the commitment rules, plans and other elements required to fulfil that role. This allows agents to switch between roles and also facilitates the reuse of previously defined roles when necessary.

\begin{figure}[!hb]
   \includegraphics[width=\textwidth]{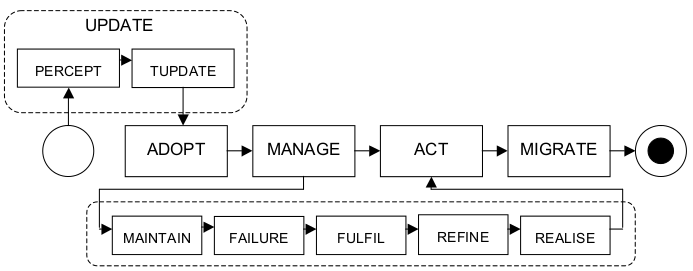}
   \caption{Agent Factory interpreter cycle (taken from~\cite{Collier2009}).}
   \label{fig:aop:agent-factory-interpreter-cycle}
\end{figure}

The AFAPL interpreter operates cyclically, and is illustrated in Figure~\ref{fig:aop:agent-factory-interpreter-cycle}, beginning each iteration by updating the agent's belief base. This is done in two stages: firstly perception occurs so beliefs from the agent's sensors can be added (PERCEPT). This is followed by a temporal update (TUPDATE) that adds any persistent beliefs to the current belief base.

Following this, any commitment rules capable of being triggered by the current belief base are adopted (ADOPT). This is followed by the commitment management stage (MANAGE), which involves such activities as ensuring that commitments can still be achieved, handling failure and dropping successful commitments. This is followed by the execution of actions (ACT), and finally a check is performed to see if the agent has decided to move to another platform (MIGRATE). As soon as this cycle has completed, it iterates again, beginning with the update of the belief base~\cite{Collier2009}.

Agent Factory has been chosen as the platform with which the concrete reference implementation for the work in this thesis was integrated. As such, it is discussed in more detail in Chapter~\ref{chap:agent-factory}.

\subsection{GOAL} \label{sec:aop:goal}

In GOAL (Goal Oriented Agent Language), the key concept is that of declarative goals ~\cite{DeBoer2007,Hindriks2010}. This is a contrast to goals as they are used in other languages such as AgentSpeak(L) and 3APL. In those languages, a goal is in effect a plan, as it represents the desire of an agent to perform some action. In contrast, a declarative goal represents a desired state to be brought about. The use of declarative goals is seen as an opportunity to make use of modal logic to specify and verify agent programs, which is not possible with other AOP languages.

The mental state of a GOAL agent consists of a belief base and a goal base. A third basic concept allows the capabilities of an agent to be defined. Basic actions can be used to add and remove beliefs and goals. Conditional actions can be specified by users that can be executed only when the mental state of the agent satisfies certain conditions.

%Actions are also specified to represent actions that an agent can undertake. Non-communicative actions are assumed to only affect the mental state of the agent itself. Built-in actions only relate the creation and removal of beliefs and goals, along with the ability to send a message to another agent. User-defined actions are also possible.

The agent program consists of conditional action rules in the form \textbf{if}~$\psi$~\textbf{then}~\texttt{a}, where $\psi$ is a mental state condition (i.e. a goal or belief, or a combination of these using boolean operators) and \texttt{a} is an action. This specifies the conditions under which the agent will perform actions.

GOAL is not only presented as a concrete programming language, but is also built upon formal operational semantics and a theory to prove the correctness of programs based on temporal logic. Correctness is considered to be when the agent program successfully realises the goals of the agent.

% TODO: This might be better in the AFAPL2 section: also need to reference an AFAPL2 section
%This is similar to the notion of a \textit{commitment rule}, as seen in AFAPL, without the temporal aspect of commitment rules.

%Cited by Novak as a state-of-the-art BDI language~\cite{Novak2009}

\begin{figure}[!ht]
\begin{Verbatim}[numbers=left,frame=single,fontsize=\footnotesize]
beliefs{ hold(fork,left). }
goals{ hold(fork,left), hold(fork,right). }
program{
  if true then think.
  if true then eat.
  if bel(hungry) 
    then adopt(hold(fork,left), hold(fork,right)).
  if goal(hold(fork,_)), bel(not(forksAvailable),neighbours(X,Y))
    then send(new:{X,Y},!hold(fork)).
  if bel(neighbour(X,D), not(hold(fork,D))), bel(X,on(fork,table))
    then ins(on(fork,table,D)).
  if bel(conversation(Id,i))
    then pickUp(fork,D) + send(Id:X, .hold(fork)).
  if bel(conversation(Id,i), hold(fork,left), hold(fork,right),
      neighbours(X,Y)) bel(X,not(on(fork,table))),
      bel(Y,not(on(fork,table)))
    then close(Id).
  if bel(conversation(Id,X)), goal(X, hold(fork))
    then putDown(fork,D) + 
      send(Id:X, .on(fork,table), not(hold(fork))).
  if bel(conversation(Id,X), neighbour(X,D)), bel(X, hold(fork))
    then del(on(fork,table,D)) + send(Id:X, ?on(fork,table)).
}

action-spec{
  think{
    pre{not(hungry)}
    post{hungry}
  } 
  pickUp(fork,D){
     pre{on(fork,table,D)}
     post{hold(fork,D),not(on(fork,table,D))}
  } 
  eat{
    pre{hungry,hold(fork,left),hold(fork,right)}
    post{not(hungry)}
  } 
  putDown(fork, D){
    pre{hold(fork,D)}
    post{on(fork,table,D),not(hold(fork,D))}
  }
}
\end{Verbatim}
\caption{Sample GOAL code (taken from~\cite{Hindriks2010})}
\label{fig:aop:goal-code}
\end{figure}

Figure~\ref{fig:aop:goal-code} shows sample code for a GOAL agent. This is written as part of an agent to solve the classical ``dining philosophers'' problem from concurrency theory~\cite[p. 122]{BenAri2006}. In short, this problem involves a number of philosophers sitting around a circular dining table with one fork between each person. To eat, a philosopher must have a fork in each hand. Philosophers can either think (making them hungry) or eat (which reduces hunger). They must negotiate with one another to secure access to two forks at the same time so they can eat.

The initial beliefs and goals are defined at the beginning of the program. The agent begins with the belief that it holds a fork in its left hand. It has goals to hold a fork in both its left and right hands.

At the bottom of the program (from line 25 onwards), a number of actions are specified. Each is defined in terms of its preconditions and postconditions. For example, the \texttt{eat} action (defined from line 34) can occur when an agent is hungry and holds forks in both its right and left hands. The outcome of this action is that it is no longer hungry. Other actions can be seen elsewhere in the program that are available by default. These include \texttt{ins} to insert a belief into the belief base, \texttt{del} to delete a belief from the belief base and \texttt{send} to send a message.

The program itself begins on line 3. The interaction between agents begins with the rule on line 8. If an agent has a goal to hold a fork but they are not available (\texttt{forksAvailable} is defined elsewhere in the program and is not shown), then it will initiate a new conversation with its neighbours to instruct them (the exclamation mark in the message indicates that it is an imperative message) to pick up the fork. Another informative rule begins on line 18. Here, if the agent is engaged in a conversation with a neighbour (identified by \texttt{X}), and it is aware that \texttt{X} has the goal to hold the fork, it reacts by putting the fork on the table and informing \texttt{X} (the dot in the message indicates that it is a declarative type message) that the fork is on the table and the sender no longer holds the fork. The other type of message supported by GOAL can be seen on line 22. An interrogative message is indicated by a question marks. In this message, the sender is enquiring whether the fork is on the table.

\subsection{JADE and Jadex} \label{sec:aop:jadex}

JADE (Java Agent DEvelopment framework) is a Java-based middleware platform with the aim of facilitating the development of MASs~\cite{Bellifemine2007b,Bellifemine1999}. It provides a mechanism to deploy distributed MASs, along with a suite of development and debugging tools to aid development. Its principal feature is that it provides a Java implementation of the FIPA agent architecture including such services as agent management, communication, addressing and discovery. An Agent class is provided that may be extended by developers to create agent programs in Java. Debugging tools include the provision of a Dummy Agent (which can be used to inject messages into the MAS) and a Sniffer Agent (which can intercept communication to give the developer an insight into the interaction that is occurring within the MAS). JADE does not natively support any particular AOP language. All programs running natively within JADE must be written in the Java OOP language. However, JADE is a common choice for a base on which AOP languages can be based.

An example of this is Jadex, which is a BDI layer built on the JADE middleware platform~\cite{Braubach2004,Pokahr2005}. Whereas JADE concentrates on providing a communication infrastructure and platform services, Jadex concentrates on the internal reasoning of the agents themselves. Agent definitions are written in an XML format that allows the developer to specify such things as plans, goals and initial beliefs. The implementation of the actual beliefs and plans themselves is by way of Java classes.

The principal agent-oriented constructs that Jadex makes available are beliefs, plans, goals and capabilities. Goals are contained in a goal base, which is accessible to the reasoning components, which will consult available plans in deciding how goals may be achieved. Jadex supports four types of goal:

\begin{itemize}
   \item A \textit{perform} goal is related to the performance of some action, regardless of the outcome of that action.
   \item \textit{Achieve} goals relate to a desire to bring about some state of the world.
   \item A \textit{query} goal relates to the availability of some information the agent wants to know about.
   \item An agent will continually plan to re-establish a desired state when using a \textit{maintain} goal.
\end{itemize}

The implementation of Jadex agents is done by combining an XML file that defines the beliefs, goals and plans of the agent with procedural code written in Java, built on the JADE multi-agent framework.

\begin{figure}[!htb]
   \includegraphics[width=\textwidth]{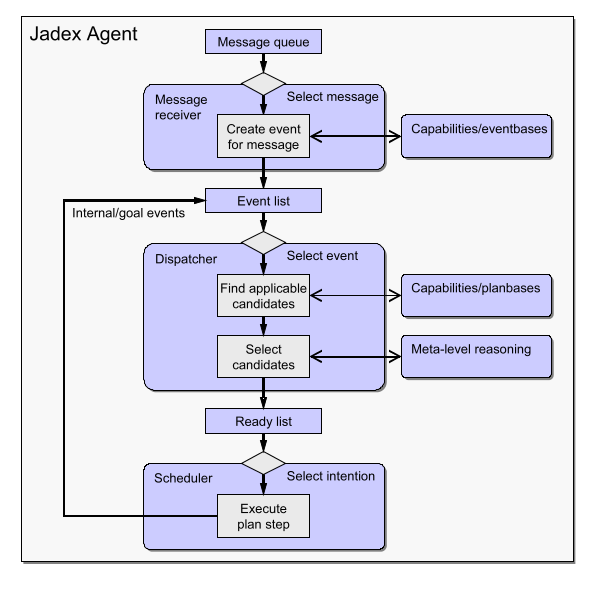}
   \caption{Jadex execution model (taken from~\cite{Pokahr2005}).}
   \label{fig:aop:jadex-execution-model}
\end{figure}

The execution model of a Jadex agent is illustrated in Figure~\ref{fig:aop:jadex-execution-model}. The internal structure consists of a number of clearly-defined, separate components. The Message Receiver is responsible for identifying appropriate capabilities to handle the message. A capability is responsible for generating an appropriate event for the message.

Events are handled by the Dispatcher. It matches the events against available plans to find those that match the event. If there are multiple plans available, it then chooses one or more to execute. This selection is performed by the Meta-level reasoning, which is extensible to allow various plan selection policies to be integrated with the system. Selected plans are considered to be ready for execution.

The Scheduler selects ready plans and executes them step-by-step. Execution continues until a plan explicitly waits or substantially alters the internal state of the agent. When this occurs, any events generated by such a change are fed back into the Event List to be processed by the Dispatcher.

An example of Jadex code is now presented. This is taken from an example of an agent for a ``blocks world'' problem~\cite{Pokahr2005}. In this type of problem, a number of stackable coloured blocks are placed on a table and an agent is required to achieve particular configurations. In the first part of the agent definition, the beliefs of the agent are defined. This is done using the following code:

\begin{Verbatim}[numbers=left,frame=single,fontsize=\footnotesize]
<beliefs>
  <belief name="table" class="Table">
    <fact>new Table()</fact>
  </belief>
  <beliefset name="blocks" class="Block">
    <fact>
      new Block(new Color(240,16,16),$beliefbase.table)
    </fact>
    <fact>
     new Block(new Color(16,16,240),$beliefbase.table.allBlocks[0])
    </fact>
    <fact>
     new Block(new Color(240,240,16),$beliefbase.table.allBlocks[1])
    </fact>
  </beliefset>
   ...
</beliefs>
\end{Verbatim}

Within the \texttt{<fact>} tags, Java objects are created to represent the entities about which the agent has beliefs. The type of object that should be contained in each is defined in the \texttt{<belief>} or \texttt{<beliefset>} tag. Each belief or beliefset is also given a name by which it can be identified elsewhere in the agent definition. For example, the new Block object created on line 7 is placed on the Table object previously created in line 3. This is accessed by means of the \texttt{\$beliefbase} variable, with which the table can be referenced by the name it was given in line 2.

The next extract shows the definition of an achievement goal. It is defined as follows:

\begin{Verbatim}[numbers=left,frame=single,fontsize=\footnotesize]
<goals>
  <achievegoal name="clear">
    <parameter name="block" class="Block" />
    <targetcondition>$goal.block.isClear()</targetcondition>
  </achievegoal>
   ...
</goals>
\end{Verbatim}

As with beliefs, goals are also given names to identify them. This goal is defined with a parameter named \texttt{block} that must be a Block object (defined in Java). This goal is intended to ``clear'' a particular block by ensuring that there are no other blocks stacked on it. The condition that the goal is intended to achieve is specified in the \texttt{<targetcondition>} tag. The \texttt{\$goal.block} variable refers to the parameter that was defined in the previous line. The \texttt{isClear()} method (invoked on the Block object) should return true when the goal has been achieved.

The final extract relates to the definitions of plans that are intended to achieve goals. Its code is as follows:

\begin{Verbatim}[numbers=left,frame=single,fontsize=\footnotesize]
<plans>
  <plan name="clear">
    <bindings>
      <binding name="upper">
         select $upper from $beliefbase.blocks where
           $upper.getLower()==$event.goal.block
      </binding>
    </bindings>
    <body>new StackBlocksPlan($upper, $beliefbase.table)</body>
    <trigger><goal ref="clear" /></trigger>
  </plan>
  ...
</plans>
\end{Verbatim}

This plan is intended to achieve the ``clear'' goal defined above (the adoption of this goal can be seen as the trigger of this plan in line 10). A new variable is created with the plan by means of a \texttt{<binding>} tag. Binding a value to this variable (called \texttt{\$upper}) is done via an SQL-style query of the agent's belief base. It is selected as a block that is above the block that the goal is intended to clear. The \texttt{getLower()} method of the Block class will return the block below the block it is invoked on and \texttt{\$event.goal.block} relates to the block that was passed as a parameter to the ``clear'' goal that triggered this event.

The body of the plan instantiates another Java class that defines the steps required to carry out the plan. The variables that were bound in this Jadex definition are passed to its constructor. This can be seen on line 9.

%
%
%Because of the use of JADE, Jadex agents make use of FIPA ACL for communication.

% Per Dastani2008: adds syntactic constructs to Java (as Jack does) but the lack of semantics for the beliefs and goals means that plans that partially help towards achieving a goal cannot be generated.

\subsection{2APL} \label{sec:aop:2apl}

2APL is a formally-specified, BDI-based AOP language, with a focus on MASs that share access to common external environments~\cite{Dastani2008}. It extends and modifies the earlier 3APL language, which was aimed more at the creation of individual agents~\cite{Dastani2005,Hindriks1999}.

For individual agents, implementation is by way of beliefs, goals, actions, plans events and rules. Beliefs and goals are declarative, with plans operating in an imperative style. Each agent is assumed to be situated in an environment (modelled by Java objects). An agent can monitor its environment either actively (through a sensing action) or passively (by means of the environment generating events).

Rules can generate plans to achieve goals, process events (including the receipt of messages) or catch and repair failed plans. Thus, 2APL agents can behave reactively if the raising of events triggers rules that result in particular actions. Proactive, deliberative behaviour is also possible through the rules that are aimed at achieving goals. Goals persist until they are achieved, meaning that if a plan fails to achieve the goal for which it was executed, another plan can be selected for execution instead. This can continue until the goal has been achieved.

A key feature of 2APL is its hybrid of declarative and imperative programming. The beliefs and goals of an agent can be queried, with the results available for use in actions and plans to modify the external environment. Information gleaned from observation of the environment can similarly be used to update the agent's beliefs and goals.

For multi-agent concerns, a separate specification language allows a developer to specify what agents need to be created and what environmental components each agent requires access to.

2APL makes use of FIPA ACL for inter-agent communication. This is done by making an action available for agents to send individual messages, with incoming messages treated as events. The distributed mode of 2APL is implemented as a layer on the JADE middleware platform (see Section~\ref{sec:aop:jadex}).

There are numerous elements to a 2APL agent, many of which are illustrated in the code samples that follow. The context for these examples is a virtual world where trash and gold are placed in particular locations. The role of the agents is to clean the trash from the world and collect the gold.

A 2APL agent typically begins with the declaration of some initial beliefs and belief rules. These exactly follow the syntax of Prolog facts and rules. In the following example, the agent believes that trash exists in a particular location, and a rule states that the world is clean if there is no trash in any location.

\begin{Verbatim}[numbers=left,frame=single,fontsize=\footnotesize]
Beliefs:
   trash(2,5).
   clean(blockWorld) :- not trash(_,_).
\end{Verbatim}

Goals are also declared at the beginning (though agents can adopt new beliefs and/or goals during execution). The following example shows an agent with two goals, separated by a comma. The first indicates that it wishes to bring about a situation whereby it possesses 5 units of gold and the world is clear of trash. The second simply desires the collection of 10 gold pieces.
\begin{Verbatim}[numbers=left,frame=single,fontsize=\footnotesize]
Goals:
  hasGold(5) and clean(blockWorld), hasGold(10)
\end{Verbatim}

A belief update rule takes the form of a Hoare triple~\cite{Hoare1969}. It indicates what the effect of an action on the goal base will be. Each belief update rule consists of three items: a precondition, an action and a postcondition. In the following example, the action becomes possible if it believes that trash is present at a position given by some co-ordinates \texttt{(X,Y)} and that its own position is the same. As with Prolog and AgentSpeak(L), capital letters indicate variables. In this context, executing the \texttt{RemoveTrash()} action will result in the outcome that trash is no longer present that that location.

\begin{Verbatim}[numbers=left,frame=single,fontsize=\footnotesize]
BeliefUpdates:
   {trash(X,Y) and pos(X,Y)} RemoveTrash() {not trash(X,Y)}
\end{Verbatim}

Plans provide a mechanism to accomplish more complex tasks by combining multiple basic actions. Plan operators include a sequence operator, conditional choice operator, conditional iteration operator and a non-interleaving operator. The following examples illustrates a 2APL plan. The brackets around the first two actions creates an atomic plan that ensures that the \texttt{ChgPos(5,5)} is executed immediately after (the semicolon is the sequential operator) the agent enters the world. The \texttt{enter} action is an external action provided by the \texttt{blockworld} environment.
\begin{Verbatim}[numbers=left,frame=single,fontsize=\footnotesize]
Plans:
  [@blockworld(enter(5,5,red),L);ChgPos(5,5)]
\end{Verbatim}

A Planning Goal rule (PG-rule) causes a plan to be generated when a particular mental state of the agent occurs. PG-rules consist of a head (consisting of goal expressions), a condition (beliefs that should be present) and a body (the actions to perform). In the following example, if the agent has the goal to clean some space (\texttt{clean(R)}) and it has a belief relating to its own position (\texttt{pos(X1,Y1)}) and the location of some trash (\texttt{trash(X2,Y2)}), it should perform the actions in the plan. In this case, that involves sequentially travelling to the location of the trash and executing the \texttt{RemoveTrash()} action.
\begin{Verbatim}[numbers=left,frame=single,fontsize=\footnotesize]
PG-rules:
  clean(R) <- pos(X1,Y1) and trash(X2,Y2) |
              {[goTo(X1,Y1,X2,Y2);RemoveTrash()]}
\end{Verbatim}

Procedure Call rules (PC-rules) are defined in the same way, except the head of the rule is not a goal query. Instead, it is used to react to the receipt of messages or other events.

Another type of plan rule, known as a Plan Repair rule (PC-rule) is used to replace a failed plan with another, if certain conditions hold. For example, a plan to move two steps to the east in a grid world can be replaced by a plan to first travel one step north, then move two steps east and then return to the south. This plan would be used to avoid an obstacle that prevents direct travel to the east.

\begin{figure}[!htb]
	\centering
	\includegraphics[width=\textwidth]{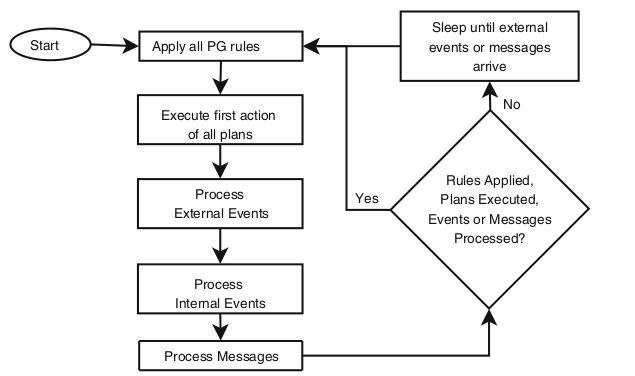}
	\caption{The deliberation cycle for 2APL agents (taken from~\cite{Dastani2008})}
	\label{fig:aop:2apl-interpreter-cycle}
\end{figure}

In order to allow for integration with various external environments, actions can be implemented as Java classes. These can be invoked from within the 2APL agent in the same way as built-in actions.

The deliberation cycle of 2APL is shown in Figure~\ref{fig:aop:2apl-interpreter-cycle}. The agent only acts whenever unhandled events or messages are present. When this is the case, plan generation rules are used to generate appropriate plans to handle those events. Next, the first action of all generated plans are executed. This is followed by the processing of external, internal and message events respectively. This cycle then operates in an iterative fashion for the lifetime of the agent.

\section{Summary}

This Chapter is the first of three that review the state-of-the art in the area of AOP. A number of potential elements of agency are considered, with the weak notion of agency proposed in~\cite{Wooldridge1995} being adopted. This defines an agent as a computational entity that has the properties of \textit{autonomy}, \textit{social ability} and exhibits both \textit{reactive} and \textit{proactive} behaviour. Additional elements that could be included in a definition are omitted due to the nature of this work.

Numerous efforts have been undertaken to create AOP languages, toolkits and frameworks. This Chapter outlines a number of these, concentrating on active projects that are currently in use. From this analysis, it is clear that a myriad of different mental models are in active use amongst the agent programming community. Most AOP languages make use of a subset of beliefs, goals, desires, intentions, commitments and capabilities as part of their mental models. For any technology seeking interoperability between diverse languages or frameworks, no particular internal model can be safely assumed.

As the key focus of this thesis is on the communication between agents, Chapter~\ref{chap:communication} surveys the efforts that have been made to date in the area of modelling agent communication. This is done both by attempting to model individual communications and their effects and also by modelling more complex communication consisting of a lengthier sequence of interactions.
\chapter{Agent Communication}
\label{chap:communication}

%TODO: add in some back-channels stuff

\section{Introduction} \label{sec:communication:introduction}

The ability of intelligent agents to interact socially is widely acknowledged as a key feature of a multi-agent system~\cite{Singh1998,Winikoff2005,Wooldridge1995}. This ability allows groups of agents to collaborate in solving problems and performing tasks that would be beyond the ability of a single agent.

To facilitate this inter-agent communication, great efforts have been made to standardise Agent Communication Languages (ACLs). These are aimed at providing a standard mechanism by which agents can interoperate and be mutually understood. ACLs tend to be based on the principles of \textit{speech act theory}, where the act of speaking is considered to be an action that has an effect on the world~\cite{Labrou2001}.

Early research tended to concentrate on the particular effects of individual messages. It is only after the semantics of these were decided upon that attention moved towards the concept of agents engaging in structured conversations consisting of multiple related messages.

This Chapter provides an overview of agent communication, from its origins in speech act theory to later advances in conversation management. This serves as the background against which the research discussed in this thesis has been conducted.

% % % % % % % % % % % % % % % % % % % % % % % % % % % % % % % % % % % % % % % % % % % % % % 

%Some stuff \cite{Braubach2007}.
% \begin{itemize}
%    \item Macro Level \begin{enumerate}
%       \item What are the objectives behind the interaction?
%       \item What are the characteristic properties of the interaction?
%       \item How can the interaction be described and analysed?
%       \end{enumerate}
%
%    \item Micro Level \begin{enumerate}
%       \item What are the objectives of the interacting agents?
%       \item How is the interaction related to the agent architectures?
%       \item How is the interaction related to domain-specific behaviour?
%       \end{enumerate}
% \end{itemize}
% 
 % % % % % % % % % % % % % % % % % % % % % % % % % % % % % % % % % % % % % % % % % % % % % % % % % % % % 

\section{Standalone Agent Communication}

A discussion of agent communication must necessarily begin with \textit{standalone} communication. In this instance, ``standalone'' refers to the fact that individual communications are unrelated to others with all messages being sent individually. There are two principal approaches to research on agent communication languages~\cite{Hindriks2010}. The first is based on speech act theory and the other is based on social semantics. These are discussed in the following Sections.

\subsection{Speech Act Theory} \label{sec:communication:speech-act-theory}

Speech act theory underpins a major strain of ACLs. This theory emerged from the work of Austin, who noted that some utterances have similarities to physical actions, in that they have an effect of the state of the environment~\cite{Austin1962}. Examples of such utterances include a declaration of war, a pronouncement of marriage or the naming of some entity.

In Austin's analysis, he identifies three aspects of any speech act:

\begin{itemize}
   \item \textit{Locutionary act}: the act of making an utterance (e.g. ``please make some tea'').
   \item \textit{Illocutionary act}: the action performed in saying something (e.g. ``he requested me to make some tea'').
   \item \textit{Perlocution}: the effect of the act (e.g. ``he got me to make tea'').
\end{itemize}

This work was expanded upon by Searle, who identified five types of illocutionary act~\cite{Searle1970}:

% http://skemman.is/stream/get/1946/8360/22297/1/Loftur_%C3%81rni_Bj%C3%B6rgvinsson_0105833579_-_BA_Ritgerd_-_Speech_Act_Theory_-_A_Critical_Overview.pdf

\begin{itemize}
	\item \textit{Representatives/Assertives}: This presents a proposition as representing the state of the world. Such assertions can be true or false. When making an utterance with an assertive force, it is not necessary for the speaker to be truthful. It is the hearer's belief that dictates whether the act is taken as true or false.
	\item \textit{Directives}: This is an attempt to get another to bring about some state of the world and includes such speech as requests, orders, instructions, etc. There is no requirement on the hearer to comply with the directive.
	\item \textit{Commissives}: This commits the speaker to future action, such as promises and verbal contracts. A speech act with commisive force is a statement of an intention to perform some action in the future. 
	\item \textit{Expressives}: This covers those speech acts that express something about mental state. Examples include expressing regret (apologising), appreciation (thanking) or condolences.
	\item \textit{Declaratives}: Speech acts with declarative power themselves bring about a change in the world. In this situation, an utterance declaring a state of the world that did not exist beforehand may have, in itself, the effect to bring this state about. 
\end{itemize}

A speech act can be further broken into two core components: a \textit{performative verb} and a \textit{propositional content}. These are illustrated by examples in Table~\ref{tab:communication:speech-acts}.

\begin{table}[!ht]
\centering
\label{tab:communication:speech-acts}
\caption{Types of Speech Acts.}
\begin{tabular}{l|l|l}
	\hline
	\textbf{Speech Act} & \textbf{Performative} & \textbf{Content} \\
	Please close the door & request & the door is closed \\
	The door is closed & inform & the door is closed  \\
	Is the door closed? & inquire & the door is closed  \\
	\hline
\end{tabular}
\end{table}

Much research on communicating agents has been based on the principles of speech act theory. The act of speaking is intended by an agent to have some effect on the world. However, a distinction can be drawn between physical actions and speech acts. For a speech act, the specific part of the world that the speaker wishes to modify is the mental state of other agents~\cite{Vieira2007}.

The speech acts themselves can be categorised according to their \textit{illocutionary force} (i.e. the type of utterance)
. Examples of this include a statement such as ``the door is open'', which is intended to ``tell'' or ``inform'' another agent of a fact. The \textit{perlocutionary force} represents the intentions of the speaker in making the utterance in terms of the effect they wish to bring about. For instance, the perlocutionary force of ``the door is open'' is that the listener would come to believe that the door is open. In agent communication, the illocutionary forces are explicit. This is typically done through the use of labels included in the messages to state what this illocutionary force is. These are known as \textit{performatives}~\cite{Vieira2007}.

Shoham's original AGENT0 language made use of speech act theory for communication. It featured only three performatives: \texttt{INFORM} (to tell another agent of a fact), \texttt{REQUEST} (to ask another agent to perform some action) and \texttt{UNREQUEST} (to cancel a prior request)~\cite{Shoham1991}.

Later researchers have identified other performatives that they consider to be useful for agent communication, and have included these in various message standards. The two most popular ACLs developed are the Knowledge Query Manipulation Language (KQML) and the ACL developed by the Foundation for Intelligent and Physical Agents (FIPA). Both of these are based on speech act theory and include performatives as a core aspect of each message. These are discussed in the following Sections.

%From~\cite{Hindriks2010}
%\begin{quotation}
%Both languages specify the semantics of messages by means of their pre-condition, expressing conditions on the mental states of the sender and receiver of the message that should hold if the message is sent, and their effect, expressing the effect of the message on the mental state of the sending and/or receiving agent.
%\end{quotation}
%FIPA ACL does not specify that there is any particular effect on the sender~\cite{FIPA00037}.

\subsubsection{KQML}

The Knowledge Query and Manipulation Language (KQML) was the first language to gain widespread adoption as an ACL~\cite{Singh1998}. However, it was not originally developed with agent programs in mind. As research in the area of Artificial Intelligence intensified in the 1980s, it was felt that standardisation was required to facilitate the storage, reuse and communication of knowledge between intelligent computational entities.

The Knowledge Sharing Effort (KSE) was a project sponsored by the Defense Advanced Research Projects Agency (DARPA) in the US. Its aim was to ``develop the technical infrastructure to support the sharing of knowledge among systems''~\cite{Neches1991}.

In the model adopted by the KSE, diverse knowledge bases require a mechanism by which they can exchange knowledge. This is done by way of \textit{propositional attitudes}. An example of a propositional attitude supplied by Labrou is $<a, fear, raining_{now}>$~\cite{Labrou2001}. This indicates that a knowledge base identified as $a$ has the propositional attitude of fear relating to the proposition that it is currently raining. A finite set of attitudes is permitted, including believing, asserting, fearing, wondering and hoping.

In order for this system to be standardised, three languages were developed. The Knowledge Interchange Format (KIF) was created to allow the propositions themselves to be represented~\cite{Genesereth1992}. Ontolingua was designed to specify the ontology of the prepositions~\cite{Farquhar1997}. Finally, the language developed to model the propositional attitudes was the Knowledge Query and Manipulation Language (KQML). Using this, an intelligent entity could express to another its attitude towards a particular proposition.

As the area of intelligent agents gained more traction, KQML was adopted as a suitable language for communication~\cite{Labrou2001,Labrou1994}. A KQML message contains three principal layers:
\begin{itemize}
   \item The \textit{content layer} contains the actual content of the message. The KQML specification does not restrict this to any particular representation language, although many users of KQML tended to make use of KIF for message content.
   \item The \textit{communication layer} includes the communication parameters such as the identity of the sender and receiver along with an identifier for the message.
   \item The \textit{message layer} encodes a message that is to be exchanged. This includes information about the content language used and the ontology that describes the content. It also includes a \textit{performative}, which specifies the speech act for the message. This indicates whether the message is intended as an assertion of fact, a query, a command or another type of speech act chosen from the set of primitive KQML performatives.
\end{itemize}

KQML features a predefined set of reserved performatives. However, this was not intended to be a closed set, with customisation being permitted. Agents are not required to be capable of responding to every possibly performative. 

According to Singh, attempts to standardise agent communication using KQML were unsuccessful~\cite{Singh1998}. This was due to the emergence of a variety of incompatible KQML dialects, which occurred for two principal reasons:
\begin{enumerate}
\item The sending agent may use application-specific terms that are not understandable to the recipient.
\item Basic communication components are not uniformly understood. This may be attributed to a lack of formal semantics.
\end{enumerate}

The semantics of KQML performatives were not included in the original specification and these were added later by the community (for example, in~\cite{Labrou1994,Labrou1998}). These semantics were defined in terms of the mental states of the sender and receiver of the communication. These take the form of preconditions (regarding the mental state of the participants before a communication is sent) and postconditions defining the effects of the communication. An example of the semantics for the \textit{tell} performative (used to inform another agent of a fact) is shown in Figure~\ref{fig:communication:kqml-semantics}.

\begin{figure}
\begin{itemize}
   \item Pre-conditions on the states of $S$ (sender) and $R$ (receiver):
   \begin{itemize}
      \item $Pre(S): bel(S,X) \wedge know(S,want(R,know(R,bel(S,X))))$
      \item $Pre(R): intend(R,know(R,bel(S,X)))$
   \end{itemize}
   \item Post-conditions on $S$ and $R$:
   \begin{itemize}
      \item $Pos(S): know(S,know(R,bel(S,X)))$
      \item $Pos(R): know(R,bel(S,X))$
   \end{itemize}
   \item Action completion:
   \begin{itemize}
      \item $know(R,bel(S,X))$
   \end{itemize}
\end{itemize}
\caption{Semantics for the KQML \textit{tell} performative (taken from ~\cite{Labrou1994}).}
\label{fig:communication:kqml-semantics}
\end{figure}

In this figure, agent $S$ wishes to tell agent $R$ that a proposition $X$ is true, using the construct $tell(S,R,X)$. For $S$ to send such a message, the preconditions must be satisfied. Firstly, $S$ must itself believe that $X$ is true (as such it is not permitted to lie) and also that it is aware that $R$ wants to know this information. $R$ intends to know whether or not $S$ believes $X$ to be true.

Following the sending of the message, the mental state of $S$ includes the knowledge that $R$ now knows of the belief of $S$ in $X$. The agent $R$ has satisfied its intention and now knows that $S$ does believe $X$. The ``action completion'' indicates the overall desired effect of the communication. Sometimes this may require further messages to be exchanged. In the case of the \texttt{tell} performative shown in Figure~\ref{fig:communication:kqml-semantics}, this is the same as the second post-condition. However, the overall effect of another performative may require a response. For example, an \texttt{ask-if} message is intended to be an enquiry as to the truth of a proposition. The desire is to receive a reply (with a \texttt{tell} performative) that contains this information.

This standardisation effort included an attempt to specify legal sequences of performatives that would make up a longer conversation. This takes into account those performatives that are suitable for beginning a conversation and continuing existing conversations~\cite{Labrou1994}.

\subsubsection{FIPA ACL}

% reference for ``most well-known'' is Hindriks2010 if needed

A number of criticisms of KQML resulted in the development of FIPA ACL beginning in the late 1990s. The Foundation for Intelligent and Physical Agents (FIPA) was founded in 1996 as a standards organisation for agents and Multi Agent Systems. It has been part of the IEEE Computer Society since 2005. FIPA ACL is the agent communication language developed by FIPA.

Many features of FIPA ACL are similar to those of KQML. It also draws on speech act theory, with each message specifying a performative\footnote{FIPA's specifications call these ``communicative acts''. For consistency, these are referred to as ``performatives'' in this work.} in addition to its content. The specifications also outline the effects of message exchange on the mental states and attitudes of the sending and receiving agents through formal semantics~\cite{FIPA00037}.

The FIPA ACL semantics specify \textit{feasibility preconditions} and \textit{rational effects} for each performative. Feasibility preconditions describe the conditions necessary for a message to be sent. The rational effect is the effect that an agent can anticipate occurring as a result of sending a message. As it may not be possible for the recipient of a message to make this come about, this is not a guarantee. However, it may be used by the sender in deciding when to communicate. A strictly compliant implementation of FIPA ACL requires that the relevant feasibility preconditions must be satisfied for an agent to be permitted to send a message.

Although the basic principles of FIPA ACL are very similar to those of KQML, they are not compatible on account of the differing semantics they employ for equivalent performatives~\cite{Labrou1999}. For example the KQML \texttt{tell} performative and FIPA ACL's \texttt{inform} performative both have the intention of allowing an agent make another agent aware of the truth of a proposition, though these have different semantics. However, approximate mappings for loose translations of these ACLs have been implemented. %e.g. Jason/JADE

A criticism of KQML is that it lacks organisation at the conversation level and so the messages it transmits are without context~\cite{Cost2000}. For this reason, the FIPA ACL specification allows for \texttt{conversation-id} and \texttt{protocol} fields to facilitate individual messages being grouped together into protocols~\cite{FIPA00061}.

Chapter~\ref{chap:fipa} presents a more detailed analysis of the FIPA Agent Communication standards.

\subsection{Criticisms of ACLs} \label{sec:communication:acl-criticism}

Although both KQML and FIPA ACL have gained some traction in the agents community, they are not without their critics. Hindriks notes that FIPA ACL and KQML have both been ``extensively criticised'' on the grounds of complexity and verifiability~\cite{Hindriks2010}.

The complexity issue stems from the high number of performatives that are available, combined with the fact that subtle semantic differences exist between them.

The problem of verifiability is a result of both ACLs specifying unverifiable preconditions for each message~\cite{Singh1998,Vieira2007}. For example, a FIPA message with the performative \texttt{inform} is intended to indicate to a recipient that a proposition given as the content of the message is true. The FIPA ACL standard requires that the sender of the message should itself believe the truth of the given proposition before communicating this to another agent. However, the recipient has no way of verifying whether this precondition holds.

With regard to the semantics of KQML and FIPA ACL, a number of researchers express concern regarding the suitability of the existing work in this area~\cite{Pitt1999,Singh1998,Wooldridge2000}. Although these agree that formal semantics are important, existing semantics for ACLs are overly focussed on the idea of \textit{mental agency}. Semantics of ACLs such as KQML and FIPA ACL tend to be expressed in terms of mental concepts such as beliefs, desires and an agent's knowledge. In effect, verification that an ACL has been employed correctly requires agents to be capable of ascertaining each other's mental state via some form of mind-reading. This is considered to be impractical and unrealistic, as this supposition has never been true for human interaction.

An additional criticism by Wooldridge is that the computational complexity of logically verifying whether an agent implements the semantics of an ACL is a ``major obstacle'', as the semantics of communication languages are often expressed in a multi-modal logic~\cite{Wooldridge2000}.

Some agent programming languages have taken a more pragmatic approach to agent communication in recent times. In many implementations (e.g. Agent Factory~\cite{Collier2003}, 2APL~\cite{Dastani2008} and 3APL~\cite{Dastani2005}), the semantics are ignored altogether. Here, the only effect on the mental model of the agents is that the receiver has a belief that a message has been received and the sender acquires a belief that it has been sent.

%Work has been done on providing semantics for communication in Jason's implementation of AgentSpeak(L)~\cite{Vieira2007}. Here, a small subset of KQML-based primitives were chosen for communication, with accompanying simple semantics. Commun

%TODO come back to this: find others ignoring or implementing inconsistent semantics and just using simple message notifications.

\subsection{Social Semantics} \label{sec:communication:social-semantics}
Social Semantics are proposed as an alternative method of viewing agent communication, addressing some perceived shortcomings of ACLs such as FIPA ACL and KQML~\cite{Singh2000}.

This typically models the effect of messages in terms of the \textit{commitments} that are created as a result of communication. In this context, a commitment is a directed obligation, where one agent (a \textit{debtor}) is obliged to do something for another (a \textit{creditor}).

The semantics of this model are based on the creation of commitments by way of message passing, and as such are more objective than mentalist, subjective semantics of FIPA ACL and KQML. An example of this is a message that contains a \textit{promise} performative. Objectively, sending such a message commits the sender to the accomplishment of some task. This can be verified by an independent observer that is privy to the communication. Using subjective mentalist semantics, however, the sender commits that it intends to perform the task. As discussed in the previous Section, verification of the intentions of an agent requires knowledge of its inner workings which is impossible in many circumstances.

\section{Conversation Models}

Communications frequently do not occur in isolation. Offers for service are accompanied by acceptance, calls for proposals are accompanied by bids and queries are followed by the provision of the information required. In order to allow this type of complex communication to occur, some model of a \textit{conversation} is necessary. Many different formalisations of conversations have been proposed. In analysing these, it is important to distinguish between \textit{prescriptive} and \textit{emergent} conversation policies~\cite{Nodine1998}.

A prescriptive conversation policy is based on design-time engineering of acceptable sequences of messages that may be exchanged between agents. We refer to these prescriptive conversation policies as \textit{protocols}. 

Emergent approaches allow agents to decide for themselves how interaction should occur. This is frequently based on their understanding of the effects of communicative acts, as specified by their respective semantics.

A number of competing models are discussed in the following Sections, including discussion of their benefits and disadvantages. This informs the choice of a model for the work presented in this thesis, which is introduced in  Chapter~\ref{chap:acre-introduction}.

%Discrete Finite-state Automata (DFA), Finite State Machines (FSMs). Used in \cite{Barbuceanu1995,Cost1998,Dinkloh2004,Bradshaw1996}.

%
%% taken from Agent UML section
%These interaction diagrams deal with agent \textit{roles} rather than concrete agents, meaning that an individual agent may play multiple roles in different interactions during its lifetime.

\subsection{Finite State Machines} \label{sec:communication:fsm}

Finite State Machines (FSMs) are one of the most popular models used for handling interaction~\cite{Yolum2001}. State transition diagrams were being used as far back as 1986 to model conversations~\cite{Winograd1986}.

The principal advantage of such a model is that they allow a variety of behaviours to be expressed, while remaining conceptually simple. Simplicity and ease of implementation are gained at the cost of the expressiveness of the protocol~\cite{Cost1999,Cost1998}.

Criticisms of this model include the observation that while FSMs dictate the syntax of a conversation, they tend to omit semantics, which can lead to rigid executions~\cite{Cost1999,Yolum2001}. Poorly-designed FSMs are also subject to problems with concurrency in systems where a consistent message ordering is not guaranteed~\cite{Cost1999}. It is further argued that specifying interaction protocols as a permitted sequence of communications is over-constrained and that a higher level of abstraction is more suitable for modelling conversations amongst autonomous agents~\cite{Singh2010}.

%The choice of FSMs to represent conversations in Jackal stems from their expressiveness and simplicity. Following from this, the authors have this to say about semantics of message contents:
%\begin{quotation}
%DFAs and roles dictate the syntax of a conversation, but say nothing about the conversation's semantics. The ability of an agent to read a description of a conversation, then engage in such a conversation, demands that the description specify the conversation's semantics. To be useful though, such a specification must not rely on a full-blown, highly expressive knowledge representation language. We believe that a simple ontology of common goals and actions, together with a way to relate entries in the ontology to the roles, states, and transitions of the conversations specification, will be adequate for most purposes. The approach sacrifices expressiveness for simplicity and ease of implementation. It is nonetheless perfectly compatible with attempts to relate conversation policy to the semantics of underlying performatives, as proposed for example by~\cite{Bradshaw1996,Bradshaw1997}.
%\end{quotation}

A typical view of a conversation as an FSM involves a conversation being in some state, with actions taken by the agents resulting in a transition between states. In most situations, this action involves the sending of a message, although other actions (such as silence) are capable of causing changes in state in some models.

A competing approach models each participating agent's view of the conversation as an FSM, with their own reactions to the receipt of messages being an integral part of the model.

There is some diversity in the literature with regard to the naming of this type of model. In some works they are described as ``Finite State Machines''~\cite{Cost1999,Yolum2001}. Elsewhere, they can be described as ``Deterministic Finite State Automata''~\cite{Cost1999} or ``Finite State Diagrams''~\cite{Pitt1999}. For consistency, the phrase ``Finite State Machine'' (FSM) will be used throughout this work.

%\cite{Salomaa1969} referenced in \cite{Huget2003} for Finite State Machine/DFA

%Per Yolum and Singh~\cite{Yolum2001}, the use of DFAs (they call them DFSMs - Deterministic Finite State Machines) is one of the most common methods of specifying agent protocols, and that they are also used extensively for e-commerce protocols. Their model features a single start state, with a set of final states.

%Definition (from \cite{Yolum2001}) ``A finite state machine is deterministic if any action has at most one transition from a state)''. This would suggest that a protocol designer is required to be careful in order to ensure that any one message cannot trigger multiple transitions from one state.

%Disadvantages:
%\begin{itemize}
%   \item Dictates syntax of a conversation but says nothing about the semantics~\cite{Cost1999,Yolum2001}.
%   \item Insufficient support for concurrency and verification~\cite{Cost1999}.
%   \item Hides the contents of states and leads to rigid executions~\cite{Yolum2001}.
%\end{itemize}

Different researchers have implemented a variety of approaches to FSM conversation modelling, each of which has its own features. The following Sections outline a number of related, yet distinct, approaches.

\subsubsection{COOrdination Language(COOL)}

The COOrdination Language (COOL) is a language that provides a structured conversation framework for agents~\cite{Barbuceanu1995,Barbuceanu1997,Barbuceanu2000}. A conversation between two agents is modelled as an FSM, where transitions between states occur as a result of ACL messages (either FIPA ACL or KQML) being sent between the participating agents. The transition that has occurred may be identified by reference to the performative of the message that triggered it. When participating in a conversation, each agent follows its own ``conversation plan'' (referred to as a ``conversation class'' in~\cite{Barbuceanu1995}), which defines the states that the conversation may be in, its start and end states, any variables that may be used during the execution of the conversation, and the ``conversation rules'' necessary to cause the conversation to advance. Each conversation rule represents an action (or set of actions) that an agent may take in a particular state. The choice between available rules is made using a controller component. Mechanisms exist to suspend conversations (pending completion of related conversations) and enforce timeouts. Additionally, protocols may be marked as ``friends'', meaning that they are related and may share common variables. An example of this would be if a situation arises during one conversation that requires further information that a new conversation of another type may help to provide.

A key feature of COOL is that each agent follows its own conversation plan, since the actions it takes are inherently linked to the rules contained within this plan. Thus, communicating agents have different models of the conversation they are following, which means that additional work is required to determine whether the conversation plans of two agents are compatible.

The messages exchanged by participating agents contain a conversation identifier that aids agents in linking related communications. However, no conversation plan is specified, as agents use separate plans. This has the consequence that whenever a new conversation is begun, the plan that should be followed is not immediately apparent to the recipient of the initial communication. To help in solving this problem, initiating agents may communicate their intentions in beginning a conversation, with the respondent matching this against the conversation plans it possesses, in order to identify a suitable candidate plan to follow.

%Per Labrou~\cite{Labrou2001}, ``Each arc in a COOL state transition diagram represents a message transmission, a message receipt, or both. One consequence of this policy is that two different agents must different automata to engage in the same conversation''.

%\cite{Barbuceanu1999} provides an overview of the progress of the COOL coordination programming language by 1999. The approach taken in COOL is to define ``Conversation Plans'' (CPs) that specifies the process by which an individual agent may exchange messages with others and change its state. An actual conversation is conducted by two or more agents each executing their own CP and interacting with one another accordingly.

\begin{figure}[!htb]
\begin{verbatim}
(def-conversation-plan 'customer-conversation
  :content-language 'list
  :speech-act-language 'kqml
  :initial-state 'start
  :final-states '(rejected failed satisfied)
  :control 'interactive-choice-control-ka
  :rules '((start cc-1)
           (proposed cc-13 cc-2)
           (working cc-5 cc-4 cc-3)
           (counterp cc-9 cc-8 cc-7 cc-6)
           (asked cc-10)
           (accepted cc-12 cc-11)))
\end{verbatim}
\caption{A COOL conversation plan (taken from~\cite{Barbuceanu2000}).}
\label{fig:communication:cool}
\end{figure}

Figure~\ref{fig:communication:cool} shows an example of a COOL conversation plan. This includes certain information about how the conversation should be conducted, such as the ACL to be used (KQML in this case) and also defines the conversation states that act as the initial and final states. The \texttt{:rules} slot indicates a list of conversation rules that can be applied for each state in the conversation. These named rules define actions that are appropriate for the agent to perform in a given conversation state. For example, when the \texttt{working} state is reached, rules named \texttt{cc-5}, \texttt{cc-4} and \texttt{cc-3} become available. The choice of which rule to execute depends on the control mechanism chosen (in the \texttt{:control} field).

These conversation rules will typically include the sending of a message to advance the conversation, along with any other actions that the agent needs to undertake and updating any relevant variable bindings within the agent itself. From Figure~\ref{fig:communication:cool} it can therefore be seen that the definition of a conversation is inherently integrated with the behaviour of the agent itself.

% Prior work on conversation management (prior to use of petri nets):
% Y. Peng, T. Finin, Y. Labrou, R.S. Cost, B. Chyu, J. Long, W.J. Tolone and A. Boughannam. An agent-based approach for manufacturing integration - the CIIMPLEX experience. Internation Journal of Applied Artificial Intelligence, 1999

%However, COOL requires each agent to use a separate finite state automota in order to engage in a conversation.

%``IN COOL, a \textit{conversation} is the finite state machine representation of an agent's plan to achieve some goal, based on interactions with other agents''~\cite{Chauhan1998}

The Java-based Agent Framework for Multi-Agent Systems (JAFMAS) is another example of a system that makes use of COOL-like conversation plans to make interaction a core aspect of the system~\cite{Chauhan1998}.

\subsubsection{AgenTalk}
AgenTalk is designed as a programming language capable of implementing protocols and agents that behave according to these protocols. Its focus is more on the interaction between agents than on their internal complex logic. The representation of protocols is done by means of an FSM that is extended to allow variables~\cite{Kuwabara1995,Kuwabara1995b}.

%``Many coordination protocols such as the  contract net protocol have been proposed, and many application specific coordination protocols are expected to be required as soon as building of more software agents begins''.

In AgenTalk, the execution of a protocol requires a script that contains the protocol definitions along with an agent program that defines the actions associated with the agent. The script determines when each action should be invoked in order to follow the protocol. Mixing the same scripts with different agent programs can lead to different behaviours that follow the same protocol.

Each script contains the hooks for one of the agent roles involved in a conversation. This means that, like COOL, a full view of the protocol can only be found by examining all the roles involved.

\begin{figure}[!htb]
   \centering
   \includegraphics[width=200pt]{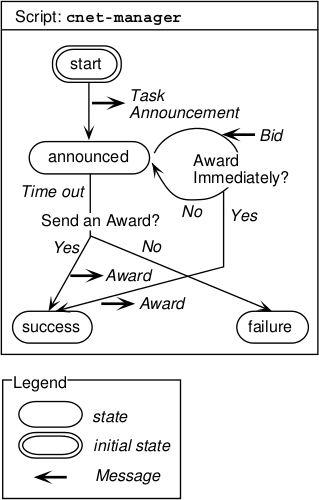}
   \includegraphics[width=200pt]{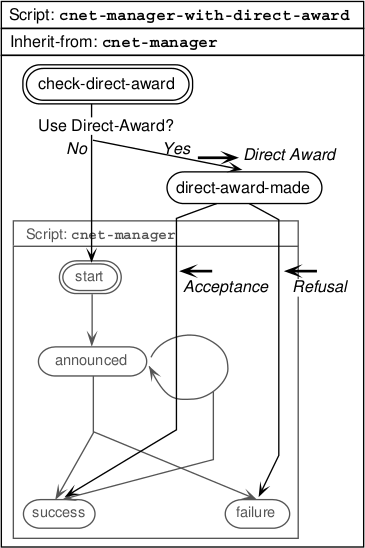}
   \caption{Inheritance in AgentTalk (taken from~\cite{Kuwabara1995}).}
   \label{fig:communication:agentalk-inheritance}
\end{figure}

A novel feature of AgenTalk is that inheritance is supported for protocols to allow for the incremental definition of protocols. An example of this can be seen in Figure~\ref{fig:communication:agentalk-inheritance}. This shows two protocols illustrating AgenTalk's inheritance. The first protocol is for a simple contract net protocol whereby a manager agent sends an announcement, receives bids and later awards the contract to the successful bidder, rejecting the others. The second protocol extends this process to allow the agent to decide against inviting bids, instead directly awarding a contract to an agent that is already known. In this situation, where a contract net is desired, the inherited protocol is invoked in exactly the same way as in the original imported protocol. This promotes the re-use of existing protocols and saves on a duplication effort that may introduce errors and/or inconsistencies.

% The initial state of a conversation is specified explicitly at the beginning of the protocol definition. The method of specifying terminal states is linked to the tasks agents are asked to perform in particular states. Terminal states can be identified by observing that the agent is told to ``exit-script'' on arrival at that state.

% % % End AgenTalk 

\subsubsection{JADE}
Although JADE does not specifically support AOP languages (see Section~\ref{sec:aop:jadex}), it does support the implementation of conversations that make use of FIPA ACL and so should be included in this Section~\cite{Bellifemine2010}.

The tasks carried out by JADE agents are implemented using one or more types of \textit{behaviour}, of which several are provided by the platform. Examples of behaviours include a \texttt{SimpleBehaviour} (simple atomic actions), \texttt{SequentialBehaviour} (where multiple sub-tasks are executed sequentially) and \texttt{ParallelBehaviour} (where multiple sub-tasks are executed concurrently). Of particular interest to the modelling of agent conversations is the \texttt{FSMBehaviour}, which schedules sub-tasks to be executed by reference to a specified FSM.

The \texttt{jade.proto} package contains role behaviours for a variety of the FIPA standard protocols (these are discussed in detail in Chapter~\ref{chap:fipa}). Roles are provided for both an initiator agent and a responder agent. This package provides the behaviours necessary for the implementation of the protocol, with the programmer required to extend the provided behaviours to encompass the application-specific handling of the various steps in a protocol. Most of the behaviours provided are based on the generic \texttt{FSMBehaviour}.

%The flexibility of the \texttt{FSMBehaviour} is limited by the fact that FSM implementations must be created at compile-time by default, although a JADE-FSM-Engine extension allows FSM Behaviours to be created at runtime~\cite{Goh2007}.

With each of these interaction protocol implementations, although the protocol is modelled as an FSM and an \texttt{FSMBehaviour} is used to implement it at the agent level, these relate to the internal state of the agent as it proceeds through a conversation, rather than the state of the conversation itself. As these are generated at compile-time, the agent lacks runtime awareness of conversations, as it is merely following precompiled behaviours.

An Eclipse plugin exists to implement conversations in JADE~\cite{Dinkloh2004}. This models interaction protocols as FSMs, ultimately generating JADE behaviours based on the \texttt{FSMBehaviour} class. 
% (FSM representation stored using cpXML~\cite{Hanson2002})

% TODO: compare with COOL - I believe this acts in a similar way: i.e. the protocol is not actually part of the reasoning process: rather each agent has its own implementation that results in a protocol being followed.

\subsubsection{Jackal}

Jackal is a Java-based package that facilitates the development of agents communicating via KQML~\cite{Cost1998}. A key feature of Jackal is its use of conversations as the unit of communication between agents.

The approach in Jackal is distinguished from that of COOL and AgenTalk in that Jackal imposes a clear separation between the messages involved in the conversation and the actions undertaken by agents as a result of communication. This can be seen in Figure~\ref{fig:communication:jackalcode}, which shows an example of a conversation template of the type used by Jackal.

%Per Cost2000, Jackal has used JDFA, a ``loose Extended Finite State Machine'' for conversation modelling.

\begin{figure}[!ht]
\begin{Verbatim}[frame=single,fontsize=\footnotesize]
// Conversation Template
// Convention: Inital and accepting states all caps,
//    other states initial caps,
//    arc-labels lower case.
(conversation
 (name kqml-ask-one)
 (author "R. Scott Cost")
 (date "3/4/98")
 (start-state START)
 (accepting-states TOLD)
 (transitions
  (arc (label ask-one) (from START) (to Asked) (match "(ask-one)"))
  (arc (label tell)    (from Asked) (to TOLD)  (match "(tell)"))
  (arc (label deny)    (from Asked) (to TOLD)  (match "(deny)"))
  (arc (label untell)  (from Asked) (to TOLD)  (match "(untell)"))
  (arc (label sorry    (from Asked) (to TOLD)  (match "(sorry)"))
  (arc (label error)   (from Asked) (to TOLD)  (match "(error)"))))
\end{Verbatim}
   \caption{A Jackal conversation template (taken from~\cite{Cost1998}).}
   \label{fig:communication:jackalcode}
\end{figure}

\begin{figure}[!ht]
   \centering
      \includegraphics[scale=0.9]{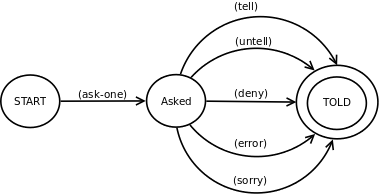}
      \caption{Finite State Machine representing the KQML ``ask-one'' conversation in Jackal (taken from~\cite{Cost1998}).}
      \label{fig:dfa-askone}
\end{figure}

As with COOL, the initial state (here described as the \texttt{start-state}) and terminal states (\texttt{accepting-states}) are explicitly specified at the beginning of the protocol definition. It can also be seen from this Figure that transitions between states are identified only by the performative in the messages exchanged, and not the content they contain.

In Jackal, conversations are used as a form of Application Programming Interface (API) for agents. An agent can describe its interface by identifying a set of conversations in which it is capable of engaging. This forms an Abstract Agent Interface (AAI) so that other agents can become aware of how, for example, a service offered by the agent may be utilised.

% in the above, Bradshaw1996 refers to:
% Bradshaw, J. M. 1996. KAoS: An open agent architecture supporting reuse, interoperability, and extensibility. In Tenth Knowledge Acquisition for Knowledge-Based Systems Workshop.

% End Jackal % % % % % % % % % %

\subsubsection{InfoSleuth} \label{sec:communication:infosleuth}
% InfoSleuth % % % % % % % % % %

% FSM with transitions consisting of sender and performative only is also used in ~\cite{Pitt1999}

\begin{figure}[!ht]
   \includegraphics[width=\textwidth]{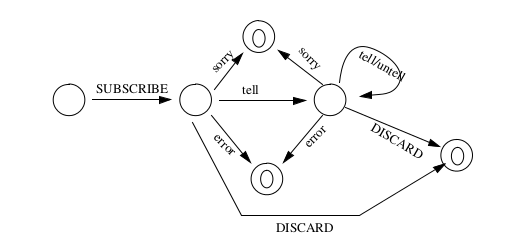}
   \caption{FSM describing an InfoSleuth ``subscribe'' protocol (taken from~\cite{Nodine1998}).}
   \label{fig:communication:infosleuth-fsm}
\end{figure}

FSMs are also used in the InfoSleuth system~\cite{Nodine1998}. An example of a graphical representation of an InfoSleuth protocol can be seen in Figure~\ref{fig:communication:infosleuth-fsm}. Transitions are defined only as KQML performatives: the message content is not used. Like Jackal's representation, the notion of a conversation is separate from the internal workings of the participating agents themselves.

Figure~\ref{fig:communication:infosleuth-fsm} illustrates the synchronisation problem associated with the use of FSMs. In this Figure, performatives sent by the conversation's initiator are shown in uppercase, with other participant's messages shown in lowercase. After the initial \texttt{SUBSCRIBE} message, either agent can send the next message from the resulting state. In this event, it is not a problem, since the initiator can only send a \texttt{DISCARD} message, which ends the conversation. However, the model does not prevent careless protocol designers from encountering this problem.

% End InfoSleuth % % % % % % % % % %

\subsubsection{KAoS}
In the KAoS multi agent framework, communication is based on speech acts and is organised into conversations~\cite{Bradshaw1997}. Although it does not use a standard ACL such as KQML, its messages are based on illocutionary acts (described as ``verbs''), which are equivalent to the performatives seen in KQML and FIPA ACL. Messages can be exchanged only within the context of conversations and do not stand alone.

\begin{figure}[!h]
   \includegraphics[width=\textwidth]{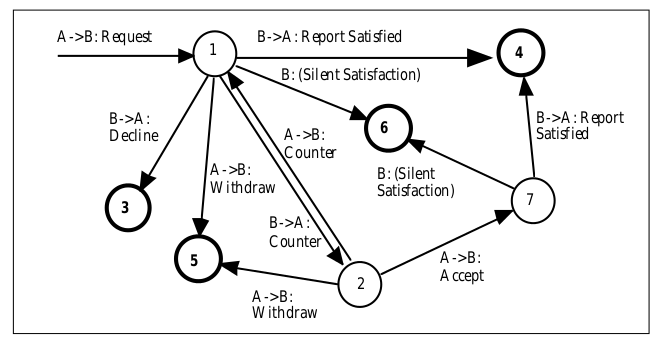}
   \caption{KAoS conversation policy (taken from~\cite{Bradshaw1997}).}
   \label{fig:communication:kaos}
\end{figure}

KAoS protocols are defined in advance as FSMs. An example of a visualisation of a KAoS FSM can be seen in Figure~\ref{fig:communication:kaos}. A number of features can be seen here, some of which are common to other FSM representations and some of which show differences in approach. Firstly, unlike the approaches of COOL and Jackal, the FSM represents the communication only and does not contain any reference to the internal reasoning of the agents themselves. Thus, this single diagram can be used to verify whether the conversation follows the specified protocol.

Each conversation begins with a transition (message) rather than commencing in some initial state. Terminal states are defined as those without any outgoing transitions. These are marked with heavier borders in Figure~\ref{fig:communication:kaos}. Transitions are labelled with the sender and recipient agents, along with the performative the message is expected to contain. Further details of the message, such as the specific content, are omitted. In addition to transitions triggered by the sending of messages, it can be seen that silence also functions as a valid transition (e.g. between states \texttt{1} and \texttt{4}). When initiating a conversation, the initiator must specify the initial verb and the conversation policy. The recipient responds to indicate whether it is capable of processing these.

\subsection{Coloured Petri Nets} \label{sec:communication:cpn}

% Reading on coloured petri-nets (not agent-based)
% K. Jensen. Coloured petri nets. Basic Concepts, Analysis Methods and Practical Use, volume 1, Basic Concepts of Monographs in Theoretical Computer Science. Springer-Verlag, 1002
% K. Jensen. Coloured Petri Nets. Basic Concepts, Analysis Methods and Practical Use, Volume 2, Analysis Methods of Monographs in Theoretcial Computer Science. Springer-Verlag, 1994
% K. Jensen. Coloured Petri Nets. Basic Concepts, Analysis Methods and Practical Use, volume 3, Practical Use of Monographs in Theoretical Computer Science, Springer-Verlag, 1997

% Some other references
%Coloured Petri Nets (CPNs): \cite{Cost1999,Mazouzi2002}.
%Also \cite{Lin1999} CPNs.

% Reading on petri-nets for agent communication
% Cost, R. S., Chen, Y., Finin, T. W., Labrou, Y., Peng, Y., 2000. Using colored petri nets for conversation modeling
% Nowostawski, M., Purvis, M., Cranefield, S., 2001. Modelling and visualizing agent conversations

% same people as behind Jackal ()Cost1998)
Cost et al. argue that FSMs are not sufficient to model complex interactions, particularly those that require concurrent communication~\cite{Cost2000}. Instead, they propose the use of Coloured Petri Nets (CPNs) to model conversations between agents. They argue that these have superior support for concurrent communication while sharing the advantages of being presentable in an intuitive graphical fashion, being simple to implement and having a variety of tools and techniques available for formal analysis and design. CPNs are another example of a prescriptive protocol definition.

\begin{figure}[!htb]
\includegraphics[width=\textwidth]{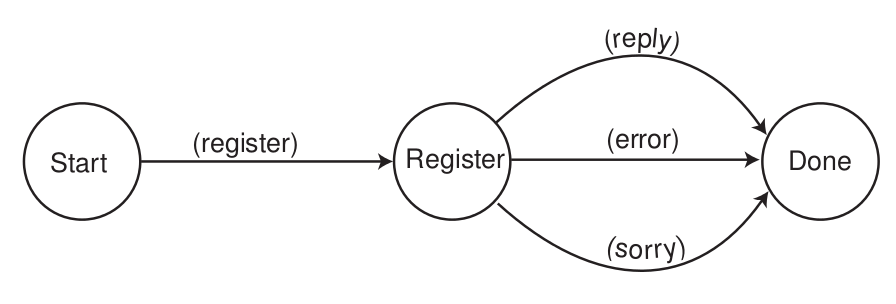}
\caption{FSM diagram of a KQML \textit{register} protocol (taken from~\cite{Cost2000}).}
\label{fig:communication:fsm-register}

\includegraphics[width=\textwidth]{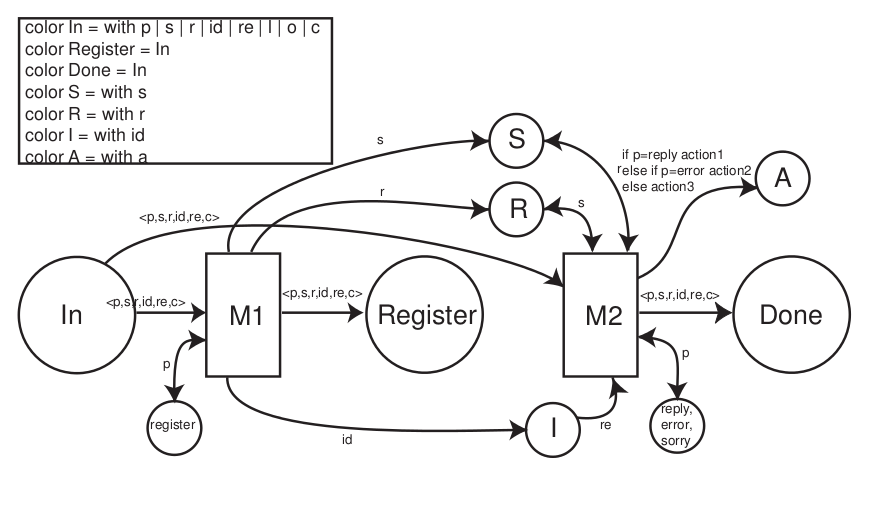}
\caption{CPN diagram of a KQML \textit{register} protocol (taken from~\cite{Cost2000}).}
\label{fig:communication:cpn-register}
\end{figure}

A Petri Net (PN) is a formal model for representing concurrency~\cite{Agerwala1979}. It is a type of directed, connected, bipartite graph. Each node in the graph is either a \textit{place} (places can be occupied by \textit{tokens}) or a \textit{transition} (which has \textit{input places} and \textit{output places}). A transition is considered to be \textit{enabled} when all places connected to it are occupied by tokens. When enabled, a transition can \textit{fire}, which removes one token from every input place and puts a token in each output place.

A CPN is an extension of this model in which tokens have data associated with them. In the case of using a CPN to represent an interaction protocol, this data includes the performative of a message, its sender and receiver, its message content and the contents of its \texttt{reply-with} and \texttt{in-reply-to} fields (these are described in more detail in Chapter~\ref{chap:fipa}). Thus the sending of messages results in the creation of tokens in the CPN. When all requisite messages have been sent, a transition can fire. In situations where multiple messages need to be sent to achieve a particular state, the ordering of messages is not important. This handling of concurrent communication is seen as the principal advantage of CPNs over FSM representations.

Figures~\ref{fig:communication:fsm-register} and~\ref{fig:communication:cpn-register} illustrate a simple KQML \texttt{register} protocol. An initiator begins by sending a  \texttt{register} message to another. The receiver has three available responses: the anticipated response (\texttt{reply}), notification that the initial message was malformed (\texttt{error}) or that some other problem has arisen (\texttt{sorry}). Figure~\ref{fig:communication:fsm-register} models this protocol as an FSM whereas Figure~\ref{fig:communication:cpn-register} shows its translation into a ``generally equivalent'' CPN~\cite{Cost2000}.

\subsection{Agent UML}

Agent UML was developed as an extension to the existing Unified Modelling Language (UML), a widely-used industry standard for the design of software systems. The intention was to bridge the gap between the academic work done in the field of MASs with existing industry standards, with a view to promoting the adoption of MAS technology in industry through the provision of familiar tools and paradigms~\cite{Bauer2001,Odell2000,Odell2001}.

A key contribution of this effort was the development of Agent UML specifications for interaction protocols. This describes the admissible sequence of messages that may be exchanged between agents fulfilling different roles and places constraints on the messages' contents. The protocols should also be consistent with the semantics of the communicative acts in the messages.

A diagram consists of an agent lifeline for each of the roles participating in the interaction, which relates to the time the agent spends playing that role rather than to the actual lifetime of the agent itself. Depending on the nature of the interaction, lifelines may split to show AND, OR or XOR parallelisms.

Messages are shown as arrows, and specify the required performative and any arguments that are necessary to provide additional information. Constraints and guards are possible also, to dictate the conditions under which messages may be sent. Parallelism is also supported for message sending. Protocols are designed to be modular, so as to support the nesting and interleaving of protocols.

The protocol specifications say nothing about the internal reasoning processing of agents in terms of how they decide to react to communications. Only the communications themselves are contained in the specifications.

The Agent UML method of representing interaction protocols was the one chosen by FIPA in defining its own library of standard agent protocols. These are discussed in more detail in Chapter~\ref{chap:fipa}. The Agent UML diagrams representing these can be seen in Appendix~\ref{app:fipa-protocols}.

An extension to Agent UML was proposed in~\cite{Ehrler2004} whereby the inputs and outputs of each operation were specified so as to allow agents to plug its application-specific code into the interaction protocol to create a fully executable implementation of the protocol.

\subsection{Dooley Graphs}

An alternative visualisation for agent interaction is a Dooley Graph~\cite{Parunak1996}. This illustrates the interaction between agents and analyses the relationship between individual communicative acts. For example, one message may be a response to another (i.e. the receipt of the first message caused the subsequent message to be sent) or a reply (where a response is communicated back to the sender of the initial message).

Dooley graphs are created from running systems or simulations and as such are most useful for analysing the behaviours that emerge from a community of agents. This can be a useful tool in the continuing development of a MAS, but is not intended to act as a formalisation of a protocol that is specified in advance of the runtime of a system.

\subsection{Global Session Types} \label{sec:communication:global-session-types}

Global session types are designed to specify multiparty interactions between distributed components. This is done to verify the correctness of these components in the way they interact~\cite{Carbone2007,Honda2008}. Although this representation was not developed with MASs in mind, it has been applied in this domain~\cite{Ancona2012}. Here, global session types were used to verify that ongoing conversations were consistent with an interaction protocol they represented. This was done in the context of a move towards implementing a form of unit testing for MASs.

Like an FSM, a global session type is described in terms of states and transitions, though there are differences in how these are treated. Transitions take the form of messages to be exchanged, each of which contains a performative and some message content. The model used in~\cite{Ancona2012} includes the ability for message content to be typed, and for these types to be verified at run-time. Content types can be simple atoms such as strings and integers, or more complex Prolog-style terms consisting of functors and arguments.

There are four principal mechanisms that facilitate the definition of transitions between states. A \texttt{seq} type allows for the definition of a valid sequence of interactions. Where \texttt{choice} is used, any number of possible interactions are specified, only one of which can be followed. A set of interactions that can be performed in any order can be specified using \texttt{fork} and finally, recursion can be used to allow for interaction loops. Care is required when using \texttt{choice} so as to avoid the synchronisation issues that are also relevant to FSMs.

This representation was used in the context of Jason agents written using a version of AgentSpeak(L) (see Section~\ref{sec:aop:jason}). Instead of sending a message in the usual way (which simply requires the specification of the intended recipient, performative and message content), a message is instead sent to a monitor agent. The monitor replies to state whether the communication is consistent with the protocol. Errors in the implementation of a protocol can be detected by the monitor.

As this work is based on a prototype, many additional interactions are required for verification, as each message is required to be sent to the monitor and the monitor must reply with its approval. It is envisaged that this may be reduced by closer integration into Jason by overriding its existing message sending action. It is also intended that this approach be extended to cover other logic-based AOP languages such as GOAL (see Section~\ref{sec:aop:goal}) and 2APL (see Section~\ref{sec:aop:2apl}).

\subsection{State Charts}

% in case: Moore uses FLBC ACL but as it also uses performatives (and isn't used by anyone else, it seems!), it's easier not to mention it here.
Another representation is a state chart visualisation for agent protocols~\cite{Moore2000}. This formulation is designed to be modular so that various protocol definitions can be combined together in order to build conversations.

\begin{figure}[!ht]
   \includegraphics[width=200pt]{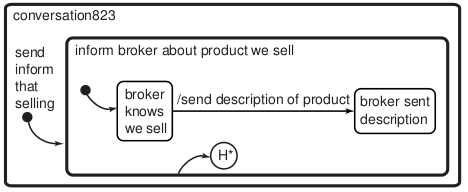}
   \includegraphics[width=200pt]{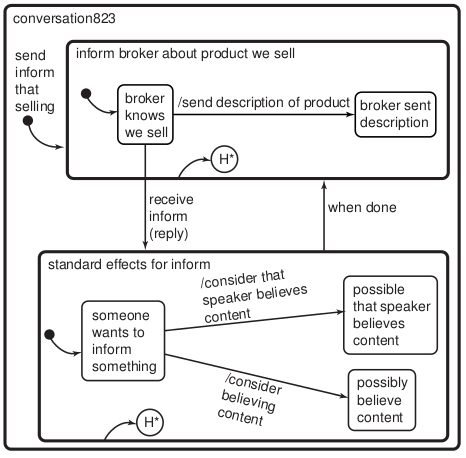}
   \caption{State Chart representation of an agent protocols (taken from~\cite{Moore2000}).}
   \label{fig:communication:moore-state-chart}
\end{figure}

An example of this can be seen in Figure~\ref{fig:communication:moore-state-chart}. On the left is a running conversation that is presented from the point of view of an initiator agent, intended to follow a protocol to inform a broker agent about a product that it sells. The inner box shows the protocol that is to be followed. This anticipates that the initiator first inform the broker that it is a seller, following this with a subsequent message to send a description of the product being sold. From the Figure, it can be seen that these are described in a high-level fashion.

The flexibility of the model allows for unexpected circumstances to be handled, however. The conversation on the right of Figure~\ref{fig:communication:moore-state-chart} illustrates a situation where the broker replies to the initial message with an \texttt{inform} message to acknowledge receipt. As this is not part of the protocol that the initiator was following, some adjustment is required. The system features a database of protocol definitions, amongst which are generic protocols for responding to messages with particular performatives. These are intended for use when a more specific protocol definition is unavailable. In this case, the initiator agent loads the standard protocol for handling \texttt{inform} performatives. Having done so, it must consider both whether the sender believes what it has sent and whether the initiator itself should believe it also. In this, the state chart becomes quite different from an FSM, as both of these transitions are used when an \texttt{inform} is received. Another noteworthy point is that not all transitions relate to message passing, including internal deliberation of the agent also. For this reason, these state chart definitions are required to be specified separately for each intended participant in a conversation (although a mechanism is available to combine two single agent protocol definitions into a multi agent definition).

Once this sub-procedure has completed, the initiator may return to the original protocol and continue by sending the description of the product that is being sold.

\subsection{Commitment Machines} \label{sec:communication:commitment-machines}

Yolum and Singh perceive weaknesses in the use of FSMs for conversation modelling~\cite{Yolum2001,Yolum2002}. They see FSMs and other prescriptive protocol definitions as being overly restrictive in that they lead to rigid executions (where out-of-order messages result in fatal errors) and obscure the contents of the states themselves. Instead of this, they proposed the use of a Commitment Machine (CM). A CM is similar to an FSM in the sense that it consists of states and transitions. Unlike an FSM, however, the key aspect is how a conversation's state is identified. Whereas an FSM concentrates on matching the communications between agents to transitions, a CM identifies the states by examining any social commitments that the participating agents have created in their communication (see Section~\ref{sec:communication:social-semantics}).

% Social Commitments relate to the undertakings agents make to each other to obtain a desired condition~\cite{Singh1999}. 

Within a CM, each state can be logically inferred by examining the social commitments in place at that point in time. This inference provides a fundamental difference from an FSM in that a CM has no particular start state, since the current state of the social commitments can be used to infer any conversation state. Actions that affect the commitments are, in effect, transitions between states. The state of the commitments present in the MAS after any action can then be used to infer the state of the conversation after any actions have occurred. Thus transitions are not triggered by examination of the messages passed between agents themselves, but rather their effect on the commitment structure of the system.

This approach is very flexible when considered in the context of protocol composition. Combining several protocols into one is easily accommodated, on condition that the final state in one protocol can be accommodated by another protocol with which it is combined. This facilitates a path that continues from the end state of the initial protocol being followed.

%They argue that FSMs hide the contents of states and lead to rigid execution. ``The content associated with each state specifies which commitments are in force in that particular state, and the content associated with each action defines how the commitments are affected by that action (thereby leading to a state change)''.

States can be derived from the commitments that are in existence at any point in time. No explicit transitions are defined: actions change the commitments so the new state can be logically inferred at run-time.

One advantage of a CM representation is that the protocol can be reconfigured by finding alternative paths between states. A conversation may be started in the middle of a protocol if an agent is satisfied that the commitments that should be present at the desired start state are indeed already in effect.

Reasoning about CMs at runtime can be a computationally expensive process, so the authors have also developed a method whereby CMs can be translated into FSMs at compile time. Although they argue that this restricts the flexibility of the protocol, it is a quicker representation to deal with, as the transitions are calculated beforehand.

%``A CM specification of a protocol emphasises that the aim of executing the protocol is not merely to perform certain sequences of actions, but the reach a state that represents the result of performing these sequences of actions''. 

It is argued that CMs change the process of protocol specification from defining how an interaction is executed to describing what interaction is to take place on a higher level~\cite{Winikoff2005b,Yolum2001}. 

%Agents' actions in CMs can change the value of state variables but may also initiate new commitments and/or discharge existing commitments. 

%Criticism of CM model from Winikoff et al. under the heading ``Communication mode assumptions not clear'':   ``... there is no explicit specification of how the conversation should be carried out between the two parties, i.e., whether it should follow a synchronous mode or an asynchronous mode.''

A number of extensions to the basic CM model have also been proposed. Winikoff et al. extend the CM model to allow specification of undesirable states~\cite{Winikoff2005b}. Fornara and Colombetti extend the model of commitments to include a temporal aspect, in order to allow deadlines to be modelled~\cite{Fornara2004}.

%\cite{Johnson2003} developed a method of verifying whether two protocols based on commitments are equivalent to one another. (criticism from \cite{Singh2010}: simplistic as it classifies )

%This model of communication was later leveraged in~\cite{Chopra2008}, which sought to determine the interoperability of two agents. That is, whether two agents can enter into, maintain and discharge commitments to each other.

A drawback of CMs is that where semantic mismatches occur, these are very difficult to debug~\cite{Nodine1998}. Verifying the interoperability of agents that make use of social semantics models is an ongoing research problem~\cite{Chopra2008,Singh2010}.

\subsection{Interaction based on Expectations} \label{sec:communication:expectations}

An alternative model of using social semantics to model agent interaction is to use expectations rather than commitments~\cite{Torroni2009}. Semantics based on expectations specify the link between observed events and possibly observable, yet unknown, events. In this context, expectations are independent of the agents involved in communication. Events are generally expected either to happen or not to happen.

Social Integrity Constraints (SICs) are used to state that if some events occur, it is expected that some other set of events have occurred previously. An example is that if a quote for a price is sent by an agent to another, it is expected that it has been preceded by a request for that quote. Protocols may be specified using SICs to outline the expectations of how the system should function. It is argued that this, like commitment-based interaction, allows more flexibility to agents than the more rigid representations such as FSMs, CPNs and Agent UML.

\subsection{Mental Models} \label{sec:communication:mental-models}

The use of communication modelled in terms of social semantics has been subject to the criticism that its semantics are overly objective, in that they say nothing about how communication affects the communicating agents themselves. An alternative semantics has instead been proposed that is based on the idea that communication affects the receiver's mental model of the sender~\cite{Hindriks2010}. This is fundamentally different to the semantics of FIPA ACL and KQML, which place unverifiable restrictions on the mental state of the sender prior to sending a message.

The GOAL AOP language contains a concrete implementation of this to add support for the grouping of messages into conversations. The motivation behind this is the use of conversations as a mechanism by which actions and communication may be synchronised in a multi-agent system~\cite{Hindriks2010}.

% analysis is mine
Conversations are tracked on the multi-agent level, so it is not the responsibility of each agent to reason about its own conversations. It is this centralised control that allows for the synchronisation that is desired: by allowing an agent to participate in only one conversation at a time, it acts in a similar way to a semaphore.

Agents may only participate in a limited number of active conversations at any time. Attempts to initiate a new conversation are placed in a queue of pending conversations if any of the participants are already engaged in an active conversation. Pending conversations may be activated when an active conversation ends (by being closed by the initiator of the conversation). Messages are linked to a particular conversation by means of a conversation ID. There is no restriction to the number of agents that may participate in a conversation.

%NOTE: analysis of drawbacks is mine. It does not have any protocol awareness
As with the other models of conversations based on semantics, there is no concept of a pre-existing protocol that should be followed by agents. When sending communications, agents tag the messages they send with conversation identifiers so that they can be matched to related messages.

%so the reasoning that is possible about a conversation is limited to identifying a) what messages are part of the conversation, b) who the participants of the conversation are and c) whether the conversation is pending, active or finished. It also makes no attempt to analyse either the type or content of messages. Once an agent tags a message with a conversation ID it is part of the conversation.

%\subsection{Decomposition into Goals and Plans}
%
%By identifying the aims of a protocol, Hutchison and Winikoff ~\cite{Hutchison2002}

% other minor representations
%\item \cite{Wagner2000} used a multi-level state machine (Abstract Task Model). Same for \cite{Elio1999}
% \item \cite{Martin2000}: Push Down Transducers

\section{Implementations} \label{sec:communication:implementation}

This Section examines the support for communication amongst those platforms and languages discussed in detail in Section~\ref{sec:aop:aop}. Table~\ref{tab:communicaion:support} summarises the support each has for ACLs, how it implements the semantics of its chosen ACL (if applicable) and what support for conversation management is present.

\begin{table}[!htb]
\centering
\caption{ACL Support in AOP}
\label{tab:communicaion:support}
\small
\begin{tabular}{|l|l|l|l|}   
\hline
Language & ACL & Semantics & Conversations \\
   /Framework & & & \\
\hline
2APL & FIPA-ACL & None & None \\
Agent Factory & FIPA-ACL & None & None \\
GOAL & Custom & Mental model of sender & Semantics-based \\
Jadex/JADE & FIPA-ACL & With JSA Add-on & Basic \\
Jason & KQML Subset & Basic & None \\
\hline
\end{tabular}
\end{table}

\subsection{FIPA ACL}

2APL and Agent Factory both use FIPA-ACL. However, they do not implement the semantics of the language. In both cases, the only effect of receiving a message is that the agent becomes aware of the message receipt. 2APL supports message events that can be used to trigger procedure call rules. Agent Factory can either use a belief that a message has been received or raise an event similar to that of 2APL. The treatment will vary depending on the AOP language being used. Neither offer default support for conversation handling.

As Jadex is built on the FIPA-compliant JADE middleware platform, it makes use of FIPA ACL for its communication also (Jadex and JADE are discussed in Section~\ref{sec:aop:jadex}). Incoming messages are placed in an agent's message queue, after which they are matched to a capability that will handle the message. This capability will receive an event if the message is part of an ongoing conversation.

Some rudimentary conversation handling is supported, via manipulation of the FIPA \texttt{conversation-id} and \texttt{reply-with}/\texttt{in-reply-to} parameters (these are discussed in Section~\ref{sec:fipa:fipa-acl}). Using a \texttt{createReply} action will result in the creation of a message whose \texttt{conversation-id} or \texttt{in-reply-to} are set automatically by consulting the appropriate values in the initial message. An action is also provided to create a unique identifier for the initiation of a conversation.

In addition to this basic conversation handling, Jadex includes default implementations of a subset of the FIPA interaction protocols (Namely \texttt{fipa-request}, \texttt{fipa-contract-net}, \texttt{fipa-iterated-contract-net}, \texttt{fipa-english-auction} and \texttt{fipa-dutch-auction}: FIPA interaction protocols are discussed in Section~\ref{sec:fipa:fipa-interaction-protocols}). These implementations include goals and beliefs relating to the relevant protocol for both the initiator and the participant. As such, the specification of the protocol is separated according to the roles the agents play.

By default, JADE does not apply any particular semantics to messages that are received by agents. However, an extension known as the Jade Semantics Add-on (JSA) adds support for FIPA's semantics to the platform~\cite{Bellifemine2007,Pautret2006}.

\subsection{KQML}

Jason (which was discussed in Section~\ref{sec:aop:jason}) uses a subset of KQML with simple semantics~\cite{Bordini2005, Vieira2007}. For instance, when receiving a \texttt{tell} message, the fact is automatically added to the belief base. Similarly, \texttt{achieve} results in the adoption of a goal. In each case, the new belief or goal that arises as a result of the message is annotated with its source, so that models of trust can be integrated into the implementation. Some undesirable consequences of automatic goal and belief adoption can be combated by the use of the social acceptance function. Messages with performatives that are not socially acceptable are discarded. For example a request for action should only be received from an agent with the social power to do so. It is left to the programmer to ensure that whatever appropriate actions are necessary are taken upon the receipt of messages.

\subsection{Custom}

GOAL (discussed in Section~\ref{sec:aop:goal}) uses three forms of message, based on natural language~\cite{Hindriks2010}. The types used are declaratives, interrogatives and imperatives. In each case, the semantics of the language involve the recipient updating its mental model of the sender and no more. Further modifications to its own mental state are left to the agent programmer to decide. For instance, given a declarative message (where agent $a$ informs agent $b$ of some fact), the recipient will believe that the sender believes the truth of what has been communicated. The aspect left to the programmer is the decision on whether or not the recipient chooses to believe the fact itself. This allows issues of trust to be built into the system. The same principle applies to interrogatives and imperatives, where the recipient only adopts beliefs about the beliefs and/or goals of the sender. GOAL's support for conversations has previously been discussed in Section~\ref{sec:communication:mental-models}.

%Communication is conducted by sending \texttt{Go!} data values via asynchronous point-to-point messages between agents. This type of communication is not based on speech act theory and interoperability with other agent programming languages is not a goal of the project.~\cite{Clark2004}

%\section{Support for Communication in MAS Frameworks}
%

%
%Agent Factory ignores semantics (FIPA ACL). Belief that a message has been received is all that happens.
%
%Some agent programming languages have taken a more pragmatic approach to agent communication in recent times. Examples include 2APL~\cite{Dastani2008} (which simply uses a mailbox with no semantics regarding agents' mental states) and the work on providing semantics for communication in AgentSpeak presented in~\cite{Vieira2007}. Here, a small subset of KQML-based primitives were chosen for communication, with accompanying simple semantics.

\section{Summary}

From the above outline of agent communication, it can be seen that there is clearly some disagreement about how communication should be modelled and handled. Two principal strains of research can be identified: the first based on prescriptive protocol definitions (e.g. FSMs, CPNs and Agent UML), with a competing approach concentrating on a more flexible run-time model based on semantics.

Much of the debate centres on the semantics used by each approach. The decision to base FIPA ACL and KQML semantics on internal mental aspects of the participating agents has been criticised, as it makes it impossible to verify whether these are being followed (this was discussed in Section~\ref{sec:communication:acl-criticism}). In addition, the motivation for the utilisation of social semantics stems from the argument that prescriptive models are overly restrictive. By specifying sequences of permissible messages, they do not permit agents to use their autonomy to deviate from protocols to take advantage of additional information or cater for exceptions. An additional criticism is that they outline how an interaction should take place, but not the meaning behind that interaction.

In contrast, approaches such as FSMs or CPNs can easily and quickly be used to verify whether agents are exchanging the correct messages, according to a protocol definition that has been engineered by a developer \textit{a priori}. Simplicity and ease of implementation are gained at the cost of the expressiveness of the protocol. Additionally, the use of semantics-based conversations in an open system is not without its own complications. Agents should have a common understanding of the semantics being used, and problems can arise where this is not the case.

The following Chapter concentrates on the FIPA standards for agent communication. This includes a more in-depth discussion of the FIPA ACL and also the library of standard interaction protocols FIPA has specified. 
\chapter{FIPA Communication Standards}
\label{chap:fipa}

%AUML originally Odell2000, latest draft is FIPA2003a

\section{Introduction}
\label{sec:fipa:introduction}

Chapter~\ref{chap:communication} outlined the history of Agent Communication Languages (ACLs) and presented some of the other research that has been conducted in the area of agent communication to date. Of the major standardisation efforts, FIPA-ACL has gained the most traction amongst modern multi agent frameworks. As such, it has been chosen as the base upon which the work in this thesis is built. This Chapter therefore gives a more detailed discussion of these communication standards.

% FIPA-compliant framework references
% Those marked with a * are from the FIPA-JACK thesis
% - eXAT http://citeseerx.ist.psu.edu/viewdoc/summary?doi=10.1.1.61.610
% - JIAC (according to wikipedia)
% - April Agent Platform (AAP) (according to wikipedia)
% Agent Factory~\cite{Collier2003}
% JACK (via plugin)~\cite{Winikoff2005}
% Comtec Agent Platform (http://fipa-comtec.ec.co.jp/ap)*
% JADE (http://jade.cselt.it) (and, by extension, Jadex?)*
% ZEUS (http://193.113.209.147/projects/agents.htm)*
% Tryllian (http://www.tryllian.com/)
% 2APL (built on JADE base) Dastani2008

The Foundation for Intelligent Physical Agents (FIPA) is a standards organisation that exists within the IEEE Computer Society. It is aimed at promoting the adoption of multi agent technologies. In particular, it is concerned with the development and promotion of standards for agent interoperability.

%TODO systems using FIPA ACL should go here
The most widely-adopted of the FIPA standards are those relating to agent communication. FIPA-ACL outlines the structure and semantics of inter-agent communications. This is discussed in Section~\ref{sec:fipa:fipa-acl}. Building on this, FIPA has also published a set of interaction protocols that dictate in what sequence FIPA ACL messages should be exchanged in order to achieve certain tasks. These are discussed in Section~\ref{sec:fipa:fipa-interaction-protocols}.

\section{FIPA ACL Message Structure} \label{sec:fipa:fipa-acl}

The structure of FIPA ACL messages is outlined in~\cite{FIPA00061}. This defines a FIPA ACL message as consisting of a set of message parameters, some of which are mandatory and others of which are optional. In addition to the parameters set out by FIPA, the standard allows for user-defined, non-standard parameters that must be prefixed with ``\texttt{x-}''.

In all, FIPA proposes 13 standard message parameters, grouped into 5 categories, relating to types of communicative acts, communication participants, message content, description of message content and the control of conversations.

The \texttt{performative} parameter indicates the type of communicative act represented by the message (e.g. providing information, requesting action, making a proposal, etc.). A full list of FIPA standard performatives is outlined in~\cite{FIPA00037} and are discussed in detail in Section~\ref{sec:fipa:performatives}. The performative is the only parameter that the standards require for all ACL messages. The use of performatives in FIPA ACL indicates that it, like the Knowledge Query Manipulation Language (KQML), is based on the principles of speech act theory.

The \texttt{sender} and \texttt{receiver} parameters refer to the agents that are participating in the communication. Each message may have only one sender but may have multiple receivers. Every agent is required to have a unique identifier by which it can be referred in ACL messages. In the case of multicast messages, the identifiers of the receiving agents must be explicitly included in the \texttt{receiver} parameter. Both of these parameters will be present in the majority of messages, but some situations exist where either may be absent (e.g. where the sender wishes to remain anonymous or where a recipient can be inferred from context).

The \texttt{reply-to} parameter is also concerned with the participants in a communication. Ordinarily, replies to messages are addressed to the agent named in the \texttt{sender} parameter. However, when the \texttt{reply-to} parameter is set, responses should instead be addressed to the agent so named.

The \texttt{content} parameter contains the actual content of the message. Interpretation of this content is to be performed by the receiver alone, meaning that the interpretation by the sender and receiver may be different. The \texttt{content} field is present in most FIPA ACL messages, though it is not compulsory. One example of a message where it is not necessary is in the case of a message with a \texttt{cancel} performative, which is intended to end a conversation. Here, if the message uses a \texttt{conversation-id} parameter (see below) to indicate the conversation that is to be cancelled, the content is implicit.

The description of the message content is performed by a combination of the \texttt{language}, \texttt{encoding} and \texttt{ontology} parameters. The \texttt{language} parameter refers to the formal language in which the content is expressed. Optionally, the \texttt{encoding} parameter indicates the method used to encode the content, and the \texttt{ontology} parameter may specify an ontology to use in interpreting the content.

A number of parameters are intended to aid in the control of conversations. The \texttt{protocol} parameter denotes the protocol that the sender is using to guide its interaction. A non-null \texttt{protocol} parameter indicates that the message is part of a conversation, for which a \texttt{conversation-id} should also be provided. Each message in a conversation must have the same \texttt{conversation-id}, which uniquely identifies a single conversation. Several conversations may occur that follow the same protocol.

The \texttt{reply-with} and \texttt{in-reply-to} message parameters are also related. In a situation where multiple interactions are occurring simultaneously, these are used to allow an agent to refer to a prior conversation. If one agent sends a message that includes ``\texttt{reply-with <expr>}'' then the receiver would reply with a message including ``\texttt{in-reply-to <expr>}''.

Finally, the \texttt{reply-by} parameter allows an agent to specify a time by which a response should be received. The FIPA standards allow the agent implementer to decide what constitutes a reply, for example the next message within a protocol or by matching the conversation specified in the initial communication. It is also open to the agent to decide what the consequences of a late reply may be, as this is also not specified in the standards.

\section{FIPA ACL Standard Performatives} \label{sec:fipa:performatives}

The \texttt{performative} parameter is the only compulsory parameter in a FIPA ACL message. Performatives play an important role in specifying the nature of the message that has been sent, and its intended effect. A list of the performatives that are available for use with FIPA-compliant
messages can be found in~\cite{FIPA00037}. Each of these has associated semantics to determine the feasibility condition and rational effect of each action.

\begin{itemize}
\item \emph{Accept Proposal} (\texttt{accept-proposal}): Used to communicate the
acceptance of a proposal that was previously submitted (usually by means of a
\texttt{propose} message). The understanding is that the recipient of the message will perform some action in the future.

\item \emph{Agree} (\texttt{agree}): Used to indicate that an agent is agreeing
to perform some action (frequently as a result of receiving a \texttt{request} to perform it).

\item \emph{Cancel} (\texttt{cancel}): This is used in a situation where the
sender previously desired that the recipient perform a particular action. The \texttt{cancel} performative is intended to indicate that this no longer holds, and that the recipient's action is no longer desired. This is distinct from a specific request to stop an action, in which case a \texttt{request} message is more appropriate. The \texttt{cancel} performative is most often seen within interaction protocols, where it is used to terminate a conversation.

\item \emph{Call for Proposal} (\texttt{cfp}): This is generally used by a sender that desires that another agent will perform some action. In this situation, a \texttt{cfp} is distributed to a number of agents to initiate a negotiation process whereby the recipients are invited to submit \texttt{propose} messages to indicate their willingness to undertake the requested action. Although this may be used to initiate an auction process, this is not necessarily the case. A \texttt{cfp} message may also be used to check the availability of a single particular agent to perform the specified action. The receipt of a proposal in return indicates that the agent is capable of performing the specified action, and is willing to do so.

\item \emph{Confirm} (\texttt{confirm}): This performative is used to indicate that the sender believes a particular proposition to be true. The sending of such a message indicates that the sender intends the recipient to come to believe that the proposition is true.

\item \emph{Disconfirm} (\texttt{disconfirm}): This performative is used to indicate that the sender believes a particular proposition to be false. The sending of such a message indicates that the sender intends the recipient to come to believe that the proposition is false.

\item \emph{Failure} (\texttt{failure}): This is used to indicate to the recipient that an action was attempted but has failed to complete successfully. The message should also include (where possible) an explanation of why the action failed to complete.

\item \emph{Inform} (\texttt{inform}): The \texttt{inform} performative is used for the sender to inform the recipient that a particular proposition is true. In sending an \texttt{inform} message, the sender also indicates its intention that the recipient should also believe the truth of the given proposition. This is closely related to the \texttt{confirm} and \texttt{disconfirm} performatives. The choice of which to use depends on the sender's knowledge of the recipients beliefs. If the sender believes that the recipient has no knowledge of the proposition, an \texttt{inform} performative should be used. If, on the other hand, the sender is uncertain about the proposition, a \texttt{confirm} (or \texttt{disconfirm} if the proposition is false) is the appropriate mechanism to clarify the recipient's beliefs.

There is some ambiguity regarding this performative, however. Amongst FIPA's interaction protocols (see Section~\ref{sec:fipa:fipa-interaction-protocols}), several refer to \texttt{inform-done} and \texttt{inform-result} in the same way as other performatives, though these are not included in~\cite{FIPA00037}. In some situations (e.g. in the \texttt{fipa-request} protocol), the protocol specification uses \texttt{inform-done} to indicate that a request has been completed and \texttt{inform-result} to indicate the result of the request being completed. Additionally, an \texttt{inform-t/f} performative appears in the \texttt{fipa-query} protocol, to communicate a boolean value.

Despite this distinction between a simple \texttt{inform} message and the variations thereon, no guidelines are supplied either within the interaction protocols themselves or within the FIPA Communicative Act Library Specification~\cite{FIPA00037} to indicate whether these should be treated as separate performatives or whether the nature of the message should be contained in the content of an \texttt{inform} message.

Some clarity may be sought by consulting systems that aim to follow the FIPA interaction standards. The MadKit~\cite{Gutknecht2000} system includes these additional performatives in the same way as those outlined in~\cite{FIPA00037}. In contrast, JADE~\cite{Bellifemine1999} and Agent Factory~\cite{Collier2003} do not include these additional performatives, preferring to implement only those that appear in the standards.

\item \emph{Inform If} (\texttt{inform-if}): This is a macro act that is not directly sent as the performative in a message. The \texttt{inform-if} performative is equivalent to informing an agent whether a proposition is true or false. An example of this usage is where an \texttt{inform-if} message is included as the content of a \texttt{request} message that asks the receiver to reply with an \texttt{inform} message that indicates whether or not Paris is the capital of France.

\item \emph{Inform Ref} (\texttt{inform-ref}): In a similar way to \texttt{inform-if}, \texttt{inform-ref} is also a macro act that is not directly sent in an ACL message. Whereas \texttt{inform-if} is designed for communicating the truth of a proposition, \texttt{inform-ref} is used to provide details about some entity. For example, one agent sends a \texttt{request} message that asks the receiver (via an \texttt{inform-ref}) to reply to say what the capital of France is. The receiver should reply (using an \texttt{inform}) to indicate that Paris is the capital of France.

\item \emph{Not Understood} (\texttt{not-understood}): This is used to indicate that the sender did not understand some action that was previously performed by the recipient. This action is typically a communicative act, meaning that the sender of the \texttt{not-understood} message previously received a message from the recipient that it was unable to understand.

\item \emph{Propagate} (\texttt{propagate}): This allows an agent to send a message to another agent, and also request that it be forwarded to some set of agents. The contents of a \texttt{propagate} message should contain two distinct parts. Firstly, a description is provided that will be used to select the agents to which to forward the message. Secondly, the message itself should also be embedded.

\item \emph{Propose} (\texttt{propose}): This is typically sent as a response to a \texttt{cfp} to indicate that the sender of the \texttt{propose} message is prepared to perform a particular action, subject to certain preconditions. A typical precondition would be to specify the price of a bid in an auction or negotiation protocol.

\item \emph{Proxy} (\texttt{proxy}): This is similar to the \texttt{propagate} performative, with the exception that the recipient is only asked to forward the message to other agents. As such, the recipient should not interpret the embedded message content as being addressed to itself and is merely for the attention of the agents to which the message is forwarded.

\item \emph{Query If} (\texttt{query-if}): This is used by the sender of the message to ascertain whether or not the recipient believes a particular proposition to be true. It is anticipated that the recipient will \texttt{inform} it of the truth of the proposition.

\item \emph{Query Ref} (\texttt{query-ref}): This is used for the sender of the message to request a particular object that is described in the content of the message. The recipient is expected to respond with an \texttt{inform} message containing details of the object or set of objects that correspond to the specified descriptor.

\item \emph{Refuse} (\texttt{refuse}): This is used to indicate that the sender is refusing to perform a particular action on the grounds that the action in question is unfeasible.

\item \emph{Reject Proposal} (\texttt{reject-proposal}): Following the receipt of a \texttt{propose} message, an agent may respond with a \texttt{reject-proposal} message to indicate that it does not desire the sender of the proposal to carry out the relevant action contained in the proposal.

\item \emph{Request} (\texttt{request}): This is used by a sender to request that the recipient of the message performs some action (which is described in the content of the message). The FIPA standards explicitly state that the action to be performed may be another communicative act, such as sending an \texttt{inform} message.

\item \emph{Request When} (\texttt{request-when}): This is similar to the \texttt{request} performative, with the exception that the requested action is not required to be performed immediately. Instead, the content of the message should also indicate some precondition for the action. This message indicates that the requested action should be performed when the specified precondition becomes true.

\item \emph{Request Whenever} (\texttt{request-whenever}): Unlike \texttt{request-when}, the request to perform an action does not lapse as soon as the action has been performed. Instead, the recipient of the message is requested to perform this action every time the specified precondition becomes true. This persistent request may be terminated by the sending of a \texttt{cancel} message by the agent that originally sent the \texttt{request-whenever} message.

\item \emph{Subscribe} (\texttt{subscribe}): This is similar to \texttt{query-ref}, in that it is used to ask an agent for some value. The principal difference between \texttt{query-ref} and \texttt{subscribe} is that the latter also indicates that an agent wishes to be informed of any change in the specified value in the future, rather than being a one-off request for information. The subscription initiated by this message may be terminated by a \texttt{cancel} message being sent to indicate that the sender is no longer interested in receiving updates on the value in question.
\end{itemize}

\section{FIPA Interaction Protocol Specifications} \label{sec:fipa:fipa-interaction-protocols}

FIPA recognised that defining a communication for individual messages is not sufficient for the demands of the MAS community. As such, a number of standards relating to common interaction protocols were also developed. These are specified by means of Agent UML interaction diagrams (based on extensions to UML 1.x~\cite{Odell2001}) and informal semantic descriptions of the interaction protocols. The UML diagrams for all of the FIPA interaction protocols are contained in Appendix~\ref{app:fipa-protocols}.

The majority of the protocols involve communication between two agents. The agent that sends the first message in a conversation is referred to as the ``initiator'', with the other agent involved being referred to as the ``participant''.

The FIPA standards documents tend not to differentiate between protocols and conversations, referring to each as ``Interaction Protocols''. For clarity and consistency, some quotes from the standards have been edited to refer to a ``protocol'' (a specification of a permissible sequence of messages) or a ``conversation'' (a running instance of a protocol). In each case, the replacement is indicated by the use of brackets.

\subsection{Common Elements}

For each of the FIPA interaction protocols, there are two situations where the conversation flow may be interrupted by messages that are not specified in the protocol itself. The first situation is where the recipient of a message does not understand the message that it has received. This may occur for any message that forms part of a conversation. In this case, the recipient may respond with a \texttt{not-understood} message to indicate its lack of understanding. According to the FIPA specifications, this ``may terminate the entire [conversation] and termination of the interaction may imply that any commitments made during the interaction are null and void''~\cite{fipa-request,fipa-query}. However, no definitive guidance is provided on how a \texttt{not-understood} message is processed.

The second exception allows the initiator of a conversation to cancel the conversation at any point. Unlike the \texttt{not-understood} case, the cancellation of a conversation must be acknowledged by the other participant in the conversation, using a meta-protocol provided in each of the interaction protocol specifications discussed below. This meta-protocol is outlined in Section~\ref{sec:fipa:fipa-cancel}.

Each interaction protocol in the following Sections is accompanied by the reference number by which it is referred within FIPA. These have an ``SC'' prefix where they have become confirmed standards, with experimental standards featuring an ``XC'' prefix.

\subsection{Bipartite Protocols}

The following Sections outline those protocols that are intended for use by exactly two communicating agents.

\subsubsection{FIPA Request Interaction Protocol Specification (SC00026)}
The \texttt{fipa-request} protocol provides a mechanism by which the initiator agent can request another to perform a particular action~\cite{fipa-request}. It is illustrated in Figure~\ref{fig:fipa-request} in Appendix~\ref{app:fipa-protocols}.

There is no requirement on the participant agent to agree to perform this request, so it may be refused by means of a \texttt{refuse} message. Alternatively, if the participant accedes to the request, it may respond with an \texttt{agree} message to indicate this. The protocol explanation requires that this occur ``if conditions indicate that an explicit agreement is required'', although it is not specified exactly how this is to be communicated or agreed. When this does occur, in addition to communicating its agreement, the participant should also communicate the outcome of performing the requested task. This can either be to notify the initiator of the failure of the task (using a \texttt{failure} message) or its success via an \texttt{inform} message.

The nature of this \texttt{inform} message can either be simply to communicate that the requested action was performed successfully, or else to communicate a particular result of the action. These are referred to as \texttt{inform-done} and \texttt{inform-result} respectively. The decision as to which of these messages to send is solely at the volition of the participant.

The specification also notes that if the requested action is quick, the participant may skip the agreement phase in favour of simply performing the action and communicating its result. This may be indicated by using the \texttt{reply-by} parameter in the initial \texttt{request} message. If the requested action is capable of being performed before the timeout, the agreement stage may be skipped in favour of only communicating the result of the action. For longer-running actions, the initial agreement is necessary in order to ensure that the conversation does not time out and in order to indicate to the initiator that it intends to carry out the requested action.

\subsubsection{FIPA Query Interaction Protocol Specification (SC00027)}

This protocol serves two related functions. One option for the initiator is to query the participant for the truth value of a particular proposition (via a \texttt{query-if} message). Alternatively, the initiator may query for a particular object by means of a \texttt{query-ref} message~\cite{fipa-query}. The agent UML diagram for this protocol can be seen in Figure~\ref{fig:fipa-query} of Appendix~\ref{app:fipa-protocols}. The structure of the protocol is very similar to the \texttt{fipa-request} protocol: the participant can communicate its agreement or refusal to participate in the conversation and the ultimate result is the sending of either a \texttt{failure} or \texttt{inform} message.

In addition to the initial message, it is the nature of these final messages that differs from \texttt{fipa-request}. Here, there are also two types of \texttt{inform} messages (referred to as \texttt{inform-t/f} and \texttt{inform-result}). Unlike the \texttt{fipa-request} protocol, however, guidance is given as to which type of \texttt{inform} message should be used to end the conversation. In the event that the conversation began with a \texttt{query-if} message, the final \texttt{inform} should merely indicate the truth value of the proposition that was contained in the original message (this is the \texttt{inform-t/f} message). If the conversation began with a \texttt{query-ref} message, the final message should contain the answer to the query. This is referred to in the same way as in \texttt{fipa-request}, namely as an \texttt{inform-result} message.

Given that the eventual terminating message depends on the performative used in the initial message, it could be argued that this protocol is logically two different protocols rolled into one. Given that the final communication is dependent on the nature of the initial communication, this protocol could be split into separate \texttt{query-if} and \texttt{query-ref} protocols.

The \texttt{fipa-query} protocol shares the same ambiguity as \texttt{fipa-request} with regard to the optional communication of agreement prior to the initiator being informed of the final result. A \texttt{fipa-query} is also terminable by sending a \texttt{not-understood} message or using the \texttt{fipa-cancel} meta-protocol.

\subsubsection{FIPA Request When Interaction Protocol Specification (SC00028)}

The Request When Protocol allows the initiator to request that a task be carried out by another agent whenever some specified precondition becomes true~\cite{fipa-request-when}. The recipient of the initial message is entitled to refuse to perform this action but in the event that it agrees, then it is expected to communicate the result of its attempt to carry out the requested task. The protocol is illustrated by Figure~\ref{fig:fipa-request-when} in Appendix~\ref{app:fipa-protocols}.

Other than the performative of the initial message being \texttt{request-when} rather than \texttt{request}, the structure of the \texttt{fipa-request-when} protocol is identical to that of \texttt{fipa-request}. The differences lie in the interpretation of the protocol. Unlike in a \texttt{fipa-request} conversation, the requested action is not to be performed immediately and so it is necessary for the participant to communicate its agreement to perform the task. Once the precondition expressed in the original message is satisfied, the action is performed and the outcome communicated to the initiator agent. Again, this can be done via two different types of \texttt{inform} message (\texttt{inform-done} and \texttt{inform-result}).

The \texttt{fipa-request-when} protocol also allows for termination of the conversation by means of a \texttt{not-understood} message or using the \texttt{fipa-cancel} meta-protocol.

\subsubsection{FIPA Subscribe Interaction Protocol Specification (SC00035)}
The \texttt{fipa-subscribe} interaction protocol allows the initiator to request that an action be performed immediately, but also that this action be performed again whenever a referenced object changes~\cite{fipa-subscribe}. It is illustrated by Figure~\ref{fig:fipa-subscribe} of Appendix~\ref{app:fipa-protocols}.
 
This protocol is similar in structure to the \texttt{fipa-request} and related protocols. Beginning with a \texttt{subscribe} message (rather than a \texttt{request}), the participant must either \texttt{agree} or \texttt{refuse} to participate further. The original message must describe some object in which the initiator is interested.

Once it has agreed to participate, the participant must send \texttt{inform} messages to the initiator whenever the object referred to in the initial message changes (this is described as an \texttt{inform-result} in the protocol specification). As with similar protocols, a \texttt{failure} message terminates the conversation if a failure is experienced by the participant.

As with the other interaction protocols, the initiator may end the conversation by invoking the \texttt{fipa-cancel} meta-protocol or either participant may send a \texttt{not-understood} message that may also result in termination. Other than the participant refusing to participate or experiencing a failure, this is the only recognised way to formally end the conversation and its associated subscription.

\subsubsection{FIPA Propose Interaction Protocol Specification (SC00036)}

The \texttt{fipa-propose} interaction protocol is intended for use by an agent to send an unsolicited proposal to indicate to another agent that it proposes to perform a specified task. A relatively simple protocol, the conversation ends on receipt of a response (either \texttt{accept-proposal} or \texttt{reject-proposal})~\cite{fipa-propose}. Unlike some other protocols (such as \texttt{fipa-contract-net} and \texttt{fipa-request}), no further communication is specified in the protocol to allow the initiator to communicate the result of the action to the other participant. The protocol is illustrated in Figure~\ref{fig:fipa-propose} in Appendix~\ref{app:fipa-protocols}.

As with the other interaction protocols, either the \texttt{not-understood} or the \texttt{fipa-cancel} mechanisms may be used to terminate the conversation at any time.

\subsection{Group Protocols}

Two of the FIPA standard interaction protocols can be considered to be group protocols, as they allow an initiator to converse with an arbitrary number of participants (described as 1:N protocols). In each case, the standard leaves the choice of a \texttt{conversation-id} to the discretion of the initiator agent. The initiator has the option either to use the same \texttt{conversation-id} for every participant with whom it has begun a conversation, or else to use a separate \texttt{conversation-id} with each participating agent. Using separate conversation identifiers effectively treats the entire interaction as being composed of a number of separate one-to-one conversations.

\subsubsection{FIPA Contract Net Interaction Protocol Specification (SC00029)}

The \texttt{fipa-contract-net} protocol (illustrated in Figure~\ref{fig:fipa-contract-net} in Appendix~\ref{app:fipa-protocols}) is intended as a mechanism by which the initiator can find another agent to perform a particular task~\cite{fipa-contract-net}. It does this by soliciting proposals from other agents using a \texttt{cfp} message. This specifies the task, and also outlines any conditions that are placed on the execution of the task. The participants outline the conditions or constraints under which they are willing to perform the requested task via a \texttt{propose} message. Alternatively, an agent may \texttt{refuse} to submit a proposal.

Having received the proposals from all willing participants, the initiating agent can choose between them and indicate that it has accepted one or more of the received proposals (using an \texttt{accept-proposal} message), rejecting the others (with a \texttt{reject-proposal} message).

The initiator may also set the \texttt{reply-by} parameter in the initial message to set a deadline by which participants must respond. Failure to respond results in rejection, on the grounds that the proposal was too late.
 
 Participants whose proposals are accepted then perform the specified task. As with the \texttt{fipa-request} and \texttt{fipa-request-when} protocols, the result of this execution should be communicated back to the initiator using either a \texttt{failure} or an \texttt{inform} message. Again, the \texttt{inform} can take the form of either \texttt{inform-done} or \texttt{inform-result}.

\subsubsection{FIPA Iterated Contract Net Interaction Protocol Specification (SC00030)}

The \texttt{fipa-iterated-contract-net} interaction protocol is closely related to the \texttt{fipa-contract-net} protocol, with the principal difference being that the initiator is not limited to sending only a single initial \texttt{cfp}~\cite{fipa-iterated-contract-net}. This protocol is illustrated in Figure~\ref{fig:fipa-iterated-contract-net} of Appendix~\ref{app:fipa-protocols}.

With the iterated version of the protocol, once a number of proposals have been received the initiator has the option of accepting or rejecting those proposals as in \texttt{fipa-contract-net}. With this protocol, it also has the option to re-issue a modified \texttt{cfp} to some or all of the participants who submitted a proposal. Thus the iterative nature of the protocol is created, with re-invited participants once again deciding whether or not to make a proposal in response to the modified \texttt{cfp}. Participants who had previously refused to submit a proposal, or whose proposals were rejected, are not sent a modified \texttt{cfp} as they have left the process (either their refusal to make a proposal or the rejection of a proposal will end the conversation).

It is at the discretion of the initiator whether or not to instigate a new iteration each time it has received replies from all invited participants. In the final iteration, all remaining proposals must be accepted or rejected.

Any agents whose proposals have been accepted are required to communicate the result of performing the required task back to the initiator. Notably, in contrast with the \texttt{fipa-request}, \texttt{fipa-request-when} and \texttt{fipa-contract-net} protocols, no distinction is made in this protocol specification between the type of \texttt{inform} message that may be sent to communicate the successful completion of the task.

\subsection{Macro Protocols}

The FIPA standards include definitions of two \emph{macro protocols}. These are categorised as such because they make use of embedded messages which may be part of a conversation that follows a different underlying sub-protocol.

\subsubsection{FIPA Brokering Interaction Protocol Specification (SC00033)}
The \texttt{fipa-brokering} protocol is a macro protocol for use in systems where a broker agent is required~\cite{fipa-brokering}. A broker agent is an intermediary through which communications must be routed in order for two other agents to interact. As part of its function, the broker may be required to identify agents that are suitable for the interaction (possibly based on knowledge of the capabilities of the agents in the system).

The intention of the protocol is to provide a wrapper around some other protocol that the interacting agents are following. Thus the messages sent to the broker (using a \texttt{proxy} performative) themselves wrap another message that will typically identify a different underlying sub-protocol. Because the structure of the interaction depends on this sub-protocol, the brokering protocol is intentionally written generically.

The protocol allows for the broker agent to agree to forward messages, to perform the actual forwarding of those messages in both directions and to make the interacting agents aware of any failures in the brokering process.

This Agent UML diagram of this protocol can be seen in Figure~\ref{fig:fipa-brokering} of Appendix~\ref{app:fipa-protocols}.

\subsubsection{FIPA Recruiting Interaction Protocol Specification (SC00034)} \label{sec:fipa-recruiting}

The \texttt{fipa-recruiting} protocol (shown in Figure~\ref{fig:fipa-recruiting} of Appendix~\ref{app:fipa-protocols}) is very similar to the \texttt{fipa-brokering} protocol with the principal difference that instead of the broker agent acting as a proxy for messages in both directions, it is involved only in the identification of suitable agents with which to interact~\cite{fipa-recruiting}. Once the initial message is sent to a suitable recipient, the interacting agents can then contact each other directly for the remainder of the conversation.

\subsection{Experimental Standards}

Two of the specifications did not reach the stage of being declared as ``Standards'' and remain as ``Experimental'' specifications. As such, the style of these specifications differs from the standard protocols in two significant ways:
\begin{itemize}
\item The sending of a \texttt{not-understood} message is explicitly included in the protocol itself, rather than being permitted at any time.
\item The \texttt{fipa-cancel} meta-protocol is not included in the specification.
\end{itemize}

These are also features of the experimental drafts of those protocols that did become standards. As such, it is reasonable to assume that a modern implementation of these protocols would make use of these standard methods of terminating the conversation.

\subsubsection{FIPA English Auction Interaction Protocol Specification \\ (XC00031)}

An English Auction consists of an auctioneer attempting to find the market price of some item. It does this by initially proposing a price lower than its notion of what the ultimate price will be and then continually incrementing the price until agents are no longer willing to pay the proposed price. When this happens, then the highest price for which proposals were received is considered to be the market price~\cite{fipa-english-auction}. The auctioneer may decide to accept this market price, depending on what its own private reservation price is. FIPA's Agent UML diagram for the English Auction is shown in Figure~\ref{fig:fipa-english-auction} of Appendix~\ref{app:fipa-protocols}.

\subsubsection{FIPA Dutch Auction Interaction Protocol Specification \\ (XC00032)}

A Dutch Auction is one where the auctioneer attempts to find the market price by starting the bidding at a much higher price, with the asking price being progressively reduced until a bidder is found. As with the English Auction, the FIPA Dutch Auction Interaction Protocol also remains in the ``experimental'' state~\cite{fipa-dutch-auction}. The Agent UML diagram contained in the experimental document can be seen in Figure~\ref{fig:fipa-dutch-auction} of Appendix~\ref{app:fipa-protocols}. The protocol is designed so as to allow for partial acceptance also (e.g. when an auctioneer is selling a quantity of some good, it may sell part of the goods to different buyers).

\subsection{Others}

In addition to those interaction protocols that were published as standards, a meta-protocol was included in each of the standard protocols in order to allow the interaction to be cancelled before it had otherwise completed.

\subsubsection{FIPA Cancel Meta Protocol (\texttt{fipa-cancel})} \label{sec:fipa:fipa-cancel}

The FIPA Cancel Meta-Protocol (\texttt{fipa-cancel}\footnote{The ``\texttt{fipa-cancel}'' label is not mentioned anywhere in the relevant FIPA standards. It has been added here so that it can be referenced in a way that is consistent with the top-level protocols for which FIPA has provided labels.}) is not specified in a standard of its own, but is included in all of the other standards (it is not included in the experimental standards) to provide a mechanism by which the initiator of the conversation may cancel the interaction. Because this may happen at any time, it is not embedded directly into the protocols to which it is relevant \cite{fipa-brokering,fipa-contract-net,fipa-iterated-contract-net,fipa-propose,fipa-query,fipa-recruiting,fipa-request,fipa-request-when,fipa-subscribe}. The initiator of the conversation that is to be cancelled indicates its desire to cancel by sending a \texttt{cancel} message to another participant. The recipient of the \texttt{cancel} message must indicate whether or not it found it possible to cancel the interaction (it may be the case that a task that it was previously asked to perform is already in progress, making it too late to cancel the procedure). If the conversation can be cancelled, this is indicated by means of an \texttt{inform-done} message. Failure to cancel the conversation is communicated with a \texttt{failure} message. The protocol is illustrated by Figure~\ref{fig:fipa-cancel} in Appendix~\ref{app:fipa-protocols}.

As with all of the FIPA standard protocols, the contents of the messages exchanged as part of a conversation following the cancel meta-protocol are not specified in standards themselves. This has the potential to cause ambiguity.

Within the FIPA standards, the \texttt{conversation-id} parameter of the messages in a \texttt{fipa-cancel} conversation must be the same as that of the conversation it attempts to cancel. Thus the \texttt{inform-done} or \texttt{failure} message that concludes the cancellation has the potential to be confused with similar messages that may be due as part of the original conversation, with the agent having ignored the request for cancellation.

The FIPA standards state that ``[e]laboration on this pattern will almost certainly be necessary in order to specify all cases that might occur in an action agent interaction. Real world issues such as the effects of cancelling actions, asynchrony, abnormal or unexpected [conversation] termination, nested [conversations], and the like, are explicitly not addressed here.'' As such, a great deal of interpretation is required of agent programmers in implementing the cancel meta-protocol. This is reflected in the decision of the JADE developers, who have decided not to support \texttt{fipa-cancel} on the grounds that it is insufficiently specified in the standards~\cite{Bellifemine2010}.

\subsection{Problems with FIPA Interaction Protocols} \label{sec:fipa:problems}

A number of criticisms can be identified with the FIPA standard interaction protocols that make them difficult to produce concrete implementations for. These are as follows:

\begin{itemize}
   \item \textit{Generic:} The FIPA interaction protocols only specify what the performatives of the messages exchanged should be. They do not state what the message content should be, as they are designed to be sufficiently generic to be used in a variety of situations.
   \item \textit{Inconsistent:} Many of the interaction protocols specified by FIPA refer to \texttt{inform-done} and \texttt{inform-result} performatives. However, these are not part of the ACL message standard.
   \item \textit{Incomplete:} The \texttt{fipa-english-auction} and \texttt{fipa-dutch-auction} IPs were never formalised into standards after being published as draft proposals. As such, the last published versions are quite different in style to the other IPs.
   \item \textit{Ambiguous:} The \texttt{fipa-brokering} IP states that ``parts of this protocol are written very generically''. The \texttt{fipa-request} IP states that ``the \texttt{agree} may be optional depending on circumstances''. Because of ambiguities like this, it can be difficult to correctly implement some of the FIPA standard IPs in a manner consistent with other implementations.
\end{itemize}

\section{Summary}
This Chapter, along with Chapters~\ref{chap:aop} and~\ref{chap:communication} outline the state-of-the-art in terms of the development of MASs, with particular emphasis on issues of communication. The following Chapters describe the Agent Communication Reasoning Engine (ACRE), which forms the core of this work. This has FIPA ACL at its core and so it is particularly important to outline the features of this ACL in detail prior to the discussion of ACRE.

Like KQML, FIPA ACL is based on speech act theory, with each message specifying a performative that represents its effect. Although semantics have been defined for each of the standard performative types, concrete implementations of FIPA ACL systems rarely enforce these. As such, these semantics are typically understood to be advisory in nature rather than strict rules that are required to be followed.

In addition to the basic communication language, FIPA has also defined a number of standard interaction protocols that are intended to facilitate agents to engage in more complex interaction using a series of messages. Underlying this is the facility to embed identifiers for protocols and conversations in each FIPA ACL message. Although these serve as a useful indicator of the types of common interactions that the community requires, the way in which these are presented is not without problems.

\part{ACRE: Agent Conversation Reasoning Engine}
\chapter{Introduction to ACRE}
\label{chap:acre-introduction}

\section{Introduction}

The Agent Conversation Reasoning Engine (ACRE) is aimed at providing conversation-level communication capabilities to agents. As noted in Chapter~\ref{chap:communication}, the majority of multi agent platforms, languages and toolkits support inter-agent communication on a message-level basis only. However, in reality messages are rarely sent in isolation and will frequently be related to other messages that have previously been sent or received, This occurs, for example, as part of an auction or other negotiation.

ACRE models conversations between agents as deterministic Finite State Machines (FSMs), where messages exchanged between agents trigger transitions between states. These FSMs are defined externally to the agents, so as to be independent of the choice of agent platform and Agent Oriented Programming (AOP) language employed.

This Chapter introduces the definitions, concepts, advantages and limitations of ACRE and also provides explanatory examples of how ACRE is used in managing conversations between agents. Subsequent chapters provide more detailed discussion including a formal model of ACRE's conversation handling (Chapter~\ref{chap:semantics}), an abstract architecture describing how ACRE may be integrated into an existing MAS framework (Chapter~\ref{chap:acre-architecture}) and an example of a concrete architecture in which ACRE was integrated into the Agent Factory framework (Chapter~\ref{chap:agent-factory}).

\section{Aims and Features}

\begin{description}
   \item[Automatic conversation management:] The principal aim of ACRE is to facilitate agents to engage in complex interaction in the form of conversations. To achieve this aim, it must be possible for messages to be grouped together in an automated fashion while providing the agents with sufficient capabilities to reason about conversations.

   To this end, a Conversation Manager component is provided, which is capable of matching messages to conversations. The mechanism by which this is done is discussed informally in Section~\ref{sec:acre-introduction:informal-model} and formally in Chapter~\ref{chap:semantics}.

   \item[Compatibility:]  ACRE is intended for use with multiple AOP languages and frameworks. Its design is not tied to any one AOP approach. This has the consequence that ACRE makes very few assumptions about the capabilities or mental models of the agents that employ it. Instead, it offers convenient conversation modelling that aids agents in acquiring knowledge about their communication and engaging in new interactions. How these are used is a decision for the developers of these agents according to the tools available to them.

   \item[Simplicity and Usability:] ACRE is intended to be a practical tool that will be used by developers of Multi Agent Systems (MASs). For this reason, Finite State Machines (FSMs) were chosen as the model on which to base protocol definitions. Although this has some limitations when compared with some alternative approaches (see Section~\ref{sec:acre-introduction:limitations} for a deeper discussion of ACRE's limitations), this model was chosen for its simplicity and intuitive understandability.

   To aid with the usability of the system, a number of development tools are provided with ACRE also, including a graphical protocol designer and a real-time conversation viewer that allows developers to view the progress of conversations while in the debugging phase of development.

   \item[Formal model:] In addition to the informal discussion of ACRE's conversation handling presented in this Chapter, the full operational semantics of the conversation manager are presented in Chapter~\ref{chap:semantics}. As the current implementation is written in the Java programming language, it is not immediately integrable with multi agent frameworks written in other languages. The availability of operational semantics facilitates the development of equivalent implementations in other languages that will reflect the reference implementation.

   \item[Generic and reference architecture:] To aid with the integration of ACRE into a multi agent framework, a generic architecture is provided in Chapter~\ref{chap:acre-architecture}. This divides the components supplied with ACRE into those that relate to specific agents and those that are shared amongst agents as platform services. This is intended to act as a guide to any developer aiming to integrate ACRE with a multi agent framework. Following this, the concrete integration with the Agent Factory framework is discussed in Chapter~\ref{chap:agent-factory}.

   \item[Complex conversations:] By default, ACRE's representation of conversation supports bipartite conversations. However, more complex conversations involving more than two parties are desirable in many situations.

   To cater for this situation, ACRE also provides a Group Reasoner component, which is intended to allow an agent to define groups of related conversations (e.g. multiple bipartite conversations related to the same auction) and declare events in which the agent is interested. This gives the agent the capability and flexibility to handle more complex communication patterns. This is discussed in more detail in Section~\ref{sec:acre-introduction:limitations}.

   \item[Conversations independent of agents:] In ACRE, the definition of protocols and conversations is entirely independent of the reasoning process the agent goes through in deciding what interactions to engage in. The FSM that represents a conversation neither dictates nor depends on the participating agents' mental states. This is in contrast to some other approaches where the actions of the agent are intertwined with the protocol definitions (e.g. COOL and JADE).

   Having protocols defined in this independent manner has the side-effect that an external monitoring agent that has access to messages exchanged between participating agents can be used to verify that interaction protocols are being followed correctly, which can be used as part of the debugging process (verifying compliance by observation~\cite{Torroni2009}).

   \item[Shared protocol repositories:] Agents can make use of shared repositories of protocols that can be accessed remotely through HTTP or alternatively can be stored on a local filesystem. Additionally, protocols are given version numbers in order to ensure that agents can verify that they are making use of the same protocol. %Bradshaw1996 had this in KAoS

   \item[Re-use of protocols:] Following the lead of~\cite{Kuwabara1995}, the definition of ACRE protocols is designed with inheritance in mind. More complex protocols can import from simpler protocols and extend them.

   \item[Standards compliant:] ACRE follows the FIPA communication standards (discussed in detail in Chapter~\ref{chap:fipa}) both in its support for FIPA ACL performatives and also its use of FIPA ACL's support for protocol and conversation identifiers.

   \item[Support for interaction with non-ACRE systems] The key focus of ACRE is for use by all communicating agents in a system. However, interoperability with systems that are not ACRE-enabled has also been included. An ACRE-enabled agent may communicate with any agent that is capable of sending and understanding FIPA ACL messages. The process of matching messages to conversations is not fully dependent on the \texttt{protocol} and \texttt{conversation-id} fields being populated. It is described in more detail in Section~\ref{sec:acre-introduction:informal-model} and Chapter~\ref{chap:semantics}.

   \item[Agent API:] In a MAS, the principal method of interacting with agents is by ACL message passing. As noted by Cost et al.~\cite{Cost1999}, sets of supported conversations can be used as a form of agent `API' so it becomes clear how one should go about interacting with a particular agent. This is aided by preserving the independence of agent deliberation and protocol definitions, along with the maintenance of shared protocol repositories.
\end{description}

\section{Definitions}

Before a more in-depth discussion of how ACRE represents and reasons about complex communication, it is necessary to provide definitions of those elements that will be under discussion in the following sections.

\begin{description}
\item[Protocol] A \textit{protocol} is a definition of a sequence of communications in which two agents may engage. Each protocol is a description of a deterministic FSM that may be used by agents in their communications.
\item[Conversation] A \textit{conversation} is a specific instance of two agents engaging in a sequence of communications that follow a specific protocol. There is no limit to the number of conversations that may take place that have the same underlying protocol. Conversations are executing FSMs.
\item[State] At any time, a conversation is described as being in some \textit{state}. The state of a conversation will dictate what form the next message should take, in accordance with the underlying protocol definition. In this definition, the state is identified only by name. However, as an FSM, the state of the conversation is a function of both this state name and also the values bound to any variables defined in the protocol.
\item[Transition] A \textit{transition} links two states, in order to allow a conversation to move between states. Each transition is a description of a message that must be exchanged between the participants of a conversation. Whenever a message is communicated that matches this transition, the transition is ``triggered'' and the state of the conversation will change. Variable bindings (whereby variables are assigned associated values) may be created or altered as variables are matched against message fields.
\item[Current State] Any active conversation is, at all times, in some state, known as the \textit{current state}.
\item[Start State] A \textit{start state} is defined as one that has no incoming transitions. Each protocol must define exactly one start state. A new conversation that follows that protocol is initially in the start state, before any messages have been sent or received.
\item[End State] Any state from which no transitions are defined is known as an \textit{end state}. Upon reaching an end state, a conversation is considered to have terminated. There is no restriction on the number of end states a conversation may have.
\item[Active Transition] An \textit{active transition} is a transition that begins at the current state. At any time, only active transitions are candidates to be triggered by a message being communicated.
\end{description}

\section{Interaction Protocols as Finite State Machines}

ACRE protocols are represented as Finite State Machines (FSMs), which are discussed in Section~\ref{sec:communication:fsm}. FSMs are a popular method of representing interaction protocols and offer an intuitive representation of a conversation that is easy to represent and model.

The following Sections describe how the states and transitions of a FSM related to message exchange in ACRE.

\subsection{States}

Within an ACRE protocol, each state is defined using only its name. However, during the execution of the FSM to model a single conversation, the state of the FSM is determined both by this name and by any variable bindings that have arisen as a result of previous state changes. At the beginning of a conversation, no bindings are present as no communication has yet occurred. Variable bindings will be acquired as messages are exchanged.

\subsection{Transitions}

Within a conversation FSM, a transition between states is triggered by an ACL message that satisfies certain criteria. Each transition has exactly one \textit{from state} and exactly one \textit{to state}. The from state is the state from which the transition may be triggered. If this is the current state of the conversation, then the transition is considered to be \textit{active}. The end state is the state in which the conversation will be upon activation of the transition.

Transitions specify values for a number of message fields, and the transition is triggered if all of the specified fields match against the corresponding parameters in the message. Full details of how this matching is done is provided in Chapter~\ref{chap:semantics}. This is described more informally in the following sections.

The message parameters that ACRE allows a transition to specify are as follows:
\begin{description}
   \item[Performative:] The \texttt{performative} message parameter of a FIPA ACL message. In an ACRE transition, this must be provided as a constant value that must match the message's performative exactly.
   \item[Sender:] The \texttt{sender} message parameter, which contains the unique identifier of the agent that sent the message. In an ACRE transition, this may either be the actual identifier of the agent that is expected to send the message or, more commonly, a variable that is intended to acquire the name of the sender, so that it can match the same agent identifier for future messages in the conversation. A full description of how variables are handled by ACRE is given in Section~\ref{sec:acre-introduction:mutability}.
   \item[Receiver:] The \texttt{receiver} message parameter. In contrast to the \texttt{sender} parameter, the FIPA standards allow multiple agents to receive the same message~\cite{FIPA00061}. Thus the \texttt{receiver} parameter may contain multiple agent identifiers. As ACRE only facilitates conversations between two agents, a message that is sent to $n$ different receivers is treated as $n$ distinct messages with a single receiver specified. As with the \texttt{sender} field, the \texttt{receiver} field may contain the actual identifier of an agent or a variable.
   \item[Content:] This is the actual content of the message, contained in the FIPA \texttt{content} parameter.
\end{description}

\section{Content Language} \label{sec:acre-introduction:content-language}

The content language used in ACRE transitions is modelled on the principles of first-order logic, featuring predicates, functions and variables. As such, it is intended that its operation will be intuitive to users of AOP languages that use similar language (e.g. AgentSpeak(L), AFAPL). It is also intended that converting between ACRE's content language and any other logic-based language used by agents will not be difficult. The full formal semantics of the content language are defined in Chapter~\ref{chap:semantics}: this section serves as an informal introduction.

In the definition of a protocol, the content language is used to define the properties of messages that will successfully advance a conversation that follows that protocol. The mechanism by which this is done is discussed later in Section~\ref{sec:acre-introduction:informal-model}. Each field in the message is compared against its definition in the protocol to check if it matches and if so the conversation will advance.

A variable may appear as an argument within a predicate or function. When variables are used in the definition of conversations, they can match any predicate or function in a message. Additionally, the definition of a conversation allows for the maintenance of \textit{variable bindings}. These record the values that have been previously matched against variables during the process of advancing the conversation. In some cases variable bindings are fixed for the duration of the conversation whereas in others they may be redefined at specified points in the process. This is dictated by whether the protocol designer has used these variables in a mutable or immutable context, as discussed in the following Section.

\subsection{Mutable and Immutable Context for Variables} \label{sec:acre-introduction:mutability}

The context in which a variable is used will dictate whether its value can be changed or not. From the programmer's point of view, when a variable is used in a mutable context, it may have its value overwritten in the conversation's set of bindings. If used in an immutable context, this cannot occur. An immutable-context variable is indicated by a `\texttt{?}' sigil, whereas a variable prefixed by `\texttt{??}' indicates that it is used in a mutable context. Thus, `\texttt{?name}' and `\texttt{??name}' refer to the same variable. The difference in how it is treated when matching against values in messages and generating bindings.

Thus when a variable is used in an immutable context, its behaviour depends on whether a binding previously exists for that variable. If the variable has no previous bound value, then it is free to match against any value and a binding is then created between the variable and that matched value. However, if a binding does previously exist, the variable can only be matched against a value that in turn matches this bound value. No new binding can be created by this process.

On the other hand, when a variable is used in a mutable context, it is possible for a new binding to be associated with it, which replaces the previous bound value. This means that the previous bound value is not relevant when matching, since the variable is capable of matching any value (and acquiring that value as its new binding).

From this discussion, we can define ``matching'' as a variable being paired against some value with which it is (in the current context) capable of being bound.

In the formal model, bindings are \textit{applied} to terms before they are matched with other terms. For a variable used in an immutable context, this results in the variable being replaced with any value with which it is associated in the bindings. If it has no previous bindings, it remains as a variable.

When the variable is used in a mutable context, however, this approach would prevent it from acquiring a replacement binding. Thus when applying bindings to a term, it is necessary to avoid replacing a variable that is used in a mutable context with another value. This means that during the matching stage, it can be associated with another term, which results in a replacement binding.

From this analysis, we can see that the only difference in treatment of a variable is in the application of bindings:
\begin{itemize}
   \item Variables used in a mutable context are not replaced with their values before matching against a message. This leaves them free to match against any value and acquire these as new bindings, which replace any bound value that existed previously.
   \item Variables used in an immutable context are replaced with their values, if a binding exists, before matching against a message. Thus the attempted match is between this bound value and the value in the message, meaning that no new binding can be generated for the variable. Where no previous binding exists, no replacement is made and the variable is free to acquire a new binding in the same way as if it was used in a mutable context.
\end{itemize}

\subsection{Anonymous Variable} \label{sec:acre-introduction:anonymous-variable}

The anonymous variable is a special variable (\texttt{?}) that has no name. Thus it cannot be bound to any values. It is used in situations where the programmer is allowing ``anything'' to appear in a particular location, without the desire to capture that value for further use. The anonymous variable can match against any value, but no bindings are created when it does so.

\section{Conversation Handling} \label{sec:acre-introduction:informal-model}

This Section provides an informal outline of how ACRE models and reasons about protocols and conversations. It is followed by some sample conversations for illustration. A full formal model of this process is presented in Chapter~\ref{chap:semantics}.

In the ACRE model, tuples are used to represent the elements that are required in such a system. Messages, protocols and conversations are the principal components for which a model is required. Additionally, as protocols and conversations are FSMs consisting of states and transitions, tuples are also used to represent these states and transitions.

The tuple $(s,r,c,\phi,p,x)$ is used to represent a \textit{message}. Here, $s$ is the unique agent identifier of the message's sender, $r$ is the unique agent identifier of the recipient, $c$ is the conversation identifier, $\phi$ identifies the protocol, $p$ is the message performative and $x$ is the message content.

Each \textit{protocol} is represented by a tuple $(\phi, S, T, \iota, F)$ where $\phi$ is the protocol's unique identifier, $S$ and $T$ are sets of states and transitions respectively, $\iota$ is the name of the initial state  and $F$ is a set of names of final (terminal) states. Both $\iota$ and the names in $F$ should match the names of states contained in $S$. Although the model does not specifically restrict $\iota$ from appearing in $F$, this is not particularly useful as it would represent a state that has no transitions attached.

Within these protocols, each \textit{state} is represented by the tuple $(n,s,\phi)$ where $n$ is the name of the state, $s$ is the status of the state (whether it is a start, end or intermediate state) and $\phi$ is the identifier of the protocol it belongs to. Each \textit{transition} is represented by $(\sigma,\epsilon,s,r,p,x)$. Here, $\sigma$ and $\epsilon$ are the names of the start and end states respectively, $s$ and $r$ represent the agent identifiers of the sender and receiver respectively, $p$ is the performative of the message triggering the transition and $x$ is the message content. The states $\sigma$ and $\epsilon$ may be linked with multiple distinct transitions, though care must be taken in defining these transitions so that they cannot be triggered by the same message. Failure to do so would result in the creation of a non-deterministic FSM.

Finally, a \textit{conversation} may be represented by $(\phi,A,s,c,B,\psi)$ where $\phi$ is the protocol identifier, $A$ is the set of participating agents, $s$ is the name of the conversation's current state, $c$ is the conversation identifier, $B$ is the current set of variable/value bindings and $\psi$ is the conversation status (active, completed or failed).

The values permitted in the tuples shown here are based on first-order logic, meaning that all values are constants, variables, functions or predicates.

When comparing values, the following rules apply: 
\begin{itemize}
\item Constant values match against other identical constant values.
\item Variables match against any value.
\item Functions match other functions that have the same functor, have the same number of arguments and whose arguments in turn match.
\item Predicates match other predicates that have the same predicate symbol, have the same number of arguments and whose arguments in turn match.
\end{itemize}

\begin{figure}[!htb]
	\includegraphics[width=\textwidth]{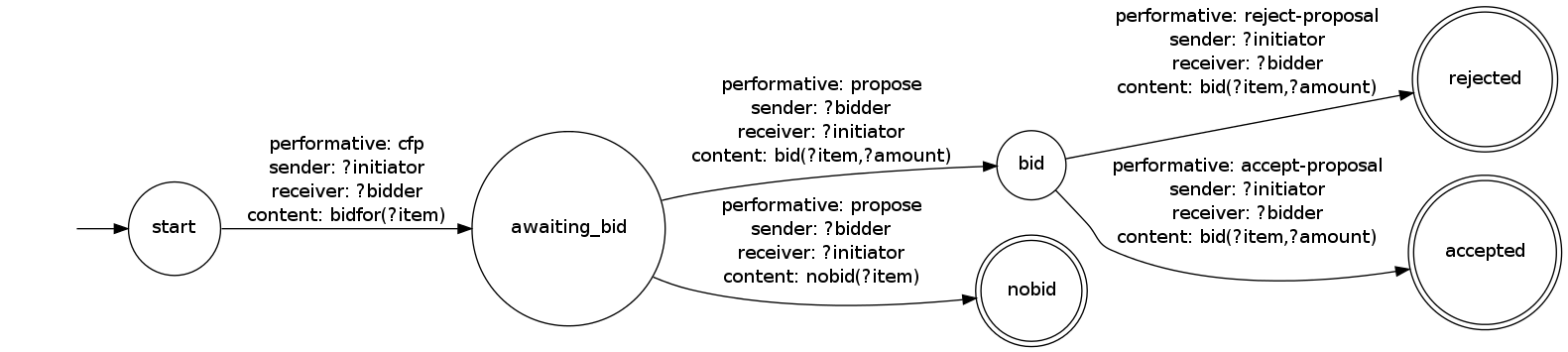}
	\caption{FSM representation of the Vickrey Auction protocol.}
	\label{fig:acre-introduction:fsm-vickrey}
\end{figure}

An advantage of using FSMs as a representation for protocols is that an intuitive graphical visualisation can easily be generated. An example of this can be seen in Figure~\ref{fig:acre-introduction:fsm-vickrey}. This figure shows an FSM for a simple, one-shot Vickrey-style auction. In this type of auction, an auctioneer sends a call for proposals (using a \texttt{cfp} performative), to which a participant may respond either by making a bid or by declining to do so. If a bid has been made, the auctioneer may either accept or reject it. 

Figure~\ref{fig:acre-introduction:fsm-vickrey} shows the states and transitions associated with this protocol. Double-lined borders indicate end states (\texttt{nobid}, \texttt{rejected} and \texttt{accepted} in this example). The state named \texttt{start} is shown with an incoming arrow that does not originate from another state. This indicates that \texttt{start} is the initial state. Transitions are triggered by comparison with messages exchanged between the participating agents. Each of the fields shown in the diagram (performative, sender, receiver, content) corresponds to an element of the tuple used to represent it.

On receipt of a message, these transitions will be compared to the values contained in the message to find whether a match can be identified. For example, the first transition in Figure~\ref{fig:acre-introduction:fsm-vickrey} can only be triggered by a message with the performative \texttt{cfp}, since this field contains a constant value. In contrast, the sender or receiver of the first message can be any agent, since no particular value has been bound to the \texttt{?initiator} and \texttt{?bidder} variables at this stage of the conversation.

The bindings associated with the conversation ($B$) is a set of key/value pairs that binds variables to constants or functions against which they have been matched in triggering a transition. Any variables that have been matched against a constant or function in a triggering message are given a binding that is stored in $B$. In the example from Figure~\ref{fig:acre-introduction:fsm-vickrey}, the sender of the initial message will have their agent identifier bound to the \texttt{?initiator} variable, so any further messages must be sent by/to that same agent, whenever the \texttt{?initiator} variable is used. This is an example of a variable being used in \textit{immutable context}. Once the variable has been bound to a value, that value may not change for the duration of the conversation if the variable is only used in immutable context. Using the same variable in \textit{mutable context} (i.e. as \texttt{??initiator}) would allow a new value to be bound during the execution of the conversation.

The following Sections outline the three key stages of the conversation management algorithm. By convention, elements of tuples are denoted by using subscripts (e.g. the initial state ($\iota$) of a protocol ($p$) is shown as $p_\iota$).

\subsection{Identifying Candidate Conversations}

The first stage of the conversation management algorithm is carried out whenever a message is exchanged and is shown in Figure~\ref{fig:acre-introduction:candidate}. Whenever a message is sent, it must be checked against existing active conversations to identify any it is capable of advancing. A set of \textit{candidate conversations} is generated, which consists of all active conversations that can be advanced by the message. An active conversation is considered a candidate whenever it has an active transition that can be triggered by the message.

\begin{figure}[!h]
\begin{algorithmic}
\STATE $C \gets \emptyset$ to store candidate conversations
\STATE $m \gets$ message sent/received
\FOR{\textbf{each} active conversation ($c$)}
   \IF{($m_c = c_c$ \AND $m_\phi = c_\phi$) \OR $m_c = \bot$}
     \FOR{\textbf{each} transition ($t$) \textbf{where} $t_\sigma = c_s$}
        \IF{$matches(m_s, apply(c_B,t_s))$ \AND $matches(m_r,apply(c_B,t_r))$ \\ \hspace{0.5cm}\AND $matches(m_x,apply(c_B,t_x))$ \AND $matches(m_p,t_p)$}
           \STATE Add $c$ to $C$
        \ENDIF
      \ENDFOR
   \ENDIF
   \IF{$m_c = c_c$ \AND $c \notin C$ }
      \STATE $c_\psi \gets failed$
   \ENDIF
\ENDFOR
\end{algorithmic}
\caption{Identifying candidate conversations.}
\label{fig:acre-introduction:candidate}
\end{figure}

The \texttt{apply(B,a)} function is used to apply a set of bindings (\texttt{B}) to a term (\texttt{a}). If \texttt{a} is a variable used in an immutable context for which a binding exists in \texttt{B}, then the bound value is returned. Otherwise, \texttt{a} is returned unaltered. This has the effect of differentiating between the two contexts that can be associated with a variable. Because a variable used in immutable context is replaced with its bound value, it will only match if its bound value matches the value in the message. However, when a variable is used in mutable context (or when it is in immutable context but has no previously bound value), it is not replaced as it is free to match against any value. This matched value will later become bound to the variable if it successfully advances the conversation later in the process.

Amongst active conversations, conversation identifiers must be unique. Thus in situations where the message to be processed contains a conversation identifier, a maximum of one candidate conversation will be identified. However, ACRE is intended to be used by agents communicating with others that lack conversation management capabilities. For this reason, it is necessary to make provision for messages received without an explicit conversation identifier. In this case, the message is compared against all active conversations to ascertain if any can be advanced by it. Here, the set of candidate conversations may have multiple elements, being all those candidate conversations that have an active transition capable of being triggered by the message. The handling of situations with multiple candidate conversations is outlined in Section~\ref{sec:acre-introduction:advancing}.

If the message contains a defined conversation identifier, but that conversation cannot be advanced by the message, the status of the conversation must be changed to \textit{failed}.

Although this approach allows for interaction with agents that are not conversation-aware while still allowing for conversation management, it is important to note that certain undesirable side-effects can result. For example, an agent may make use of protocols that have been defined with similar transitions, or multiple conversations following the same protocol may arrive in the same state. In these cases, multiple candidate conversations will be identified and none can be definitively advanced. Dealing with this type of situation may require further negotiation or clarification between the parties to the conversation, or re-visiting previously matched messages in case they have been matched to the wrong candidate conversation. Handling these issues is outside the scope of the work presented in this thesis.

The possibility of such multiple candidate conversations being identified when communicating with a system that is not conversation-aware something that should be borne in mind by protocol designers where the protocols are intended for use for communicating with such systems.

\subsection{Identifying Candidates for New Conversations}

If a message cannot be associated with an active conversation, the second stage is to find whether it is possible for the message to initiate a conversation that follows a known protocol. This procedure is shown in Figure~\ref{fig:acre-introduction:candidate_protocol}.

\begin{figure}[!ht]
\begin{algorithmic}
\IF{$|C|$ = 0}
  \FOR{\textbf{each} protocol ($p$)}
     \IF{$m_\phi=p_\phi$ \OR $m_\phi = \bot$}
        \FOR{\textbf{each} transition ($t$) where $t_\sigma = p_\iota$}
           \IF{$matches(m_s, t_s)$ \AND $matches(m_r,t_r)$ \AND $matches(m_x,t_x)$}
              \IF{$m_c = \bot$}
                 \STATE Add $(p_\phi, \{m_s,m_r\}, p_\iota, nextid(), \emptyset, active)$ to $C$
              \ELSE
                 \STATE Add $(p_\phi, \{m_s,m_r\}, p_\iota, m_c, \emptyset, active)$ to $C$
              \ENDIF
           \ENDIF
        \ENDFOR
     \ENDIF
  \ENDFOR
\ENDIF
\end{algorithmic}
\caption{Identifying candidate protocols for new conversations.}
\label{fig:acre-introduction:candidate_protocol}
\end{figure}

In cases where the message contains a protocol identifier, then only the protocol with that identifier is considered. Otherwise, the message is compared against the initial transition of every available known protocol. If a suitable protocol is found, a new conversation is created and added to the set of candidate conversations (\texttt{C}). If the message contained a conversation identifier, this is used as the identifier for the new conversation. Otherwise, a new unique conversation identifier is generated (by means of the \texttt{nextid()} function).

\subsection{Advancing the Conversation} \label{sec:acre-introduction:advancing}

Having identified new or existing conversations that match against the given message,  a conversation must be advanced as appropriate. This process is shown in Figure~\ref{fig:acre-introduction:advance}. At this stage, events are raised that can be seen by the agent to inform it of the outcome of the conversation reasoning process. If the message was not capable of advancing an existing conversation or initiating a new one, an ``unmatched'' event is raised. If there were multiple candidate conversations (which cannot be the case if conversation identifiers are defined for all messages), an ``ambiguous'' event is raised.

\begin{figure}[!ht]
\begin{algorithmic}
\IF{$|C| = 1$}
   \STATE $c \gets$ the matched conversation in $C$
   \STATE $t \gets$ the transition matched by the message $m$
   \STATE $c_s \gets t_\epsilon$
   \STATE $c_B \gets mergeBindings(c_B, getBindings(m,apply(c_B,t)))$
      \IF{$c_s$ is an end state}
         \STATE $c_\psi \gets completed$
         \STATE $raiseEvent(completed,c)$
      \ELSE
          \STATE $raiseEvent(advanced,c)$
      \ENDIF
   \ELSIF{$|C|=0$}
      \STATE $raiseEvent(unmatched,m)$
   \ELSE
      \STATE $raiseEvent(ambiguous,m)$
   \ENDIF
\end{algorithmic}
\caption{Advancing the conversation.}
\label{fig:acre-introduction:advance}
\end{figure}

If one candidate conversation was identified, this is advanced to the next appropriate state by setting its current state to be the end state of the transition that was triggered by the message. Its bindings must also be updated to include bindings for variables in the transition that were matched against values in the message. The anonymous variable does not acquire a binding. 

The generation of these bindings depends on the context in which any variables were used. Firstly, each of the fields in the definition of the transition have the current conversation bindings applied to them (using the \textit{apply} function). As discussed in Section~\ref{sec:acre-introduction:mutability}, variables with previous bindings that are used in an immutable context will be replaced, with unbound variables or those used in mutable context remaining unchanged.

Once this transformation occurs, the \textit{getBindings} function generates a set of bindings that arises from comparing the message fields against the appropriate fields in the transition definition. Where the transition field contains a variable other than the anonymous variable, a binding between that variable and its matched value is generate.

Finally, once these new bindings are generated, these are merged with the original bindings that were present in the conversation. This merging is slightly more complex than a simple union operation, since there can be circumstances where a binding exists for the same variable in both sets of bindings. This occurs when a variable with a previous binding is used in a mutable context. In this case, the variable will not be replaced in the \textit{apply} step, but will then match against a value in the message and generate new bindings in the \textit{getBindings} stage. Thus a binding for that variable will exist both in the returned bindings from \textit{getBindings} as well as in the original set of bindings ($c_B$). For this reason, the \textit{mergeBindings} function gives priority to bindings contained in the second set of bindings, which can override the previous binding if one exists.

This can be represented procedurally in the following pseudocode, in which $M$ is the set of merged bindings to be returned. $M$ is initialised as the original set of bindings ($B1$), with bindings being replaced with bindings from the newly-generated bindings ($B2$) if they relate to the same variable. Bindings from $B2$ that relate to different variables are simply added to $M$.

\begin{figure}[!ht]
   \begin{algorithmic}
      \STATE $B1 \gets$ a set of variable name/value pairs in the form $(n,v)$
      \STATE $B2 \gets$ a set of variable name/value pairs in the form $(n,v)$
      \STATE $M \gets B1$
      \FORALL{$(n,v) \in B2$}
         \IF{$(n,x) \in M$}
            \STATE remove $(n,x)$ from $M$
         \ENDIF
         \STATE add $(n,v)$ to $M$
      \ENDFOR
      \STATE return $M$
      
   \end{algorithmic}
   \caption{Procedural description of the \textit{mergeBindings} function.}
   \label{fig:acre-introduction:mergeBindings}
\end{figure}

\section{Limitations} \label{sec:acre-introduction:limitations}

The choice of FSMs as a model for inter-agent conversations does have a number of limitations that must be addressed. These are discussed in the following sections.

\subsection{Ordering of Actions} \label{sec:acre-introduction:ordering}

In a conversation where actions can happen in any order, this can be difficult to model in an FSM. For example, consider a system that models the delivery of and payment for goods. We assume that it is permissible for payment to occur in advance of delivery or alternatively may happen afterwards.

\begin{figure}[!ht]
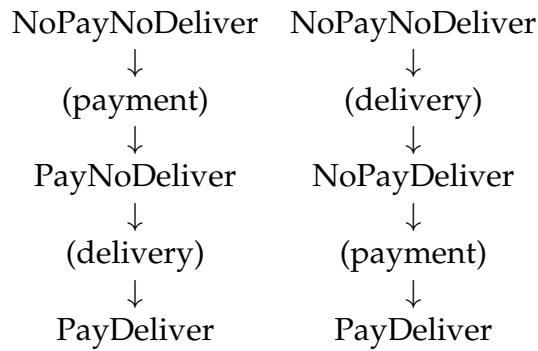

   \centering
   \begin{tabular}{cc}
      NoPayNoDeliver & NoPayNoDeliver \\
      $\downarrow$ & $\downarrow$ \\
      (payment) & (delivery) \\
      $\downarrow$ & $\downarrow$ \\
      PayNoDeliver & NoPayDeliver \\
      $\downarrow$ & $\downarrow$ \\
      (delivery) & (payment) \\
      $\downarrow$ & $\downarrow$ \\
      PayDeliver & PayDeliver \\
   \end{tabular}
   \caption{Ordering of operations.}
   \label{fig:acre-introduction:ordering}
\end{figure}

This issue is illustrated in Figure~\ref{fig:acre-introduction:ordering}, which shows two sequences of actions (enclosed in parentheses) to get from a situation where goods have neither been paid for nor delivered to a final state where both actions have occurred. Where these are permitted to happen in any order, two intermediate states are required if both possibilities are to be modelled within a single protocol. This becomes more unwieldy if there are more than two actions that can take place in any order.

In this case, it may become necessary to separate a large protocol such as this into separate protocols that each allow some stricter ordering of operations. This would require the participating agents to agree on the ordering beforehand (e.g. they would agree to use the protocol that insists on prepayment for goods that are to be delivered afterwards).

\subsection{Synchronisation}

A significant limitation of using FSMs to represent agent interaction protocols is the issue of synchronisation. This problem arises any time a conversation is in a state from which either participant may send a message. In this situation, one agent may send a message to move the conversation to a particular state, while the other agent does likewise before the receipt of this message. Where this arises, each agent will then have a different view of the state of the conversation, leading to confusion and misunderstanding in respect of further messages.

\begin{figure}[!hbt]
\begin{center}
\includegraphics[scale=0.5]{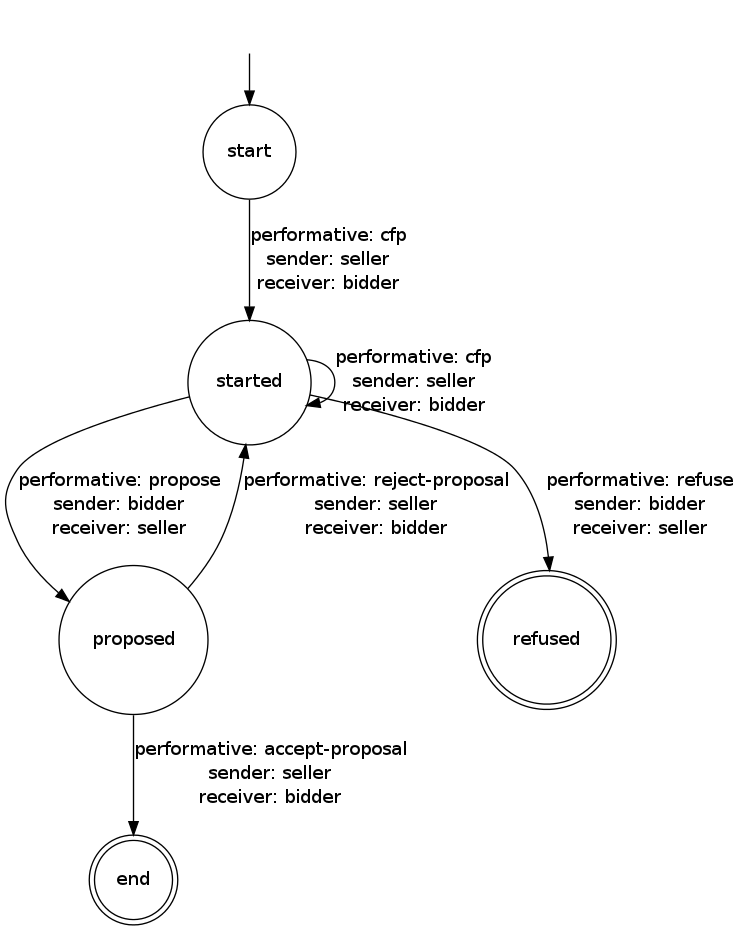}
\end{center}
\label{fig:acre-introduction:sync}.
\caption{English Auction Protocol with Synchronisation Problems.}
\end{figure}

Figure~\ref{fig:acre-introduction:sync} shows a protocol that has been designed without taking synchronisation into account. This diagram represents a protocol designed to implement a form of English Auction~\cite{fipa-english-auction}\footnote{For clarity, the message content fields used by ACRE have been omitted from this example.}. In an English Auction, the seller proposes a price to the bidders at which an item is to be sold. If a bid is made for the item at this price, the price is raised and the seller invites new bids at the higher price. Eventually, a price is reached for which no bids are received. When this occurs, the bidder that agreed to the highest price is considered the winner of the auction (subject to that bid being higher than the seller's privately-known reserve price).

In Figure~\ref{fig:acre-introduction:sync}, the initiator of the conversation initially sends a \texttt{cfp} message to invite bids for some item. This brings the conversation to the ``started'' state, at which point the bidder is permitted to make a proposal (via the \texttt{propose} performative). If the seller has found no higher bid, it will accept this bid (using an \texttt{accept-proposal} message) and the auction ends with this bidder being considered the winner. If a higher bid is found elsewhere, the bid is rejected (using \texttt{reject-proposal}) and the conversation is returned to the ``started'' state. In effect, a bidder has three options on receiving a call for proposals:

\begin{enumerate}
   \item Make a proposal at the advertised price, bringing the conversation to the ``proposed'' state.
   \item Refuse to bid, taking the conversation to the ``refused'' state and terminating the conversation. This removes the bidder from the auction, as refusal to bid at one price implies refusal to bid at a higher price.
   \item Remain non-committal and await future calls for proposals. This may be the case if an agent is waiting for external factors to facilitate a bid being made (e.g. the availability of resources with which to pay the relevant price).
\end{enumerate}

The synchronisation issue occurs in this ``started'' state, since in addition to a bidder being permitted to make or refuse to make a proposal, the seller is also permitted to send further \texttt{cfp} messages to solicit bids at increasing prices as bids are received from other participants in the auction.

If a bidder sends a proposal, it will consider the conversation to be in the ``proposed'' state. As such, a further \texttt{cfp} message from the seller will appear to be an out-of-turn message. This is despite the fact that both agents are using the same protocol and that, from the seller's point of view, the message was a perfectly valid one to send (as it was sent before its receipt of the proposal from the bidder).

For this reason, care must be taken when writing protocols to ensure that this type of situation does not occur. In practice, this will generally mean that at any particular state of a conversation, the onus is on only one agent to act. This is, however, not enforced by ACRE as the aim is to enable programmers to do more with conversations rather than place restrictions upon them.

One example of where it may be desirable to allow this situation to occur would be in a case where one agent is acting as a proxy between two other agents who cannot communicate directly. In this case, the two conversing agents are engaged in a conversation that follows some agreed interaction protocol. In doing so, they send messages to the proxy agent, which is expected to forward it to the other conversing agent. In this case, the design of a generic proxying protocol would require messages to be sent in the order specified in the underlying protocol being followed by the conversing agents. Once the agreement is made for one agent to act as a proxy, the proxying protocol cannot insist on any particular ordering of the message exchange, as to do so would restrict the underlying protocols that can be used. The alternative to this, if ACRE were to enforce synchronisation restrictions, would be to write separate proxying protocols for each underlying protocol. This would result in unnecessary additional effort to create near-duplicate protocols. Thus the enforcement of synchronisation is left as a task to the protocol developer.

\subsection{Two Participants in Each Conversation}

In the ACRE model of conversations, only two agents may participate in any one conversation. This is in contrast to other representations such as Agent UML or Coloured Petri Nets, which can cater for interactions featuring more than two participants. This is done in order to keep the complexity of protocol definitions as simple and intuitive as possible.

In the event that an agent wishes to correspond with multiple agents, it must begin separate conversations with each and manage any relationship between these at the deliberative layer. An example of this would be where an agent conducts an auction by sending a call for proposals to multiple participants and receives bids in reply. Each of these interactions represents a separate conversation. However, these conversations are indelibly linked, in that they relate to the same auction. ACRE addresses this limitation by allowing agents to manage and reason about groups of conversations.

The group management facilities of ACRE (described in more detail in Section~\ref{sec:acre-architecture:group-reasoner}) are designed to support the raising of events relating to multiple conversation groups instead of on the individual conversation level. In the auction example, a suitable event would be that all participating agents have submitted a bid (and in doing so have all advanced their respective conversations to the same state) or that some reply has been received from all of the participating agents (including refusals to bid).

Developers are given the facility to define their own custom events and so are offered a great deal of flexibility regarding the management of multiple related conversations.

\subsection{Counting Iterations}

It is possible to define a protocol that includes a cycle (e.g. the example protocol used in Section~\ref{sec:acre-introduction:process-documents} below). Here, a number of iterations of this cycle may be desired before the conversation comes to an end by triggering a transition that exits the cycle. Because of the choice of an FSM for modelling protocols, it is not possible to enforce a specific number of iterations that must be performed before breaking the cycle. Similarly, it is not possible to set upper or lower bounds on the number of iterations that must be performed. Where a specific number of exchanges of similar messages are required, these must be represented by separate transitions that do not actually form a cycle in the FSM.

\section{ACRE Protocol Examples} \label{sec:acre-introduction:examples}

This Section presents examples of ACRE protocol implementations. Initially, a simple request/response protocol is defined and a sample conversation is followed to illustrate how ACRE matches messages as part of its automated conversation handling. More complex examples are used later in order to show other common use-cases and illustrate further features of the system.

The examples in this Section assume normal operating conditions. Thus events such as the cancellation or timeout of a conversation are not considered at this time. These are discussed below in Section~\ref{sec:acre-introduction:exceptions}.

\subsection{Basic Variable Use: The Request/Response Protocol} \label{sec:acre-introduction:request-response-example}

\begin{figure}[!ht]
\begin{center}
\includegraphics[scale=0.8]{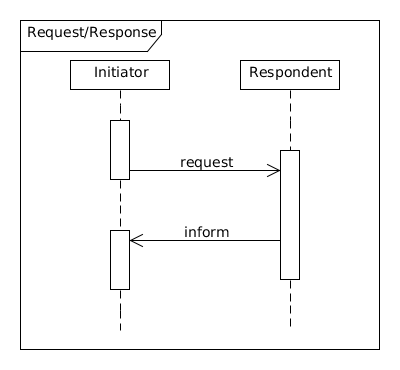}
\end{center}
\caption{Agent UML representation of Request/Response protocol.}
\label{fig:acre-introduction:request-response-auml}
\end{figure}

Figure~\ref{fig:acre-introduction:request-response-auml} shows a visual representation of a simple request/response protocol using Agent UML~\cite{Odell2001}. This is designed to allow one agent (the ``Initiator'') to ask for information from another (the ``Respondent'') on a one-off basis. The initial message is required to use the \texttt{request} performative, with the response using \texttt{inform}. At this point, only the performatives are used to identify messages, with the contents being ignored. For clarity, the \texttt{content} fields of the messages are omitted, although their inclusion would not affect the functioning of the example. Message content is used in later examples.

\begin{figure}[!ht]
\includegraphics[scale=0.53]{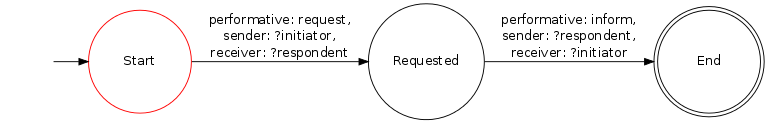}
\caption{Request/Response Protocol in the \textit{Start} state.}
\label{fig:acre-introduction:request-response-fsm-start}
\end{figure}

This protocol can be represented by the Finite State Machine (FSM) shown in Figure~\ref{fig:acre-introduction:request-response-fsm-start}. Here, the protocol consists of three states: \textit{Start} at the beginning of the conversation when no messages have yet been sent, \textit{Requested} after the initial \texttt{request} message has been sent by the Initiator agent, and \textit{End} after the Respondent has replied with the required information. The \textit{Start} state is highlighted in red as this is the initial state of any conversation that is to follow the Request/Response protocol.

Each of the transitions is labelled with the details of the message that is required to trigger the transition. In each case, the performative of the message is a fixed constant term (\texttt{request} for the initial message, \texttt{inform} for the reply). Any message that does not contain the correct performative cannot trigger a transition.

The participants in the protocol, however, are indicated by variables that are matched against the names of the agents. If variables were not used here, then only agents of specific names would be capable of participating in a conversation of this type, which would be very restrictive\footnote{This is still permitted by the model, however, as it may be desirable in some systems to have only one agent that is allowed to send particular messages (an agent named ``ManagerAgent'', for example)}.

In this example, both the \texttt{?initiator} and \texttt{?respondent} variables are used in an immutable context at all times, meaning that their values cannot change once set. This means that the agent that sends the initial message (at which point its name is bound to the \texttt{?initiator} variable) must be the recipient of the response that follows.

Given this protocol, suppose the following FIPA ACL message is sent from an agent named ``agent1'' to another named ``agent2''. This is a fully valid FIPA ACL message as it contains a performative, which is the only mandatory parameter~\cite{FIPA00061}.

\begin{Verbatim}[samepage=true]
(request
   :sender    agent1
   :receiver  agent2
)
\end{Verbatim}

This message will match the first transition, as it satisfies all three of the matching rules:
\begin{enumerate}
   \item The performative of the message is \texttt{request}.
   \item The \texttt{?initiator} variable has not previously been bound and will as such match any term: in this case ``agent1''.
   \item Similarly, the \texttt{?respondent} variable is given the value ``agent2''.
\end{enumerate}

\begin{figure}[!ht]
\includegraphics[scale=0.53]{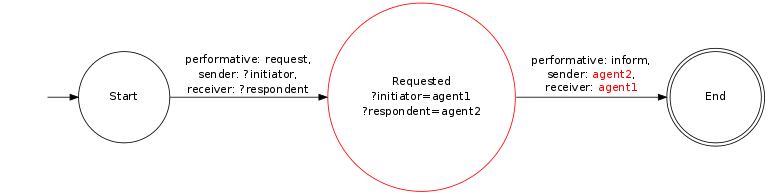}
\caption{Request/Response Protocol in the \textit{Requested} state.}
\label{fig:request-response-fsm-requested}
\end{figure}

The state of the conversation is now shown in Figure~\ref{fig:request-response-fsm-requested}. The \textit{Requested} state is now highlighted, since the first message triggered a transition. The variable bindings are also shown.

If the second transition is now examined, it can be seen that since the two variables used were in an immutable context, they must only match against terms that are equal to their values. Thus this transition has, in effect, changed from its original definition and no longer contains variables that can match any message content.

At this point, a second message is sent, in the following form:
\begin{Verbatim}[samepage=true]
(inform
   :sender    agent2
   :receiver  agent1
)
\end{Verbatim}

If this is compared with the relevant transition, it also matches:
\begin{enumerate}
   \item The message's performative is \texttt{inform}.
   \item The message was sent by ``agent2'' (which is the value bound to the \texttt{?respondent} variable)
   \item The message was received by ``agent1'' (which is the value bound to the \texttt{?initiator} variable).
\end{enumerate}

\begin{figure}[!htb]
\includegraphics[scale=0.53]{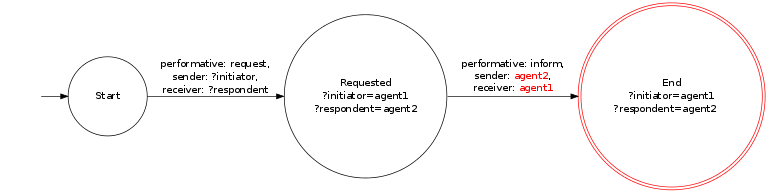}
\caption{Request/Response Protocol in the \textit{End} state.}
\label{fig:request-response-fsm-end}
\end{figure}

Having triggered this second transition, the conversation enters its final \textit{End} state, as illustrated in Figure~\ref{fig:request-response-fsm-end}. Again, the current state is highlighted in red. As this is marked as a terminal state, the conversation has now ended.

\subsection{Anonymous Variables: The Status Report Protocol}

In the previous example, all variables were used in immutable context. This section presents a slightly more complex protocol, which involves the following additional features:
\begin{itemize}
   \item Use of the anonymous variable;
   \item Multiple optional transitions;
   \item Use of message content.
\end{itemize} 

\begin{figure}[!htb]
\begin{center}
\includegraphics[scale=0.75]{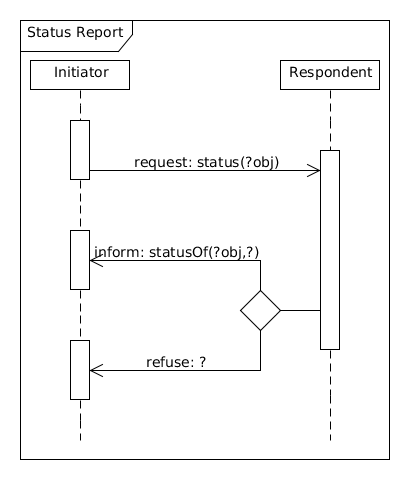}
\end{center}
\caption{Agent UML representation of a Status Report protocol.}
\label{fig:acre-introduction:status-auml}
\end{figure}

The Agent UML diagram for the status report protocol is shown in Figure~\ref{fig:acre-introduction:status-auml}. This protocol is designed so that one agent (the ``Initiator'') has a method by which it can ask another agent (the ``Respondent'') about the status of some object. As with the request/response protocol in the previous section, a conversation must begin by the Initiator sending a \texttt{request} message, which may be replied to by way of an \texttt{inform} message. However, in this case the Respondent also has the option of refusing the request (by means of a \texttt{refuse} ACL message).

In addition to this, the message content is now important. Whereas in the previous example, any \texttt{request} message with any (or no) content would be capable of triggering the first transition, in this case it is necessary that the content of the message match the predicate \texttt{status(?obj)}. Here, \texttt{?obj} is a variable in the immutable context seen in Section~\ref{sec:acre-introduction:request-response-example}. It may initially match against any term but its value is fixed thereafter. The \texttt{status(...)} portion, however, is fixed, and can only match predicates of that type.

\begin{figure}[!htb]
\includegraphics[scale=0.53]{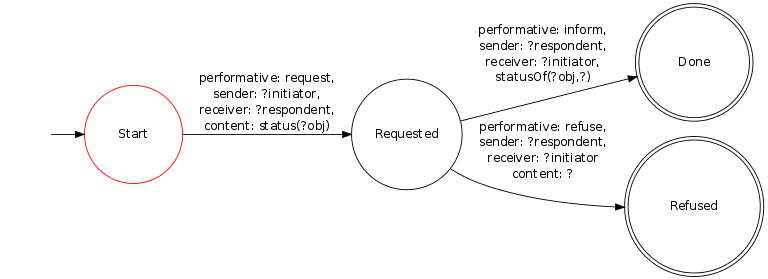}
\caption{Status Report Protocol in the \textit{Start} state.}
\label{fig:acre-introduction:status-fsm-start}
\end{figure}

When represented by an FSM, the status report protocol is as shown in Figure~\ref{fig:acre-introduction:status-fsm-start}. As the conversation has not yet begun, any conversation following this protocol is initially in the \textit{Start} state.

\begin{figure}[!htb]
\includegraphics[scale=0.53]{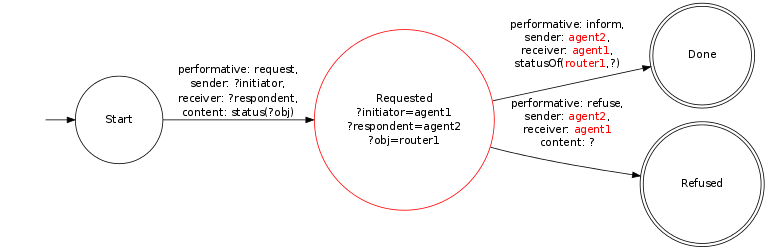}

\caption{Status Report Protocol in the \textit{Requested} state.}
\label{fig:acre-introduction:status-fsm-requested}
\end{figure}

From this initial state, suppose the initial \texttt{request} message is sent by an agent named \texttt{agent1} to another named \texttt{agent2}. This message takes the following form:

\begin{Verbatim}[samepage=true]
(request
   :sender    agent1
   :receiver  agent2
   :content   status(router1)
)
\end{Verbatim}

In this example, \texttt{agent1} is asking for an update on the status of a router (identified as \texttt{router1}). The intention of this interaction is to discover whether the router is currently functional or not. According to the protocol being followed, this message matches the transition from the \textit{Start} state to the \textit{Requested} state. As with the previous example, the variables \texttt{?initiator} and \texttt{?respondent} are bound to the names of the participating agents. On this occasion, however, the message content is also taken into account. The message content \texttt{status(router1)} is matched against the transition rule \texttt{status(?obj)} and the variable \texttt{?obj} is bound with the value \texttt{router1} in the same way as the variables relating to the conversation participants. Thus future transitions containing the variable \texttt{?obj} must match the specific string \texttt{router1}. The state of the conversation after this initial transition is shown in Figure~\ref{fig:acre-introduction:status-fsm-requested}.

At this point, the \textit{Requested} state has two possible outgoing transitions, which will be triggered in accordance with the next message to be exchanged. The following message will be sufficient to trigger the transition to the \textit{Done} state:

\begin{Verbatim}[samepage=true]
(inform
   :sender    agent2
   :receiver  agent1
   :content   statusOf(router1,up)
)
\end{Verbatim}

\begin{figure}[!hbt]
	\includegraphics[scale=0.53]{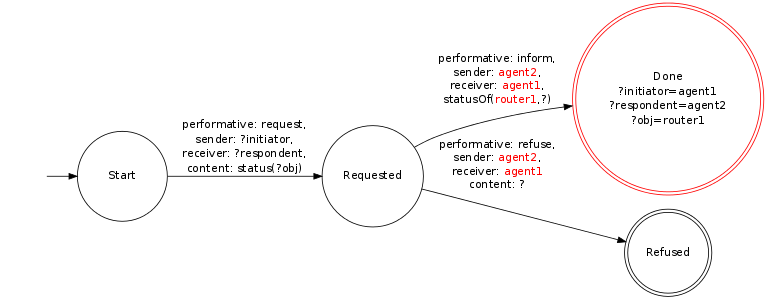}
	
	\caption{Status Report Protocol in the \textit{Done} state.}
	\label{fig:acre-introduction:status-fsm-done}
\end{figure}

The participant names match the \texttt{?initiator} and \texttt{?respondent} variables as illustrated in the previous example. However, this transition makes use of the anonymous ``\texttt{?}'' variable in the message content matching rule. Whereas \texttt{?obj} must match the string it was previously bound to (i.e. \texttt{router1}), the anonymous variable may match any string (in this case, it is the term \texttt{up} that is matched) but does not cause any variable binding to occur. Having matched that transition, this message causes the conversation to enter the terminal \textit{Done} state, as shown in Figure~\ref{fig:acre-introduction:status-fsm-done}.

\begin{figure}[!h]
	\includegraphics[scale=0.53]{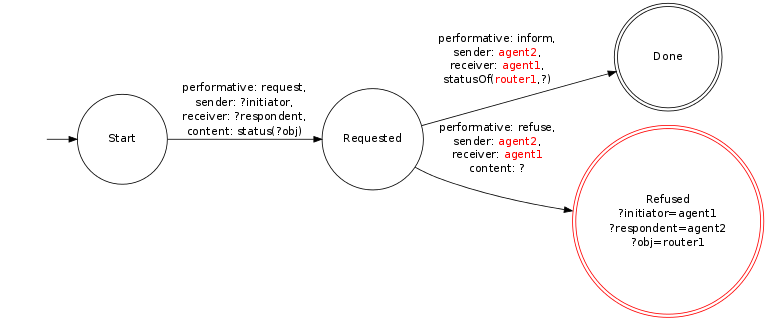}
	
	\caption{Status Report Protocol in the \textit{Refused} state.}
	\label{fig:acre-introduction:status-fsm-refused}
\end{figure}

From the \textit{Requested} state, the following message would cause the transition to the \textit{Refused} state to be triggered:

\begin{Verbatim}[samepage=true]
(refuse
   :sender    agent2
   :receiver  agent1
   :content   status(router1)
)
\end{Verbatim}

This message is intended to communicate that the sender is unwilling (or unable) to report on the status of \texttt{router1}. This is similar to the previous example in that all of the variables in this transition are ordinary named variables that already have bindings used in an immutable context. Thus, only the values \texttt{agent1}, \texttt{agent2} and \texttt{router1} could possibly match, as is the case in this message. After transitioning to the \textit{Refused} state, the conversation is as shown in Figure~\ref{fig:acre-introduction:status-fsm-refused}.

\subsection{Rebinding Variables: Process Documents Example} \label{sec:acre-introduction:process-documents}

\begin{figure}[!htb]
\centering
\includegraphics[scale=0.8]{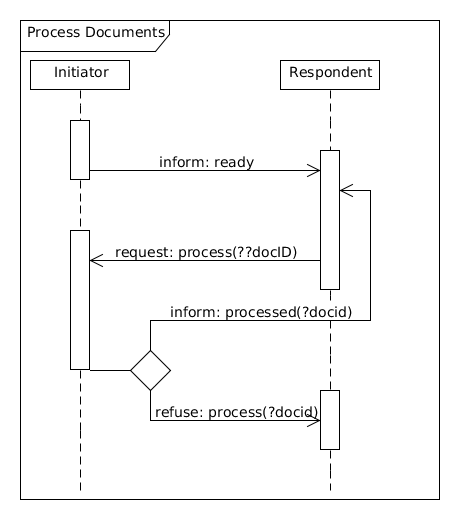}
\caption{Agent UML representation of a Process Documents protocol.}
\label{fig:acre-introduction:process-auml}
\end{figure}

\begin{figure}[!htb]
\includegraphics[width=\textwidth]{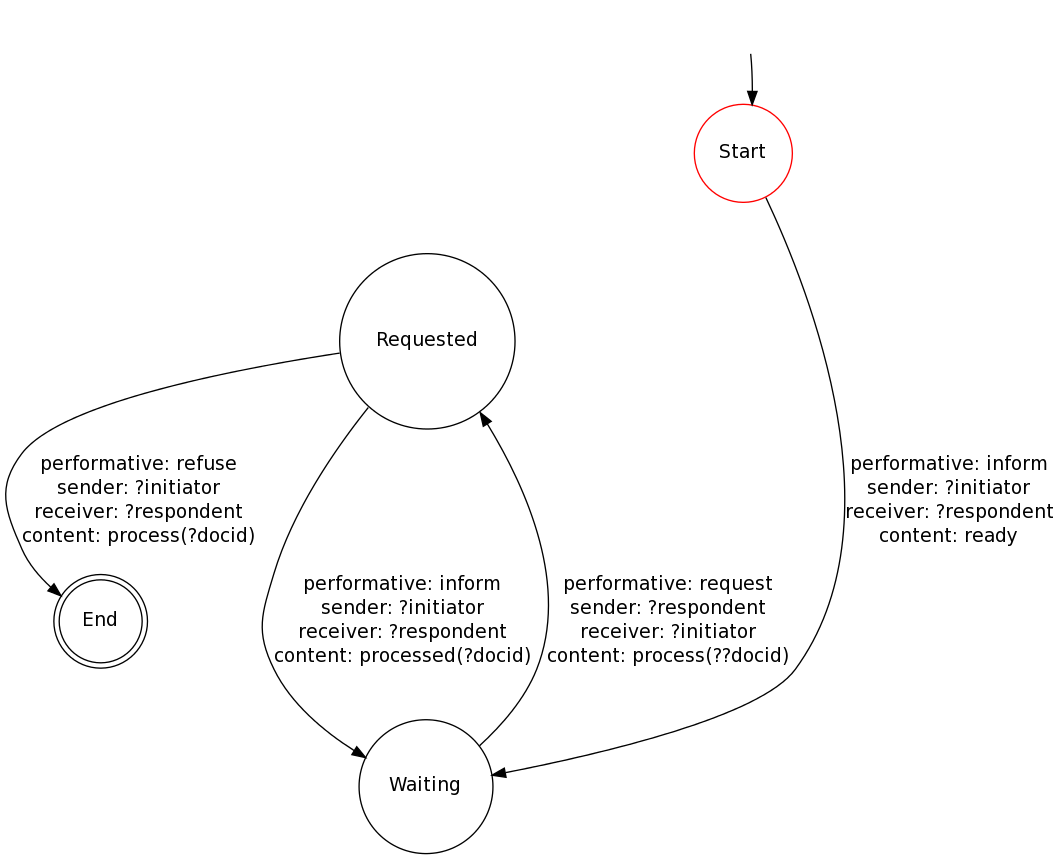}

\caption{Process Documents Protocol in the \textit{Start} state.}
\label{fig:acre-introduction:process-fsm-start}
\end{figure}

In the two previous examples, once a variable has had a value bound to it, that binding persists for the duration of the conversation. Where a variable is used in a transition, the transition can only be fired if the variable's previous bound value is matched. However, in some circumstances it may be desirable that the values that are bound to the conversation's variables be permitted to change in specified places. One example is the Process Documents protocol illustrated in the Agent UML diagram in Figure~\ref{fig:acre-introduction:process-auml} and the FSM shown in Figure~\ref{fig:acre-introduction:process-fsm-start}.

In this example, it is assumed that the participating agents are involved in a production line that processes text documents. One agent acts as the ``Manager'' that controls which documents are to be processed by the processing agents. In the Process Documents protocol, the initiator of the conversation is a ``Processor'' agent that is seeking information on which document(s) it should process. Initially, it tells the respondent (the Manager agent) that it is ready to perform such processing. Once this has been done, the Manager agent will request that the Processor agent perform the processing of a document with a specified unique identifier. The Processor can react to this either by processing the document and informing the Manager of this, or by refusing to process the document. Refusal ends the conversation. For as long as the Processor agent continues to  process the documents it has been asked to handle, the Manager will continue to request that more documents be processed. The use of variables in this example ensures that the documents referred to in the replies from the Processor have the same identifiers as those it has been asked to process. However, it is not desirable for the Manager to be restricted from requesting different documents each time.

\begin{figure}[!htb]
\includegraphics[width=\textwidth]{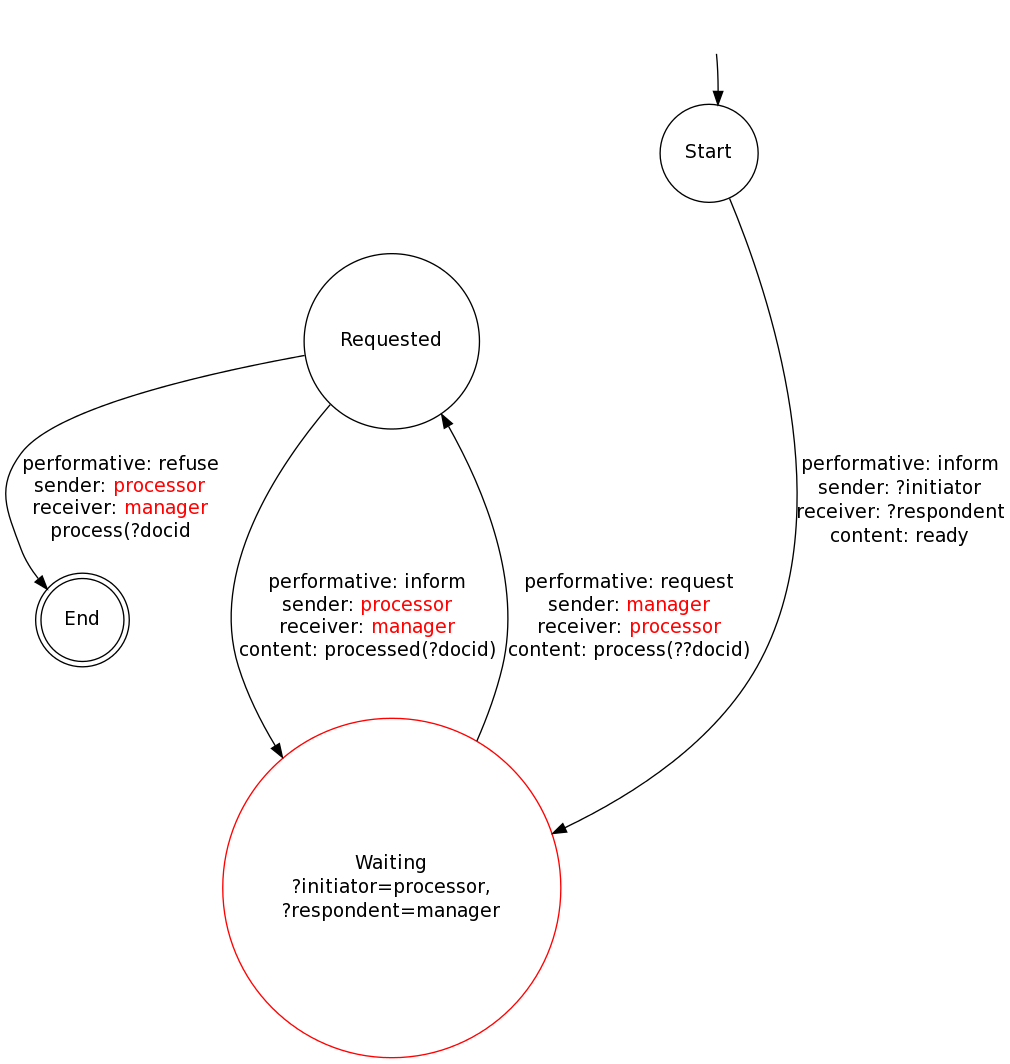}
\caption{Process Documents Protocol in the \textit{Waiting} state.}
\label{fig:acre-introduction:process-fsm-waiting1}
\end{figure}

Initially, the conversation is in the \textit{Start} state, as illustrated in Figure~\ref{fig:acre-introduction:process-fsm-start}. The conversation is advanced by the Processor agent sending an \texttt{inform} message, as follows:

\begin{Verbatim}[samepage=true]
(inform
   :sender    processor
   :receiver  manager
   :content   ready
)
\end{Verbatim}

The transition this message triggers insists that the performative of the message be \texttt{inform} and the content be the exact term \texttt{ready}, both of which are present in the message. Additionally, the variables \texttt{?initiator} and \texttt{?respondent} bind to the values \texttt{processor} and \texttt{manager} respectively, in a similar way to the previous examples. Once this message has been sent, the conversation enters the \textit{Waiting} state, as illustrated in Figure~\ref{fig:acre-introduction:process-fsm-waiting1}. As with the previous examples, these bound variables are replaced by their values in further transitions.

From this state, the next available transition is triggered by the following message:

\begin{Verbatim}[samepage=true]
(request
   :sender    manager
   :receiver  processor
   :content   process(doc123)
)
\end{Verbatim}

\begin{figure}[!hb]
	\includegraphics[width=\textwidth]{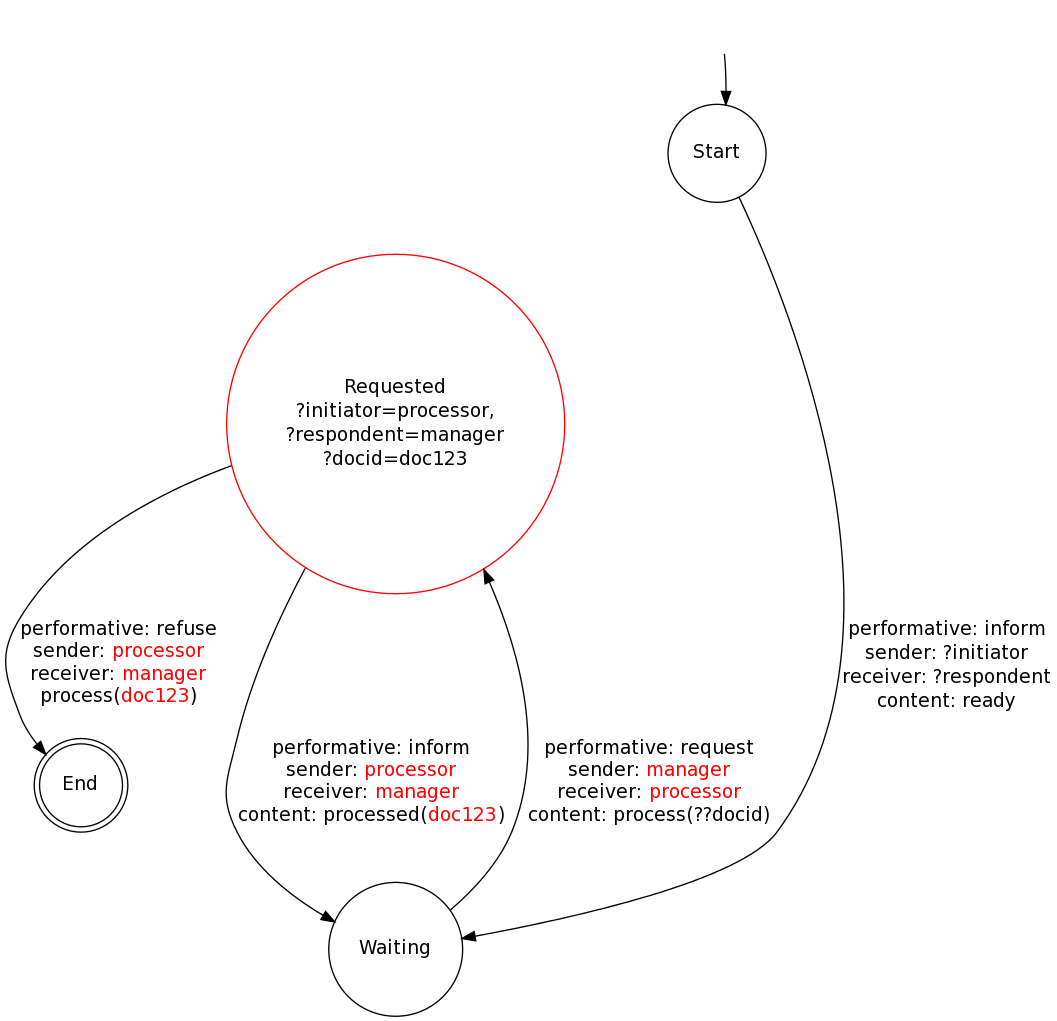}
	\caption{Process Documents Protocol in the \textit{Requested} state.}
	\label{fig:acre-introduction:process-fsm-requested1}
\end{figure}

This message, sent by the Manager agent, indicates that it wishes the processor to process the document with the identifier \texttt{doc123}. The participants match the transition rule in the same way as with the previous examples. With regard to the content, this protocol uses the mutable named variable \texttt{??docid}. As this variable does not yet have any bindings, it acts in the same way as the other variables to date. It matches the term \texttt{doc123} and this binding is added to the conversation. The result of this transition to the \textit{Requested} state is shown in Figure~\ref{fig:acre-introduction:process-fsm-requested1}.

As with the Status Report protocol, there are two possible transitions available from the \textit{Requested} state. Both contain the variable \texttt{?docid} in their content matching rule. As this has now been associated with a binding following the previous transition, these variables can only match the term \texttt{doc123}.

\begin{figure}[!hb]
	\includegraphics[width=\textwidth]{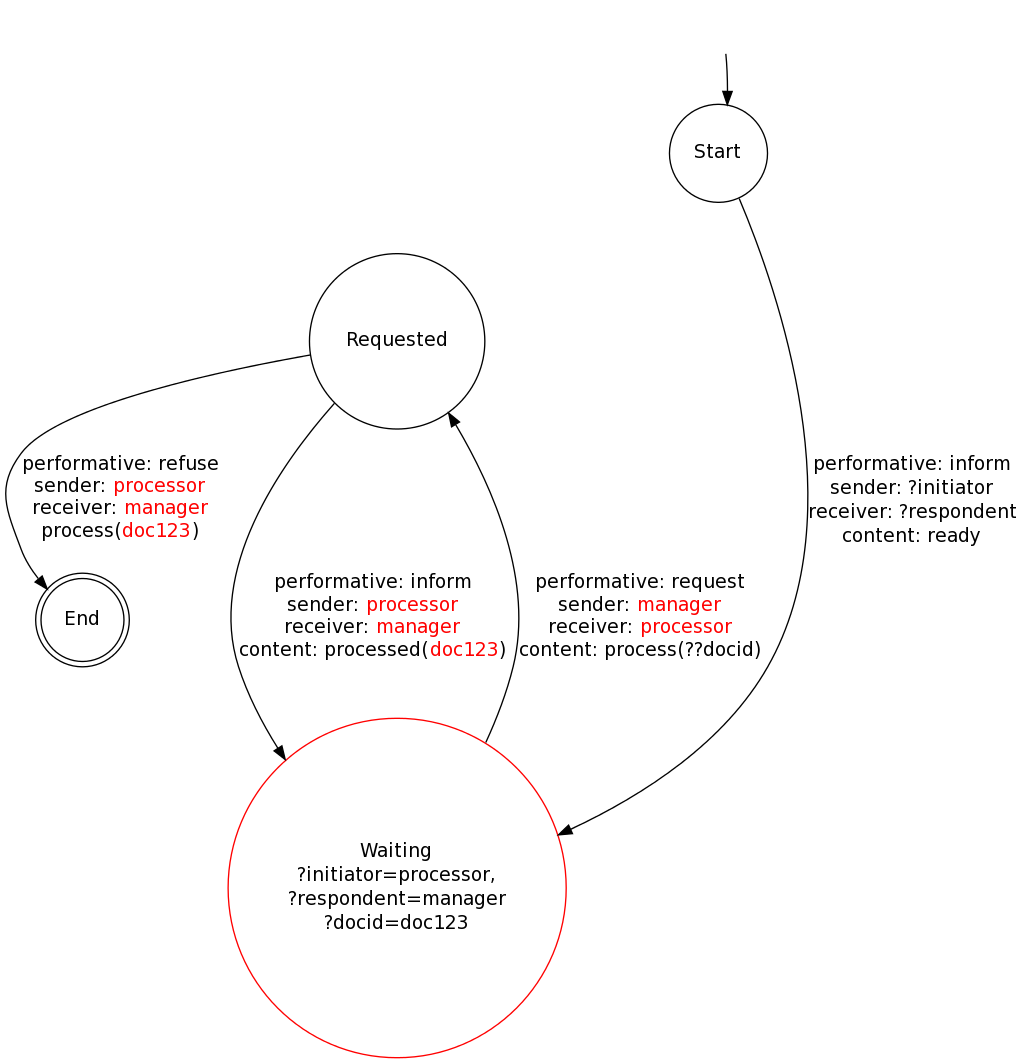}
	\caption{Process Documents Protocol having returned to the \textit{Waiting} state.}
	\label{fig:acre-introduction:process-fsm-waiting2}
\end{figure}

If the Processor agent decides to proceed with the processing of the specified document, it communicates its success to the Manager agent by means of an \texttt{inform} message as follows:

\begin{Verbatim}[samepage=true]
(inform
   :sender    processor
   :receiver  manager
   :content   processed(doc123)
)
\end{Verbatim}

As this matches the transition to the \textit{Waiting} state, the conversation state is now as shown in Figure~\ref{fig:acre-introduction:process-fsm-waiting2}.

The conversation has now returned to the \textit{Waiting} state, although on this occasion the \texttt{?docid} variable has a binding associated with it, which was not the case in Figure~\ref{fig:acre-introduction:process-fsm-waiting1}. It is at this stage that the necessity for a mutable context for variables becomes apparent. Had the transition between the \textit{Waiting} and \textit{Requested} states used the variable \texttt{?docid} in its immutable form, then the conversation could only proceed from this stage if the manager agent was to request the processor to process the same document again (i.e. document \texttt{doc123}). Thus, as mutable context was used, \texttt{??docid} can once again match any string in the same way as anonymous or unbound variables. If it does so, it will cause the binding for the \texttt{?docid} variable to be replaced (recall that these refer to the same named variable and it is only the context that is altered by the use of the different ``\texttt{??}'' sigil). This will occur in the event of a message such as the one below being sent.

\begin{Verbatim}[samepage=true]
(request
   :sender    manager
   :receiver  processor
   :content   process(doc1024)
)
\end{Verbatim}
\begin{figure}[!hbt]
\includegraphics[width=\textwidth]{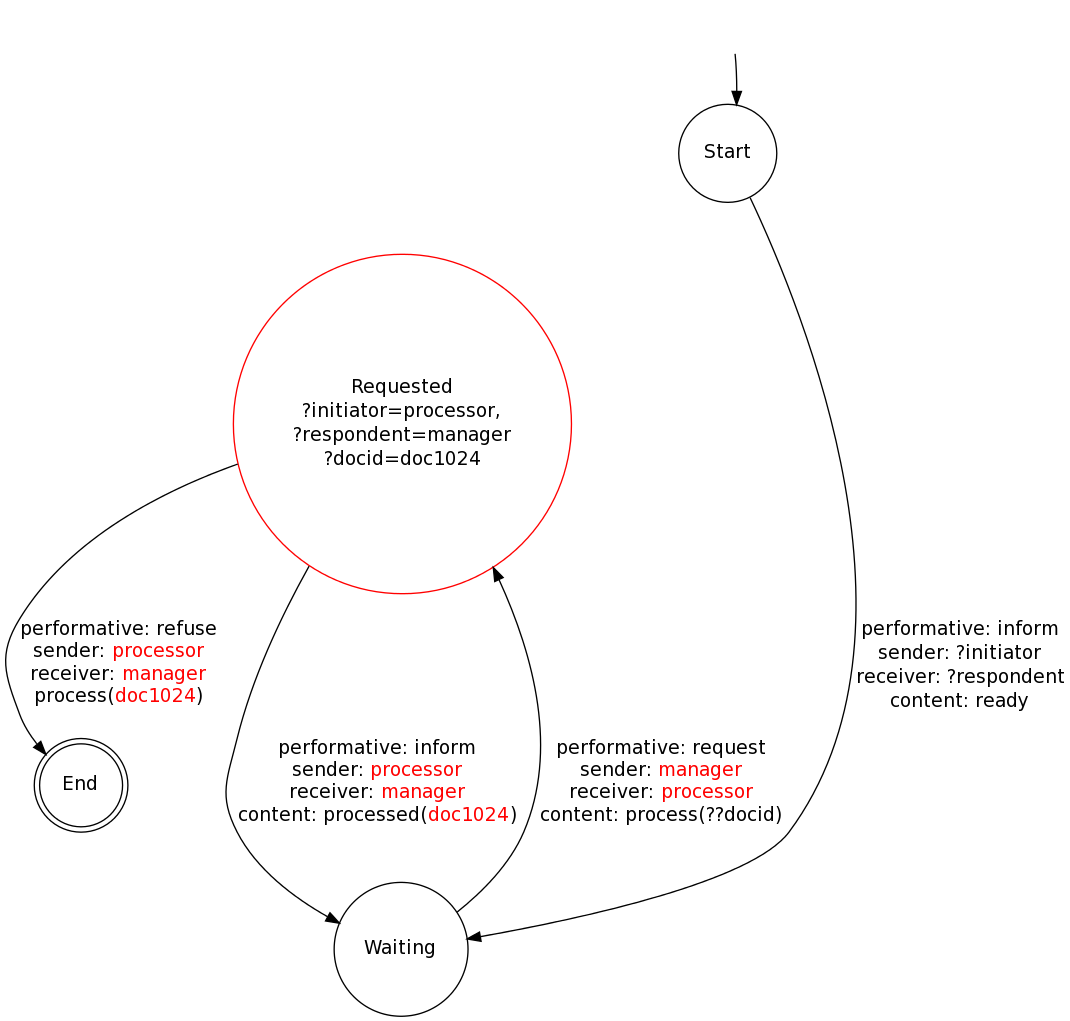}
\caption{Process Documents Protocol having returned to the \textit{Requested} state.}
\label{fig:acre-introduction:process-fsm-requested2}
\end{figure}

Because of the behaviour of the mutable \texttt{??docid} variable, this message will trigger the transition back to the \textit{Requested} state, causing the value of the \texttt{?docid} variable to be rebound to \texttt{doc1024}. From Figure~\ref{fig:acre-introduction:process-fsm-requested2}, it can be seen that because of this rebinding, the messages necessary to move from the \textit{Requested} state have changed from the first time the conversation was in this state (as shown in Figure~\ref{fig:acre-introduction:process-fsm-requested1}.

The mixture of the mutable and immutable usages of the \texttt{?docid} variable are essential to ensure the correct functioning of this protocol. As previously stated, the mutable use of the variable in the transition from the \textit{Waiting} state allows the manager to request the processor to process any document, regardless of what has previously occurred in the conversation. However, once such a request has been made, the processor must reply with reference to the last document it was asked to process. This is enforced by the immutable use of the variable in the transitions that require the processor to send the message.

In order to finish the example, it can be seen by reference to the examples thus far that once the conversation is in the \textit{Requested} state, a \texttt{refuse} message from the processor to the manager with the content \texttt{process(doc1024)} will cause the conversation to reach the terminal \textit{End} state.

\section{Exceptions to Interaction Protocol Flow} \label{sec:acre-introduction:exceptions}

The preceding examples illustrated situations where conversations proceeded from the underlying protocol's initial state to some terminal state. There are, however, many situations where a conversation may not proceed in this way. For example, one participant may wish to cancel an interaction before the completion of the conversation (e.g. having begun an auction, the agent decides not to proceed as another agent has made a one-time offer for an immediate agreement). Alternatively, an agent may wish to add a timeout condition to a conversation so a failure (or simply a failure to act) of a respondent is handled correctly and does not cause the conversation to hang. A third option occurs when a message is exchanged that does not fit the protocol definition. The mechanism by which ACRE deals with these situations is outlined in the following Sections.

\subsection{Intentional Termination} \label{sec:acre-introduction:cancel}

In a system using one-off message passing, there is no explicit expectation of any particular message being sent at any given time (although such an expectation may be implied from the agent code itself). In a conversation-aware system, however, a conversation only ends once a terminal state is reached, according to the underlying interaction protocol. One consequence of this is that if an agent does not wish to continue with the interaction, it must either take the conversation through to completion regardless, or else leave the conversation stuck in the same state indefinitely after it declines to proceed any further.

One solution to this is to allow agents the ability to explicitly cancel a conversation. The FIPA standard interaction protocols go some way towards providing such a facility. Each standard protocol specification also includes the \texttt{fipa-cancel} meta-protocol, which allows the initiator of a conversation to cancel that conversation at any point. This meta-protocol was previously discussed in Section~\ref{sec:fipa:fipa-cancel}.

In terms of adapting this meta-protocol to ACRE, a few issues arise:
\begin{itemize}
   \item The FIPA standards state that the meta-protocol may be used to cancel an existing conversation and must be initiated by the original initiator of the conversation that is to be cancelled. ACRE does not place this restriction on cancellation. As the effect of a cancellation is that a conversation is prematurely terminated, it is not important at what point this occurs. Thus any issues with synchronisation are moot (particularly as the cancel meta-protocol cannot be done unilaterally and requires agreement from the other agent).
   \item The meta-protocol specifies that the content of the \texttt{cancel} message should be the ``cancelled communicative act''. However, the meta-protocol is not intended to cancel individual communicative acts: its intention is to cancel a conversation. Elsewhere in the FIPA standards~\cite{FIPA00061}, it is noted that the \texttt{cancel} performative has implicit content when used in conjunction with the \texttt{conversation-id} parameter. As ACRE uses \texttt{cancel} to cancel conversations, the cancellation messages it sends include a \texttt{conversation-id} but have no content.
   \item The cancel meta-protocol includes an \texttt{inform-done} message. However, \texttt{inform-done} is not a performative listed in~\cite{FIPA00037} (as discussed in Section~\ref{sec:fipa:problems}). In any event, it is potentially dangerous to send a message as part of a cancellation meta-protocol that may be confused with communications in the underlying conversation (as, according to FIPA, the \texttt{conversation-id} parameter values should be the same. For this reason, ACRE sends a standard acknowledgement in response to a \texttt{cancel} message. This uses the \texttt{inform} performative and makes use of the \texttt{in-reply-to} parameter of FIPA ACL. The initial \texttt{cancel} message has no content, but contains a \texttt{reply-with} value of ``\textit{cancel}''.
   \item In the same way, \texttt{failure} messages may be used in the context of other interaction protocols also, so ACRE also seeks to use a well-defined message content so such confusion may be avoided. Communicating that the cancellation of a conversation has failed is done by means of a \texttt{failure} message with the content and with ``\textit{cancel}'' as the \texttt{in-reply-to} parameter value.
\end{itemize}

\subsection{Timeouts} \label{sec:acre-introduction:timeout}

In the context of conversation management, it is important for an agent to be capable of dealing with late responses and expected responses that are never received. The latter issue can result in agents waiting indefinitely for events to occur before acting, for example when awaiting bids from all bidders as part of an auction. For this reason, ACRE facilitates the setting of timeouts on all messages that are sent as part of a conversation. This is done by making use of the FIPA \texttt{reply-by} parameter~\cite{FIPA00061}.

This parameter is required to hold the time ``expressed according to the sender's view of the time on the sender's platform''. Thus the FIPA standard ignores the issue of cross-platform time synchronisation and timezone issues. Additionally, it does not specify what format this parameter should take. ACRE does not seek to address these issues at present. The value used for the parameter is the Unix epoch time\footnote{This is defined as the number of seconds that have elapsed since midnight on 1st January 1970 Coordinated Universal Time (UTC), excluding leap seconds.} according to the sender's host. This is capable of being converted to a local time on the receiving platform if desired.

The timeout can be set by the agent at any point during a conversation (including prior to the first message being sent). Once set, the \texttt{reply-by} parameter is set for all messages sent as part of that conversation. It may be changed for future messages at any time. Once a conversation has reached the timeout time without a response being received, ACRE raises an event to inform the agent of this fact. The mechanics of this are described in more detail in Chapter~\ref{chap:acre-architecture}.

\subsection{Inadvertent Termination}

In systems depending on one-off message passing, a message that does not conform to the expectations of the recipient can be difficult to identify and may frequently go unnoticed. This is because many agent programs are implemented by way of rules that are triggered whenever an anticipated situation comes about (e.g. that a message of a particular type has been received). Where rules of this type are inadequate, however, is in identifying situations where an unanticipated situation occurs (e.g. that a message has been received but it does not trigger any of the existing rules).

ACRE aids this situation in matching all incoming messages against known conversations and protocols: the latter to see if the message is capable of initiating a new conversation. If the message fails to match any of these, ACRE will raise an event to the agent to make it aware of this situation.

There are numerous reasons why a message may be received that cannot be successfully matched to a new or existing conversation:
\begin{itemize}
   \item The message specified the \texttt{conversation-id} of an existing conversation but the content or performative did not match an active transition.
   \item The message specified an unknown \texttt{conversation-id} but the message was not capable of beginning a conversation following the specified \texttt{protocol}.
   \item The message specified a \texttt{protocol} that was not known to the recipient.
   \item The message did not specify a \texttt{conversation-id} but could neither be matched against an active conversation nor begin a conversation following any known protocol (unmatched message).
   \item The message did not specify a \texttt{conversation-id} but it is capable of advancing or beginning multiple conversations (ambiguous message).
\end{itemize}

The FIPA standards provide for a \texttt{not-understood} performative to be used whenever a message is received that is not recognised by the recipient~\cite{FIPA00061,FIPA00037}. ACRE also provides the facility to send such a message whenever an unmatched message is received. This message contains a \texttt{conversation-id} parameter that matches the message that was not understood.

\section{Comparison with Related Systems}

Having outlined how ACRE represents protocols and handles conversations, it is important to draw comparisons with other approaches to interaction. The following Sections discuss some of the design decisions involved in the development of ACRE, particularly where they differ from those taken by other researchers in tackling the problem of structured agent communication.

\subsection{Other Finite State Machine Representations}

As discussed in Section~\ref{sec:communication:fsm}, the use of FSMs has been a common method of modelling agent interaction in a number of systems. ACRE's use of FSMs is, though rooted in a similar theoretical model, quite different to several of these other implementations.

One key difference is the decision to model a conversation itself as an FSM independent of the internal workings and reasoning of the participating agents. This is in contrast to systems such as COOL, AgenTalk and JADE. These use separate FSMs for each of the participants in a conversation, mixing agents' own actions with the messages exchanged. This was not considered to be a suitable approach for ACRE for two principal reasons. Firstly, because ACRE is intended for use with a variety of AOP languages and frameworks, it is important to maintain a clear distinction between the actual exchange of messages and the reasoning surrounding this. Agent developers should be left free to implement their agents' decision-making as they see fit, in accordance with the features and capabilities of their chosen tools. Secondly, the separation of a protocol into multiple roles increases the difficulty of verifying whether two agents are compatible and whether a protocol is being followed correctly by its participants. Using a single protocol definition means that an outside observer can monitor the validity of a protocol implementation in any AOP language that supports the sending of FIPA-compliant messages.

Another distinction can be drawn between ACRE and the KaOS model of FSMs, which allows silence to constitute a valid transition between states. Silence is a difficult concept to model, particularly as it is indistinguishable from a failure of a participating agent or the communications channel. Additional complexity arises from needing to decide the time lapse required to constitute silence. Certain delays in the communication of messages or in the deliberation of agents are unavoidable. Thus there is a danger that in setting too short a time period, a transition may incorrectly be triggered despite another participant intending to send an explicit reply. ACRE's approach is that any transition can only be triggered by active communication, which leads to greater clarity of intention. The support for conversation timeouts allows an agent to be notified when an expected reply has not been received, so that it may react according to a developers wishes (e.g. cancelling the conversation, verifying that an error has not caused the other participant to crash).

The other principal distinction between ACRE and other FSM models is that ACRE allows the content of messages to be specified in the protocol in addition to the performative and the participants. Representations of interaction protocols (including those defined by FIPA) are typically created to be generic and to be widely applicable to numerous situations. These types of protocols are possible to define in ACRE if the protocol developer decides against specifying any particular constraints on permissible message contents. However, the internal handling of messages within agents frequently relies on the content of those messages. For this reason, it is desirable to permit developers of MASs to create clearly-defined protocols that can be used to verify that every aspect of agent communication is being performed correctly. If an agent is created to react to messages with a particular content, this can be represented within the protocol definition.

In allowing message content to be specified, it is important to be mindful of the fact that the choice of a content language will affect the compatibility of ACRE with existing AOP languages. A balance must be struck between remaining compatible with as many AOP languages as possible and allowing clear protocols to be defined. As the vast majority of AOP languages are logic-based, the choice of a similar content language allows for wide compatibility. It should be noted that, in contrast, many previous FSM representations are inherently incompatible with other systems as they integrate directly into a specific AOP language.

\subsection{Global Session Types}

Global session types (introduced in Section~\ref{sec:communication:global-session-types}) are similar to FSMs in the sense that they consist of transitions triggered by the sending of messages.

The concept of a \texttt{fork} construct in a global session type (where a number of communications can occur in any order) is one feature that is not easily captured by an FSM. This is a limitation outlined in Section~\ref{sec:acre-introduction:ordering}.

The integration of global session types into Jason is one of the few approaches to cater for the message content in addition to performatives and participants. This allows typed data to be present in messages, with the correct type being checked at run-time. However, verification that related messages refer to the same item is not currently possible. ACRE's bound variables allow for this type of checking to occur.

In addition, the current prototype implementation only supports successful, verified communications. The agent is not aware if its proposed message does not match the protocol specified by the global session type and so a miscommunication will cause silent failure of the conversation. In contrast, ACRE can make an agent aware when exceptions or cancellations occur, to facilitate the agent in attempting recovery.

\subsection{Coloured Petri Nets}

In choosing an FSM as the representation of ACRE protocols, it was necessary to balance the superior support for concurrency offered by Coloured Petri Nets (CPNs) against the increased complexity this incurs (CPNs are introduced in Section~\ref{sec:communication:cpn}). Equivalent representations of a simple KQML \texttt{register} protocol have previously been shown in Figures~\ref{fig:communication:fsm-register} (FSM) and~\ref{fig:communication:cpn-register} (CPN) (both diagrams are taken from~\cite{Cost2000}). As can be seen from these illustrations, the CPN representation has much greater complexity, even for a simple protocol.

A key aim of the ACRE system is to formulate an intuitive system that developers can easily become familiar with and design interaction protocols for. For this reason, it was decided that the greater simplicity of an FSM-based model was preferable to the more complex CPNs. This involves the tradeoff that for complex interactions, some protocols may need to be split into multiple sub-protocols with the agent reasoning capabilities being used to manage the interplay between these. Additionally, some adjustments are required to handle concurrency where asynchronous communication is used. In its simplest form, this requires that only one agent is permitted to act at any particular time.

It is believed that imposing the use of complex CPN-based protocols for even simple interactions would act as a significant barrier to the uptake of a conversation-handling framework for those developers more accustomed to communication via individual unrelated messages.

\subsection{Approaches based on Semantics}

A number of approaches to agent interaction are based on various models of semantics, including Commitment Machines (discussed in Section~\ref{sec:communication:commitment-machines}), Expectations (Section~\ref{sec:communication:expectations}) and Mental Models (Section~\ref{sec:communication:mental-models}). These solutions tackle interaction from a different approach, where agents make use of their autonomy and reasoning capabilities to proactively negotiate communication patterns based on their understanding of the meaning of the communication.

ACRE aims to be as widely applicable as possible, and so efforts have been made to define it in such a way as to be compatible with a variety of existing AOP languages and frameworks. As noted in the previous comparison with other FSM-based solutions, the ACRE approach does not attempt to interfere with the mental model or reasoning process of the agents that use it. Additionally, it makes minimal assumptions about the nature of agents' mental models and the concepts they are based on. An agent may not have any concept of a social commitment, or of expectations. Even for the semantics of individual messages based on speech act theory, as defined for FIPA ACL and KQML, support varies widely amongst AOP toolkits. As discussed in Section~\ref{sec:communication:acl-criticism}, this ranges from systems that do not support semantics at all, to those with partial support.

Because of this treatment of the semantics of communication, the approach of ACRE is to focus on a more traditional approach to communication where a developer engineers an interaction protocol at design time. This then informs the development of the agents that are expected to follow it, as well as allowing for runtime verification that specific protocols are being followed by conversation participants. Although this is a less flexible approach, it does provide a framework within which interaction can occur. The autonomy of the participating agents is still respected, as their own deliberation will dictate how, when and if they engage in interaction.

\section{Summary}
This Chapter has served as an informal introduction to the ACRE system. In particular it concentrated on how interaction protocols are defined, how they relate to the messages that are exchanged between agents, and how the ACRE system deals with these in grouping messages into conversations. In addition, it considers situations where exceptional circumstances occur (e.g. cancellation, timeout, etc.) that cause conversations to terminate abnormally.

The following Chapters provide more detail on ACRE. Chapter~\ref{chap:semantics} shows the operational semantics that underlie ACRE's automated conversation management. As part of this illustration, the informal examples from this Chapter (seen in Section~\ref{sec:acre-introduction:examples}) are re-used in a more formal setting to show how the same conversations can be modelled using the operational semantics supplied. Following this, Chapters~\ref{chap:acre-architecture} and~\ref{chap:agent-factory} show two architectures for the assimilation of ACRE into an existing MAS framework. These are a generic architecture and a specific concrete implementation respectively. The generic architecture is intended to be a general architecture suitable for integration with a variety of MAS systems whereas the concrete implementation illustrates the result of integration with the Agent Factory MAS framework.

\chapter{ACRE Formal Model}
\label{chap:semantics}

% NOTE: some protocols (e.g. FIPA's subscribe protocol) do not have any end states but continue indefinitely until a cancel meta-protocol is invoked.

\section{Introduction}
\label{sec:semantics:introduction}

It has become common for creators of AOP languages to specify the semantics of their languages using operational semantics~\cite{Plotkin1981}. Following the informal description of ACRE in Chapter~\ref{chap:acre-introduction}, this chapter outlines the formal model of the ACRE system in a similar way to that of Jason~\cite{Bordini2005,Vieira2007}.

\section{Assumptions of the Model} \label{sec:semantics:assumptions}

In order for the model in this Chapter to be applicable, a number of assumptions are required to be made in advance. These assumptions are as follows:\textbf{}

\begin{itemize}
  \item \textit{Messages are sent between only two agents.} The FIPA ACL Message Structure Specification allows for a message to have multiple recipients although these must be explicitly declared as a set of agent identifiers~\cite{FIPA00061} so the recipients are specifically enumerated in the message\footnote{Some agent frameworks have extended this model to allow wildcard matching of agent identifiers within ACL messages (e.g.~\cite{Lillis2007c}) This is a non-standard extension and as such it is not supported by ACRE.}. In the ACRE model, these are treated as separate communications that happen to have the same content.
  \item \textit{Variables cannot appear in the messages themselves.} It is possible that a term that looks like a variable may be included, but this is not treated as such. Thus, the recursive application of bindings is not required, as the value of a binding cannot itself contain a variable that may have its own associated binding.
  \item \textit{The protocol developer has avoided concurrency issues}, having ensured that no state exists from which either participant may advance the conversation.
  \item \textit{The protocol developer has created a valid deterministic finite state machine} by ensuring that multiple transitions from the same state cannot be triggered by the same message. If such transitions exist within the protocol definition, a non-deterministic finite state machine has instead been created, which is outside the scope of this formal model.
  \item \textit{The content of a message is a predicate}, and so more complex formulae are not supported.
\end{itemize}

\section{Notation}

Uppercase letters are used to indicate data structures holding multiple items (lists, sets). Lowercase letters indicate individual items.

Parentheses are used to indicate tuples. Each entity modelled in this Chapter is represented by a fixed-length tuple of values that may be of various types. Square brackets are used to represent variable-length lists of similar items (e.g. a queue of messages). Angle-brackets are used for expressions written in Eindhoven Quantifier Notation~\cite{Backhouse2006}.

For convenience, where a symbol is used to denote some entity that is normally represented by a tuple, a non-numeric subscript is used to refer to an individual element in that tuple. For instance if $a$ is defined as the tuple $(b,c,d)$ then $a_b$ refers to the named element $b$, which is the first element in the tuple representing $a$. This is the same notation that is used in~\cite{Bordini2005,Vieira2007}.

Numeric subscripts are used to donate the position of an item within a function, predicate or list. E.g. $a_4$ is the fourth element in the list $[a_1,a_2,a_3,a_4,a_5]$. The only non-numeric subscripts used for list positions are $n$, relating to the length of the list, and $i$, which refers to a particular list index.

Where a value is undefined (e.g. a message does not have a conversation identifier included), $\bot$ is used to indicate this undefined value.

The valence of a function is defined as the number of arguments it has. The notation $|f|$ represents the valence of the function $f$, i.e.:

$|f| = n \impliedby f = f'(t_1,\ldots,t_n)$

\subsection{Use of Ellipses} \label{sec:semantics:ellipses}

Ellipses ($\ldots$) are used to indicate ranges of values in lists, predicates and functions. As the use of ellipses can be ambiguous in some situations, this section clarifies some properties of ellipse-based ranges as they are used in this Chapter.

Sample usages are as follows:

\begin{itemize}
   \item $f(t_1,\ldots,t_5) = f(t_1,t_2,t_3,t_4,t_5)$
   \item $p(t_1,\ldots,t_3) = p(t_1,t_2,t_3)$
   \item $[t_1,\ldots,t_7] = [t_1,t_2,t_3,t_4,t_5,t_6,t_7]$
\end{itemize}

In each case, the subscript of $t$ indicates its index within the function, predicate or list. The use of ellipses indicates that all integer subscripts between the first and last values are included. More formal definitions of each of these are presented in Section~\ref{sec:semantics:language}.

The ellipses are typically used to indicate a function, predicate or list with variable length. This type of declaration is as follows (declaring a function and a predicate with $n$ arguments, and a list with $n$ elements):

\begin{itemize}
   \item $f(t_1,\ldots,t_n)$
   \item $p(t_1,\ldots,t_n)$
   \item $[t_1,\ldots,t_n]$
\end{itemize}

Two special cases must be defined for situations where the end of the range is not greater than the beginning. Firstly, the one-point range is defined as follows:

\begin{itemize}
   \item $f(t_1,\ldots,t_n) = f(t_1) \impliedby n=1$
   \item $p(t_1,\ldots,t_n) = p(t_1) \impliedby n=1$
   \item $[t_1,\ldots,t_n] = [t_1] \impliedby n=1$
\end{itemize}
Finally, an empty range is defined as follows:

\begin{itemize}
   \item $f(t_1,\ldots,t_n) = f() = f \wedge f \in \textit{Const} \impliedby n<1$
   \item $[t_1,\ldots,t_n] = [] \impliedby n<1$
\end{itemize}

\section{Language} \label{sec:semantics:language}

The language of ACRE (denoted by $L_{ACRE}$) is represented using statements of predicate logic. The principal types of term that are used in this logic are constants, variables and functions. \textit{Const}, \textit{Var} and \textit{Funct} are defined to be the set of constants, variables and functors respectively. $\Theta_{ACRE}$ is the set of all terms allowed in the language.

% variables themselves are not actually part of the language - only variable instances are
% functions are included later
%All variables and functors are included in the set of allowable terms:
%\begin{itemize}
%   \item $v \in \textit{Var} \Rightarrow v \in \Theta_{ACRE}$
%   \item $f \in \textit{Funct} \Rightarrow f \in \Theta_{ACRE}$
%\end{itemize}

The set of valid terms begins with the inclusion of constant terms. The set of constant terms is denoted by \textit{Const}. All constants are valid terms in the language.

\begin{itemize}
   \item $c \in \textit{Const} \Rightarrow c \in \Theta_{ACRE}$
\end{itemize}

Any time a variable is used within the language, it must be accompanied by an associated \textit{context}. This context may either be \textit{mutable} (i.e. it may override an existing binding) or \textit{immutable} (it may not override an existing binding). Thus the set of variable instances (\textit{VarInst}) is defined as a set of tuples of the form $(v,c)$ where $v$ is a variable and $c$ indicates the context in which it has been used. The effect of mutable and immutable variables is discussed in Section~\ref{sec:acre-introduction:mutability}. These variable instances are also considered valid terms.
\begin{itemize}
   \item $\textit{VarInst} = \textit{Var} \times \{\textit{mutable},\textit{immutable}\}$
   \item $v \in \textit{VarInst} \Rightarrow v \in \Theta_{ACRE}$
\end{itemize}

Any function consisting of a functor and a number of arguments that are valid terms of the language is itself also a valid term:
\begin{itemize}
\item $f \in \textit{Funct} \wedge \langle\forall i : 1 \le i \le n : t_i \in \Theta_{ACRE}\rangle \Rightarrow f(t_1,\ldots,t_n) \in \Theta_{ACRE}$
\end{itemize}

% Old definition: taken out as per Koen's corrections
% Constants are defined as functions with 0 valence, as follows:
%\begin{itemize}
%\item $\textit{Const} = \{f \in Funct : |f| = 0\}$
%\end{itemize}

The language $L_{ACRE}$ may now be defined. Firstly, \textit{Pred} is defined as the set of predicate symbols. Predicates whose arguments are valid terms of $\Theta_{ACRE}$ comprise the language, as follows:
\begin{itemize}
   \item $p \in Pred \wedge \langle\forall i:1\le i\le n:t_i \in \Theta_{ACRE}\rangle \Rightarrow p(t_1,\ldots,t_n) \in L_{ACRE}$
\end{itemize}

\subsection{Grounded Language}

Some entities modelled in this language do not permit the use of variables. To facilitate this, a separate but related language is defined. $G_{ACRE}$ is the grounded language of ACRE, which is similar to $L_{ACRE}$ with the exception that neither variables nor variable instances are permitted.

Firstly, it is necessary to define a subset of $\Theta_{ACRE}$ to be the set of grounded terms permissible in the grounded language. This set of grounded terms is denoted by $\Gamma_{ACRE}$. 
\begin{itemize}
\item $c \in \textit{Const} \Rightarrow c \in \Gamma_{ACRE}$
\item $f \in \textit{Funct} \wedge \langle\forall i : 1 \le i \le n : t_i \in \Gamma_{ACRE}\rangle \Rightarrow f(t_1, ..., t_n) \in \Gamma_{ACRE}$
\end{itemize}

The set of valid formulae in the grounded language is now defined as follows:
\begin{itemize}
   \item $p \in \textit{Pred} \wedge \langle\forall i:1\le i \le n : t_i \in \Gamma_{ACRE}\rangle  \Rightarrow p(t_1,\ldots,t_n) \in G_{ACRE}$
\end{itemize}

From these definitions, it can be seen that the following relationships hold:
\begin{itemize}
   \item $\Gamma_{ACRE} \subset \Theta_{ACRE}$
   \item $G_{ACRE} \subset L_{ACRE}$
\end{itemize}
Thus any functions that operate on elements of $\Theta_{ACRE}$ will also be applicable to grounded terms in $\Gamma_{ACRE}$.

\section{Entities}

Within the model, there are a number of entities that must be represented, so that operations can be performed on them. These are represented by means of tuples and are outlined in the following sections.

\subsection{Bindings}

A \textit{binding} is the association of a value to a variable. The use of bindings within ACRE is described in Section~\ref{sec:acre-introduction:mutability}. In this model, \textit{Bindings} is the set of all possible bindings that can be made. It is described as follows: 

\begin{equation}
Bindings = Var \times \Gamma_{ACRE}
\end{equation}

This results in a set of bindings in which each element is a pair $(v,c)$ where $v$ is a variable (i.e. $v \in Var$) and $c$ is a grounded function that is bound to it (i.e. $c \in \Gamma_{ACRE}$). Each variable should be unique within a set of bindings. This means that the following property should always hold for any set of bindings $B \subset \textit{Bindings}$:

\begin{equation}
\forall (v,c) \in B : (\neg \exists (v',c') \in B : v = v' \wedge c \ne c')
\end{equation}

This restriction requires care in adding a new binding into an existing set of bindings. When a variable is used in a mutable context, it may match against some function despite the fact that it already has a binding associated with it. When combining existing bindings with the new bindings arising from unifying this variable with its new value, care must be taken to ensure that the above restriction is not violated. This can be done by using the $combine(s_1,s_2)$ function presented in Section~\ref{sec:semantics:combine}.

\subsection{Performative}

One key aspect of ACL communication is the \emph{performative} (see Section~\ref{sec:fipa:fipa-acl}), which must be present in each message exchanged between agents. We define \textit{Perf}, which is the set of valid FIPA performatives, as follows:

\begin{eqnarray}
\textit{Perf} & = & \{ \textit{accept-proposal}, \textit{agree}, \textit{cancel}, \textit{cfp}, \textit{confirm},  \textit{disconfirm}, \textit{failure}, \nonumber \\ 
& & inform, \textit{inform-if}, \textit{inform-ref}, \textit{not-understood}, \textit{propagate}, \textit{propose}, \nonumber \\
& & \textit{proxy}, \textit{query-if}, \textit{query-ref}, \textit{refuse}, \textit{reject-proposal}, \textit{request}, \nonumber \\
& & \textit{request-when}, \textit{request-whenever}, \textit{subscribe}\}
\end{eqnarray}

This definition arises from the FIPA Communicative Act Library Specification~\cite{FIPA00037}. However, nothing in this formal model prevents the use of an alternative definition of \textit{Perf} (for example for use with a different Agent Communication Language such as KQML). In either case, a performative will always be a single constant value, thus the principal restriction on the domain of \textit{Perf} is as follows:
\begin{equation}
\textit{Perf} \subset \textit{Const}
\end{equation}

\subsection{Message} \label{sec:semantics:message}

Messages are the means by which agents participate in conversations. This definition is based on the FIPA ACL message standard~\cite{FIPA00061}. The set of messages is denoted by \textit{Messages}, which is a set of 6-tuples of the form $(s,r,c,\phi,p,x)$ where:
\begin{itemize}
   \item $s \in \textit{Const}$ is the unique identifier of the agent that was the sender of the message. This is the value in the \texttt{:sender} field of a FIPA ACL message.
   \item $r \in \textit{Const}$ is the unique identifier of the agent that was in receipt of the message. This is a value from the \texttt{:receiver} field of a FIPA ACL message. Although the FIPA standard allows multiple recipients to be specified, ACRE's approach to this is outlined in Section~\ref{sec:semantics:assumptions}.
   \item $c \in \textit{Const}$ is the unique identifier of the conversation to which the message belongs. This may be undefined ($\bot$) if the sender does not support conversation management. This represents the \texttt{:conversation-id} field of a FIPA ACL message.
   \item $\phi \in \textit{Const}$ is the unique identifier of the protocol that the conversation is following. This may also be undefined if the sender does not support conversation management. This is the \texttt{:protocol} field of a FIPA ACL message.
   \item $p \in \textit{Perf}$ is the performative of the message. This is the \texttt{:performative} field in a FIPA ACL message.
   \item $x \in G_{ACRE}$ is the actual content of the message, as contained in a FIPA ACL message's \texttt{:content} field.
\end{itemize}

\subsection{State} \label{sec:semantics:state}

As illustrated in Chapter~\ref{chap:acre-introduction}, ACRE Protocols are represented as Finite State Machines. In order to use such a model, it is necessary to define the states a conversation may be in, along with the transitions that allow it to move between states.

States are represented in the set \textit{States}, which contains 2-tuples of the form $(\nu,\phi)$, where:

\begin{itemize}
\item $\nu \in \textit{Const}$ is the name of this state.
\item $\phi \in \textit{Const}$ is the unique identifier of the protocol this state is related to.
\end{itemize}

% This is now defined in the protocol

% The status of the state depends on the presence of incoming and outgoing transitions to and from the state. It is defined as follows:

% \begin{equation}
% s = \left\{ \begin{array}{ll}
%    start & \neg \exists t \in p_T : t_\epsilon = n \\
%    end &  \neg \exists t \in p_T : t_\sigma = n \\
%    intermediate & otherwise
% \end{array} \right.
% \end{equation}
% where $p$ is a protocol such that $p_\phi = \phi$

\subsection{Transition} \label{sec:semantics:transition}

At any point in time, a conversation will be in one particular state. The state of a conversation can only change in the event of a message being sent between the participating agents that matches a transition originating from that state.

The majority of the elements of a transition are terms that are to be matched against elements of messages that are sent and received by participating agents. In each case, variables may be used so that data from messages may be stored for re-use later in the conversation.

\textit{Transitions} is a set of transitions represented by the tuples of the form $(\phi,\sigma,\epsilon,s,r,p,x)$, where:

\begin{itemize}
\item $\phi \in \textit{Const}$: The unique identifier of the protocol to which this transition belongs.
\item $\sigma \in \textit{Const}$: The name of this transition's start state.
\item $\epsilon \in \textit{Const} $: The name of this transition's end state.
\item $s \in (\textit{Const} \cup \{v \in \textit{VarInst} : v_c = \textit{immutable}\})$: The name of the sending agent. This may be a constant value (which will only match against a specific agent identifier) or an instance of a variable used in an immutable context.
\item $r \in (\textit{Const} \cup \{v \in \textit{VarInst} : v_c = \textit{immutable}\})$: The name of the receiving agent. The same restrictions apply to this field as to the message sender.
\item $p \in \textit{Perf}$: The performative of the message.
\item $x \in L_{ACRE}$: The content of the message that will trigger this transition.
\end{itemize}

A transition represents a mechanism by which the state of a conversation may be altered. Transitions are intended to be matched against ACL messages, with a conversation advancing whenever a message matches an active transition.

As the names of states are defined as constant terms in Section~\ref{sec:semantics:state}, they must also be represented as such in a transition between states. The names of participating agents may be defined as either constants or variable instances. A constant value in these fields indicate that only messages to and from specific agents may match the transition. Variables can match against any agent names, thus allowing any agent to send or receive a triggering message. However, these variables may be replaced with constants according to the bindings associated with the appropriate conversation. Variables used in the sender or receiver fields may not be in mutable context (see Section~\ref{sec:acre-introduction:mutability} for an explanation of what is meant by ``mutable context'').

\subsection{Protocol} \label{sec:semantics:protocol}

The set of protocols is denoted by \textit{Protocols}, which is a set of 5-tuples of the form $(\phi,S,T,\iota,F)$, where:

\begin{itemize}
\item $\phi \in \textit{Const}$: The unique identifier of this protocol.
\item $S = \{s \in States : s_\phi=\phi\}$: The set of the states that are used in the protocol's finite state machine. The protocol identifier in each state must match the unique identifier of this protocol ($\phi$).
\item $T = \{t \in Trans : t_\phi=\phi \wedge (\exists s \in S : t_\sigma=s_\nu) \wedge (\exists s \in S : t_\epsilon=s_\nu)\}$: The set of transitions that are part of this protocol. These transitions must all begin and end at states that are contained in $S$.
\item $\iota \in \{s_\nu : s \in S \wedge (\neg \exists t \in T : t_\epsilon=s_\nu)\}$: The name of the initial state of any conversation following this protocol. This must be the name of a state that is contained in $S$, with the additional restriction that no transitions may end at that state. A well-formed protocol may not have multiple states that fulfil this criterion.
\item $F  = \{ s_\nu : s \in S \wedge \neg \exists t \in T : t_\sigma = s_\nu \}$: The set of names of states that are considered to be final states for any conversation following this protocol. In a similar way to the initial state, these states must be contained in $S$ and no transitions should begin at a final state. Unlike the initial state, there is no restriction on the number of final states a protocol may have. This includes the possibility of a protocol having no defined final states, in which case it will not reach a natural conclusion and must be interrupted in some other way.
\end{itemize}

\subsection{Conversation} \label{sec:semantics:conversation}

It is a conversation that acts like a functioning finite state machine, rather than a protocol itself. Protocols define the finite state machines that are executed as conversations. A conversation represents an instance of two agents communicating in a way that follows a particular protocol. The set of all conversations is denoted by \textit{Conversations} and contains 7-tuples of the form $(\phi,A,s,H,c,B,\psi)$, where:

\begin{itemize}
\item $\phi \in Const$ The unique ID of the protocol that this conversation follows.
\item $A \subset \textit{Const}$: Set of participating agent names.
\item $s \in \{p_S : p \in Protocols \wedge p_\phi=\phi\}$: The name of the current state of the conversation.
\item $H = [m_1,...,m_n] \wedge \langle\forall i : 1 \le i \le n : m_i \in Messages\rangle$: A list of past messages in this conversation.
\item $c \in Const$: The unique identifier of this conversation.
\item $B \subset Bindings$: A set of variable bindings that apply to this conversation.
\item $\psi \in \{active,completed,failed\}$: The conversation status.
\end{itemize}

\subsection{Event}
Events represent information that may be accessed by the intentional layer of an agent, in order to support reasoning about conversations. They represent the fact that errors have occurred in matching messages to conversations, or that conversations may be begun, advanced or completed.

An event is contained in the set \textit{Events} and defined as the tuple $(d,o)$, where:

\begin{itemize}
\item $d \in Const$: A description of the event that has occurred.
\item $o \in Messages \cup Conversations$: The message or conversation to which this event relates.
\end{itemize}

\subsection{Conversation Manager} \label{sec:semantics:conversation-manager}

The key element in this model is the Conversation Manager that is charged with identifying messages that are associated with particular conversations, advancing conversations, raising events to the agent's intentional layer, and other associated tasks. Within an ACRE-enabled Multi Agent System, each agent has its own Conversation Manager to take care of reasoning about the conversations in which the agent is engaged. The operational semantics for the operation of a Conversation Manager are presented in Section~\ref{sec:semantics:semantics}.

A Conversation Manager is represented by the tuple $(n,M,C,P,E,s,\mu)$, where:

\begin{itemize}
   \item $n \in \textit{Const}$ is the name of the agent to which this conversation manager belongs.
   \item $M = [m_1,...,m_n] \wedge \langle\forall i : 1 \le i \le n : m_i \in Messages\rangle $ is a Message queue, which is a list of messages that have been sent or received by the agent since the last iteration of message handling. These are assumed to be stored in chronological order of receipt, with the earlier messages at the beginning of the list.
   \item $C \subset Conversations$ is a Conversation Store, which is a set that contains all the conversations in which the agent is participating.
   \item $P \subset Protocols$ is a Protocol Store, which is a set of protocols about which the agent is aware.
   \item $E \subset Events$ is a set of Events that may be accessed by the intentional layer of the agent. These events are raised by the Conversation Manager during its reasoning process, to be consumed by the agent afterwards.
   \item $s \in \{start, initialise, match, fail, new, update, done\}$ is the state of the conversation management system. A description of what is meant by each of these states, along with their semantics, is given in Section~\ref{sec:semantics:semantics}.
   \item $\mu$ is the memory of the conversation management system. Its purpose is to record which conversations a given message can potentially advance. It is a tuple of the form $(m,C)$ where:
   \begin{itemize}
   \item $m \in Messages$ is the message that is currently being matched.
   \item $C \subset Conversations$ is a set of conversations that have been matched to $m$ as a result of identifying that $m$ is capable of advancing them.
   \end{itemize}
\end{itemize}

\section{Predicates}

In the defined logic, a number of predicates are defined that are used in reasoning about the entities described above.

\subsection{Matching}

Before any predicates can be defined that deal with the entities described above, it is necessary to define what it means for two items to match. Because of the nature of ACRE, one of the operands in each case will be a grounded term or predicate. This is as a result of the fact that these are sourced in messages that are exchanged, which cannot contain variables. Variables can only occur in protocol definitions (and by extension conversations and transitions also). For the purposes of this model, three types of matching are necessary:
\begin{itemize}
   \item \textit{tmatches} is used to test if individual terms match (i.e. one term is an element of $\Theta_{ACRE}$ and the other is an element of $\Gamma_{ACRE}$).
   \item \textit{lmatches}: is used to test if a list of terms matches (i.e. to match two lists of terms, one of which contains elements of $\Theta_{ACRE}$ and the other of which contains elements of $\Gamma_{ACRE}$).
   \item \textit{pmatches} is used to match predicates (i.e. one that is an element of $L_{ACRE}$ and one that is an element of $G_{ACRE}$).
\end{itemize}

In each case, variables are taken into account, so as to ensure that the same variable cannot match against two different values in the ground argument.

\subsubsection{TMatches} \label{sec:semantics:tmatches}

This predicate equates to true if two given terms match. As mentioned above, the second term is a grounded term, since it originates in a message.

Arguments:
\begin{itemize}
\item $s_1 \in \Theta_{ACRE}$
\item $s_2 \in \Gamma_{ACRE}$
\end{itemize}

$tmatches(s_1, s_2) \equiv $ \\
\begin{equation}
\left\{ \begin{array}{ll}

% first argument is a variable
         true & \mbox{$s_1 \in \textit{VarInst}$} \\

% both are constants and they're equal
         true & \mbox{$s_1 \in \textit{Const} \wedge s_2 \in \textit{Const}$} \wedge s_1 = s_2 \\

% Both functions
         \mbox{$lmatches([t_1,\ldots,t_n],[u_1,\ldots,u_n])$} & \mbox{$f(t_1,...,t_n) = s_1 \wedge f(u_1,...,u_n) = s_2$} \\
         false & otherwise \\
         
         \end{array} \right.
\end{equation}

%Two terms are considered not to match (i.e. this predicate will evaluate to false) unless one of the following conditions holds:

% explain about applying bindings to previous elements in funcions/predicates

\begin{itemize}
\item The first argument is a variable. Prior to testing if terms match, the \textit{tapply} function (discussed in Section~\ref{sec:semantics:tapply}) should be applied to $s_q$. The different handling of variables depending on context thus arises as a combination of \textit{tapply} and \textit{tmatches}. A variable with a previous bound value that is used in an immutable context will already have been replaced with its bound value (either a constant value or a function). As such, these are never evaluated for matching with \textit{tmatches}. Only variable instances that are not replaced with bound values can be evaluated as $s_1$. This only includes variables used in a mutable context (which are free to match against any value) or variables for which no previous bindings exist.
\item Both terms are constants and are equal.
\item Both terms are functions whose functors are equal and whose arguments match as a list, using \textit{lmatches}.
\end{itemize}

\subsubsection{LMatches}

This operates on lists of terms and indicates whether those lists match each other. Again, the second argument contains grounded terms (i.e. its contents are elements of $\Gamma_{ACRE}$) whereas the first argument may contain any ACRE terms (i.e. its contents are elements of $\Theta_{ACRE}$). The length of each list must be equal for this to apply.

Arguments:
\begin{itemize}
   \item $[t_1,\ldots,t_n] : \langle\forall i : 1 \le i \le n : t_i \in \Theta_{ACRE}\rangle$
   \item $[u_1,\ldots,u_n] : \langle\forall i : 1 \le i \le n : u_i \in \Gamma_{ACRE}\rangle$
\end{itemize}

$lmatches([t_1,\ldots,t_n],[u_1,\ldots,u_n]) \equiv $
\begin{equation}
\left\{  \begin{array}{ll}
% empty lists
true & n = 0 \\

% non-empty lists
\mbox{$lmatches([t_1,\ldots,t_{n-1}],[u_1,\ldots,u_{n-1}])$} & n > 0 \\
    \mbox{$\wedge~tmatches(tapply(B,t_n),u_n)$} & \\
\end{array} \right.
\end{equation}

where $B = lbind([t_1,\ldots,t_{n-1}],[u_1,\ldots,u_{n-1}])$

Empty lists are always considered to match (this relies on the notion of an empty list outlined in Section~\ref{sec:semantics:ellipses}). Where the lists are not empty, \textit{lmatches} acts as a recursive function. For lists of length $n$, it is first necessary to test if the lists match up to position $n-1$, recursively. If so, the $n$th terms are compared, after applying the bindings arising from the earlier part of the lists. The \textit{tapply} function is defined below in Section~\ref{sec:semantics:tapply}. Its purpose is to apply a set of bindings to a term.

\subsubsection{PMatches}

This tests to see if two predicates match. The first predicate may contain variable terms, so any predicate of the ACRE language $L_{ACRE}$ is permissible here. Once again, the second argument must be grounded, and as such it must be part of $G_{ACRE}$.

Arguments:
\begin{itemize}
   \item $s1 \in L_{ACRE}$
   \item $s2 \in G_{ACRE}$
\end{itemize}

\begin{equation}
\begin{split}
pmatches(s1,s2) \equiv~ &   p(t_1,\ldots,t_n) = s1 \wedge p(u_1,\ldots,u_n) = s2 \\ & \wedge lmatches([t_1,\ldots,t_n],[u_1,\ldots,u_n])
\end{split}
\end{equation}

Two predicates will match if they have the same predicate identifier ($p$) and their arguments match as a list, using \textit{lmatches}.

\subsection{Triggers}

The \textit{triggers} predicate indicates whether a message will trigger a specific transition, given a set of bindings.

Arguments:
\begin{itemize}
   \item $m \in \textit{Messages}$: a message
   \item $t \in \textit{Transitions}$: a transition
   \item $B \subset \textit{Bindings}$: the set of bindings already present in the conversation
\end{itemize}

\begin{equation}
	\begin{split}
triggers(m,t,B) \equiv 	& 	m_p = t_p  \\
					& \wedge tmatches(tapply(t_s, B),m_s) \\
					& \wedge tmatches( tapply(t_r, B'),m_r) \\
					& \wedge pmatches( papply(t_x, B''),m_x)
	\end{split}
\end{equation}
where $B'= combine(B, tbind(tapply(t_s,B), m_s))$ \\
and $B'' = combine(B',tbind(tapply( t_r, B'), m_r))$

A message will trigger a transition if all of the following criteria are met:
\begin{itemize}
\item The performative of the message ($m_p$) and the transition ($t_p$) are equal.
\item The message's sender ($m_s$) matches the transition's `sender' field ($t_s$).
\item The message's recipient ($m_r$) matches the transition's `recipient' field ($t_r$).
\item The message's content ($m_x$) matches the transition's `content' field ($t_x$).
\end{itemize}
In each of the above cases, with the exception of the performative (which cannot be a variable), the conversation's bindings are applied before matching, so as to avoid re-binding variables that have already been bound. The functions to apply these bindings are defined below in Section~\ref{sec:semantics:applying}. The \textit{combine} function, for combining sets of bindings, is defined in Section~\ref{sec:semantics:combine}.

\subsection{Initiates}

This predicate indicates whether a specific message is capable of initiating a conversation to follow a specific protocol. A message is considered to be capable of initiating a conversation if it matches a transition from the initial state of the given protocol.

Arguments:
\begin{itemize}
\item $m \in \textit{Messages}$: a message
\item $p \in \textit{Protocols}$: a protocol
\end{itemize}

\begin{equation}
initiates(m,p) \equiv (m_\phi = p_\phi \vee m_\phi = \bot) \wedge (\exists t \in p_T : p_\iota = t_\sigma \wedge triggers(m,t,\emptyset))
\end{equation}

If the message has a defined protocol identifier ($m_\phi$), it may only be matched against a protocol if this is the same as the protocol identifier of the protocol itself ($p_\phi$). If this is the case, the message may initiate the protocol if the protocol contains a transition that begins at the protocol's initial state ($p_\iota$) and can be triggered by the message. The `bindings' argument to \textit{triggers} is the empty set in this case, as there are no bindings yet associated with the conversation that is to be initiated.

\subsection{Advances}

Given a specific conversation, the \textit{advances} predicate indicates whether a given message is capable of advancing that conversation to a different state. This involves checking the transitions of the underlying protocol to find one that is active and can be triggered by the message.

Arguments:
\begin{itemize}
\item $m \in Messages$: the message to be checked.
\item $c \in Conversations$: a conversation against which the message is checked.
\end{itemize}

\begin{equation}
advances(m,c) \equiv (m_c = \bot \vee m_c = c_c) \wedge (\exists t : t \in p_T \wedge c_s = t_\sigma \wedge triggers(m,t,c_B))
\end{equation}
where $p \in P \wedge p_\phi = c_\phi$

A message advances a conversation if a transition exists for which all of the following criteria are met:
\begin{itemize}
   \item If the message contains a conversation identifier ($m_c$), it is the same as that of the conversation ($c_c$) 
   \item The transition ($t$) is part of the protocol that the conversation is following. This is indicated by the transition being contained within the protocol's set of transitions ($p_T$). The protocol is identified as that whose protocol identifier ($p_\phi$) matches that of the conversation ($c_\phi$).
   \item The conversation's current state ($c_s$) is the same as the transition's start state ($t_\sigma$).
   \item The message triggers the transition, given the conversation's current bindings ($c_B$).
\end{itemize}

\section{Functions}

In addition to the predicates defined above, a number of functions (that do not evaluate to boolean values) must also be defined to facilitate ACRE's conversation management.

\subsection{Head}
The \textit{head} function is a standard list operation that returns the first element of a list. For an empty list, this will return the undefined value $\bot$.

\begin{equation}
head(L) = \left\{ \begin{array}{ll}
   i_1 & n > 0 \\
   \bot & n = 0
   \end{array} \right.
\end{equation}
where $L = [i_1,\ldots, i_n]$

\subsection{Tail}
The \textit{tail} function is a standard list operation that returns all elements in a list other than the first element. If the list already has fewer than two elements, an empty list will be returned.
\begin{equation}
tail(L) = \left\{ \begin{array}{ll}
   [i_2,\ldots, i_n] & n > 1 \\
   \left[ \right] & n \le 1
   \end{array} \right.
\end{equation}

where $L = [i_1,\ldots, i_n]$

\subsection{Append}
The \textit{append} function is used to append an element $e$ to the end of a list $L$.

\begin{equation}
append(L,e) = [i_1,\ldots,i_n,e]
\end{equation}
where $L = [i_1,\ldots,i_n]$

\subsection{New Conversation}

The \textit{newConversation} function is used to create a new conversation. As its arguments, it takes a message that initiates a conversation ($m$) and the protocol that the conversation should follow ($p$).

Arguments:
\begin{itemize}
\item $m \in \textit{Messages}$: a message used to initiate the conversation.
\item $p \in \textit{Protocols}: initiates(m,p)$: a protocol for which $m$ is capable of beginning a conversation.
\end{itemize}
  
\begin{equation}
newConversation(m,p) = (p_\phi,\{m_s,m_r\},p_\iota,[],c,\emptyset,active)
\end{equation}
where:
\begin{equation*}
c = \left\{ \begin{array}{ll}
   m_c & m_c \ne \bot \\
   nextid() & m_c = \bot
\end{array} \right.
\end{equation*}
where $nextid()$ is a function that generates a unique conversation identifier

This function returns a tuple representing a conversation. The protocol identifier is the same as that of the protocol ($p_\phi$). The participants in the conversation are the sender and recipient of the initial message ($m_s$ and $m_r$ respectively). The conversation's state is the initial state of the protocol ($p_\iota$).

Initially, the message history relating to a conversation is an empty list, as the conversation has yet to be advanced from its initial state (this occurs via the \textit{advance} function defined below). 

For the conversation identifier, a function named \textit{nextid()} is assumed. This will create a unique identifier for a conversation in the event that a conversation identifier is not specified already in the message initiating the conversation. If the message does contain a conversation identifier ($m_c$), this is used as the identifier for the new conversation ($c$). c

A new conversation does not yet have any bindings associated with it, and is in the \textit{active} state.

% Not used in new operational semantics
% \subsection{Trigger}
% 
% Arguments:
% \begin{itemize}
% \item $m \in \textit{Messages}$: A message
% \item $t \in \textit{Transitions}$: A transition
% \item $B \subset \textit{Bindings}$: A set of bindings
% \end{itemize}
% 
% \begin{equation}
% \begin{split}
% trigger(m,t,B) = & bindings( m_s, apply( t_s,B ) ) \\
% & \cup bindings(m_r, apply( t_r,B' ) ) \\
% & \cup bindings( m_x, apply( t_x, B'')
% \end{split}
% \end{equation}
% where $B'= B \cup bindings( m_s, apply( t_s,B ) )$ \\
% and $B'' = B' \cup bindings(m_r, apply( t_r,B' ) )$
% 
% Triggering a transition takes place when a message is matched against it. If the transition contains variables in its fields, these must be bound against the values contained in the message. This function equates to the bindings that are created during this process (inclusive of the set of bindings that are provided as an argument to the function). The bindings in question relate to:
% \begin{itemize}
% \item The sender of the message
% \item The recipient of the message
% \item The content of the message
% \end{itemize}
% In each case, the transition's values for each of these fields firstly has any existing bindings applied, so as not to bind variables that have already been bound. This function should only be used in cases where $triggers(m,t,B)$ is true.

\subsection{Combine} \label{sec:semantics:combine}

The \textit{combine} function is designed to merge two sets of bindings. The second set of bindings takes precedence, in that if there are any variable/value pairs in the two sets that share the same variable, it is the value from the second set of bindings that is used in the combined result. Thus, this function can be considered to add the bindings of $B'$ to $B$, overriding any bindings for common variables.

Arguments:
\begin{itemize}
   \item $B \in \textit{Bindings}$
   \item $B' \in \textit{Bindings}$
\end{itemize}

\begin{equation}
combine(B,B') = \{ (v,c) \in B : (v,c') \notin B'\} \cup B'
\end{equation}

Support for overriding bindings is included because of the possibility of variables used in a mutable context. This is particularly important when a variable used in mutable context already has a previous value bound to it. In this situation, it is free to acquire a new binding that will override the previous value. In practice, the \textit{combine} function will be used with the older bindings as the first argument and new bindings as the second, thus causing the older binding to be overridden. Because of how the \textit{tapply} function (discussed in Section~\ref{sec:semantics:tapply}) is defined, a variable used in an immutable context will be replaced by its bound value before any matching occurs, which means that it cannot cause a new binding to arise that will override a previous bound value.

\subsection{Generating Bindings}

The following sections define three functions for generating a set of bindings after comparing two terms, predicates or lists. The \textit{tbind} function applies to terms, \textit{lbind} applies to lists, \textit{pbind} is for predicates.

\subsubsection{TBind} \label{sec:semantics:tbind}

The \textit{tbind} function compares two terms (one of which is grounded) and returns the bindings that arise.

Arguments:
\begin{itemize}
\item $t \in \Theta_{ACRE}$
\item $g \in \Gamma_{ACRE}$
\end{itemize}

% Original version
%\begin{equation}
%tbind(t,g) = \left\{ \begin{array}{ll}
%\emptyset & \neg tmatches(t,g) \\
%(t_v,g) & t \in VarInst \\
%lbind([t_1,\ldots,t_n],[u_1,\ldots,u_n]) & l = f(t_1,\ldots,t_n) \\
%& \wedge~g = f(u_1,\ldots,u_n) \\
%& \wedge lmatches([t_1,\ldots,t_n],[u_1,\ldots,u_n]) \\
%
%\end{array} \right.
%\end{equation}

\begin{equation} \label{eqn:semantics:tbind}
tbind(t,g) = \left\{ \begin{array}{ll}
(t_v,g) & t \in VarInst \wedge t_v \ne \bot \\
lbind([t_1,\ldots,t_n],[u_1,\ldots,u_n]) & tmatches(t,g) \\
& \wedge~l = f(t_1,\ldots,t_n) \\
& \wedge~g = f(u_1,\ldots,u_n) \\
\emptyset & otherwise \\
\end{array} \right.
\end{equation}

If the first argument $t$ is a variable instance, a single binding of that variable to the other argument $g$ is returned. In practice, this will only arise where the variable has either been used in a mutable context or where it has no value previously bound to it. For a variable used in an immutable context that previously had a bound variable, it will have been replaced with that bound value by means of the \textit{tapply} function (discussed in Section~\ref{sec:semantics:tapply}) prior to this function being used. Thus it will not come to \textit{tbind} as a variable and will instead be handled by one of the other cases.

A special case is when the anonymous variable is used. This is modelled by the variable name being undefined (i.e. $t_v = \bot$). As discussed in Section~\ref{sec:acre-introduction:anonymous-variable}, the anonymous variable does not acquire any bindings.

For functions, \textit{lbind} is used to generate bindings between the functions' respective arguments (when converted to lists). Constant terms or terms that do not match will not produce any bindings and are handled by the third case in Equation~\ref{eqn:semantics:tbind}.

\subsubsection{LBind}

The \textit{lbind} function is intended to generate the bindings that arise when comparing two lists. In common with similar functions, the first argument is a list of terms (elements of $\Theta_{ACRE}$) whereas the second argument consists of grounded terms (elements of $\Gamma_{ACRE}$).

Arguments:
\begin{itemize}
   \item $[t_1,\ldots,t_n] : \langle\forall i : 1 \le i \le n : t_i \in \Theta_{ACRE}\rangle$
   \item $[u_1,\ldots,u_n] : \langle\forall i : 1 \le i \le n : u_i \in \Gamma_{ACRE}\rangle$
\end{itemize}

\begin{equation}
lbind([t_1,\ldots,t_n],[u_1,\ldots,u_n]) = \left\{ \begin{array}{ll}
\emptyset & n = 0 \\
combine(B, tbind(tapply(B,t_n),u_n)) & n > 0 \\
\end{array} \right.
\end{equation}
where $B = lbind([t_1,\ldots,t_{n-1}],[u_1,\ldots,u_{n-1}])$

No bindings are generated when two empty lists are compared. For non-empty lists, the function acts recursively. It first calls \textit{lbind} on the lists up to element $n-1$ (where $n$ is the length of each list). These bindings are then applied to element $n$ and combined with any bindings that arise from comparing the terms at position $n$ in the lists.

Notably, \textit{tapply} is used to replace terms with their bound values prior to attempting to generate new bindings. As discussed in Section~\ref{sec:semantics:tapply}, this will replace any variables used in an immutable context with their previous bindings, if these exist. Hence these are not capable of generating new bindings when passed to \textit{tbind}, as they have been replaced with functions or constant values. A variable used in a mutable context (or that has no previous binding associated with it) will not be replaced by \textit{tapply} and as such is free to acquire a new bound value.

It is important to note that this function does not check to see if the two lists match. As such, undesirable results may arise if \textit{lmatches(t,g)} is not evaluated prior to attempting to make use of the bindings between them.

\subsubsection{PBind}

The \textit{pbind} function generates a set of bindings from comparing two predicates. As with similar functions, the second argument may only contain grounded terms as its arguments.

Arguments:
\begin{itemize}
\item $l \in L_{ACRE}$
\item $g \in G_{ACRE}$
\end{itemize}

\begin{equation}
pbind(l,g) = \left\{ \begin{array}{ll}
lbind([t_1,\ldots,t_n],[u_1,\ldots,u_n]) & pmatches(l,g) \\
& \wedge~l = p(t_1,\ldots,t_n) \\
& \wedge~g = p(u_1,\ldots,u_n) \\
\emptyset & otherwise
\end{array} \right.
\end{equation}

Where the two predicates have the same predicate identifier ($p$) and the same number of arguments, the bindings are generated by using the \textit{lbind} function to bind the argument lists. Where this is not the case, no bindings are generated.

\subsection{Applying Bindings} \label{sec:semantics:applying}

Applying bindings to a term or predicate is the process of replacing variables with the values to which they are bound. Two functions are defined for this process: \textit{tapply} for applying bindings to terms and \textit{papply} to apply bindings to predicates.

% We don't ever use LApply
%\subsubsection{LApply}

% Arguments:
% \begin{itemize}
%    \item $[t_1,\ldots,t_n] : \langle\forall i : 1 \le i \le n : t_i \in \Theta_{ACRE}\rangle$
%    \item $B \in \textit{Bindings}$
% \end{itemize}

% \begin{equation}
% lapply([t_1,\ldots,t_n],B) = [u_1,\ldots,u_n]
% \end{equation}
% where $\langle\forall i : 1 \le i \le n : u_i = tapply(t_1,B)\rangle$

\subsubsection{TApply} \label{sec:semantics:tapply}

The \textit{tapply} function is to apply a set of bindings to a single term (i.e. an element of $\Theta_{ACRE}$). The result is the same term, with any variable arguments replaced by their associated value from the set of bindings, if one exists and if the variable was used in an immutable context.

Arguments:
\begin{itemize}
   \item $t \in \Theta_{ACRE}$
   \item $B \in \textit{Bindings}$
\end{itemize}

There are three possible outcomes when using the \textit{tapply} function, each of which is reflected in Equation~\ref{eqn:semantics:tapply}.
\begin{enumerate}
   \item If the term $t$ is a function, the functor is left unchanged, while the \textit{tapply} function is applied to each of the function's arguments.
   \item The treatment of variable instances is key to the operation of mutable and immutable variable contexts. Clearly, a variable cannot be replaced by a bound value if none yet exists, thus it is a requirement that $B$ contains a binding that relates to the variable instance in question. When such a binding exists, however, it is only used when the variable was used in an immutable context. In this situation, the variable should only be capable of matching the value against which it was previously bound, so it is replaced by this value before any matching takes place. For a variable used in a mutable context, it is not required to match its previous value. This means that a mutable-context variable is not replaced with its bound value, as it remains free to match against any value. This matching is encapsulated by the complementary \textit{tmatches} predicate discussed in Section~\ref{sec:semantics:tmatches}.
   \item The final situation applies for anything other than the previous two cases. This includes any variables used in mutable context, as discussed above. It also includes variables used in an immutable context for which bindings have not yet been created, in addition to constant values. In each of these cases, no change is made and the term is returned unchanged. 
\end{enumerate}

\begin{equation} \label{eqn:semantics:tapply}
tapply(t,B) = \left\{ \begin{array}{ll}
f(u_1,\ldots,u_n) & t = f(t_1,\ldots,t_n) \\
v & t \in VarInst \wedge t_c = \textit{immutable} \wedge \exists(t_v,v) \in B \\
t & otherwise
\end{array} \right.
\end{equation}
where $\langle\forall i : 1 \le i \le n : u_i = tapply(t_i,B)\rangle$

% Arguments:
% \begin{itemize}
% \item $s \in \Theta_{ACRE}$
% \item $B \subset \textit{Bindings}$: a set of bindings
% \end{itemize}

% \begin{equation}
% apply(s,B) = \left\{ \begin{array}{ll}
%  v 				& \mbox{ $s \in Var \wedge \neg mutable(s) \wedge \exists (s,v) \in B$ }	\\
%  f(u_1,\ldots,u_n) & \mbox{ $s \in Func \wedge f(t_1,...,t_n) = s $}			\\
% 				& \mbox{ $\wedge~u_i = apply(t_i, B)$ } 			\\
% p(u_1,\ldots,u_n)	& \mbox{ $s \in Pred \wedge p(t_1,...,t_n) = s $}			\\
% 				& \mbox{ $\wedge~u_i = apply(t_i, B)$ }				\\
% s 				& \mbox{ $otherwise$ }
% \end{array} \right.
% \end{equation}

% If $s$ is a variable and $B$ contains a binding for $s$, return the binding for $s$.
% If $s$ is either a function or a predicate then return that function or predicate, with the bindings of $B$ applied to its arguments.

\subsubsection{PApply}

This function applies a set of bindings to a predicate, replacing any variables amongst its arguments with their associated values in the set of bindings, if one exists.

Arguments:
\begin{itemize}
   \item $l \in L_{ACRE}$
   \item $B \in \textit{Bindings}$
\end{itemize}

\begin{equation}
papply(l,B) = p(u_1,\ldots,u_n)
\end{equation}
where $p(t_1,\ldots,t_n) = l$ \\
and $\langle\forall i : 1 \le i \le n : u_i = tapply(t_i,B)\rangle$

The return value of this function is the same predicate, with \textit{tapply} applied to its arguments.

\subsection{Advance} \label{sec:semantics:advance}

The \textit{advance} function is used to advance the state of a specified conversation in response to a given message. The protocol that the conversation is following must also be provided as an argument.

Arguments:
\begin{itemize}
   \item $m \in Messages$
   \item $c \in Conversations$
   \item $p \in Protocols$
\end{itemize}

\begin{equation}
advance(m,c,p) = (c_\phi,c_A,t_\epsilon,append(c_H,m),c_c,B'',\psi)
\end{equation}
where $t \in p_T \wedge t_\sigma = c_s \wedge triggers(m,t,c_B)$ \\
and $B = combine(c_B,tbind(tapply(t_s,c_B),m_s))$ \\
and $B' = combine(B,tbind(tapply(t_r,B),m_r))$ \\
and $B'' = combine( B', pbind(papply(t_x,B'),m_x))$ \\
and $\psi =  \left\{ \begin{array}{ll}
   completed & t_\epsilon \in p_F \\
   active & otherwise
   \end{array} \right.$

When advancing an existing conversation, the conversation's existing protocol identifier ($c_\phi$), participants ($c_A$) and conversation identifier ($c_c$) are unchanged. Additionally, the conversation's status will always be \textit{active} after a successful advancement, unless the conversation has come to an end. This occurs when the new conversation state $t_\epsilon$ is a final state according to the underlying protocol ($p_F$).

In order to ascertain the new current state of the conversation, a transition must be identified that begins at the conversation's current state (i.e. $t_\sigma=c_s$) and is capable of being triggered by the message, given the conversation's current set of bindings ($c_B$). The current state of the conversation after it is advanced is the end state of the triggered transition ($t_\epsilon$).

The bindings associated with the conversation after it has been advanced is a result of comparing several aspects of the message against the transition that is triggered. The bindings are generated in the following sequence:
\begin{enumerate}
   \item The existing conversation bindings ($c_B$) are applied to the `sender' field of the transition $t$. Any further bindings arising from comparing this to the message sender are added to these bindings to be applied in the next step.
   \item The bindings from the above step are applied to the `recipient' field of the triggered transition before it is compared with the message recipient. Again, any further bindings from this comparison are included for the next stage.
   \item Finally, the bindings accumulated thus far are applied to the `content' field of the transition. This is then compared against the message content in order to generate further bindings. All of these bindings are then part of $B''$, which becomes the current conversation bindings after it has been advanced.
\end{enumerate} 

It is important to note that this function does not check if the conversation in question is capable of being advanced by the given message. As such, it is necessary to check that \textit{advances(m,c)} holds true before applying this function.

\section{Operational Semantics} \label{sec:semantics:semantics}

\begin{figure}[!ht]
\centering
\includegraphics[width=10.0cm]{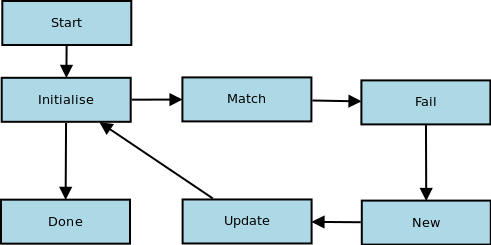}
\caption{Stages in the conversation reasoning process.}
\label{fig:semantics:stages}
\end{figure}

This Section describes the operational semantics of the Conversation Manager that forms part of each agent. Figure~\ref{fig:semantics:stages} shows the stages that the conversation reasoning processes goes through on each iteration of the agent interpreter. The Conversation Manager begins each iteration in the \textit{Start} state and ends in the \textit{Done} state, having iterated through all the messages then available. The Conversation Manager does not itself begin its next iteration: this is left to the agent itself, in accordance with the policy of the agent interpreter and scheduler.

In the following sections, the Conversation Manager is represented by the tuple $(n,M,C,P,E,s,\mu)$, as described in Section~\ref{sec:semantics:conversation-manager}. For the purposes of these formal semantics, it is assumed that the Conversation Manager has separately been made aware of the name of the agent to which it belongs ($n$) and the set of Protocols of which the agent is aware ($P$). Additionally, as the agent perceives incoming messages and decides to send outgoing messages, these are added to the message queue ($M$).

\subsection{Start}

The \textit{Start} state is the initial state in the process and is shown in Equation~\ref{eqn:semantics:start}. The principal function of this state is to prepare the Conversation Manager for the reasoning process it must undertake. This is done by initially emptying the event set $E$ so that by the end of the process, all events in the set have been triggered on this iteration of the agent interpreter cycle.

It is assumed that any events remaining in the event set from the previous interpreter cycle can be safely discarded, as they will have been read by an external entity following that iteration.

\begin{equation} \label{eqn:semantics:start}
\frac{true}{(n,M,C,P,E,start,\mu) \longrightarrow (n,M,C,P,E',initialise,\mu)}
\end{equation}
where $E' = \emptyset$

At this stage, emptying this set is the only work to be done. No processing of a message has yet occurred, meaning that no changes to the active conversations or other elements has taken place at this stage. The Conversation Manager may now enter the next state: \textit{initialise}.

\subsection{Initialise}

The \textit{initialise} state, shown in Equation~\ref{eqn:semantics:initialise}, is where a message is removed from the message queue so that it can be processed by the Conversation Manager. In the case where there is at least one message in the message queue (\textit{M}), this is removed and moved into the memory of the Conversation Manager ($\mu$). In this situation, some further processing will be required in order to match this message to an appropriate conversation (or to create an appropriate new conversation). Thus the state of the Conversation Manager changes to \textit{match} in this situation.

\begin{equation} \label{eqn:semantics:initialise}
\frac{head(M) \ne \bot}{(n,M,C,P,E,initialise,\mu) \longrightarrow (n,M',C,P,E,match,\mu')}
\end{equation}
where $M' = tail(M)$ \\
and $\mu' = (head(M), \emptyset)$

If there are no messages in the message queue, the Conversation Manager has no further processing to do on this iteration of the agent, so its state changes to \textit{done}. This situation may occur either because no messages were sent or received since the last iteration of the agent, or because the Conversation Manager has already processed all the messages that were in the queue. In the latter situation, it will have entered this \textit{initialise} state via the \textit{update} state.

\begin{equation}
\frac{head(M) = \bot}{(n,M,C,P,E,initialise,\mu) \longrightarrow (n,M,C,P,E,done,\mu')}
\end{equation}
where $\mu' = (\bot,\emptyset)$ 

\subsection{Match} \label{sec:semantics:match}

The \textit{match} state is designed to attempt to match the message that was previously stored in memory against the conversations that are already active (stored in $C$). Any conversations that can be advanced by the message in question are considered to be candidate conversations and are stored in the memory of the Conversation Manager ($\mu_C$). Further reasoning will take place later in the \textit{update} state (discussed in Section~\ref{sec:semantics:update}) to ensure that the message is matched against the correct conversation.

\begin{equation}
\frac{true}{(n,M,C,P,E,match,\mu) \longrightarrow (n,M,C,P,E,fail,\mu')}
\end{equation}
where $\mu' = (\mu_m,\{c \in C : advances(\mu_m,c)\})$

Once this stage is complete, the Conversation Manager enters the \textit{fail} state.

\subsection{Fail}

The \textit{fail} state checks for failed conversations. A conversation is considered to have failed if its conversation identifier is specified in the message, but the message cannot advance that conversation because it does not match against any active transition. Either one of two situations may occur in this state.

In the first situation, an active conversation $c$ exists (in the set of active conversations $C$) that can be considered to have failed, as the message $m$ is incapable of advancing it, despite specifying its conversation identifier (i.e. $m_c = c_c$). When this occurs, the status of the conversation must be changed to \textit{failed}. An event to indicate the failure of conversation $c$ is added to the event set ($E$) so that the agent will become aware of the failure of the conversation. The Conversation Manager moves to be \textit{new} state.

\begin{equation}
\frac{\exists c \in C : m_c = c_c \wedge \neg advances(m,c)}{(n,M,C,P,E,fail,\mu) \longrightarrow (n,M,C',P,E',new,\mu)}
\end{equation}
where $m = \mu_m$ \\
and $C' = (C \setminus \{c\}) \cup \{c'\}$ \\
and $c' = (c_\phi,c_A,c_s,c_H,c_c,c_B,failed)$ \\
and $E' = E \cup \{(failed,c)\}$

The second situation arises where no failed conversations are identified. This will occur for one of two reasons:
\begin{enumerate}
   \item The message does not specify any conversation identifier (i.e. $m_c = \bot$). A message can only cause a conversation to fail if it explicitly references its conversation identifier.
   \item Each active conversation (in $C$) either has a different conversation identifier to that of the message, or is capable of being advanced by the message. In the latter case, this can only happen when the conversation identifiers of the conversation and the message are equal.
\end{enumerate}

In this situation, the Conversation Manager's state changes to \textit{new}, with no other changes being made. In the case where a conversation is identified by the identifier contained in the message and can be advanced by it, that conversation will previously have been added to the conversation manager's memory as a candidate for advancement during the \textit{match} state. Thus no action with this conversation is necessary at this stage.

\begin{equation}
\frac{m_c = \bot \vee (\forall c \in C : m_c \ne c_c \vee  advances(m,c))}{(n,M,C,P,E,fail,\mu) \longrightarrow (n,M,C,P,E,new,\mu)}
\end{equation}
where $m = \mu_m$

\subsection{New}

In the \textit{new} state, the Conversation Manager checks for new conversations that could be started by the message being processed. Because conversation identifiers must be unique, this does not happen if the message contained a conversation identifier that uniquely identifies a conversation that is already active. For the purposes of creating new conversations, it is not important whether or not the message is capable of advancing its matching active conversation. The fact that the identifiers match is considered to be a clear indication that the intent of the message is to advance that message. If it cannot do so, that conversation will have been considered to have failed during the \textit{fail} state. In either case, the message will not be used to create a new conversation and no change is made to the Conversation Manager other than to move it to the \textit{update} state.

A message will only generate a new conversation if it is capable of initiating a conversation that follows a known protocol. Moving the Conversation Manager to the \textit{update} state is also the only change required if there is no protocol that the message is capable of initiating.

Equation~\ref{eqn:semantics:new2} applies to both of the above situations, where either the message has a conversation identifier that matches an active conversation or where it cannot initiate a conversation that follows any known protocol. The only alteration is the change of state in the Conversation Manager.

\begin{equation} \label{eqn:semantics:new2}
\frac{(\exists c \in C : m_c = c_c) \vee (\neg \exists p \in P : initiates(\mu_m, p))}{(n,M,C,P,E,new,\mu) \longrightarrow (n,M,C,P,E,update,\mu)}
\end{equation}

% Original thesis submitted version
%\begin{equation} \label{eqn:semantics:new2}
%\frac{(m_c \ne \bot \wedge |\mu_c|>0 ) \vee \neg \exists p \in P : initiates(\mu_m, p)}{(n,M,C,P,E,new,\mu) \longrightarrow (n,M,C,P,E,update,\mu)}
%\end{equation}

% new conversation only added to candidate conversations
% can't make it active unless we know it's unambiguous

% This is an older version that will check for new protocols even if the message has a conversation id that has matched an active conversation.
%\begin{equation}
%\frac{\exists p \in P : initiates(\mu_m, p)}{(n,M,C,P,E,new,\mu) \longrightarrow (n,M,C,P,E,update,\mu')}
%\end{equation}
%where $\mu' = (\mu_m, \mu_C \cup \{c'\})$ \\
%and $c'= newConversation(\mu_m,p)$

Two conditions must be satisfied in order for a message to be capable of beginning a new conversation. Firstly, it must not contain a conversation identifier that matches an active conversation, as discussed above. This condition also takes into account the situation where the message does not contain any conversation identifier, as all active conversations must contain a unique identifier meaning that no match can occur. Secondly, the message must be capable of initiating a conversation that follows some known protocol (in $P$).

When a new conversation is to be initialised, it is created (using the \textit{newConversation} function) and added to the Conversation Manager's memory ($\mu$). At this stage, the new conversation is not added to the set of active conversations ($C$), as further reasoning is required before the Conversation Manager can activate the conversation. This is done in the \textit{update} state, to which the Conversation Manager now moves.

\begin{equation}
\frac{(\neg \exists c \in C : m_c = c_c) \wedge (\exists p \in P : initiates(\mu_m, p))}{(n,M,C,P,E,new,\mu) \longrightarrow (n,M,C,P,E,update,\mu')}
\end{equation}
where $\mu' = (\mu_m, \mu_C \cup \{c'\})$ \\
and $c'= newConversation(\mu_m,p)$

The reason why the new conversation is not yet activated is because of the possibility of messages not specifying a conversation identifier. When this occurs, a message may previously have been matched against one or more active conversations during the \textit{match} state discussed in Section~\ref{sec:semantics:match}. However, if this message is also capable of initiating a new conversation, it is unclear what the intent of the message is and so it will be considered to be an ambiguous message during the \textit{update} state discussed in Section~\ref{sec:semantics:update}.

% Version submitted in first thesis submission
%\begin{equation}
%\frac{(m_c = \bot \vee |\mu_c|=0 ) \wedge \exists p \in P : initiates(\mu_m, p)}{(n,M,C,P,E,new,\mu) \longrightarrow (n,M,C,P,E,update,\mu')}
%\end{equation}
%where $\mu' = (\mu_m, \mu_C \cup \{c'\})$ \\
%and $c'= newConversation(\mu_m,p)$

\subsection{Update} \label{sec:semantics:update}

When in the \textit{update} state, the active conversations must be updated to reflect any changes that may have occurred as a result of sending or receiving this message. 

In this state, one of three situations may have occurred, depending on how many conversations could be matched against the message. In each situation, appropriate events must be added to the event set $E$ so that the agent to which the Conversation Manager is attached can gain knowledge about the status of its communication. The types of event that can be raised are the same as those previously discussed in Section~\ref{sec:acre-introduction:advancing}.

The first case is where the message was matched to exactly one conversation. Here, the existing conversation ($c$) is first removed from the list of active conversations ($C$). If this is a new conversation that is not already in $C$, this will have no effect. Once the conversation has been advanced (creating $c'$), it is added back to $C$ .

Three events are available in this scenario: an `advanced' event will always be raised when a conversation is successfully advanced, a `completed' event indicates that the conversation was advanced to a final state and a `started' event arises when the conversation was previously in its initial state (i.e. this is the first time it has been advanced).

Multiple events may be raised, depending on the state of the conversation. For any protocol, a `started' or `completed' event will be accompanied by an `advanced' event. All three may arise for the the simplest protocol consisting of only two states and one transition. Here, the conversation may be begun, advanced and finished by the same message. 

\begin{equation}
\frac{|\mu_C| = 1}{(n,M,C,P,E,update,\mu) \longrightarrow (n,M,C',P,E',initialise,\mu)}
\end{equation}
where $E' = E \cup E''$ \\
and $ (completed,c') \in E'' \Leftarrow  c'_s \in p_F$ \\
and $(advanced, c') \in E''$ \\
and $(started, c') \in E'' \Leftarrow c_s = p_\iota$ \\
and $p \in P : p_\phi = c_\phi$ \\ % was originally p_\psi = c_\psi but that doesn't make sense.
and $c \in \mu_C$ \\
and $C' = (C \setminus \{c\} ) \cup \{ c' \}$ \\
and $c' = advance(\mu_m,c,p)$ \\

In the second case, the message has been matched with multiple candidate conversations. This situation can only occur where the message does not contain a conversation identifier, as conversation identifiers must be unique. Because it is impossible to be certain as to which conversation the message was intended to relate to, or whether it was instead intended to begin a new conversation, no alteration is made to any active conversation and an `ambiguous' event is raised to indicate that this situation has occurred.

\begin{equation}
\frac{|\mu_C| > 1}{(n,M,C,P,E,update,\mu) \longrightarrow (n,M,C,P,E',initialise,\mu)}
\end{equation}
where $E' = E \cup \{(ambiguous,\mu_m)\}$

The final situation to be handled is where no conversation was found that could advanced by the message in question and where the message was also incapable of initiating a new conversation that followed a known protocol. In this case, an `unmatched' event is raised to inform the agent of this. Unlike an ambiguous message, a message that does contain a specific conversation identifier may still be unmatched, either because of a previous problem with that conversation, or because the details of the message itself do not correctly match the specified protocol definition.

\begin{equation}
\frac{|\mu_C| = 0}{(n,M,C,P,E,update,\mu) \longrightarrow (n,M,C,P,E',initialise,\mu)}
\end{equation}
where $E' = E \cup \{(unmatched,\mu_m)\}$

\subsection{Done}
The \textit{done} state indicates the end of the reasoning process. At this point, the Conversation Manager will perform no further tasks until the agent resets it to the \textit{start} state on its next iteration.

\begin{figure}[!htb]
	\centering
	\includegraphics[scale=0.5]{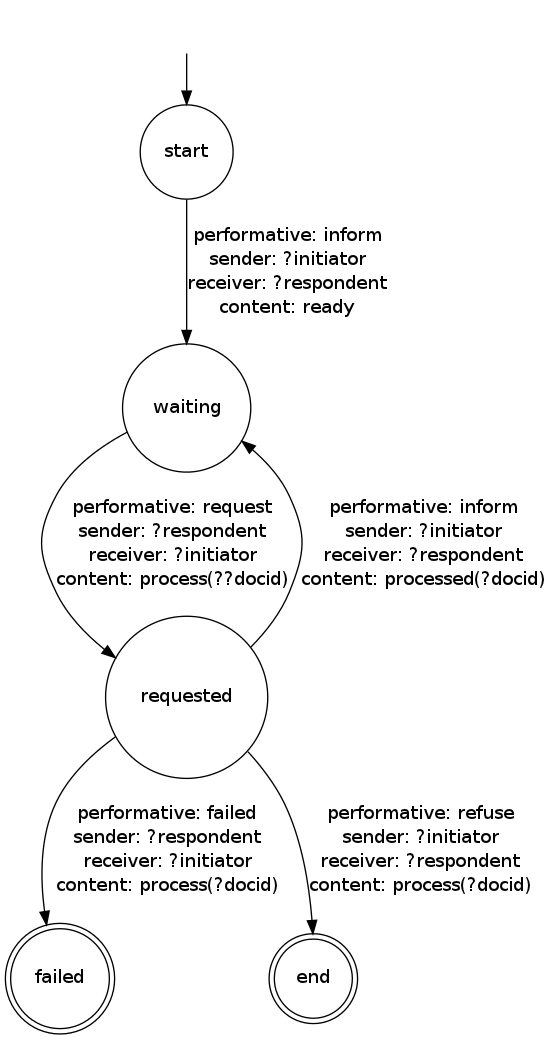}
	\caption{Process Documents Protocol.}
	\label{fig:semantics:processdocuments}
\end{figure}

\section{Example}
This section presents a slightly modified version of the Process Documents protocol originally presented the example in Section~\ref{sec:acre-introduction:process-documents}. Similar messages are used as in that example, with the formal semantics being used to show how the conversation is modelled by ACRE.

This example uses a sample conversation between two agents (named \textit{agent1} and \textit{agent2}), of which one supports conversation management and the other does not. The lack of support for conversation management means that \textit{agent2} sends messages that do not contain \texttt{conversation-id} or \texttt{protocol} parameters.

The protocol used is illustrated in Figure~\ref{fig:semantics:processdocuments}.

\subsection{Modelling}

Firstly, the protocol is given a unique identifier so that it can be referred to by the various elements within the conversation. In this case, we use \texttt{proc\_docs} as the unique protocol identifier.

The \texttt{proc\_docs} protocol consists of five states, which are stored in a set named $S$. These are defined as the following 2-tuples, in accordance with the definition of a state found in Section~\ref{sec:semantics:state}.

\begin{math}
S = \{(start,proc\_docs),(waiting,proc\_docs),(requested,proc\_docs),\\\hspace*{0.5cm}(end,proc\_docs),(failed,proc\_docs)\}
\end{math}

Having modelled the protocol's states, the transitions must now be modelled according to the definition set out in Section~\ref{sec:semantics:transition}. The set of states to be used in the protocol is named $T$ and is defined as follows:

\begin{math}
T = \{ \\
\hspace*{0.5cm}(proc\_docs,start,waiting,(?initiator,immutable),\\
\hspace*{1cm}(?respondent,immutable),inform,ready), \\
\hspace*{0.5cm}(proc\_docs,waiting,requested,(?respondent,immutable),\\
\hspace*{1cm}(?initiator,immutable),request,process((?docid,mutable))),\\
\hspace*{0.5cm}(proc\_docs,requested,waiting,(?initiator,immutable),\\\hspace*{1cm}(?respondent,immutable),inform,processed((?docid,immutable))),\\
\hspace*{0.5cm}(proc\_docs,requested,end,(?respondent,immutable),\\\hspace*{1cm}(?initiator,immutable),refuse,process((?docid,immutable))),\\
\hspace*{0.5cm}(proc\_docs,requested,failed,(?respondent,immutable),\\\hspace*{1cm}(?initiator,immutable),failure,process((?docid,immutable)))\\\}
\end{math}

The next stage is to model the protocol itself ($p$). This is done following the model set out in Section~\ref{sec:semantics:protocol}.

\begin{math}
p = (proc\_docs, S, T, start, \{end,failed\})
\end{math}

where $S$ and $T$ are the previously-defined sets of states and transitions respectively. This protocol's initial state is named \textit{start}, and ends if either the \textit{end} or \textit{failed} state is reached.

Finally, the conversation manager must be modelled, according to Section~\ref{sec:semantics:conversation-manager}. Initially, this is as follows:

\begin{math}
(agent1,[],\emptyset,\{p\},\emptyset,start,(\bot,\emptyset))
\end{math}

In this example, it is  assumed that the name of the parent agent is \textit{agent1} and that $p$ is the only protocol of which the agent is aware. Initially, the conversation manager has no messages in its queue, has processed no conversations, has raised no events and has not loaded anything into its memory. As such, all of these begin as empty sets, empty lists or undefined values.

\subsection{Start}
The \textit{start} state involves only the emptying of the conversation manager's event queue. On the first iteration, this is already empty and so it remains:

\begin{math}
(agent1,[],\emptyset,\{p\},\emptyset,start,(\bot,\emptyset))
\end{math}

\subsection{First Iteration}

\subsubsection{Initialise}
The \textit{initialise} state takes the first message in the message queue and adds it to memory (if such a message exists). The message queue is initially assumed to contain the following message (the \texttt{conversation-id} and \texttt{protocol} fields have been added for the purposes of illustrating the formal model): 

\begin{Verbatim}[samepage=true]
(inform
   :sender            agent1
   :receiver          agent2
   :content           ready
   :conversation-id   c1
   :protocol          proc_docs
)
\end{Verbatim}

For convenience, this message is referred to as $m_1$ and is represented in the formal model according to the definition in Section~\ref{sec:semantics:message} as follows:

\begin{math}
m_1=(agent1,agent2,c1,proc\_docs, inform,ready)
\end{math}

In the \textit{initialise} state, this message is removed from the message queue and loaded into the conversation manager's memory and the state is changed to \textit{match}. Thus the state of the conversation manager becomes:

\begin{math}
(agent1,[],\emptyset,\{p\},\emptyset,match,(m_1,\emptyset))
\end{math}

\subsubsection{Match and Fail}
The \textit{match} and \textit{fail} states both depend on the set of active conversations being non-empty. As this is not the case, the only effect these will have on this iteration is to change the state of the conversation manager. Thus at the end of the \textit{fail} state, the conversation manager is represented as follows:

\begin{math}
(agent1,[],\emptyset,\{p\},\emptyset,fail,(m_1,\emptyset))
\end{math}

\subsubsection{New}
The \textit{new} state is the phase where the conversation manager attempts to begin new conversations that are based on known protocols. It checks the message stored in memory against each known protocol in turn and checks each to see if the message is capable of initiating a new conversation that follows the protocol in question.

In this example, $p$ is the only protocol of which the conversation manager has knowledge, so a new conversation will be created if $initiates(m_1,p)$ evaluates to true.

One aspect of the \textit{initiates} predicate is to check that the protocol identifiers in the message and the protocol are equal (or that the message does not specify a particular protocol). In this case, both $m_1$ and $p$ have a protocol identifier of $proc\_docs$.

Additionally, it compares the message with all transitions that begin in the protocol's initial state. In this case, the only transition that satisfies this criterion is:

\begin{math}
t = (proc\_docs,start,waiting,(?initiator,immutable),\\(?respondent,immutable),inform,ready)
\end{math}

This must be compared with the message using the \textit{triggers} predicate. As this would be a new conversation, no bindings are yet in effect. Thus, $triggers(m_1,t,\emptyset)$ predicate in this situation evaluates to true, as follows:
\begin{itemize}
   \item The performative in the message is the same as the performative in the transition.
   \item The message sender is specified in the transition as a variable, which matches anything using the \textit{tmatches} predicate.
   \item Similarly, the transition's specification of the message receiver is also a variable, which also results in a match.
   \item Finally, the contents match, as they are the same predicate without any parameters.
\end{itemize}

This is illustrated in Table~\ref{tab:semantics:newtable}. In this table, ``Pre'' shows an element of the transition definition before any bindings are applied to it. ``Post'' refers to the same elements after applying bindings. ``Message'' is the same field as it appears in the message. ``Match'' indicates whether or not the post-bindings transition element matches the corresponding message element. Finally ``New'' shows any new bindings that arise from matching these.

\begin{table}[!hbt]
\centering
\caption{Initial message matched against available transition in the \textit{new} state.}
\begin{tabular}{|l|l|l|l|l|}

\hline
\multicolumn{5}{|l|}{\textbf{Transition}} \\
\hline
\multicolumn{5}{|c|}{$t = (proc\_docs,start,waiting,(?initiator,immutable),$} \\
\multicolumn{5}{|c|}{$(?respondent,immutable),inform,ready)$} \\
\hline
\textbf{Pre} & \textbf{Post} & \textbf{Message} & \textbf{Match} & \textbf{New} \\
\hline
inform & inform & inform & Yes & \\

?initiator & ?initiator & agent1 & Yes & ?initiator $\rightarrow$ agent1 \\

?respondent & ?respondent & agent2 &  Yes & ?respondent $\rightarrow$ agent2 \\

ready & ready & ready & Yes & \\
\hline
\end{tabular}
\label{tab:semantics:newtable}
\end{table}

As the message is capable of beginning a conversation that follows the \textit{proc\_docs} protocol, a new conversation is created and added to the conversation manager's memory. This conversation takes the following form, and is referred to as $c$ for convenience:

\begin{math}
c = (proc\_docs,\{agent1,agent2\},start,[],c1,\emptyset,active)
\end{math}

This conversation has the same protocol identifier as the protocol, has agents \textit{agent1} and \textit{agent2} as its participants, is initially in the protocol's \textit{start} state, has had no messages yet associated with it, can be identified by the conversation identifier \textit{c1} (taken from the message), has no initial bindings and is itself in an \textit{active} state.

Having created this new conversation, the conversation manager can progress to the \textit{update} state. At this point, it is modelled as follows:

\begin{math}
(agent1,[],\emptyset,\{p\},\emptyset,update,(m_1,\{c\}))
\end{math}

\subsubsection{Update}

The \textit{update} state examines any candidate conversations that have been identified during the previous states and advances a conversation if the message has been matched against one. In this case, only one candidate conversation is present in the conversation manager's memory, and so no event is raised to notify of either an ambiguous or unmatched message. The conversation $c$ is advanced using the \textit{advance} function specified in Section~\ref{sec:semantics:advance}.

The mechanism of the \textit{advance} function is similar to the \textit{triggers} predicate in terms of how the bindings are created, and so the procedure shown in Table~\ref{tab:semantics:newtable} is also relevant to this step. This creates a new representation of a conversation as it appears following the sending of the first message. This conversation is as follows:

\begin{math}
c' = (proc\_docs,\{agent1,agent2\},waiting,[m_1],c1,B,active)
\end{math} \\
where $B =\{(?initiator,agent1),(?respondent,agent2)\}$

Following this process, the conversation manager is now:

\begin{math}
(agent1,[],\{c'\},\{p\},\{(started,c'),(advanced,c')\},done,(m_1,c))
\end{math}

\subsection{Second Iteration}

On the second iteration of the agent, the Conversation Manager is again placed in the \textit{initialise} state. On this iteration, it is assumed that a reply has been received, as follows:

\begin{Verbatim}[samepage=true]
(request
   :sender    agent2
   :receiver  agent1
   :content   process(doc123)
)
\end{Verbatim}

This message has been sent by an agent that does not support conversation management and as such it lacks the \texttt{conversation-id} and \texttt{protocol} parameters. It is left to the Conversation Manager to match this message against the available protocols and active conversations.

\subsubsection{Start and Initialise}
As with the first iteration, the \textit{start} and \textit{initialise} states are concerned with initially emptying the Conversation Manager' event queue, and then loading the first message from the message list into its memory. In this iteration, the incoming message is modelled as $m_2$, which is defined as follows:

\begin{math}
m_2=(agent2,agent1,\bot,\bot,request,process(doc123))
\end{math}

In this instance, the elements representing the conversation and protocol identifiers contain undefined values ($\bot$). After the \textit{initialise} state, the Conversation Manager may be modelled as follows:

\begin{math}
(agent1,[],\{c\},\{p\},\emptyset,match,(m_2,\emptyset))
\end{math}

\subsubsection{Match}
In the \textit{match} state, the conversation manager examines the stored message and attempts to match it against existing active conversations. The conversation manager now contains one active conversation, against which the message must be compared. The set of candidate conversations in memory will include any conversation $c$ for which $advances(m_2,c)$ evaluates to true.

The \textit{advances} predicate compares the message against any of the protocol's transitions that begin at the current state of the conversation. In this case, there is only one transition that satisfies this criterion, namely:

\begin{math}
t = (proc\_docs,waiting,requested,(?respondent, immutable),\\\hspace*{1cm} (?initiator,immutable),request,process((?docid,mutable)))
\end{math}

This transition is compared to the message using the \textit{triggers} predicate, in the same way as in the first iteration. However, on this occasion, there are already bindings associated with the conversation. The process of matching the transition against the message is set out in Table~\ref{tab:semantics:matchtable}.

\begin{table}[!hbt]
\centering
\caption{Second message matched against available transition in the \textit{match} state.}
\begin{tabular}{|l|l|l|l|l|}
\hline
\multicolumn{5}{|l|}{\textbf{Transition}} \\
\hline
\multicolumn{5}{|c|}{$t = (proc\_docs,waiting,requested,(?respondent, immutable),$} \\ \multicolumn{5}{|c|}{$(?initiator,immutable),request,process((?docid,mutable)))$} \\
\hline
\textbf{Pre} & \textbf{Post} & \textbf{Message} & \textbf{Match} & \textbf{New} \\
\hline
request & request & request & Yes & \\

?respondent & agent2 & agent2 & Yes & \\

?initiator & agent1 & agent1 &  Yes &  \\

process(??docid) & process(??docid) & process(doc123) & Yes & ?docid \\

&&&& $\rightarrow$ doc123 \\
\hline
\end{tabular}
\label{tab:semantics:matchtable}
\end{table}

Following the first iteration, bindings had been created for the \texttt{?initiator} and \texttt{?respondent} variables, and so these are replaced with their bound values before a comparison is made to the message. Thus for this conversation, \texttt{?initiator} must always match against \texttt{agent1} (as it is always used in an immutable context) and \texttt{?respondent} may only match against \texttt{agent2}.

Since the transition can be triggered by the message, this conversation can be considered to be a candidate for advancement, and so is added to the Conversation Manager's memory as such. This results in the Conversation Manager at the end of the \textit{match} state being modelled as follows:

\begin{math}
(agent1,[],\{c\},\{p\},\emptyset,fail,(m_2,\{c\}))
\end{math}

\subsubsection{Fail}
In the \textit{fail} state, messages that include \textit{conversation-id} parameters are checked against any existing conversation with the same identifier and mark the conversation as having failed if the message is not capable of advancing it. Although the set of active conversations is now non-empty, the message being processed does not contain a \texttt{conversation-id} parameter, and so the \textit{fail} state will have no effect other than to transition the conversation manager to the next, \texttt{new}, state.

\subsubsection{New}

As the message does not contain a conversation identifier, the \textit{new} state will cause each protocol to be checked to see if the message is capable of initiating a conversation that follows that protocol. As there is only one loaded protocol, this procedure is identical to that in the first iteration. The process of matching $m_2$ to the transition is set out in Table~\ref{tab:semantics:newtable2}. However, on this occasion, the performative required by the transition does not match the performative in the message, and so this protocol is not a candidate match for this message.

\begin{table}[!hbt]
\centering
\caption{Second message matched against available transition in the \textit{new} state.}
\begin{center}
\begin{tabular}{|l|l|l|l|l|}
\hline
\multicolumn{5}{|l|}{\textbf{Transition}} \\
\hline
\multicolumn{5}{|c|}{$t = (proc\_docs,start,waiting,(?initiator,immutable),$} \\
\multicolumn{5}{|c|}{$(?respondent,immutable),inform,ready)$} \\
\hline
\textbf{Pre} & \textbf{Post} & \textbf{Message} & \textbf{Match} & \textbf{New} \\
\hline
inform & inform & request & No & \\
\hline
\end{tabular}
\label{tab:semantics:newtable2}
\end{center}
\end{table}

\subsubsection{Update}

In the \textit{update} state, there is once again only one conversation that matched the message. This conversation will be advanced, and appropriate events generated.

The message $m_2$ is added to the message history and the bindings are updated to reflect the fact that the \texttt{?docid} variable has gained a bound value.

\begin{math}
c' = (proc\_docs,\{agent1,agent2\},requested,[m_1,m_2],c1,B',active)
\end{math} \\
where $B' =\{(?initiator,agent1),(?respondent,agent2),(?docid,doc123)\}$

Following this process, the conversation manager is now modelled as follows:

\begin{math}
(agent1,[],\{c'\},\{p\},\{(advanced,c')\},done,(m_2,c))
\end{math}

On this occasion, the conversation has neither started nor ended, so the only event raised is that the conversation was advanced.

\subsection{Third Iteration}

On the third iteration of the Conversation Manager, a further message has been sent, as follows:

\begin{Verbatim}[samepage=true]
(inform
   :sender            agent1
   :receiver          agent2
   :content           processed(doc123)
   :conversation-id   c1
   :protocol          proc_docs
)
\end{Verbatim}

This message is modelled as before and is named $m_3$, as follows:

\begin{math}
m_3=(agent1,agent2,c1,proc\_docs, inform,processed(doc123))
\end{math}

\subsubsection{Start, Initialise and Match}

The \textit{start} and \textit{initialise} stages occur in an identical fashion to the second iteration. During the \textit{match} state, the message is compared against each of the three available transitions in the active conversation. This matching is presented in Table~\ref{tab:semantics:matchtable2}.

\begin{table}[!hbt]
\centering
\caption{Third message matched against available transition in the \textit{match} state.}
\begin{tabular}{|l|l|l|l|l|}
\hline
\multicolumn{5}{|l|}{\textbf{Transition}} \\
\hline
\multicolumn{5}{|c|}{$t=(proc\_docs,requested,waiting,(?initiator,immutable),$} \\
\multicolumn{5}{|c|}{$(?respondent,immutable),inform,processed((?docid,immutable)))$} \\
\hline
\textbf{Pre} &
\textbf{Post} &
\textbf{Message} &
\textbf{Match} &
\textbf{New} \\

\hline
inform & inform & inform & Yes & \\

?initiator & agent1 & agent1 &  Yes &  \\

?respondent & agent2 & agent2 & Yes & \\

processed(?docid) & processed(doc123) & processed(doc123) & Yes & \\ 

\hline
\end{tabular}

%\begin{math} t=(proc\_docs,requested,end,(?respondent,immutable),\\
%\hspace*{1cm}(?initiator,immutable),refuse,process((?docid,immutable)))
%\end{math}

\begin{center}
\begin{tabular}{|l|l|l|l|l|}
\hline
\multicolumn{5}{|l|}{\textbf{Transition}} \\
\hline
\multicolumn{5}{|c|}{$t=(proc\_docs,requested,end,(?respondent,immutable),$} \\
\multicolumn{5}{|c|}{$(?initiator,immutable),refuse,process((?docid,immutable)))$} \\
\hline
\textbf{Pre-bindings} & \textbf{Post-bindings} & \textbf{Message} & \textbf{Match} & \textbf{New} \\
\hline
refuse & refuse & inform & No & \\
\hline
\end{tabular}
\end{center}

%\begin{math}
%t=(proc\_docs,requested,failed,(?respondent,immutable),\\
%\hspace*{1cm}(?initiator,immutable),failure,process((?docid,immutable)))
%\end{math}

\begin{center}
\begin{tabular}{|l|l|l|l|l|}
\hline
\multicolumn{5}{|l|}{\textbf{Transition}} \\
\hline
\multicolumn{5}{|c|}{$t=(proc\_docs,requested,failed,(?respondent,immutable),$} \\
\multicolumn{5}{|c|}{$(?initiator,immutable),failure,process((?docid,immutable)))$} \\
\hline
\textbf{Pre-bindings} & \textbf{Post-bindings} & \textbf{Message} & \textbf{Match} & \textbf{New} \\
\hline
failure & failure & inform & No & \\
\hline
\end{tabular}

\label{tab:semantics:matchtable2}
\end{center}
\end{table}

The message $m_3$ is only capable of triggering one of the available transitions in the active conversation. As such, the conversation can be added to the set of candidate conversations, as follows:

\begin{math}
(agent1,[],\{c\},\{p\},\emptyset,fail,(m_3,\{c\}))
\end{math}

\subsubsection{Fail}
The third iteration is the first in which the conditions necessary for the \textit{fail} state to be relevant are present. This occurs whenever the message being processed includes a conversation identifier (as is the case with $m_3$), but the conversation that this refers to cannot be advanced by the message.

In this case, message $m_3$ is capable of advancing conversation $c$ (as illustrated in Table~\ref{tab:semantics:matchtable2}) and so the conversation is not marked as having failed.

This means that the conversation manager can enter the \textit{new} state and be modelled as follows:

\begin{math}
(agent1,[],\{c\},\{p\},\emptyset,new,(m_3,\{c\}))
\end{math}

\subsubsection{New}
In the \textit{new} state, one of the conditions in which no new conversation will be initiated is if the message contains a conversation identifier and the corresponding candidate conversation has already been identified (according to Equation~\ref{eqn:semantics:new2}). In this instance, $m_3$ does contain a conversation identifier and it has already been established that conversation $c$ is a candidate conversation, meaning that no check is performed to establish whether a new conversation may be initiated.

Thus, the conversation manager can move to the \textit{update} state, at which point it is modelled as:

\begin{math}
(agent1,[],\{c\},\{p\},\emptyset,update,(m_3,\{c\}))
\end{math}

\subsubsection{Update}

The \textit{update} state on this iteration is similar to that in the second iteration. Once again, there is only one conversation that matched the message. 

Message $m_3$ is added to the message history, although on this occasion, no additional bindings were generated by the matching shown in Table~\ref{tab:semantics:matchtable2} so these are unchanged.

The conversation can now be advanced to the waiting state, and is modelled as follows:
\begin{math}
c' = (proc\_docs,\{agent1,agent2\},waiting,[m_1,m_,m_3],c1,B,active)
\end{math} \\
where $B =\{(?initiator,agent1),(?respondent,agent2),(?docid,doc123)\}$

Following this process, the conversation manager is now modelled as follows:

\begin{math}
(agent1,[],\{c'\},\{p\},\{(advanced,c')\},done,(m_3,c))
\end{math}

\subsection{Fourth Iteration}

On the fourth iteration of the agent, it is assumed that a further message has been exchanged, taking the following form: 

\begin{Verbatim}[samepage=true]
(request
   :sender            agent2
   :receiver          agent1
   :content           process(doc234)
)
\end{Verbatim}

Following previous iterations, this message is named $m_4$ and is modelled as follows:

\begin{math}
m_4=(agent2,agent1,\bot,\bot,request,process(234))
\end{math}

Once again, it can be seen that messages sent by \texttt{agent2} lack the \texttt{protocol} and \texttt{conversation-id} fields as that agent does not support conversation management.

\subsubsection{Start, Initialise and Match}

The \textit{start} and \textit{initialise} states are carried out in the same way as the previous iteration.

In the \textit{match} state, the message is compared with the one active conversation. As that conversation is in the \textit{waiting} state, there is only one available transition against which to match the message. The matching of the message against this transition is outlined in Table~\ref{tab:semantics:matchtable3}.

\begin{table}[!hbt]
\caption{Fourth message matched against available transition in the \textit{match} state.}
\begin{tabular}{|l|l|l|l|l|}
\hline
\multicolumn{5}{|l|}{\textbf{Transition}} \\
\hline
\multicolumn{5}{|c|}{$t = (proc\_docs,waiting,requested,(?respondent, immutable),$} \\ \multicolumn{5}{|c|}{$(?initiator,immutable),request,process((?docid,mutable)))$} \\
\hline
\textbf{Pre} & \textbf{Post} & \textbf{Message} & \textbf{Match} & \textbf{New} \\
\hline
request & request & request & Yes & \\

?respondent & agent2 & agent2 & Yes & \\

?initiator & agent1 & agent1 &  Yes &  \\

process(??docid) & process(??docid) & process(doc234) & Yes & ?docid \\ 
&&&& $\rightarrow$ doc234 \\
\hline
\end{tabular}
\label{tab:semantics:matchtable3}
\end{table}

One key difference between this match and that done for message $m_2$ is in the content field. Although the \texttt{?docid} variable already has a binding in the active conversation, it is used in a mutable context in this transition.

According to Equation~\ref{eqn:semantics:tapply}, since the variable is used in a mutable context, applying bindings to it does not replace it with a bound value. As such, it is free to acquire a new bound value (i.e. ``doc234'') when it is matched against a new term.

As the message matches an available conversation, conversation $c$ is added to the set of candidate conversations, leaving the conversation manager in the following situation entering the \textit{fail} state:

\begin{math}
(agent1,[],\{c\},\{p\},\emptyset,fail,(m_4,\{c\}))
\end{math}

\subsubsection{Fail, New and Update}

As the message did not contain a conversation identifier, the \textit{fail} state will have no effect on any active conversation, in the same way as in the second iteration. 

However, an attempt must be made to check if any new conversation could be initiated by this message, so in the \textit{new} state, the message is compared against the initial transitions of known protocols. In this example, only one protocol is known, and so the message must be matched only against that protocol. This is illustrated in Table~\ref{tab:semantics:newtable3}.

\begin{table}[!hbt]
\caption{Fourth message matched against available transition in the \textit{new} state.}
\begin{center}
\begin{tabular}{|l|l|l|l|l|}
\hline
\multicolumn{5}{|l|}{\textbf{Transition}} \\
\hline
\multicolumn{5}{|c|}{$t = (proc\_docs,start,waiting,(?initiator,immutable),$} \\
\multicolumn{5}{|c|}{$(?respondent,immutable),inform,ready)$} \\
\hline
\textbf{Pre} & \textbf{Post} & \textbf{Message} & \textbf{Match} & \textbf{New} \\
\hline
inform & inform & refuse & No & \\
\hline
\end{tabular}
\label{tab:semantics:newtable3}
\end{center}
\end{table}

As with the message sent in the second iteration, message $m_4$ is also incapable of starting a new conversation and so no change is made to the candidate conversations.

Thus the only further change to the conversation manager on this iteration is in the \textit{update} state, where the only candidate conversation that was identified is actually advanced.

Because of the use of the mutable variable in the transition that was triggered, the bindings associated with the conversation are altered by the update in this case.

The conversation returns to the ``requested'' state, and is modelled as follows:
\begin{math}
c' = (proc\_docs,\{agent1,agent2\},requested,[m_1,m_,m_3,m_4],c1,B,active)
\end{math} \\
where $B =\{(?initiator,agent1),(?respondent,agent2),(?docid,doc234)\}$

The conversation manager is also changed to reflect this change, along with the event that the conversation has been advanced.

\begin{math}
(agent1,[],\{c'\},\{p\},\{(advanced,c')\},done,(m_4,c))
\end{math}

\subsection{Fifth Iteration}

The final iteration of this conversation consists of the initiator agent refusing to process the requested document, thus terminating the conversation by bringing it to a terminal state. The message associated with this action is:

\begin{Verbatim}[samepage=true]
(refuse
   :sender            agent1
   :receiver          agent2
   :content           process(doc234)
   :conversation-id   c1
   :protocol          proc_docs
)
\end{Verbatim}

As before, this message is modelled as a tuple called $m_5$ as follows:

\begin{math}
m_5=(agent1,agent2,c1,proc\_docs,refuse,process(234))
\end{math}

\subsubsection{Start, Initialise and Match}

The functioning of the \textit{start} and \textit{initialise} states is the same as for previous iterations. 

In terms of matching this message against available conversations, it is similar to the process conducted on the third iteration, where the available conversation was also in the ``requested'' state.

In this state, there are three available transitions against which the message should be compared. This process is shown in Table~\ref{tab:semantics:matchtable4}.

\begin{table}[!hbt]

\caption{Fifth message matched against available transition in the \textit{match} state.}
\begin{center}
\begin{tabular}{|l|l|l|l|l|}
\hline
\multicolumn{5}{|l|}{\textbf{Transition}} \\
\hline
\multicolumn{5}{|c|}{$t=(proc\_docs,requested,waiting,(?initiator,immutable),$} \\
\multicolumn{5}{|c|}{$(?respondent,immutable),inform,processed((?docid,immutable)))$} \\
\hline

\hline
\textbf{Pre} &
\textbf{Post} &
\textbf{Message} &
\textbf{Match} &
\textbf{New} \\

\hline
inform & inform & refuse & No & \\
\hline
\end{tabular}
\end{center}

\begin{center}
\begin{tabular}{|l|l|l|l|l|}
\hline
\multicolumn{5}{|l|}{\textbf{Transition}} \\
\hline
\multicolumn{5}{|c|}{$t=(proc\_docs,requested,end,(?respondent,immutable),$} \\
\multicolumn{5}{|c|}{$(?initiator,immutable),refuse,process((?docid,immutable)))$} \\
\hline
\textbf{Pre} & \textbf{Post} & \textbf{Message} & \textbf{Match} & \textbf{New} \\
\hline
refuse & refuse & refuse & Yes & \\
?respondent & agent2 & agent2 & Yes & \\
?initiator & agent1 & agent1 & Yes & \\
process(?docid) & process(doc234) & process(doc234) & Yes & \\
\hline
\end{tabular}
\end{center}

\begin{center}
\begin{tabular}{|l|l|l|l|l|}
\hline
\multicolumn{5}{|l|}{\textbf{Transition}} \\
\hline
\multicolumn{5}{|c|}{$t=(proc\_docs,requested,failed,(?respondent,immutable),$} \\
\multicolumn{5}{|c|}{$(?initiator,immutable),failure,process((?docid,immutable)))$} \\
\hline
\textbf{Pre} & \textbf{Post} & \textbf{Message} & \textbf{Match} & \textbf{New} \\
\hline
failure & failure & refuse & No & \\
\hline
\end{tabular}

\label{tab:semantics:matchtable4}
\end{center}
\end{table}

On this occasion, one transition can be triggered by the message. This transition is the second that is compared, where the state of the conversation would be brought to the ``end'' state, which is a terminating state.

As the message is capable of advancing an active conversation $c$, this conversation is added to the set of candidates, leaving the conversation manager as follows:

\begin{math}
(agent1,[],\{c\},\{p\},\emptyset,fail,(m_5,\{c\}))
\end{math}

\subsubsection{Fail, New and Update}

As with the previous iterations, the \textit{fail} and \textit{new} states do not have any effect on the conversation manager (as the message is capable of successfully advancing the conversation it identified).

In the \textit{update} state, the conversation is once again advanced using the message $m_5$. No new bindings were generated by the triggering of the active transition, so the bindings remain as they were after the previous iteration. The conversation can now be modelled as:

\begin{math}
c' = (proc\_docs,\{agent1,agent2\},end,[m_1,m_,m_3,m_4,m_5],c1,B,completed)
\end{math} \\
where $B =\{(?initiator,agent1),(?respondent,agent2),(?docid,doc234)\}$

It is notable here that since the ``end'' state is a terminating state, the conversation status now becomes $completed$.

Having advanced the conversation, the conversation manager is now as follows:

\begin{math}
(agent1,[],\{c'\},\{p\},\{(advanced,c'),(completed,c')\},done,(m_5,c))
\end{math}

At this point, the conversation has now been completed, as it has reached a state that has no outgoing transitions.

\section{Summary}
This Chapter outlines the formal model underpinning the ACRE system, including the operational semantics for the conversation manager, which is the most crucial component of the system. 

The formal model includes modelling of conversations, protocols, messages and the other associated entities that are necessary for such a system. In addition to this, a number of other predicates and functions are necessary in order to be able to reason about and manipulate the entities that have been modelled.

An example is also presented where the operational semantics are used on a sample conversation, indicating the transitions that occur at each state.

Chapter~\ref{chap:acre-architecture} follows this work in providing a generic architecture to demonstrate how the ACRE system may be integrated into a Multi Agent platform. A concrete implementation that integrates with a specific Agent Oriented Programming framework then follows in Chapter~\ref{chap:agent-factory}.
\chapter{Generic Architecture}
\label{chap:acre-architecture}

\section{Introduction} \label{sec:acre-architecture:introduction}

The preceding Chapters have shown how the Agent Conversation Reasoning Engine (ACRE) models agent conversations, leading to the formal model of agent conversation presented in Chapter~\ref{chap:semantics}. Following from this work, it is necessary to show how ACRE may be used in a practical sense, within a Multi Agent System (MAS) framework.

This Chapter presents the generic ACRE architecture, which refers to an abstract architecture that is designed to fit into a variety of MAS frameworks. More concrete details of a specific implementation are given in Chapter~\ref{chap:agent-factory}.

The generic architecture consists of a number of components that provide the core services of ACRE, combined with existing elements of the MAS framework with which these must be integrated. The ACRE components are designed to be, to the greatest extent possible, independent of platform, framework and AOP language. The generic architecture aims to keep the amount of framework-specific integration to a minimum, with only one ACRE component interacting directly with the MAS framework and its agents.

In order to be compatible with ACRE, a MAS framework is required to provide a number of pre-existing features, such as message sending capabilities. This Chapter focuses on these existing components that the framework is expected to have, along with describing the ACRE components and the role they play within the overall system.

\section{Overview}

\begin{figure}[!htb]
\includegraphics[width=5.7in]{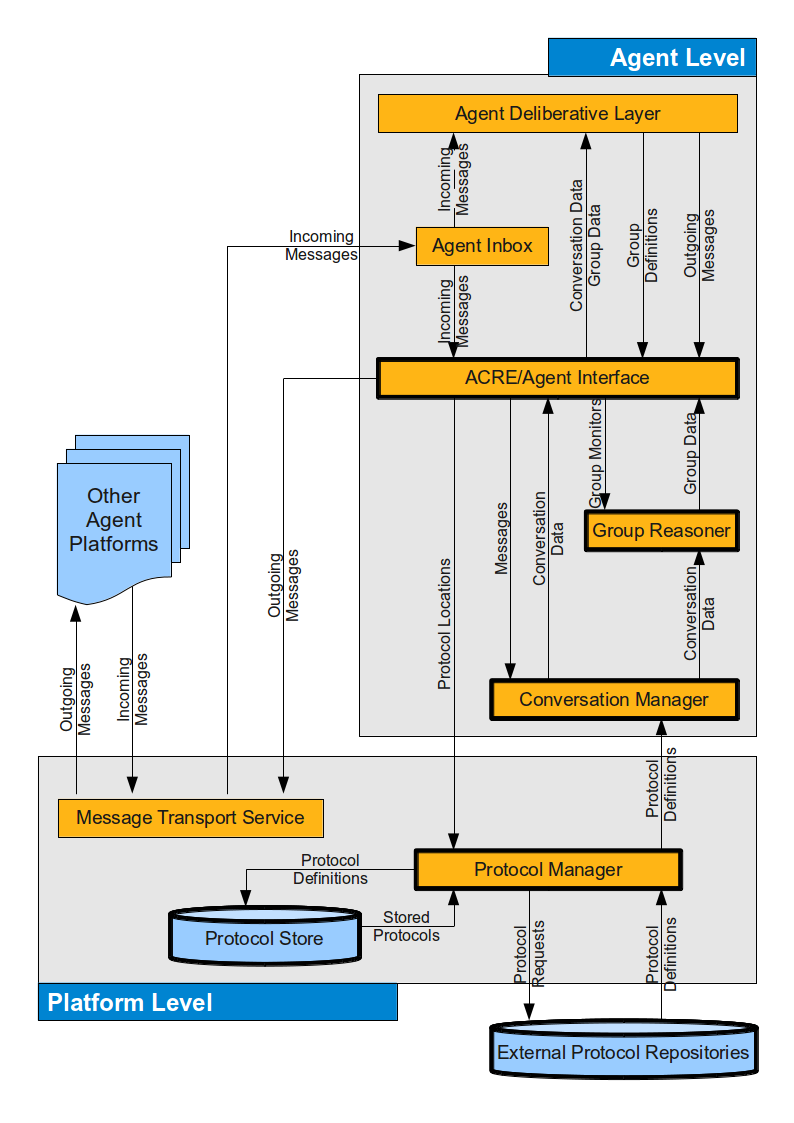}
\caption{Generic ACRE Architecture.}
\label{fig:acre-architecture:generic-acre-architecture}
\end{figure}

The Generic Architecture is shown in Figure~\ref{fig:acre-architecture:generic-acre-architecture}. In this diagram, components with thicker borders indicate those that are part of ACRE, whereas those with thin borders are required as pre-existing components of the framework within which the agents reside. Arrows between components indicate the flow of data through the overall system. Each of the components is outlined in detail in the following sections, along with details of how they should be integrated into an existing MAS system. 

The system's components are separated into groups based on whether they operate externally, at the platform level or the individual agent level.  Those that are grouped in the \textit{platform level} are shared between agents, and only one of each of these components is present within one agent platform at any point in time. Components at the \textit{agent level} belong only to a single agent, and each ACRE-enabled agent in the system should have its own instances of these.

The following sections outline the components in the architecture that are part of ACRE, along with the data that flows between them. The architecture is designed so that the ACRE/Agent Interface is the only component that interacts directly with the existing components of the agent platform on which ACRE is running. As such, this is the only component for which a framework-specific version needs to be created in order to integrate with a new MAS framework. Java implementations of all other components have been created, as Java is a popular language in the creation of MAS frameworks. Thus, these are capable of being deployed in their present form within any Java-based MAS framework (e.g. Agent Factory~\cite{Collier2003}, Jason~\cite{Bordini2005}). Where relevant, details of how these should be integrated into a MAS framework are supplied.

\section{External Components}

%TODO: add in zoomed portion of architecture diagram

This Section discusses aspects of ACRE that are entirely independent of any framework within which ACRE is used. These are accessed remotely using standard networking protocols (such as HTTP) and so no work is required to tailor them to specific frameworks.

\subsection{External Protocol Repositories} \label{sec:acre-architecture:external-protocol-repositories}

A Protocol Repository is a resource from which protocol definitions can be sourced. Access to a Protocol Repository can currently be gained either via HTTP or by reading from a local filesystem, although the structure of repositories does not preclude the implementation of access mechanisms that use other protocols such as SSH or FTP.

Each Protocol Repository is identified by a URL that identifies its base directory. Within that base directory, it is necessary to have a \texttt{repository.xml} file that describes the contents of the repository, and a \texttt{repository} directory that holds the individual protocol definitions. This layout is loosely based on the package repositories that are used by the Advanced Packing Tool (APT) to install programs on the Debian Linux Operating System\footnote{\url{http://www.debian.org}} and its derivatives.

A sample of a valid \texttt{repository.xml} file is shown in Figure~\ref{fig:acre-architecture:repository-xml}. This file follows the XML schema document (XSD) for \texttt{repository.xml} files shown in Appendix~\ref{app:schemas}.

\begin{figure}[!htb]
\begin{Verbatim}[fontsize=\footnotesize]
<?xml version="1.0" encoding="UTF-8"?>
<repository xmlns="http://acre.lill.is" 
   xmlns:xsi="http://www.w3.org/2001/XMLSchema-instance" 
   xsi:schemaLocation="http://acre.lill.is
    http://acre.lill.is/repository.xsd">
   <base>http://acre.lill.is</base>
   <namespaces>
      <namespace name="is.lill.fipa">
         <protocol name="fipa-request" version="1.0"/>
         <protocol name="fipa-contractnet" version="1.0"/>
         <protocol name="fipa-iterated-contractnet" version="1.0"/>
      </namespace>
      <namespace name="is.lill.acre">
         <protocol name="processdocuments" version="1.0"/>
         <protocol name="english-auction" version="1.0"/>
         <protocol name="vickreyauction" version="1.0"/>
      </namespace>
   </namespaces>
</repository>
\end{Verbatim}
\caption{Example of a valid \texttt{repository.xml} file.}
\label{fig:acre-architecture:repository-xml}
\end{figure}

In this file, the \texttt{<base>} tag specifies the location of the repository (where the \texttt{repository.xml} file itself is located).

Following this, a number of namespaces are declared. Each protocol must exist within a declared namespace, so as to allow for the possibility that different protocol designers may use the same name for their protocols.

In order to avoid conflicts, it is recommended that namespaces take a form similar to the established convention amongst the Java programming community for creating unique package names~\cite{Gosling2005}. This takes the components of an Internet domain name (separated by dots) and reverses them, component by component. Further components may then be appended to this, in accordance with the convention used within the organisation that controls the domain name. Thus the domain \texttt{lill.is} becomes a namespace prefix \texttt{is.lill}, with ACRE protocols stored in that domain under the \texttt{is.lill.acre} namespace.

The XSD specifies that the valid characters in a namespace are the same as for a valid domain name, namely: lowercase letters, numbers, hyphens (but not as a starting or ending character of a component of the namespace) and dots (as separators between the components of the namespace).

Within each namespace is a list of protocols, including the name and the version number. The version is to ensure that a correction to a previous protocol will not cause confusion amongst agents that are under the impression that they are using a protocol they both understand, but in reality are not.

In addition to the \texttt{repository.xml} file, which lists the protocols available, each protocol must be present in its own file within the repository. These are stored in a directory named \texttt{repository} that is located under the same directory that holds the \texttt{repository.xml} file (as referenced in the \texttt{repository.xml} file in the \texttt{<base>} tag). These protocol definitions are also contained in XML files. The filenames follow the format ``\textit{namespace}\_\textit{name}\_\textit{version}.acr''. With this system, the URLs of individual protocols can be constructed by combining the base of the repository with the known data about the protocol itself. Thus in the \texttt{repository.xml} example in Figure~\ref{fig:acre-architecture:repository-xml}, the \texttt{process-documents} protocol is to be found at the location \url{http://acre.lill.is/repository/is.lill.examples_process-documents_1.0.acr}.

An XSD file is also available for protocol files, which is contained in Appendix~\ref{app:schemas}. An example of a valid protocol definition is shown in Figure~\ref{fig:acre-architecture:protocol-xml}.

\begin{figure}[!htb]
\begin{Verbatim}[fontsize=\footnotesize]
<?xml version="1.0" encoding="UTF-8"?>
<protocol xmlns="http://acre.lill.is" 
  xmlns:xsi="http://www.w3.org/2001/XMLSchema-instance" 
  xsi:schemaLocation="http://acre.lill.is
  http://acre.lill.is/protocol.xsd">
   <namespace>is.lill.examples</namespace>
   <name>process-documents</name>
   <version>1.0</version>
   <description>
     Example protocol to illustrate use of ACRE
   </description>
   <states>
      <state name="requested"/>
      <state name="start"/>
      <state name="waiting"/>
      <state name="end"/>
   </states>
   <transitions>
      <transition content="process(??docid)" from-state="waiting"
        performative="request" receiver="?initiator" 
        sender="?respondent" to-state="requested"/>
      <transition content="process(?docid)" from-state="requested"
        performative="refuse" receiver="?respondent" 
        sender="?initiator" to-state="end"/>
      <transition content="processed(?docid)" from-state="requested" 
        performative="inform" receiver="?respondent" 
        sender="?initiator" to-state="waiting"/>
      <transition content="ready" from-state="start" 
        performative="inform" receiver="?respondent" 
        sender="?initiator" to-state="waiting"/>
   </transitions>
</protocol>
\end{Verbatim}
\caption{Example of a valid protocol definition file.}
\label{fig:acre-architecture:protocol-xml}
\end{figure}

The protocol definition begins by specifying the namespace, name and version of the protocol contained within the file. This must match the corresponding entries in the \texttt{repository.xml} file and also the filename in which this protocol definition was found.

This is followed by an optional \texttt{<description>} tag, which contains an informal description of the purpose of the protocol. This is intended to aid developers in choosing between available protocols. It is recommended that this, in conjunction with descriptive names for the states of the protocol, be used in a way that helps to clarify the situations in which the protocol will be suitable.

The remainder of the file consists of declarations of the states and transitions that form the protocol, as described in Chapter~\ref{chap:acre-introduction} (and formally in Chapter~\ref{chap:semantics}).

Each \texttt{<state>} tag has one mandatory attribute: the name of the state. This is used in the definitions of the transitions to refer to this state. State names must be unique within a protocol in order for it to function correctly. There is no requirement to explicitly mark the initial and final states. These are identified dynamically as that state with no incoming transition and those states with no outgoing transitions respectively.

Within a \texttt{<transition>} tag, a number of attributes may be defined. The \texttt{from-state}, \texttt{to-state} and \texttt{performative} attributes are mandatory. The former two attributes refer to the name of the state at which the transition begins and ends, respectively. The latter is the performative that must be contained in a message that is to match the transition.

The other attributes are optional, defaulting to the anonymous variable \texttt{?} if they are not defined. This variable will match against any value in the corresponding message fields but will not cause any bindings to be created. It is thus treated in a similar way to a wildcard match.

The \texttt{sender} and \texttt{receiver} attributes are typically variables that, during the execution of a conversation, bind to the unique identifiers of the agents engaged in the conversation. They may, however, be restricted to specific agent names if so desired.

The \texttt{content} attribute is a predicate or variable that is matched against the actual content of the message.

One additional tag available in protocol definitions that is not shown in Figure~\ref{fig:acre-architecture:protocol-xml} is the \texttt{<import>} tag. This is used to extend an existing protocol by importing all of its states and transitions into the protocol being defined. This is inspired by the approach taken in~\cite{Kuwabara1995}.

The current mechanism for extending protocols is somewhat basic, given that it simply directly imports a set of states and transitions from another protocol. This has restrictions in terms of extending multiple protocols that may have identical names for states that do not represent the same situation. A more sophisticated mechanism of protocol reuse is a possibility for further work and is discussed in Section~\ref{sec:conclusions:further-work}.

%TODO: explain more..... based on Kuwabara, limited re-use, refer to future work section on this later on.
%TODO: find original ref for imported protocols
% it works for commitment machines using a slightly different mechanism in Yolum2001
% Kuwabara1995 (and Vitteau2004)

\begin{figure}
\includegraphics[width=\textwidth]{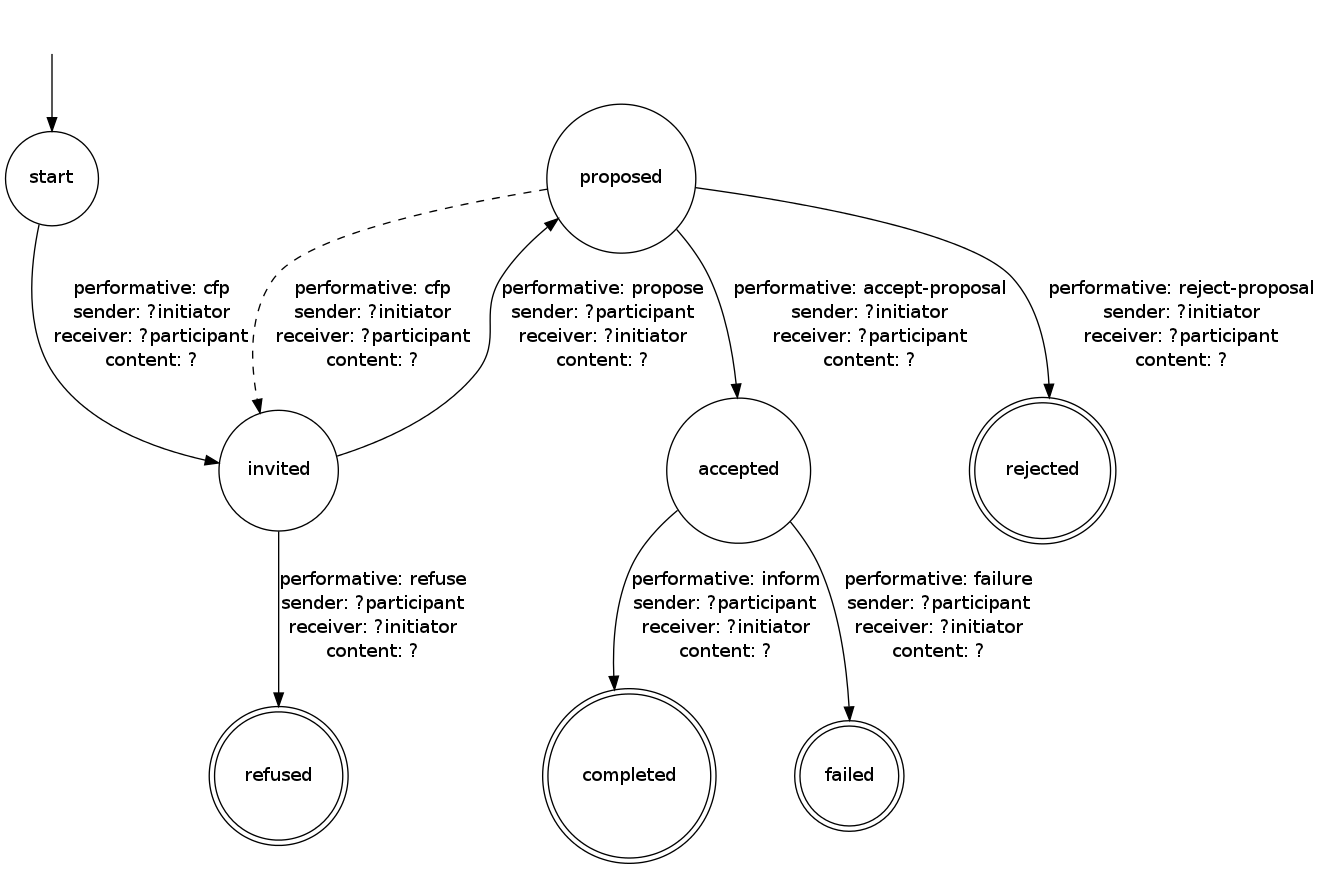}
\caption{Finite State Machine illustrating the use of imports in a protocol definition.}
\label{fig:acre-architecture:protocol-import-fsm}
\end{figure}

Transitions defined in the protocol containing the \texttt{<import>} tag may refer to states that are defined in the protocol that is being imported. Figure~\ref{fig:acre-architecture:protocol-import} shows an example of this tag in use. This example is illustrated by the protocol visualisation shown in Figure~\ref{fig:acre-architecture:protocol-import-fsm}.

\begin{figure}
\begin{Verbatim}[fontsize=\footnotesize]
<?xml version="1.0" encoding="UTF-8"?>
<protocol 
   xmlns="http://acre.lill.is"
   xmlns:xsi="http://www.w3.org/2001/XMLSchema-instance"
   xsi:schemaLocation="http://acre.lill.is
   http://acre.lill.is/protocol.xsd">
   <namespace>is.lill.fipa</namespace>
   <name>fipa-iterated-contract-net</name>
   <version>1.0</version>
   <import>
      <namespace>is.lill.fipa</namespace>
      <name>fipa-contract-net</name>
      <version>1.0</version>
   </import>
   <states/>
   <transitions>
      <transition
       from-state="proposed" to-state="invited" performative="cfp"
       receiver="?participant" sender="?initiator" />
   </transitions>
</protocol>
\end{Verbatim}
\caption{Example of the use of the \texttt{<import>} tag in a protocol definition.}
\label{fig:acre-architecture:protocol-import}
\end{figure}

In these Figures, an implementation of the \texttt{fipa-iterated-contract-net} protocol~\cite{fipa-iterated-contract-net} can be seen, which is one of the interaction protocols specified by FIPA. This protocol is very similar to the \texttt{fipa-contract-net} protocol~\cite{fipa-contract-net}, with the addition that a call for proposals may be sent multiple times, causing multiple iterations of the process.

This protocol definition indicates that version 1.0 of the \texttt{fipa-contract-net} protocol, located in the \texttt{is.lill.fipa} namespace is to be imported into this protocol. The states \texttt{proposed} and \texttt{invited} that are used in the sole transition contained in this XML definition are defined in the imported protocol. Similarly, the \texttt{?participant} and \texttt{?initiator} variable names are chosen to match those used in the imported protocol to refer to the participants in the conversation.

The Finite State Machine illustrating this protocol is shown in Figure~\ref{fig:acre-architecture:protocol-import-fsm}. This shows the states and transitions from the imported \texttt{fipa-contract-net} as solid lines, with the additional transition declared in the \texttt{fipa-iterated-contract-net} shown as a dashed line. From this it can be seen that the extra transition fits seamlessly into the pre-existing protocol, thus promoting reuse in the case of protocols with common bases.

\begin{figure}
   \includegraphics[width=\textwidth]{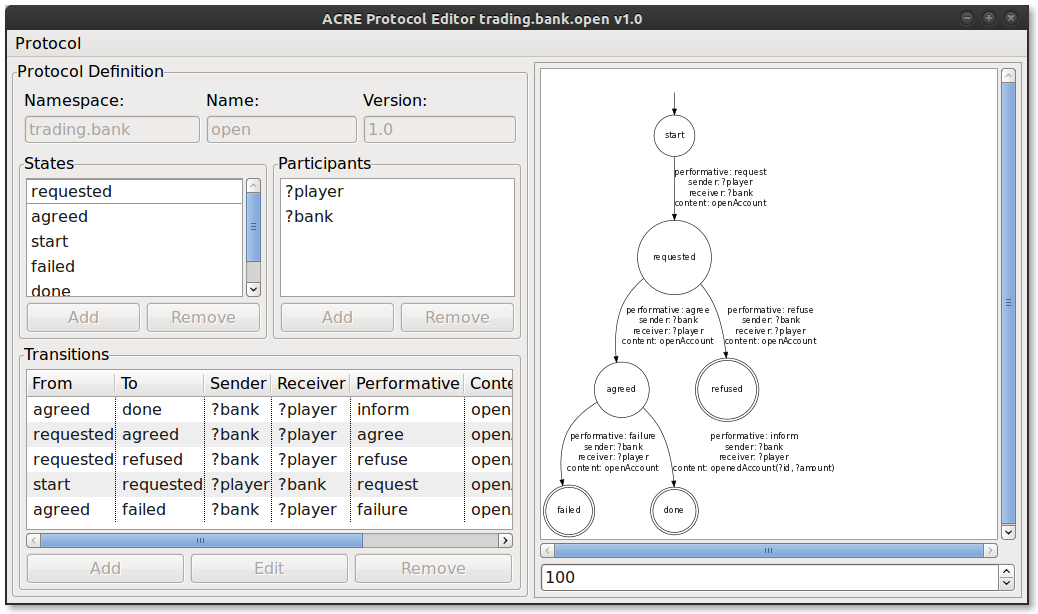}
   \caption{Screenshot of the Protocol Editor.}
   \label{fig:acre-architecture:protocol-editor}
\end{figure}

Although the underlying format is XML, a graphical Protocol Editor has also been created to allow system designers to develop protocols in an easier and more intuitive manner. A screenshot of this tool can be seen in Figure~\ref{fig:acre-architecture:protocol-editor}. This tool also includes the ability to manage a protocol repository automatically. In doing so, it creates the correct directory structure, maintains the \texttt{repository.xml} file and generates the appropriate filenames for each protocol in the repository.

\section{Platform Level Components}

\begin{figure}[!ht]
   \includegraphics[width=\textwidth]{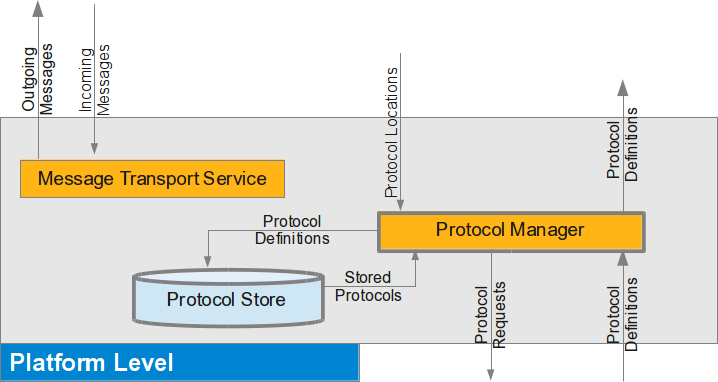}
   \caption{Platform Level of the Generic ACRE Architecture.}
   \label{fig:acre-architecture:platform-level}
\end{figure}

Figure~\ref{fig:acre-architecture:platform-level} shows the platform level section of the generic ACRE architecture. This includes all the components of the system that are shared between all agents on the agent platform. The following Sections discuss these three components: two of which are provided by ACRE and the other of which is expected to be previously available in any framework with which ACRE is integrated.

\subsection{Message Transport Service}

A Message Transport Service (MTS) is a key requirement of FIPA-compliant MAS systems~\cite{FIPA00067}. Even in systems to do not purport to be FIPA-compliant, a conversation handling framework such as ACRE is only relevant where there are existing facilities for the sending and receipt of ACL messages.

The generic architecture requires that some mechanism be pre-existing in any system into which ACRE is to be integrated. The diagram in Figure~\ref{fig:acre-architecture:platform-level} shows this as a single MTS component, but there is no reason why this cannot be made up a number of MTS-type components (possibly each communicating using ACL over different underlying transport protocols such as HTTP, UDP or local message passing).

The key requirement is simply that there is a method available by which an agent may send FIPA ACL messages to others, and also receive messages from other agents.

The platform-level ACRE components do not interact with this directly, as it is the individual agents that are involved in the sending and receiving of messages.

\subsection{Protocol Manager} \label{sec:acre-architecture:protocol-manager}

The Protocol Manager is a shared component that is available to all ACRE-enabled agents on an agent platform. Its task is to allow all agents on the platform to share access to any protocols that have been previously fetched from external protocol repositories, regardless of which agent discovered the repository's location.

The ACRE/Agent Interface (see Section~\ref{sec:acre-architecture:acre-agent-interface}) provides a link between the agent itself and the Protocol Manager, so giving the agent the capability of requesting access to definitions of protocols that it intends to use. Having requested such a protocol, the Protocol Manager will download its definition if it is available in any of the External Protocol Repositories of which it is aware.

In addition to this, the Protocol Manager will make agents aware (via the ACRE/Agent Interface) of any protocols that have been downloaded at the request of other agents on the platform.

The implementation of the Protocol Manager is a Java object. When integrating, it will be necessary to wrap this in a suitable manner so that it is discoverable and accessible by other system components, particularly the ACRE/Agent Interface. Its API allows new single protocols or external protocol repositories to be added to the system, which will then be available to all agents on the platform as this is a shared component. In the case of external repositories, the Protocol Manager will scan the repository and download all its protocols. An agent can be made aware of new protocols being added, as the Protocol Manager raises an event whenever one is added. In practice, this is done by using Java's \texttt{Observable} mechanism, whereby an \texttt{Observer} can be attached to the Protocol Manager and gain access to these events. Finally, it offers a mechanism whereby an agent can read the details of the Protocol Store (see Section~\ref{sec:acre-architecture:protocol-store}) in order to find which protocols are currently available.

The specific form in which agents gain information about the Protocol Manager will be different according to the framework and AOP language being used. The specific details of one concrete implementation may be seen in Chapter~\ref{chap:agent-factory}. Although this deals with a specific implementation, the data that can be made available to agents is not platform-dependent.

\subsection{Protocol Store} \label{sec:acre-architecture:protocol-store}

When an agent requests a protocol definition from the Protocol Manager, it is downloaded and saved locally in a Protocol Store. This has a number of advantages. Firstly, it improves access times to protocol definitions, as it is not necessary to download these from online repositories each time they are required. Additionally, restarting the agent platform will not result in this data being lost in the event of an online repository becoming unavailable. This facilitates the system's recovery from crashes, as remote protocol repositories do not need to be discovered and explored each time. The Protocol Store is structured in the same way as an external repository (which means that it could be read by another agent platform if this was desired). The Protocol Manager is responsible for creating the requisite protocol definitions and \texttt{repository.xml} file to accompany them.

Where downloaded protocols import other protocols, the imported protocol will be incorporated directly when the original protocol is being stored in the Protocol Store. As the \texttt{<import>} tag refers to specific versions of protocols that are to be imported, the imported protocols should not change without a corresponding change in version number. This has the effect that alterations to the imported protocols will not change those protocols that import them.

Protocol definitions are designed to be static (changes to a protocol should be reflected in an altered version number), so refreshing the Protocol Store after checking for updates to existing protocol definitions is not necessary. If a bug is identified in an existing protocol definition, a new definitions (with an incremented version number) should be created and released to supersede it. This policy also prevents the situation arising where two agents are erroneously using different versions of the same protocol.

The Protocol Store is automatically loaded by the Protocol Manager and so no additional integration effort is required, aside from setting a single parameter to indicate where the Protocol Store should be created.

\section{Agent Level Components}

\begin{figure}[!ht]
   \includegraphics[width=\textwidth]{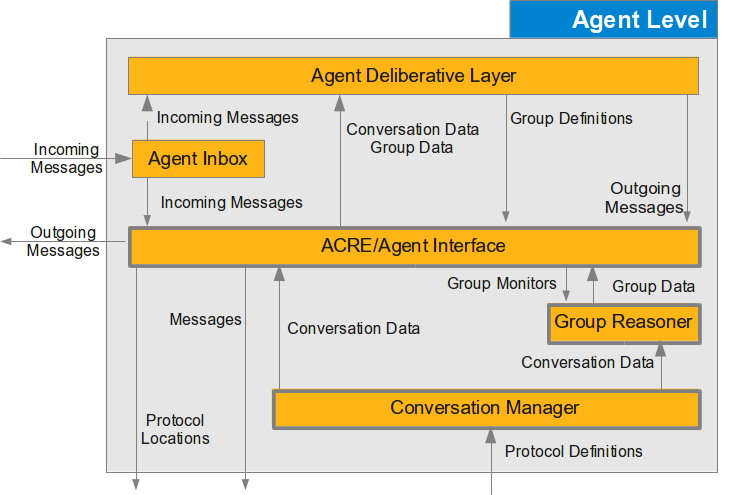}
   \caption{Platform Level of the Generic ACRE Architecture.}
   \label{fig:acre-architecture:agent-level}
\end{figure}

The next group of components that require discussion are those that operate on the agent level. These are components for which a separate instance is required for each agent in the system. Figure~\ref{fig:acre-architecture:agent-level} shows these components (for presentation reasons these have been reorganised from Figure~\ref{fig:acre-architecture:generic-acre-architecture}).

As with the platform level components, some of these (with the thicker borders) are specific to ACRE, with a subset of these being available as concrete implementations. Those components with thinner borders are required as prerequisites of the MAS framework.

\subsection{Agent Inbox}
It is necessary that the messages being sent to the agent are available to the ACRE components. This allows them to reason about incoming messages so that they can be matched to existing conversations, or begin new ones.

For any specific integration, it will not necessarily be the case that this feature will be implemented in exactly this way, via an agent level inbox component. Any other mechanism by which the ACRE components can gain access to incoming messages (including interacting directly with the MTS if necessary) is acceptable.

\subsection{Agent Deliberative Layer}

The Agent Deliberative Layer is the agent program itself, written in some AOP language. This contains all the reasoning capabilities of the agent, and it is to this layer that the ACRE/Agent Interface provides information about ongoing communications and the facility to reason about these. The services provided by this interface to the agent are outlined below in Section~\ref{sec:acre-architecture:acre-agent-interface}.

\subsection{Conversation Manager} \label{sec:acre-architecture:conversation-manager}

The Conversation Manager is the most important element of the ACRE architecture. It represents the running implementation of the formal model shown in Chapter~\ref{chap:semantics}, along with extensions to deal with situations beyond the beginning, advancement and end of conversations that are the subject of the model. Conversation Managers operate at the agent level, meaning that each agent has its own instance of a Conversation Manager, which is independent of those belonging to other agents.

It is the responsibility of the Conversation Manager to accept outgoing messages sent to it by the agent via the ACRE/Agent Interface and incoming messages from the Agent Inbox (or equivalent), also via the ACRE/Agent Interface. Both types of messages are matched against the known protocols and conversations. The Conversation Manager is then responsible for providing information to the agent about the conversations in which it is engaged.

This information includes details such as the current state of active conversations, the other participants in conversations and events such as the conversation being begun, advanced or completed. It will also alert the agent to situations where a message is not suitable for advancing an active conversation or beginning a new conversation based on a known protocol. The specific details of the types of information that the Conversation Manager makes available to its parent agent are outlined below in Section~\ref{sec:acre-architecture:acre-agent-interface}.

In addition to providing information to the ACRE/Agent Interface for forwarding to the agent, the Conversation Manager also sends details of active conversations to the Group Reasoner (discussed below in Section~\ref{sec:acre-architecture:group-reasoner}) to enable it to reason about groups of conversations.

One extension that the Conversation Manager makes to the formal model is to expand the list of conversation statuses. As outlined in Section~\ref{sec:semantics:conversation}, a conversation's status will, according to the formal model, be one of \textit{active}, \textit{completed}, or \textit{failed}. The Conversation Manager adds the following possible statuses also:
\begin{description}
   \item[Ready:] A conversation has been created but has no messages as yet.
   \item[Stale:] A timeout has occurred.
   \item[Cancelling:] A cancellation request has been made.
   \item[Cancelled:] The conversation has been cancelled.
\end{description}

A conversation that is ``ready'' does not yet exist according to the formal model, and is only available with in the Conversation Manager for convenience. With any other status, the conversation is still technically ``active'' according to the formal model, as it has neither been successfully advanced to a final state nor caused to fail by attempting to match an inappropriate message.

Like the Protocol Manager, the Conversation Manager is implemented as a Java object. Integration of this into an existing MAS framework may require the creation of a wrapper that allows the Conversation Manager to be discoverable and accessible. It also utilises Java's \texttt{Observable} mechanics to raise events that are of interest to the associated agent. Within the ACRE subsystem, messages are passed in objects that implement an \texttt{IACREMessage} interface. This requires some integration work within the ACRE/Agent Interface, which is tasked with translating between the MAS framework's internal representation of messages and that of ACRE.

\subsection{Group Reasoner} \label{sec:acre-architecture:group-reasoner}

ACRE conversations are between two agents only. However, it is common that multiple agents may be engaged in conversations that are closely related to one another. For example, in conducting an auction, an auctioneer agent may communicate with many bidders, which engage in separate conversations that are nonetheless related to one another. Thus ACRE also allows agents to reason about groups of related conversations or agents so as to better deal with this type of situation.

The Group Reasoner is the component that facilitates the grouping of related agents and conversations into groups, providing information about these to the agent via the ACRE/Agent Interface. Like the Conversation Manager, it also operates at the agent level, meaning that each agent will have its own Group Reasoner.

The Group Reasoner allows two types of groups to be created: \emph{agent groups} and \emph{conversation groups}.

An agent group is similar to a group of contacts in an address book: a named group of agents that are linked for a particular purpose (e.g. a set of agents that have expressed an interest in participating as bidders in an auction).

Conversation groups are groups of related conversations, following the same protocol, that can be reasoned about. ``Group monitors'' can be loaded into the Group Reasoner in order to gain information about the conversations that form part of the group. These group monitors raise events when particular circumstances arise that can be seen by the agent. The nature of these events varies depending on the type of group monitors that have been loaded, with each having a different purpose.

Whereas the range of events that may be associated with an individual conversation is well defined (e.g. a conversation has begun, been advanced, ended, failed, etc.), the same cannot be said for groups of conversations. Situations may arise when some conversations are active while others have ended or failed. Conversations may be in different states or may arrive at common states.

Additionally, the policies used for the handling of conversation groups will also dictate that types of event in which an agent is interested. For example, in an auction situation, it may be important to conclude the auction quickly and so a policy may be implemented whereby the auctioneer agent will not wait for all potential bidders to respond. Instead, it may be interested in an event stating that 75\% of expected responses have been received so that it can then proceed to accept the best bid received thus far.

Another scenario involves an agent inviting others to participate in some process, but allowing preferential treatment to certain potential participants (e.g. more trusted, the participant has access to greater resources or capabilities, etc.). This may require different events to be raised regarding preferred agents, for instance to allow them more time to reply than others.

For reasons such as these, support for generating beliefs from groups of conversation groups is more flexible than for individual conversations.

Conversations that are part of groups generate the same events and provide the same information as standalone conversations. However, the use of group monitors allows more information to be gleaned from conversation groups about their progress. These are loaded into the Group Reasoner by the ACRE/Agent Interface at the request of the agent. Different group monitors are designed to raise events in different scenarios, with these being routed back to the agent through the ACRE/Agent Interface.

A number of default group monitors are packaged with ACRE that deal with common situations. In addition to these, agent programmers may implement custom group monitors also, in order to handle any scenarios that are not covered by the existing default group monitors. This facility is provided due to the far greater variety of interesting situations in which a group of conversations may find itself, when factors are considered including the state of a conversation, the participants in a conversation, whether a conversation has timed out or failed and when new conversations are added to (or removed from) a group.

Group monitors are implemented so as to allow them to be configured upon loading by means of parameters that may be passed to them by the ACRE/Agent Interface. The number of parameters required varies from monitor to monitor and is specified within the monitor code itself. Group monitors are Java objects, rather than depending on deliberative AOP code. The decision to perform this task at the lower level was made for two key reasons:

\begin{enumerate}
   \item From an agent's perspective, an event is a primitive concept. Insisting that these are generated in the agent's deliberative AOP code adds unnecessary complexity at this level. Implementing group monitors in Java with events being routed through the ACRE/Agent Interface means that group events are treated in the same way as events arising from the Conversation or Protocol Managers.
   \item ACRE is intended for use within a variety of MAS frameworks using a multitude of AOP languages. Requiring group monitors to be implemented in an AOP language would add to the burden of porting ACRE to a new platform as these would need to be translated to the target AOP language. As group monitors are an integral part of ACRE, it is more appropriate that these be implemented in the same programming language as the rest of the system.
\end{enumerate}

\subsubsection{Default Group Monitors}

A number of group monitors are included with ACRE by default. These relate to common situations in which agents may be interested.

\begin{description}
   \item[\texttt{AllInState}] This monitor will raise an event to notify the agent any time all of the conversations in the group are in a specific state. It takes one configuration parameter, the name of the state in question. On reaching the specified state, this event will be raised on every call to this group monitor for as long as the conversations remain in that state.
   \item[\texttt{AllReachedState}] The \texttt{AllReachedState} group monitor is very similar to \texttt{AllInState}, in that it raises an event an event whenever all the conversations in the group reach a specific state, provided as its only configuration parameter. Unlike its related monitor, however, once the specified state is reached no further events will be raised unless one or more conversations have left the specified state in the meantime.
   \item[\texttt{NoneInState}] This group monitor is the opposite to the \texttt{AllInState} monitor. It raises events whenever none of the conversations in the group are in the specified state (again, passed as its only parameter).
   \item[\texttt{AllFinished}] The \texttt{AllFinished} group monitor will raise an event whenever the situation arises whereby all of the conversations in the group have reached some terminal state. A conversation that has failed or timed out is not considered to be finished for the purposes of this group monitor. Failed conversations would need to be removed from the group before the \texttt{AllFinished} event could be raised.
\end{description}

\subsubsection{Custom Group Monitors}

In addition to the default group monitors, it is also possible for developers to define custom group monitors to deal with less common situations. Custom group monitors are implemented as Java classes that implement the \texttt{IGroupMonitor} interface. The recommended way to do this is by extending the \texttt{AbstractGroupMonitor} class that is also provided with ACRE.

The abstract group monitor handles the initialisation of the group monitor, including its configuration by means of parameters passed to it when it is created.

To extend the \texttt{AbstractGroupMonitor} class, two things are required. Firstly, it is necessary to set the value of the \texttt{PARAMS} attribute. This is used to specify (as an \texttt{int}) the number of parameters that instances of this group monitor should expect when they are being configured. The actual mechanism by which this parameter is passed is platform and language specific and is done via the ACRE/Agent Interface.

In addition to this, an implementation of the \texttt{event} method must be supplied. This method is called by the event sensor on each iteration of the agent in order to check whether the group monitor has detected a situation arising in which it raises an event. The method signature requires that a boolean value be returned: \texttt{true} if an event has occurred and \texttt{false} otherwise.

The ACRE/Agent Interface will bring this to the attention of the agent whenever an event occurs.

\subsection{ACRE/Agent Interface} \label{sec:acre-architecture:acre-agent-interface}

The ACRE/Agent Interface is the only component of the ACRE architecture that interacts directly with the existing elements provided by the MAS framework. Thus it is the only component that must be implemented in a platform-specific way, in order to provide an interface between the platform and the ACRE components. Operating on the agent level, an instance of an ACRE/Agent Interface is created for each agent, to link it with its own Group Reasoner and Conversation Manager, and with the platform-wide Protocol Manager.

The ACRE/Agent Interface serves three broad principal functions:

\begin{description}
   \item[Routing Messages to the Conversation Manager] Both incoming and outgoing messages must be provided to the Conversation Manager so that they can be matched against conversations and protocols. In a non-ACRE system, outgoing messages are typically sent by the agent directly to the MTS. In contrast, an ACRE-enabled system requires that the messages are either routed via the ACRE/Agent Interface or are made available to it for modification before sending. The modification is performed by the Conversation Manager, which adds appropriate \texttt{conversation-id} and \texttt{protocol} parameters to those messages that are sent as part of a conversation.

   Similarly, agents will typically read incoming messages from the Agent Inbox. In an ACRE system, the ACRE/Agent Interface must also be capable of reading the messages from this inbox in order to send these to the Conversation Manager.

   \item[Providing Additional Complimentary Features:] There are some aspects of a conversation-management system that are not directly provided by the ACRE components. If some feature is absent from both the existing ACRE components and the MAS framework within which it sits, it can be included in the ACRE/Agent Interface. One example of such a feature is the management of an address book. The FIPA standards on identifying agents are very simple, in that they require that each agent have a unique identifier. However, each agent may have a number of different addresses by which it can be reached. This may occur if its agent platform uses multiple MTSs that make use of different underlying transport protocols. If the management of a type of address book is not already handled by the system, one may be integrated within the ACRE/Agent Interface.

   \item[Providing Information] Whenever messages are matched to appropriate conversations, it is necessary for the agent to gain knowledge of this process. Thus the ACRE/Agent Interface has a major role in allowing the Protocol Manager, Conversation Manager and Group Reasoner to communicate with the agent.

   This information can take the form of one-off events (e.g. a conversation has been advanced by the receipt of a message) or ongoing knowledge (e.g. a particular protocol is available for use). This information should relate knowledge about individual conversations, protocols and conversation groups. An illustration of the specific information that can be provided by the ACRE/Agent Interface is set out as part of the concrete implementation outlined in Chapter~\ref{chap:agent-factory}.

   \item[Allowing the Agent to Act] As well as the ability to gain information about the active conversations in which it is engaged, an agent must also be able to act upon these also. For this reason, the ACRE/Agent Interface is also required to provide an API through which the agent is capable of performing actions provided by the Protocol Manager, the Conversation Manager and the Group Reasoner.

   The actions provided will depend on the specific platform on which the system runs, but will typically fall under some or all of the following categories:
   \begin{description}
      \item[Actions for Setup:] Any actions that are required to initialise the various ACRE components being used.
      \item[Actions to Manage Contacts:] If address book functionality has been included, actions will be required to allow the agent to add or remove contacts to and from this contact list.
      \item[Actions to Send Messages:] A key capability is the ability to begin and advance conversations by the sending of messages.
      \item[Actions to Manage Conversations:] By default, all conversations will cause events to be generated and will be available. It may be desirable to add the ability to remove conversations from memory, either permanently or temporarily. 
      \item[Actions to Cancel Conversations:] As discussed in Section~\ref{sec:acre-introduction:cancel}, ACRE agents may cancel conversations, or react to a request for cancellation.
      \item[Actions for Agent Groups:] Adding and removing agents to and from groups should be available through the interface.
      \item[Actions for Conversation Groups:] Beginning or advancing groups of conversations is an important ability. This heading may also include abilities such as adding and removing conversations to and from groups, or cancelling groups of conversations.
   \end{description}
\end{description}

Unlike the other components that are part of ACRE, the ACRE/Agent Interface is heavily dependent on the nature of the framework within which it runs. In implementing an ACRE/Agent Interface, it will be necessary to have it available to its parent agent and also it in turn should have access to that agent's Conversation Manager and Group Reasoner as well as the platform-level Protocol Manager.

Once the ACRE/Agent Interface can discover these components, it should attach itself to them so as to be made aware of the events that they raise. This is done by extending Java's \texttt{Observer} interface. Additionally, ongoing information about conversations can be done by proactively accessing the methods that the ACRE components make available to read data about the current state of conversations, protocols and groups. The mechanism by which this is triggered is heavily platform independent. Many agent architectures operate in a cyclical manner, with ``sensors'' or ``perceptors'' running periodically. Such an architecture, if available, is a suitable method of integration for elements of ongoing knowledge.

Another issue to be solved at this level is the passing of messages to and from the ACRE subsystem. The ACRE/Agent Interface must be capable of accessing the messages being sent to and from its agent (by connecting to the MTS and Agent Inbox) and passing these to the Conversation Manager. This will also require some translation of the messages from their existing representation into an ACRE-suitable object that extends the \texttt{IACREMessage} interface.

Another integration issue is the handling of the content language that is contained in the messages. As mentioned in Section~\ref{sec:acre-introduction:content-language}, the content language used with ACRE is, like many AOP languages, based on first-order logic. The formatting of logic elements may differ between languages, and so translation will be necessary for this also. This will require a parser to be implemented that can accept language-specific logic strings and create a set of objects that represent ACRE's view of the content language.

\section{Summary}
This Chapter describes the generic ACRE architecture, outlining the various components that are available and how they integrate into a MAS framework. The majority of ACRE components are platform-independent and can be integrated as-is into any Java-based agent platform.

The only component for which re-engineering is required for integration is the ACRE/Agent Interface, which acts as an intermediary between the agent itself and the components that make up the ACRE system.

The following Chapter describes in detail a concrete implementation of the ACRE architecture within the Agent Factory MAS framework. In particular, this requires in-depth discussion of the ACRE/Agent Interface and how it facilitates the interactions between the ACRE components outlined in this chapter and the various Agent Oriented Programming Languages supported by Agent Factory.
\chapter{Integration with Agent Factory} \label{chap:agent-factory}

\section{Introduction}

Previous chapters described how ACRE handles conversations between agents, both informally (Chapter~\ref{chap:acre-introduction}) and formally (Chapter~\ref{chap:semantics}). In the preceding chapter, the generic ACRE architecture was outlined, which shows how ACRE may be integrated into an existing Multi Agent System (MAS) framework. This Chapter discusses a concrete implementation of ACRE. This implementation fits into the Agent Factory multi agent framework~\cite{Collier2003}. It then goes on to illustrate how an agent programmer may interact with the ACRE components through the Agent Oriented Programming (AOP) languages available within Agent Factory.

\section{Agent Factory} \label{sec:agent-factory:agent-factory}

Agent Factory\footnote{The Agent Factory home page may be accessed at \url{http://www.agentfactory.com}} is a FIPA-compliant ``cohesive framework supporting a structured approach to the development and deployment of agent-oriented applications''~\cite{Muldoon2009}.

At its core, the framework features a \textit{Run Time Environment} (RTE) that serves as the container within which all other components are created. This RTE contains one or more \textit{agent platforms}, each of which in turn contains the agents that make up a multi agent application (although agents operating from different platforms on a variety of networked computers is also a common architecture). In addition to the agent platforms, the RTE will also facilitate access to a number of \textit{platform services}. These are components that are shared amongst all of the agents that inhabit an agent platform. By default, a local message transport service (to facilitate intra-platform communication between agents using an Agent Communication Language (ACL)) and an agent management service (to allow agents to be created, destroyed, started and stopped) are deployed on an agent platform. However, many other types of platform services are available (e.g. ACL communication over a network and support for various agent-augmenting technologies such as the SoSAA component layer~\cite{Dragone2009,Lillis2009}, the CArtAgO Agents \& Artifacts environment~\cite{Ricci2007} and the Environment Interface Standard (EIS)~\cite{Behrens2010}). In addition to these, custom platform services may be created by agent developers whenever a component or resource (e.g. database access) is required to be shared amongst multiple agents.

In general, each agent consists of an agent program written in one of a number of supported AOP languages and optionally one or more \textit{actions}, \textit{sensors} and/or \textit{modules}. These are discussed in the following Section.

\subsection{Common Language Framework}
The Common Language Framework (CLF) is a set of pre-written components for Agent Factory that support the building of agent interpreters on top of the Agent Factory core~\cite{Russell2011}. These operate within an agent platform to control the execution of the agent, according to an interpreted AOP language.

A key feature of the CLF is the ability of agents written in different AOP languages (and therefore using different agent interpreters) to share the same \textit{actions}, \textit{sensors} and \textit{modules}. These are the three key components in extending an agent program so that an agent may gain information about its environment and effect its intentions.

\textit{Actions} are components that enable an agent to effect some change to its environment. An interface is defined for each action that defines how it may be invoked from within an agent program (an identifier and a specific number of arguments). \textit{Sensors} support the generation of beliefs for an agent, so as to furnish it with the knowledge it requires to satisfy its goals. In the event that a group of sensors and/or actions are linked, they may be defined from within a \textit{module}, which may also provide resources that are shared amongst several sensors and actions.

To date, four AOP langagues have been implemented within the Agent Factory CLF:
\begin{description}
\item[AFAPL] The Agent Factory Agent Programming Language (AFAPL) is the original AOP language developed in conjunction with the Agent Factory framework~\cite{Ross2006}. It is based on a core of beliefs and commitments, with additional support for other features such as plans and goals.
\item[AF-AgentSpeak] is an implementation of the AgentSpeak(L) language proposed by Rao~\cite{Rao1996}. AgentSpeak(L) is a BDI-based agent programming language, which features beliefs, plans, goals, actions, intentions and events as part if its mental model.
\item[AF-TeleoReactive] is based on Nilsson's TeleoReactive model~\cite{Nilsson1994}. This model is based on circuit semantics and has a hierarchical set of action rules as its core.
\item[AF-ASTR] is a hybrid language that combines the benefits of AF-AgentSpeak with those of AF-TeleoReactive.
\end{description}

Although each of these languages features its own agent interpreter, the framework acts in such a way so that any module (with its associated actions and sensors) created with one AOP language in mind can also be used in an agent program written in any other supported language. This includes any further languages for which a CLF interpreter is added in the future. This is designed to promote the reuse of code via the creation of cross-language modules.

As with the core of the Agent Factory framework, CLF modules are written in Java, by extending existing abstract classes that are in turn designed to implement the interfaces required by a CLF component.

\section{ACRE in Agent Factory}

Agent Factory was chosen as the framework within which a concrete implementation of ACRE would be developed for a number of reasons:
\begin{itemize}
   \item Agent Factory is Java-based, which is the same language in which many other popular multi agent frameworks are implemented (e.g. Jason~\cite{Bordini2005}, JADEX~\cite{Braubach2004}, Jack~\cite{Busetta1999}). This means that the platform-independent components of the ACRE implementation may be brought to other platforms in future without difficulty.
   \item As a result of the CLF, Agent Factory supports multiple AOP languages that can all make use of an ACRE/Agent Interface created for Agent Factory with the according to CLF principles. Thus the effectiveness of ACRE may be illustrated in a number of distinctive AOP styles without the requirement of implementing separate ACRE/Agent Interfaces for each. This is intended to illustrate the wide applicability of ACRE in a variety of AOP scenarios.
\end{itemize}

\begin{figure}[!htb]
\centering
\includegraphics[width=\columnwidth]{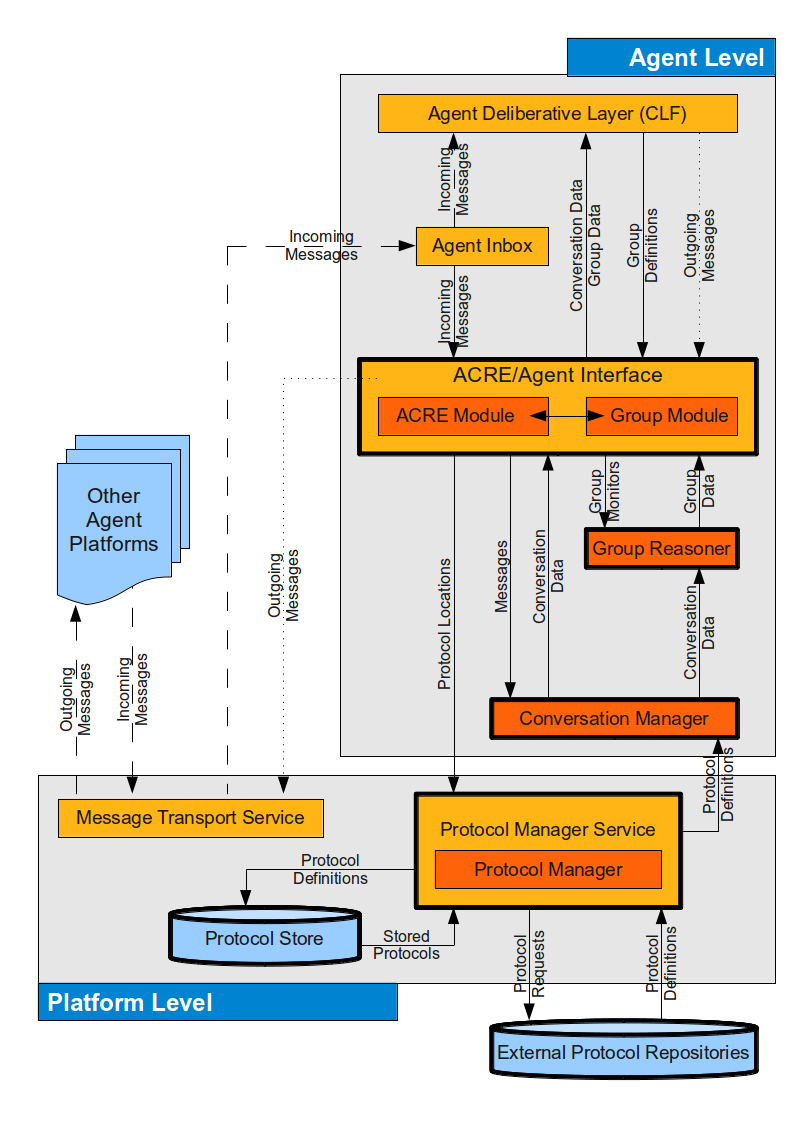}
\caption{ACRE Architecture within Agent Factory.}
\label{fig:agent-factory:architecture}
\end{figure}

Figure~\ref{fig:agent-factory:architecture} shows the architecture used in integrating ACRE into the Agent Factory Framework. In this diagram, components with thinner borders are extant within Agent Factory. Thicker borders indicate ACRE components. 

Additionally, different styles of lines are used for the data-flow arrows in order to indicate how these differ from an Agent Factory deployment that does not make use of ACRE. 
\begin{itemize}
	\item Solid lines indicate those flows of data that are specific to ACRE and are not present in a non-ACRE-enabled deployment.

	\item Dashed lines indicate existing, unaltered, data flows. The exchange of ACL messages between platforms is not affected by the adoption of ACRE. These still flow through the Message Transport Service. Incoming messages are stored in an Agent Inbox, where they are collected (via an appropriate sensor) by the addressee agent.

	\item Dotted lines show areas where the flow of data has been rerouted when compared with the standard Agent Factory deployment. Outgoing messages are normally sent by the agent by direct interaction with the Message Transport Service. However, it is necessary for such messages to be presented to the Conversation Manager and Group Reasoner so that they can be matched against active conversations (and used in the creation of new conversations). The \texttt{protocol} and \texttt{conversation-id} fields, along with recipient details, are added by the ACRE/Agent Interface (after consulting the Conversation Manager) before being sent to the Message Transport Service for sending.
\end{itemize}

\subsection{Platform Services}

Agent Factory supports the use of platform services (such as the Message Transport Services) to allow components to be shared amongst multiple agents on a platform. As the Protocol Manager operates on the platform level and is intended to be shared, it is wrapped within a platform service.

The Protocol Manager component itself is no different from the one featured in the Generic Architecture. The use of a dedicated platform service as a wrapper allows it to be discovered by other components within an Agent Factory deployment.

\subsection{ACRE/Agent Interface}

The ACRE/Agent Interface (AAI) is the key component in a concrete implementation of an ACRE system. As previously discussed in Chapter~\ref{chap:acre-architecture}, this is the only component of the system that is required to be specifically adapted for use within a particular MAS framework.

The role of the AAI is to expose the capabilities of the other ACRE components to an intelligent agent. It is also tasked with providing the agent with information regarding the status of its interactions with other agents.

The generic ACRE architecture calls for each agent to have its own AAI, which provides access to the platform-level Protocol Manager, along with the agent-specific components: the Group Reasoner and the Conversation Manager.

As a component designed to provide information and capabilities to a single agent, in the context of the CLF it is most suitable to implement the AAI as a module that can be used with any of the CFL AOP languages.

In the Agent Factory implementation, the AAI is split into two modules:
\begin{itemize}
   \item The \textit{ACRE Module} provides an interface to the Conversation Manager and Protocol Manager in order to support reasoning about individual conversations.
   \item The \textit{ACRE Group Module} provides access to the Group Reasoner and so furnishes its parent agent with the facilities to create and manage group monitors, along with providing information about these.
\end{itemize}

The separation of the AAI allows agent developers to make use of ACRE features without the additional overhead of group management if this is not desired. In a situation where the Group Module is not loaded by the agent, the creation of a Group Reasoner will also be skipped.

The ACRE Module is the key access-point through which an agent gains access to the core ACRE infrastructure. On creation, a Conversation Manager component is generated for the agent that has loaded the module. This is responsible for handling all conversation reasoning for that specific agent.

Additionally, the ACRE Module must also be made aware of the location of the Protocol Manager Service, by means of the \texttt{init} action outlined in Section~\ref{sec:agent-factory:acre-module-setup}, which must be explicitly invoked by an ACRE-capable agent. This express initialisation is a consequence of how module loading in Agent Factory is performed. A module cannot discover a platform service at load time.

As CLF modules, the ACRE Module and ACRE Group Module provide a set of actions and sensors that encapsulate the capabilities of the ACRE subsystem with which they interface. They allow the agent to effect changes to the status of the underlying ACRE components. These are outlined in the following Sections, categorised according to how they are used. Where example code is shown, it is written in the AF-AgentSpeak AOP language. The syntax of action invocation in the other CLF languages may differ slightly but the way in which the actions may be used does not differ across the various AOP languages.

\subsubsection{Actions for Setup}
\label{sec:agent-factory:acre-module-setup}

\begin{description}
	\item[\texttt{init}] Initialise the ACRE Module. Because of limitations in the underlying implementation of how modules are loaded within Agent Factory, it is not possible to automatically connect the module to the Protocol Manager Platform Service when it is initially loaded. The module binds to the platform service and gets a list of available protocols that have previously been loaded.
	\item[\texttt{addRepository(?url)}] Inform the Protocol Manager of the existence of a remote repository from which new protocols can be loaded. The \texttt{?url} parameter indicates the base of the repository.
\end{description}

\subsubsection{Actions to Manage Contacts}
\label{sec:agent-factory:acre-module-contacts}

The existing capabilities of the CLF allow agents to send FIPA messages by means of a \texttt{.send} action. This requires the agent identifier as one of its parameters, in the form \texttt{agentID(?name,addresses(?address1))} (where the ``addresses'' function can specify one or more addresses by which the agent can be contacted). Rather than requiring the address of an agent to be provided for every message, the ACRE Module provides an address book functionality that allows an agent to be referred to only by name. Agents can be added to this address book in two ways:
\begin{itemize}
	\item The initiator of a conversation will automatically be added to the address book of the other ACRE-enabled participant in the conversation.
	\item An ACRE-enabled agent may add a specific contact by using the \texttt{addContact} action.
\end{itemize}

\begin{description}
	\item[\texttt{addContact(?aid)}] Add a contact to the address book. There are two acceptable formats for the \texttt{?aid} parameter:
	\begin{enumerate}
		\item If the parameter is in the same format as an agent identifier being passed to the existing CLF \texttt{.send} action, both the name and the specified addresses of the contact agent are stored in the address book.
		\item If the parameter is a constant, this is taken to be the name of the agent, with a local address on the same agent platform assumed by default.
	\end{enumerate} 
	\item[\texttt{removeContact(?aName)}] Remove a contact from the address book. Here, only the unique agent name is required.
\end{description}

\subsubsection{Actions to Send Messages}

\begin{description}
	\item[\texttt{start(?pid, ?receiver)}] Begin a new conversation. The minimal required parameters are the unique identifier of the protocol that is to be used in the conversation (\texttt{?pid}) and the unique name of the other agent that is to be a participant in the conversation (\texttt{?receiver}). A new conversation is added to the Conversation Manager that is initially in its start state (as no messages have yet been exchanged).

	Unlike the majority of actions provided by the ACRE/Agent Interface, the \texttt{start} action is implemented within the Common Language Framework as an ``active action'', meaning that it has a return value that may be assigned to a variable. Here, the return value is the unique identifier of the conversation that has been created by the Conversation Manager as a result of the action (assuming the protocol is known).

	\item[\texttt{start(?pid, ?receiver, ?performative, ?content)}] This second version of \texttt{start} includes two further parameters and is intended for use when it is desired to also to send the first message. The performative of this message must be supplied, along with the message content. An example of how this would be used is as follows:\footnote{In these examples, ``\texttt{acre}'' is the name given to the ACRE module when it is loaded by the agent.}

	\begin{Verbatim}[fontsize=\footnotesize]
	?cid = acre.start(acre-vickrey,agent1,inform,test);
	\end{Verbatim}

	Essentially, this is shorthand for first initiating a conversation and then advancing it. As such, it is equivalent to the following (the \texttt{advance} action is discussed below):

	\begin{Verbatim}[fontsize=\footnotesize]
	?cid = acre.start(acre-vickrey,agent1),
	acre.advance(?cid,inform,test);
	\end{Verbatim}

	As with the two-argument form of the \texttt{start} action, this also returns the unique conversation identifier that has been generated by the Conversation Manager.

	\item[\texttt{start(?pid,?receiver,?performative)}] This three-argument version of \texttt{start} is identical to its four-argument counterpart except that the first message is sent with an empty \texttt{content} field.

	\item[\texttt{advance(?cid,?performative,?content)}] Sends a message that is part of a conversation (identified by its unique identifier, \texttt{?cid}). The name and address of the agent that is to receive the message have already been recorded in the Conversation Manager and as such is handled automatically. The performative and content of the message must also be specified.

	\item[\texttt{advance(?cid,?performative)}] An \texttt{advance} action is also available to send a message with an empty \texttt{content} field, given a conversation identifier and a performative.

\end{description}

\subsubsection{Actions to Manage Conversations} \label{sec:agent-factory:actions-manage-conversations}

\begin{description}
	\item[\texttt{forget(?cid)}] The conversation identified by \texttt{?cid} is permanently removed from memory.

	\item[\texttt{archive(?cid)}] In some circumstances, it may not be desirable to permanently remove all details of a conversation, yet details of all conversations need not be immediately available. The \texttt{archive} action allows details of a conversation to be temporarily removed from the belief base of an agent in a way that allows them to be recalled later. A large belief base may cause long processing times for agents, so this action allows the agent's processing to become more streamlined.

	\item[\texttt{recall(?cid)}] A previously archived conversation can be recalled.

	\item[\texttt{setTimeout(?cid,?timeout)}] May be invoked at any time during a conversation. Its effect is that any further messages that are sent as part of that conversation will have a timeout associated with them. As previously discussed in Section~\ref{sec:acre-introduction:timeout}, this is implemented by way of the FIPA \texttt{reply-by} performative. The \texttt{?timeout} parameter should be the number of seconds from the sending of the message within which the recipient is required to respond. The task of converting this to a timestamp by which a response must be received is automatically performed by the ACRE/Agent Interface.
	
	\item[\texttt{annotate(?cid,?annotation)}] Conversations may be annotated to record details about them that are not recorded by default by the Conversation Manager. This is offered as an alternative to adopting beliefs in the usual way, since these would need to be handled separately. Annotations are removed whenever a conversation is forgotten, and are archived and recalled along with their associated conversations also.
	
	\item[\texttt{deannotate(?cid)}] Removes all annotations from a specified conversation.
	
	\item[\texttt{deannotate(?cid,?annotation)}] Removes a specified annotation from a conversation if it exists.
\end{description}

\subsubsection{Actions to Cancel Conversations}

\begin{description}
	\item[\texttt{cancel(?cid)}] Initiates the cancellation meta-protocol described in Section~\ref{sec:acre-introduction:cancel} for a conversation that has not yet terminated. The actual mechanics of how the messages are formed are abstracted from the agent programmer and handled automatically by the Conversation Manager. The agent merely needs to supply the identifier of the conversation that is to be cancelled. This results in the appropriate \texttt{cancel} message being sent to the other participant. The response will likewise be handled automatically by the Conversation Manager, cancelling the conversation if confirmation is received and returning it to active status if the cancellation fails. In each scenario, an appropriate event is raised so the agent is aware of the outcome of the attempted cancellation.

	\item[\texttt{confirmCancel(?cid)}, \texttt{denyCancel(?cid)}] Two actions are provided for dealing with requests from another agent to cancel a conversation. These are convenience actions where the Conversation Manager automatically takes care of constructing appropriate messages to reflect the intention of the agent to either agree to the cancellation of a conversation or to deny this cancellation (by means of a \texttt{failure} message). In each case, the action should be used in response to an event notifying the agent of a request to cancel the conversation, otherwise the actions will fail.
\end{description}

\subsubsection{Sensors} \label{sec:agent-factory:module-sensors}

The ACRE Module provides two sensors to provide the agent with information about loaded protocols, conversation states, conversation participants and other information related to the ACRE components. Each of these sensors gathers information from all of the connected ACRE components, but they differ in the type of information they provide:

\begin{description}
\item[Knowledge Sensor] The Knowledge Sensor consistently provides information about the ongoing state of the ACRE components, along with the conversations and protocols they manage. On each iteration of the Knowledge Sensor, it generates a consistent set of beliefs about active conversations and available protocols, in order to facilitate the agent in reasoning about these. 

The beliefs that can be generated, when appropriate, by the Knowledge Sensor are as follows:
\begin{description}

\item[\texttt{conversationArchived(?cid,?pid,?aName)}] Indicates that the conversation identified by \texttt{?cid} has been archived and provides a minimal amount of information about it, namely the identifier of the protocol that conversation followed (as \texttt{?pid}) and the name of the other participant (the \texttt{?aName} parameter). This is the only percept that is generated for archived conversations. It is necessary to use the \texttt{recall} action in order to have other knowledge be available.

\item[\texttt{conversationStatus(?cid,?status)}] This indicates the status of a conversation of which the Conversation Manager is aware. The \texttt{?cid} parameter (as in all the beliefs in this section) refers to the unique identifier of the conversation to which the belief relates. The \texttt{?status} parameter can be one of ``ready'', ``active'', ``completed'', ``failed'', ``stale'', ``cancelling'' or ``cancelled'', as discussed in Section~\ref{sec:acre-architecture:conversation-manager}.

	\item[\texttt{conversationParticipant(?cid,?aName)}] This belief allows the agent to become aware of the identity of the other agent (in \texttt{?aName}) that is party to the conversation identified by \texttt{?cid}.

	\item[\texttt{conversationProtocolID(?cid,?pid)}]Each protocol is identified by a unique identifier consisting of three separate parts: a namespace, a name and a version number, as discussed in Section~\ref{sec:acre-architecture:external-protocol-repositories}. This belief contains the entire unique identifier of the protocol that the conversation follows.

	\item[\texttt{conversationProtocolName(?cid,?pname)}]
	\item[\texttt{conversationProtocolNamespace(?cid,?pnamespace)}]
	\item[\texttt{conversationProtocolVersion(?cid,?version)}]
These three percepts offer more convenient mechanisms by which an agent may reason about the individual components of the overall protocol identifier.

	\item[\texttt{conversationState(?cid,?state)}] The \texttt{?state} parameter refers to one of the states that are declared as part of the Finite State Machine defining the underlying protocol and give an indication of the progress of the conversation to date.

	\item[\texttt{conversationLength(?cid,?length)}] The \texttt{?length} parameter refers to the number of messages that have been exchanged as part of the conversation to date. This is helpful in identifying changes in state for protocols where loops exist, which allow the conversation to reach the same state on multiple separate occasions. An example of this can be seen in the \textit{Process Documents} example shown in Section~\ref{sec:acre-introduction:process-documents}.

	\item[\texttt{knownProtocol(?pid)}] This is the only belief of the Knowledge Sensor that does not come from the Conversation Manager. Instead, a belief of this type is adopted for every protocol (identified by \texttt{?pid}) that has been loaded into the Protocol Manager. This allows the agent to check whether its ACRE components are aware of a particular protocol before attempting to instigate a conversation that follows it.

	\item[\texttt{annotation(?cid,?annotation)}] Annotations added using the \texttt{annotate} action are accessed by the Knowledge Sensor to allow the agent to reason about them.

	\item[\texttt{conversationHistory(?cid,?i,?type,?p,?content)}] Details of conversations that have not been forgotten or archived are made available using this percept. A percept is generated for each message that has been part of the conversation identified by \texttt{?cid}. The information available consists of the place of the message in the conversation (the \texttt{?i} parameter, beginning at 1 for the first message in the conversation). The \texttt{?type} parameter indicates whether the message was ``sent'' or ``received'', and must contain one of those two values. The \texttt{?p} (performative) and \texttt{?content} parameters refer to the details of the message itself.
\end{description}

\item[Event Sensor] The Event Sensor causes the agent to adopt beliefs about one-off events arising as a result of activity within the ACRE components. Unlike the Knowledge Sensor, these are not repeated, as the sensor consumes any available events on each iteration. The conversation-specific events consumed by the Event Sensor are those added by the Conversation Manager in Chapter~\ref{chap:semantics}.

The events consumed by the Event Sensor are as follows:
\begin{description}
	\item[\texttt{ready(acre)}] The ACRE module has been successfully loaded and has bound itself to the Protocol Manager service.

	\item[\texttt{conversationAdvanced(?cid,?state,?length)}] A conversation identified by \texttt{?cid} has been advanced by the receipt or sending of a new message. In addition, the \texttt{?state} parameter also indicates the state the conversation has reached by advancing. The \texttt{?length} parameter indicates the number of messages that have now been part of the conversation. This ensures that each \texttt{conversationAdvanced} event is unique, even within conversations whose protocol definitions allow the same state to be reached multiple times.

	\item[\texttt{conversationMessage(?cid,?performative,?content)}] Whenever a conversation is advanced, it is important for the agent to become aware of the details of the message that caused this to occur. The performative of the message and the message contents are given as the \texttt{?performative} and \texttt{?content} parameters respectively. The sender of the message can be ascertained by cross-referencing the conversation identifier with the \texttt{conversationParticipant} belief generated by the Knowledge Sensor.

   \item[\texttt{conversationStarted(?cid)}] 
   \item[\texttt{conversationEnded(?cid)}]
   \item[\texttt{conversationFailed(?cid)}]
	A new conversation has begun, finished or failed. These events are raised by the Conversation Manager. A conversation fails when a message purports (via its \texttt{conversation-id} parameter) to advance a conversation but cannot trigger any active transition. The other events are always accompanied by \texttt{conversationAdvanced} and \texttt{conversationMessage} events, as they require normal advancement of the conversation to have occurred.

	\item[\texttt{conversationCancelRequest(?cid)}] An attempt to cancel a conversation has been made (see Section~\ref{sec:acre-introduction:cancel}).

	\item[\texttt{conversationCancelConfirmed(?cid)}]
	\item[\texttt{conversationCancelFailed(?cid)}] After a \texttt{cancel} message is sent, the receiver of that request must confirm the cancellation of the conversation or indicate that this cancellation has failed. These are indicated by the use of the \texttt{conversationCancelConfirmed} and \texttt{conversationCancelFailed} events respectively. 

	\item[\texttt{conversationTimeout(?cid)}] For conversations that have timeouts associated with them, the Conversation Manager will be responsible for checking the status of conversations so as to raise an appropriate event whenever the timeout time has been passed.

	\item[\texttt{unmatchedMessage(?performative,?sender,?content)}] An unmatched message is one that cannot be successfully matched against any active conversation so as to advance it and is incapable of initiating a conversation that follows any known protocol. This belief provides the details of the message in question.

	\item[\texttt{ambiguousMessage(?performative,?sender,?content)}] An ambiguous message is one that is capable of advancing/initiating multiple active conversations/protocols. It can only arise when a message is sent without a \texttt{conversation-id} or \texttt{protocol} parameter.
   \end{description}
\end{description}

\subsection{ACRE Group Module}

The ACRE Group Module is separate to the ACRE Module so that an agent developer may choose whether or not to load support for conversation and agent groups. The Group Reasoner is only created upon an ACRE Group Module being loaded. Without this, all conversations are treated individually.

As with the ACRE Module, the ACRE Group Module is intended to constitute part of the ACRE/Agent Interface: specifically to allow the agent to interact with the Group Reasoner component. As a CLF module, it also provides a number of actions and sensors to facilitate this interaction. These are outlined in the following Sections. The Group Reasoner supports the grouping of both agents and conversations. As these are different entities, the actions relating to each are outlined separately.

\subsubsection{Actions Relating to Agent Groups}

\begin{description}
   \item[\texttt{init}] The \texttt{init} function performs the first initialisation of the ACRE group module. This should be done after the ACRE Module has previously been loaded, as the ACRE Group Module makes use of this to send its individual messages and for access to the Conversation Manager.
   
   \item[\texttt{newAgentGroup(?agid)}] A new group of agents is created. The \texttt{?agid} parameter is the unique identifier given to the agent group\footnote{In this chapter, the name \texttt{?agid} is used for the identifiers of groups of agents, whereas conversation group identifiers are expected whenever a parameter named \texttt{?cgid} is used.}. An empty group is created with the given identifier. Agents can be subsequently added using the \texttt{addAgent} action.
   
   \item[\texttt{addAgent(?agid,?agent-list)}] The \texttt{addAgent} action allows agents to be added to an existing agent group identified by \texttt{?agid}. The \texttt{?agent-list} parameter can be either a single agent identifier or a list of agent identifiers, as follows:\footnote{In these examples, ``\texttt{acreGroup}'' is used as the name under which the ACRE group module was loaded by the agent.}
   
   \begin{Verbatim}[fontsize=\footnotesize]
      // single agent
      acreGroup.addAgent(group1,agent1)
      
      // multiple agents
      acreGroup.addAgent(group1,[agent2,agent4,agent4])
   \end{Verbatim}
      
   \item[\texttt{newAgentGroup(?agid,?agent-list)}] A second version of \texttt{newAgentGroup} allows for the creation of a new agent group whose members are specified on creation. An example of its use, to create a group named ``group1'' consisting of the agents ``agent1'' and ``agent2'' is as follows:
   
   \begin{Verbatim}[fontsize=\footnotesize]
      acreGroup.newAgentGroup(group1,[agent1,agent2])
   \end{Verbatim}
   
   The above code is equivalent to first calling the one-argument \texttt{newAgentGroup} action, followed immediately by using \texttt{addAgent} to add the requisite agents to the newly-created group, as follows:
   
   \begin{Verbatim}[fontsize=\footnotesize]
      acreGroup.newAgentGroup(group1),
      acreGroup.addAgent(group1,[agent1,agent2])
   \end{Verbatim}

   \item[\texttt{removeAgent(?agid,?agent-list)}] Remove an agent or list of agents from a specified agent group.
   
   \item[\texttt{disband(?agid)}] The \texttt{disband} action will cause an agent group to be removed from the Group Reasoner's memory so it can no longer be referenced. This means that no agent can be added or removed to or from this group, and that no new conversation groups can be initiated with this group of agents.
\end{description}

\subsubsection{Actions Relating to Conversation Groups} \label{sec:agent-factory:conversation-group-actions}

\begin{description}
   \item[\texttt{start(?pid,?agid)}] Begin a group of related conversations. Each conversation must follow the same protocol, (identified by \texttt{?pid}). This action begins a conversation with each of the members of the specified agent group. Initiating new conversations results in the creation of a conversation group, the unique identifier of which is the return value of the function.

   \begin{Verbatim}[fontsize=\footnotesize]
   ?cgid = acreGroup.start(acre-vickrey,group1)
   \end{Verbatim}

   The individual conversations are assigned their own unique identifier, of which the agent is made aware through its knowledge sensors (see Section~\ref{sec:agent-factory:module-sensors}). Each conversation is initiated in its starting state, as no messages are sent by invoking this action.
   
   \item[\texttt{start(?pid,?agid,?performative,?content)}] The four-argument version of the \texttt{start} action combines the creation of a conversation group with the sending of the first message in each of the conversations in the group. As with the two-argument \texttt{start} action, the return value is the unique identifier of the conversation group that is created.

   All messages sent using this action are identical: if it is desired to send customised messages to each recipient then the conversation group must be initially created using the two-argument \texttt{start} action, and each conversation must then be individually advanced using the \texttt{advance} action.

   \item[\texttt{start(?pid,?agid,?performative)}] A three-argument version of \texttt{start} allows the initial message in each conversation to be sent with an empty \texttt{content} field.

   \item[\texttt{remove(?cgid,?cid)}] The \texttt{remove} action removes a conversation (identified by the \texttt{?cid} parameter) from a conversation group. This is useful in circumstances such as a participant declining to bid in an auction.

   \item[\texttt{advanceAll(?cgid,?performative)}]
   \item[\texttt{advanceAll(?cgid,?performative,?content)}] This is a convenience action for situations where it is desired to send the same message as part of all of the conversations in a group. Specifying content for the message is optional.

   Depending on the state of each of the group conversations, it may be the case that such a message does not trigger any active transition of one or more conversations. If this is the case, those conversations for which the message is an appropriate next step will be advanced successfully. Other conversations will not be advanced and an \texttt{unmatchedMessage} event will be triggered and brought to the agent's attention via the event sensor (see Section~\ref{sec:agent-factory:module-sensors}).

   \item[\texttt{watch(?group-name,?monitor-list)}] Attach group monitors (see Section~\ref{sec:acre-architecture:group-reasoner}) to a conversation group. The \texttt{?monitor-list} parameter provides a list of the group monitors that are to be attached to each group. Each of these should be the class name of a Java class that implements the \texttt{IGroupMonitor} interface, along with associated parameters. If a package name is not provided, it is assumed that the monitor is to be loaded from the default package (namely \texttt{is.lill.acre.group.monitor}). Following the class name, a comma-separated list of parameters (the number of which should match the \texttt{PARAMS} attribute provided in the class) should follow within parentheses. An example of how the \texttt{AllInState} group monitor is added is as follows:
   
   \begin{Verbatim}[fontsize=\footnotesize]
   acreGroup.watch(group1,AllInState(started))
   \end{Verbatim}
   
   Here, \texttt{group1} is the unique identifier of the conversation group that is to be watched. The \texttt{AllInState} group monitor is a default monitor and so it does not require a package name. This requires just the name of the underlying protocol as its only parameter.

   \item[\texttt{unwatch(?cgid,?description-list)}] This is the opposite of the \texttt{watch} action. The \texttt{?description-list} parameter again contains descriptions of group monitors. In this case, any group monitors that were loaded with these descriptions are removed from the Group Reasoner and will no longer raise any events.

   \item[\texttt{unwatch(?cgid)}] A more general form of the \texttt{unwatch} action, this removes all group monitors from the specified conversation group.

   \item[\texttt{setTimeout(?cgid,?timeout)}]The group \texttt{setTimeout} action operates in a similar way to its counterpart for individual conversations. The same timeout value is set for all conversations in the group.

	\item[\texttt{newGroup(?cgid)}] Creates a new, empty, conversation group. Conversations can be explicitly added to and removed from this group at a later time.
	
	\item[\texttt{add(?cgid,?cid)}] Adds an existing conversation (identified by \texttt{?cid}) to a conversation group.
	
	\item[\texttt{remove(?cgid,?cid)}] Removes a specified conversation from a conversation group.
	
	\item[\texttt{annotate(?cgid,?annotation)}]
	\item[\texttt{deannotate(?cgid)}]
	\item[\texttt{deannotate(?cgid,?annotation)}] These three actions operate in exactly the same way as their counterparts for single conversations, discussed in Section~\ref{sec:agent-factory:actions-manage-conversations}. The ACRE Group module also allows annotations to be added to and removed from conversation groups.
\end{description}
%TODO: in the above, you can't archive groups.

\subsubsection{Sensors}

In the same way as the ACRE Module, the ACRE Group Module also features a Knowledge Sensor and an Event Sensor. Once again, the Knowledge Sensor makes the agent aware of ongoing situations whereas the Event Sensor consumes one-off events that occur in connection with groups of conversations.

\begin{description}
   \item[Knowledge Sensor] The ACRE Group Module's Knowledge Sensor provides information only on the Group Reasoner. Knowledge about individual conversations, whether members of a conversation group or not, is provided by the Knowledge Sensor that is part of the ACRE Module.
   \begin{description}
      \item[\texttt{groupMember(?agid,?aid)}]An agent (identified by \texttt{?aid}) is a member of the specified agent group.
      \item[\texttt{conversationGroup(?cgid,?cid)}] A conversation (identified by \texttt{?cid}) is a member of a specified conversation group.
      \item[\texttt{groupSize(?cgid,?size)}] The number of conversations that are members of the conversation group.
   \end{description}

   \item[Event Sensor] The Event Sensor provided by the ACRE Group Module consumes only those events that relate to conversation groups. No changes are made to agent groups without the direct invocation of an action by the agent (which will fail if an error occurs). Conversations operating within a group are still managed on the individual level by the Conversation Manager, so in the same way as the Knowledge Sensors, the Event Sensor that is part of the ACRE Module will adopt beliefs in response to conversations beginning, advancing and ending in the same way as for conversations that are not part of groups. The ACRE Group Module's Event Sensor only handles additional events that are not covered within the existing Event Sensor.
   \begin{description}
      \item[\texttt{groupEvent(?cgid,?monitor-description)}] Indicates that some event has been raised by an active group monitor attached to a particular conversation group. Such an event occurs whenever the \texttt{event} method of a loaded group monitor returns \texttt{true} (see Section~\ref{sec:acre-architecture:group-reasoner}). When this occurs, the \texttt{?monitor-description} is the same as that used to initially load the group monitor in the \texttt{watch} action that is also provided by the ACRE Group Monitor.

      As an example, suppose a group monitor had been loaded using the following code:
      
      \begin{Verbatim}[fontsize=\footnotesize]
acreGroup.watch(myGroup,AllReachedState(bid_submitted))
      \end{Verbatim}
      
      An event that is raised by this group monitor would be:
      \begin{Verbatim}[fontsize=\footnotesize]
groupEvent(myGroup,AllReachedState(bid_submitted))
      \end{Verbatim}
   \end{description}
\end{description}

\section{Implementation of Example Protocols}

In previous chapters, a number of protocols have illustrated the features of ACRE in both informal (in Chapter~\ref{chap:acre-introduction}) and formal (in Chapter~\ref{chap:semantics}) settings. This Section completes that series of examples by illustrating how agent programs following these sample protocols could be created using the Agent Factory implementation of ACRE.

In each case, the agent programs presented here are intended to be for illustrative purposes only. Thus some shortcuts have been taken in order to simplify the examples with the intention of highlighting the communicative aspect of the programs.

\subsection{Request/Response Protocol}

Figures~\ref{fig:agent-factory:request-agent} and~\ref{fig:agent-factory:response-agent} show implementations for the Request/Response Protocol (originally shown in Figure~\ref{fig:acre-introduction:request-response-auml}). These sample agents are intended to illustrate how ACRE integrates into AF-AgentSpeak to facilitate communication. For this reason, the protocol itself is not required to be realistic (the semantics of the \texttt{request} and \texttt{inform} performatives are not suited to messages sent without content, for example). The agents have merely been written to send messages that follow the protocol.

\begin{figure}
\begin{Verbatim}[numbers=left,frame=single,fontsize=\footnotesize]
#agent Requester

module acre -> is.lill.acre.agent.module.ACREModule;

+initialized : true <-
    acre.init;

+ready(acre) : request(?responder) <-
    acre.start(request-response,?responder,request);
   
+conversationAdvanced(?cid,end,?l) : 
  conversationProtocolName(?cid,request-response) <-
    .println( "Conversation ended: " + ?cid );
\end{Verbatim}
\caption{ACRE-enabled Requester agent for Request/Response Protocol.}
\label{fig:agent-factory:request-agent}
\end{figure}

Figure~\ref{fig:agent-factory:request-agent} shows the code for the agent that initiates the conversation. The first line is a requirement of all AF-AgentSpeak agents, specifying the type of agent the code represents. Following this, the ACRE module is loaded and initialised. This is common to all ACRE-enabled agents.

Two rules form the main body of the agent. The first (beginning on line 8) is triggered by the \texttt{ready(acre)} event, which is raised once the ACRE module has completed its setup (i.e. binding to the Protocol Manager platform service and creating a Conversation Manager for this agent). It assumed that this agent has been given an initial belief of the form \texttt{request(agentName)} from which it can become aware of the name of the other agent in the conversation. The \texttt{acre.start} action initiates a new conversation of the type \texttt{request-response}, with the first message being addressed to the other agent. The performative of this first message is \texttt{request} and it has no content.

The second rule (beginning on line 11) deals with the end of the conversation. If a conversation following the \texttt{request-response} protocol advances to the state named \texttt{end} then it simply prints that information to the console.

\begin{figure}
\begin{Verbatim}[numbers=left,frame=single,fontsize=\footnotesize]
#agent Responder

module acre -> is.lill.acre.agent.module.ACREModule;

+initialized : true <-
    acre.init;

+conversationAdvanced(?cid,requested,?l) : 
  conversationProtocolName(?cid,request-response) <-
    acre.advance(?cid,inform);
\end{Verbatim}
\caption{ACRE-enabled Responder agent for Request/Response Protocol.}
\label{fig:agent-factory:response-agent}
\end{figure}

The implementation of the other agent in this conversation is shown in Figure~\ref{fig:agent-factory:response-agent}. This agent also features the same initialisation code as the Requester agent.

The only rule for this agent that is relevant to communication is the one beginning on line 8. This responds to a conversation following the \texttt{request-response} protocol that has been advanced to the \texttt{requested} state. The response consists only of an \texttt{inform} message with no content.

\subsection{Status Report Protocol}

ACRE implementations of the Status Report Protocol (originally shown in Figure~\ref{fig:acre-introduction:status-auml}) are shown in Figures~\ref{fig:agent-factory:status-request-acre-agent} and~\ref{fig:agent-factory:status-response-acre-agent}.

\begin{figure}[!hb]
\begin{Verbatim}[numbers=left,frame=single,fontsize=\footnotesize]
#agent ACREStatusRequester

module acre -> is.lill.acre.agent.module.ACREModule;

+initialized : true <-
    acre.init;

+ready(acre) : askStatus(?responder) <-
    acre.start(status,?responder,request,status(?responder));
   
+conversationAdvanced(?cid,done,?l) :
  conversationHistory(?cid,?l,received,inform,statusOf(?o,?s)) <-
    .println("Status of " + ?o + " was " + ?s);
\end{Verbatim}
\caption{ACRE-enabled Requester agent for Status Report Protocol.}
\label{fig:agent-factory:status-request-acre-agent}

\begin{Verbatim}[numbers=left,frame=single,fontsize=\footnotesize]
#agent ACREStatusResponder

module acre -> is.lill.acre.agent.module.ACREModule;

+initialized : true <-
    acre.init;

+conversationAdvanced(?cid,requested,?l) : 
   conversationProtocolName(?cid,status) & 
   conversationHistory(?cid,?l,received,request,status(?obj)) & 
   name(?obj) &
   status(?status) <- 
     acre.advance(?cid,inform,statusOf(?obj,?status));
\end{Verbatim}
\caption{ACRE-enabled Responder agent for Status Report Protocol.}
\label{fig:agent-factory:status-response-acre-agent}
\end{figure}

As with the agents for the Request/Response protocol, both of these agents have the code necessary to load and initialise the ACRE module. The ACREStatusRequester agent is seeded with an initial belief of the form \texttt{askStatus(agentName)} that causes the rule beginning on line 8 to fire as soon as the ACRE module is ready. A new conversation is begun using the \texttt{status} protocol that asks the ACREStatusResponder agent what its status is.

Once the other participant has responded (thus bringing the state of the conversation to the \texttt{end} state), it prints the details of the message it received.

For the ACREStatusResponder agent, it responds to messages following the \texttt{status} protocol that bring the conversation to the \texttt{requested} state. By default, each AF-AgentSpeak agent has a \texttt{name} belief about what its own name is. Line 11 checks if this is the same as the entity about which the received message was enquiring (the variable is bound to the value from the message in line 10). For simplicity, the ACREStatusResponder agent is seeded with an initial belief \texttt{status(up)} that is matched in line 12. The agent continues the conversation in line 13 by informing the initiator of its status.

These simple agents illustrate how ACRE can be used within AF-AgentSpeak. However, to fully demonstrate the benefits of using ACRE, it is necessary to also draw comparisons with agent code that achieves the same task without the use of ACRE.

\begin{figure}[!hb]
\begin{Verbatim}[numbers=left,frame=single,fontsize=\footnotesize]
#agent SimpleStatusRequester

+initialized : askStatus(agentID(?responder,?addr)) <-
    .send(request,agentID(?responder,?addr),status(?responder));

+message(inform,?sender,statusOf(?o,?s)) : true <-
    .println("Status of " + ?o + " was " + ?s);   
\end{Verbatim}
\caption{Non-ACRE Requester agent for Status Report Protocol.}
\label{fig:agent-factory:status-request-simple-nonacre-agent}

\begin{Verbatim}[numbers=left,frame=single,fontsize=\footnotesize]
#agent SimpleStatusResponder

+message(request,?sender,status(?obj)) :
   name(?obj) &
   status(?status) <- 
     .send(inform,?sender,statusOf(?obj,?status));
\end{Verbatim}
\caption{Non-ACRE Responder agent for Status Report Protocol.}
\label{fig:agent-factory:status-respond-simple-nonacre-agent}
\end{figure}

Figures~\ref{fig:agent-factory:status-request-simple-nonacre-agent} and~\ref{fig:agent-factory:status-respond-simple-nonacre-agent} show code to implement simple, na\"{i}ve agents to follow the Status Report protocol. Here, agents send and receive individual messages that act as standalone communications with no formal link between them. As with the ACRE version, the SimpleStatusRequester agent is given an initial belief about the agent it needs to contact for a status update. Line 4 illustrates how the address of the receiver agent must be explicitly set every time a message is sent. ACRE assumes a local address unless specified otherwise when sending the message or when adding an agent to the address book.

Upon receipt of a reply (in the rule beginning on line 6), it simply prints the status of the responder agent to the terminal.

For the SimpleStatusResponder agent, it replies to any request for a status by sending a message back to the agent who sent the request.

Although these implementations will typically have the same effect as the ACRE-enabled agents, additional checks that are performed by ACRE are not available in this situation. A consequence of this is that the rule beginning on line 6 of the SimpleStatusRequester agent will accept a message from \textit{any} agent informing it of the status of \textit{any} item, as this rule is independent of the sending of the initial \texttt{request} message.

To combat this shortcoming, the SimpleStatusRequester must be adapted to keep track of what other status request has been made of another agent. Figure~\ref{fig:agent-factory:status-request-better-nonacre-agent} shows a superior implementation of this agent. In this code, an additional belief is adopted to record the fact that a status has been requested of another agent (this can be seen on line 4). Following this, the second rule will only fire if the agent sending the response is the agent to whom the initial request was made and the object whose status is being reported is the also the subject of the original request.

\begin{figure}[!ht]
\begin{Verbatim}[numbers=left,frame=single,fontsize=\footnotesize]
#agent BetterStatusRequester

+initialized : askStatus(agentID(?responder,?addr)) <-
    +asked(status(?responder,?responder)),
    .send(request,agentID(?responder,?addr),status(?responder));

+message(inform,agentID(?name,?addr),statusOf(?obj,?status)) :
  asked(status(?name,?obj)) <-
    .println("Status of " + ?obj + " was " + ?status);

\end{Verbatim}
\caption{Better Non-ACRE Requester agent for Status Report Protocol.}
\label{fig:agent-factory:status-request-better-nonacre-agent}
\end{figure}

In an ACRE-enabled agent, such checks are unnecessary, as the protocol definition insists that the receiver of the original message is the same as the sender of the response. The use of the \texttt{?obj} variable in the protocol definition also ensures that the same item must the subject of both the request and the response. If this is not the case then an \texttt{unmatchedMessage} event will be raised, which can then be handled separately.

Omitting these checks from a non-ACRE agent may result in unanticipated situations arising whereby rogue messages are received from malicious, deceitful or malfunctioning agents and these are being acted upon unintentionally. This problem is more pronounced in a larger agent that is expected to conduct multiple conversations (possibly following the same protocol) with many agents at once.

\subsection{Process Documents Protocol}

The final sample protocol is the Process Documents protocol seen originally in Figure~\ref{fig:acre-introduction:process-fsm-start}. This is a more complex protocol that allows the conversation to loop between two states until an agent refuses to process further documents.

In order to facilitate an implementation of this protocol, a simple scenario is proposed. After the ProcessResponder agent indicates that it is ready to process documents, the ProcessRequester agent requests it to process documents identified by integer identifiers. It is assumed that the ProcessResponder has an action available to allow it to perform this processing (not shown). 10 documents will be processed before the ProcessResponder refuses to process a subsequent document, leading to the end of the conversation.

The code to implement these protocols can be seen in Figures~\ref{fig:agent-factory:process-request-agent} and~\ref{fig:agent-factory:process-respond-agent}. The ProcessRequester agent has one additional element to its initialisation rule that is not present in previous ACRE implementations. The \texttt{last} belief is used to record the last document that has been processed. Assuming document identifiers begin at 1, the initial belief is effectively recording the fact that no documents have yet been processed.

\begin{figure}
\begin{Verbatim}[numbers=left,frame=single,fontsize=\footnotesize]
#agent ProcessRequester

module acre -> is.lill.acre.agent.module.ACREModule;

+initialized : true <-
    +last(0),
    acre.init;

+conversationAdvanced(?cid,waiting,?l) :
  conversationProtocolName(?cid,process-documents)
  & last(?last) <-
    ?next = ?last + 1,
    acre.advance(?cid,request,process(?next)),
    .replace(last(?last),last(?next));
\end{Verbatim}
\caption{Requester agent for Process Documents Protocol.}
\label{fig:agent-factory:process-request-agent}

\begin{Verbatim}[numbers=left,frame=single,fontsize=\footnotesize]
#agent ProcessResponder

module acre -> is.lill.acre.agent.module.ACREModule;
module proc -> is.lill.acre.examples.ProcessDocuments;

+initialized : true <-
    acre.init;

+ready(acre) : requester(?name) <-
    acre.addContact(?name),
    ?cid = acre.start(process-documents,?name,inform,ready),
    acre.annotate(?cid,processed(0));

+conversationAdvanced(?cid,requested,?l) : 
  conversationProtocolName(?cid,process-documents) & 
  conversationHistory(?cid,?l,received,request,process(?docID)) & 
  annotation(?cid,processed(?done)) <-
    if ( ?done < 10 ) {
       proc.process(?docID), 
       ?newdone = ?done + 1,
       acre.deannotate(?cid,processed(?done)),
       acre.annotate(?cid,processed(?newdone)),
       acre.advance(?cid,inform,processed(?docID))
    }
    else {
       acre.advance(?cid,refuse,process(?docID))
    };
\end{Verbatim}
\caption{Responder agent for Process Documents Protocol.}
\label{fig:agent-factory:process-respond-agent}
\end{figure}

Following this, the second rule in the ProcessRequester agent (beginning on line 9) is to react to a situation where any conversation following the Process Documents protocol reaches the \texttt{waiting} state by requesting that the participant processes the next document in sequence. Finally, the belief about the last document to be processed is updated using the built-in \texttt{.replace} action.

The ProcessResponder agent is more complex. The initial task to be performed is to inform the ProcessRequester that it is ready to begin processing. This is done on line 11 after adding the name of the ProcessRequester to its internal address book (it is assumed that an initial \texttt{requester} belief has been supplied to the ProcessResponder so that it will know which agent to contact). Having started the new conversation on line 11, the conversation identifier that was automatically generated by ACRE is recorded in the \texttt{?cid} variable, as the developer wished to use that information on the following line.

Line 12 illustrates how ACRE's annotation feature can be used in AF-AgentSpeak code. In this case, it is used as a mechanism to record how many documents have been processed as part of a particular ongoing conversation. The code as written will cause the agent to agree to the processing of ten documents from any conversation that it initiates, before refusing to process at the eleventh document. The use of annotations links this information to a particular conversation, ensuring that this information does not need to be explicitly removed via manually-written code once the conversation has ended.

The third rule in the ProcessResponder agent (beginning on line 14) reacts to a request to process a document. The ACRE Conversation Manager ensures that this request has been sent by an agent with which the ProcessResponder has begun a conversation. Thus another ProcessRequester agent could not have work done for it without agreeing beforehand that it may do so.

The rule will fire only if a message is received as part of a Process Documents protocol and lines 16 and 17 bind the \texttt{?docID} and \texttt{?done} variables to the identifier of the document to be processed and the number of documents previously processed as part of this conversation respectively.

If fewer than ten documents have previously been processed, the request will be satisfied, the agent will record the new total of documents processed as part of this conversation and finally use ACRE's \texttt{advance} action to send a message to indicate that the processing has been completed.

If ten documents have already been processed, a \texttt{refuse} message will be sent that brings the conversation to an end. In this simple scenario, neither agent takes any additional action once the conversation ends, but the \texttt{conversationEnded} event would be raised in both agents, allowing the developer to program the agents to act accordingly.

\section{Summary}
This Chapter has discussed a concrete implementation of the Generic ACRE Architecture, which was presented in Chapter~\ref{chap:acre-architecture}. This implementation is integrated into the FIPA-compliant Agent Factory framework, leveraging the Common Language Framework (CLF) to allow it to be accessible to developers using multiple AOP languages.

The main focus of this implementation is the platform-specific ACRE/Agent Interface that is implemented as two CLF Modules: the ACRE Module (dealing with individual conversations and protocols) and the ACRE Group Module (providing access to a Group Reasoner that supports the combination of related conversations into groups that can be reasoned about collectively).

The core role of these modules is to provide a variety of actions and sensors that form an API to the various platform-independent ACRE components. These allow an agent developer much flexibility in writing agents that are capable of reasoning about their interactions in a more structured fashion than adopting beliefs about individual communications.

This can be seen in the sample agents shown that implement example protocols using ACRE. The use of ACRE removes the responsibility for verifying aspects of the communication from the agent developer. As ACRE verifies the sender, content and order of messages, it is not necessary for developers to contend with these type of communication issues, as they would if using individual message receipt beliefs and sending actions.

The following Chapter outlines an evaluation to gauge the relative effectiveness of using ACRE when compared with creating agents using the existing message-sending capabilities of an agent framework that does not support conversation reasoning.
\chapter{Evaluation} \label{chap:evaluation}

\section{Introduction} \label{sec:evaluation:introduction}

The previous chapters set out the capabilities of ACRE and the means by which an AOP programmer can interact with the framework using the ACRE/Agent Interface. In order to demonstrate the effectiveness and usefulness of the ACRE approach, this Chapter describes a user trial that was conducted using students from two distinct classes with different levels of programming experience. Participants were tasked with writing a solution to a specially-designed problem that was posed to them. The problem was designed to require inter-agent communication while not being dependent on complex reasoning. Half of the students were requested to use ACRE in their solution, with the other half using pre-existing message-handling facilities.

The code from both groups' implementations was analysed using both objective numeric metrics and subjective analysis. This was intended to ascertain whether the use of ACRE added to the effort required by a developer to implement a solution, in addition to identifying areas where the use of ACRE may prevent certain common problems from arising.

\section{Background}

The evaluation of programming toolkits, methodologies, paradigms and languages is a matter of some discussion and debate within the wider software engineering community. For a system like ACRE, the principal aim is to make it easier for developers to perform particular tasks. Specifically, it should facilitate developers in implementing reliable, predictable, understandable and secure management of communication.

 A new proposal for a method of performing this task should not impose additional effort on the developers that use it. Examples of this include~\cite{Hochstein2008} and~\cite{Luff2009}, which related to models of parallel programming,

In a variety of software engineering disciplines, a notion of \textit{programmer effort} has been used in attempting to quantify the amount of work a programmer must undertake to complete some programming task. Numerous studies have attempted to use objective metrics to quantify programmer effort. Although not based in the agent domain, these metrics can help inform a choice of metric for AOP. Some examples include the number of non-comment, non-empty lines of code~\cite{Luff2009} and non-commented code statements~\cite{VanderWiel1997}. Another measurement of effort is the time taken to implement a solution to a standard problem~\cite{Luff2009,Hochstein2008}.

When performing an evaluation such as this, a common approach is to use two groups of participants. Each group is presented with a common problem to solve, with all factors other than the subject of the evaluation being kept equal~\cite{Hochstein2008,Luff2009,Rossbach2010}.

% Szafron1994 also uses two groups for evaluating usability of parallel programming systems

\section{Scenario Motivations} \label{sec:evaluation:motivations}

Following precedent set in other areas, it was decided to develop a scenario within which experiment participants would be required to develop solutions. Before the development of this scenario it was important to first identify its desirable features. Due to the nature of ACRE, the problem must be \textit{communication-focused}. Additionally, since communication is key, it was desirable that the scenario should not have complex reasoning as a requirement, as this would detract from the communicative aspect. Thus the problem should also be \textit{accessible}. The \textit{reproducibility} of the experiment is also important. A scoring system should be in place so that the participants' \textit{goal is clearly defined}. With this in place, participants are readily aware of what is required of them. Another important requirement was that the final score be \textit{independent} from the order in which the tasks were implemented. A clear \textit{reward for active agents} was also considered to be a necessity, so that idleness cannot result in success. Some of these motivations are relevant to any type of MAS evaluation whereas others are specific to the type of task that would be suitable for a system such as ACRE.

\begin{itemize}
	\item \textit{Communication-focused:} As ACRE is fundamentally concerned with inter-agent communication, it is essential that the scenario be inherently communication-focused. As such, it was decided that the scenario should include a number of pre-written ``core agents'' with which participants' solutions were required to interact using protocols that were supplied. This communication was considered to be an essential component without which it would be impossible for a participant to make any progress.
	\item \textit{Accessible:} The nature of the scenario should be such that it is not required to create an agent or agents that feature complex reasoning or intelligent behaviour. Though this motivation does not preclude the development of smarter agents, it should not be the main driver of success as it should be possible to succeed using simple strategies.
	\item \textit{Reproducible:} Having decided to make core agents available for communication, it is important that their behaviour should be reproducible. This means that any decisions they make, along with any non-deterministic changes to the environment, must be recorded so that any experiment can be repeated. The reason for this is twofold: 1) any experiment can be reliably re-run in order to verify its results and 2) the performance of multiple solutions can be compared using the same conditions.
	\item \textit{Clearly defined goal:} It should be clear to participants what is required of them. The tasks they are required to perform, the scoring criteria and any time constraints should be made available at the outset. It is not necessarily the case that the final score should be an accurate representation of the quality of the solution: merely that it aids the avoidance of confusion and ensures that each participant is working towards the same aims. 
	\item \textit{Independence of task ordering and final score:} In a scenario where a participant is required to perform multiple tasks, no participant should be capable of gaining an advantage merely by implementing these tasks in a different order to his/her competitors. The easiest way to achieve this independence is to fix the order in which the tasks must be performed. In the context of a communication-focused evaluation, this involves ensuring that later protocols are dependent on having previously implemented prerequisite protocols.
	\item \textit{Reward for participation:} The scenario should be set up in such a way so as to ensure that it is not possible for a participant to benefit from failing or declining to implement one or more of the tasks. Where ``participation'' consists of the implementation of protocols (as in this case), it should be impossible for an agent successfully implementing any protocol to be consequently at a disadvantage when compared to another that does not.
\end{itemize}

\section{Scenario Description} \label{sec:evaluation:description}

The scenario and instructions that were presented to the participants is contained in full in Appendix~\ref{chap:appendix-3} (with the exception of a section detailing how participants should submit their code: the only other changes to the original scenario are for reasons of typesetting and formatting). The scenario that was chosen was a simple stock and asset trading game. Each participant was required to create an agent named \textit{Player}, which was capable of interacting with a number of core agents that were provided. The Player was initially furnished with some initial capital, manifested as a quantity of virtual money, and had as its aim the buying and selling of virtual stocks and assets for profit, with the goal of maximising its capital.

The core agents with which the Player could interact were as follows:
\begin{description}
   \item[Banker:] The Banker agent is responsible for recording the amount of money each Player possesses. At the beginning of the scenario, the first task that a Player should perform is to contact the Banker to open a bank account, into which its initial capital is deposited in virtual currency. The Banker can later be contacted so an agent can ascertain its bank balance.
   \item[Stockbroker:] The Stockbroker agent allows the Player agent to buy and sell virtual stocks. The primary mechanism through which a Player can earn money is by communicating with the Stockbroker to buy stocks and later sell them at a profit.
   \item[Guru:] This agent has expert knowledge about the movements of the market, and can provide advice to Player agents about stocks that are likely to rise dramatically in value and also those that should be avoided. This allows Players to gain an advantage in terms of choosing which stocks they should buy, and when they should be sold again.
   \item[Auctioneer:] In addition to stocks, a Player may also purchase properties. The value of properties rises more quickly than stocks, making this a valuable means of increasing capital. Properties are purchased by participating in auctions conducted by the Auctioneer agent. 
   \item[Bidder:] In order to sell a property it has previously bought, a Player agent must conduct its own auction to allow a number of Bidder agents to lodge bids to buy it. There are three Bidder agents present in the system, unlike the other types of core agent where only one instance is present. It is possible for a Player to sell a property by communicating with only one Bidder agent, but in order to be guaranteed to get the best price it must consult all three Bidders.
\end{description}

The trading game's features were designed with the scenario motivations in mind. The game is heavily \textit{focused on communication}, since it is necessary to communicate with the core agents in order to succeed. A number of protocols were developed that describe the types of conversations in which the core agents are capable of participating. Each of these protocols is based on one or more of the standard FIPA interaction protocols. These protocols are summarised in Table~\ref{tab:evaluation:protocols}. An illustration of the FSM describing each protocol was also made available. An example of this is shown in Figure~\ref{fig:evaluation:broker-buy} (all FSMs can be seen in Appendix~\ref{chap:appendix-3}). This is the protocol that allows a Player to communicate with the Stockbroker in order to purchase a particular quantity or value of a specified stock.

\begin{table}
\small
	\caption{Core Agent Protocols.}
	\label{tab:evaluation:protocols}
	\centering
	\begin{tabular}{|l|l|l|l|}
		\hline
		\textbf{Agent} & \textbf{Protocol} & \textbf{Based On} & \textbf{Purpose} \\
		\hline
		Banker 			& open		& request~\cite{fipa-request} & Open a bank account \\
		& enquiry	& query~\cite{fipa-query} & Query your bank balance \\
		\hline
		Stockbroker	& listing 	& query~\cite{fipa-query} & Get a list of available stocks \\
		& price		& query~\cite{fipa-query} & Query the price of a stock \\
		& portfolio	& query~\cite{fipa-query} & Query stocks currently owned \\
		& buy		& request~\cite{fipa-request} & Buy a quantity/value of a stock \\
		& sell		& request~\cite{fipa-request} & Sell a quantity/value of a stock \\
		\hline
		Guru			& subscribe	& subscribe~\cite{fipa-subscribe} & Subscribe to Guru's stock tips \\
		\hline
		Auctioneer		& subscribe	& subscribe~\cite{fipa-subscribe}, & Subscribe to new auctions and \\ 
	&			&  english-auction~\cite{fipa-english-auction} & participate in them \\
		\hline
		Bidder			& sell		& contract-net~\cite{fipa-contract-net} & Sell a property \\
		\hline
	\end{tabular}
\end{table}

\begin{figure}[!htb]
\includegraphics[width=\textwidth]{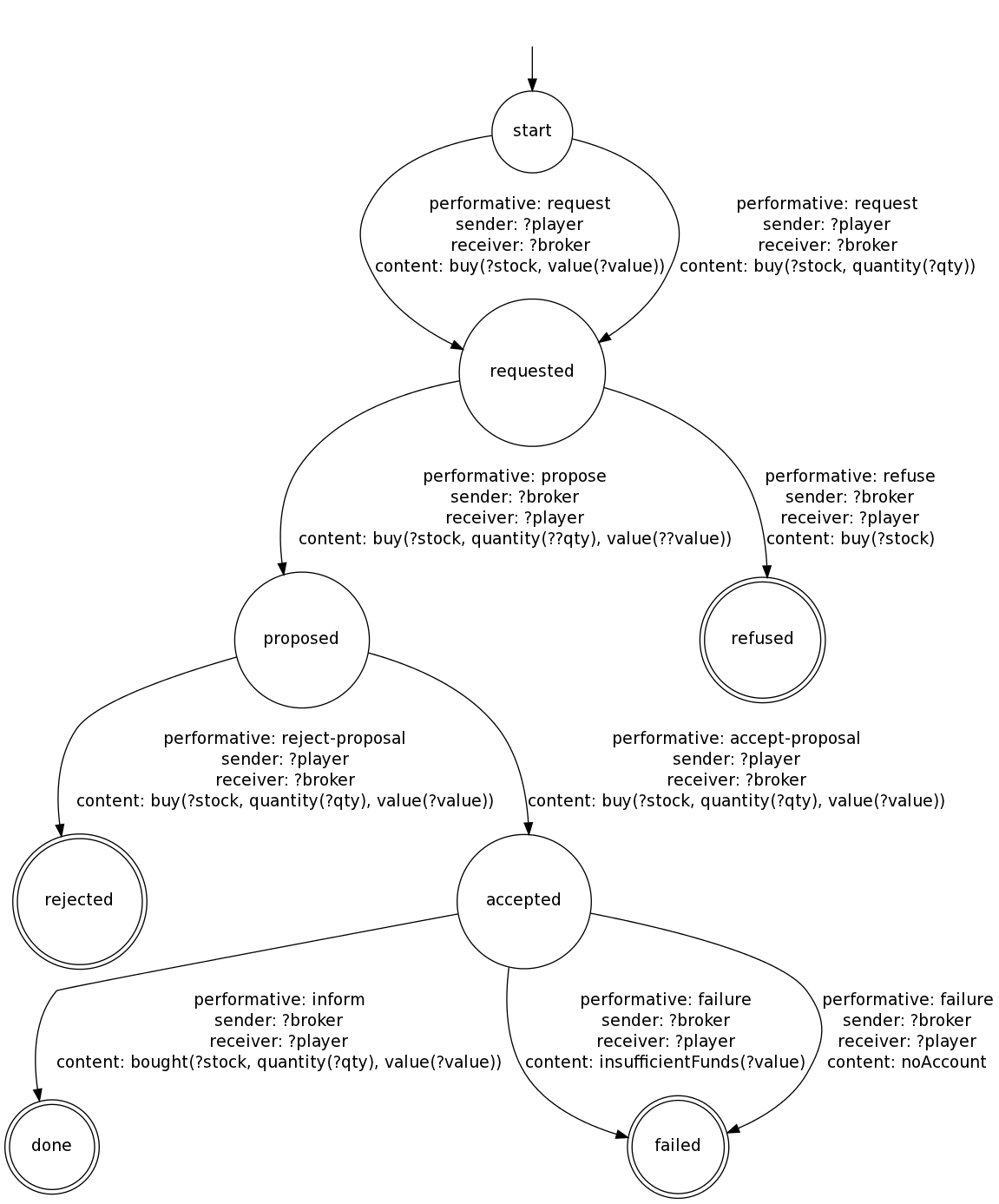}
\caption{FSM illustration of the \texttt{broker.buy} protocol.}
\label{fig:evaluation:broker-buy}
\end{figure}

The \textit{clearly defined goal} of the scenario is for a Player to maximise capital. This is an easily understandable concept, as is the buying and selling of assets for profit. Thus the participants in the evaluation were clearly aware of what was required of them.

By way of making the scenario \textit{accessible}, only a single agent is required to be developed by each participant. Additionally, this agent can achieve a reasonable level of success using a relatively simple strategy. A typical such strategy is to subscribe to the advice given by the Guru and follow it by investing all available capital in the stocks it recommends, selling it when advised. Although a more complex strategy may yield superior results, it is not a requirement for participation in the evaluation. The requirement of accessibility is especially important when participants are subject to time constraints, as they were in this case (see Section~\ref{sec:evaluation:fudan}).

The movement of stock prices was, by default, random. An internal clock was used to keep track of the time of the experiment. Stock prices may change on every ``tick'' of this clock. For experiments to be \textit{reproducible}, these stock movements may alternatively be pre-prepared (or recorded from a previous random run) and be loaded at the beginning of the experiment. This would ensure that price movements are repeated across multiple runs of an experiment.

Certain aspects of the experiment were defined in such a way so as to encourage a particular \textit{implementation sequence}. The tasks that a Player is required to perform are designed to be conducted in a particular order. Although there are no technical restrictions on the order in which participants choose to implement their solutions, most tasks are dependent on a previous task having already been completed. For example, it is impossible to buy stocks prior to opening a bank account. Stocks may be not be sold before being bought, and the advice of the Guru is also useless without a Player having the ability to act upon it by buying and selling. The protocols for buying and selling properties are more complex. As such, it is intended that participants would wait until successfully engaging with the rest of the system using simpler protocols before taking part in auctions. To enforce this, the value of all properties is set to be higher than a Player's initial capital. Thus a Player must make a profit on the stock market before they can take part in auctions. Finally, selling a property to Bidders cannot occur before it has been bought. Some protocols are for convenience and are not essential to the successful completion of tasks. For example a Player may recalculate and remember its bank balance after each transaction rather than using the \texttt{bank.enquiry} protocol to request this information from the Banker.

The requirement to \textit{reward active agents} is achieved by constraining the stock calculation mechanism so that the price of stocks only increases. Although this is an artificial restriction, it plays an important role in the development of the system by ensuring that an agent that successfully implements the protocols to buy and sell stocks must perform better than one that does not participate at all in the hope that its initial capital will suffice.

\section{Undergraduate Experiment} \label{sec:evaluation:fudan}

The evaluation experiment was first conducted with a group of final year undergraduate students in Fudan University, Shanghai, China. This evaluation was conducted as part of a module on Agent Oriented Programming. Prior to taking this module, none of the students had previous experience developing MASs or using an AOP language.

In this type of experiment, it is important to reduce to the greatest extent possibly any features of chosen programming languages or frameworks that may bias the result. For this reason, all students were required to create their implementation using the AF-AgentSpeak language within the Agent Factory multi agent framework.

The students were permitted a three hour period within which to create their solutions. The experiment was conducted within a supervised laboratory setting. The use of a fixed time period facilitates the use of quantitative comparisons between the speed at which each participant worked. Conducting the experiment in class ensured that the work submitted by each student was their own. To aid them with their task, students were permitted to access lecture notes and refer to the official manuals and user guides for Agent Factory, AF-AgentSpeak and ACRE.

The class was divided into two groups of equal size by means of random assignment. Participants in one group were requested to create their solution using ACRE whereas the other group was requested to use the pre-existing message-passing capabilities of Agent Factory.

To prepare the students for the experiment, a practical session was conducted a week before the experiment. This allowed the students to have sufficient time to become familiar with the technologies they were expected to use for the experiment, including the various message exchange mechanisms. As part of the AOP module, they had previously participated in practical sessions that exposed them to other aspects of AF-AgentSpeak and AOP in general.

The class consisted of a total of 46 students. Therefore 23 students were asked to create an ACRE solution, with the other 23 being requested to make their implementations without the use of ACRE. For the evaluation itself, one student from the non-ACRE group did not attend the evaluation, leaving 45 students in total. Additionally, one other student from this non-ACRE group instead submitted a solution using ACRE.

Of the remaining 21 students in the group not using ACRE, one submission was excluded from this research as only a single protocol had been attempted and this attempt had not been successful. This ultimately left 20 non-ACRE students in total for consideration.

In the ACRE group, 24 submissions were received. As with the non-ACRE group, one submission was excluded from this analysis as it had not successfully interacted with any core agent. Thus there are 23 submissions using ACRE that have been used for the evaluation.

The submissions were evaluated using both objective and subjective measures. Each submission was evaluated only on the source code that was submitted: additional surveys were not conducted. Initially, some simple objective measures were utilised to measure the time taken and the level of programmer effort that is required to create communication-heavy AOP programs using ACRE when compared to doing so without ACRE. This analysis is presented in Section~\ref{sec:evaluation:fudan:objective}. Following this, each implementation was examined subjectively to identify any common problems or issues that may arise during the development of such a solution. This is done in Section~\ref{sec:evaluation:fudan:subjective}.

\subsection{Objective Measures} \label{sec:evaluation:fudan:objective}

The ACRE toolkit is intended to aid developers in dealing with complex communication in an easier fashion. For this reason, it is essential to ensure that ACRE does not add to the effort required by a programmer to develop a MAS. Some simple objective measures are available to attempt to quantify the concept of \textit{programmer effort}. For this evaluation, two such metrics were used: 1) the number of protocols implemented within a specific time period (``number of protocols'') and 2) the number of non-comment, non-empty lines of code written per protocol (``lines per protocol'').

The first metric can compare two participant groups in terms of the time taken to implement protocols. Because the tasks are ordered, each participant is encouraged to implement the protocols in the same order, although this is not guaranteed. A communication protocol is considered to be the unit of work for this experiment, given that all interaction with core agents is done using protocols, and that no work can be done without interacting with these core agents. 

With regard to counting lines of code, merely counting the lines of code contained in each submission is not a useful metric, as it is affected by the overall progress of each participant's solution. As this is already measured by the ``number of protocols'' metric, the number of lines of code is taken as an average per protocol. This gives an indication of the amount of work that is required of a programmer to implement an interaction protocol.

It could be argued that a more fine-grained metric may attempt to assign lines of code to specific protocols. Although this is obvious when considering lines of code that specifically deal with the sending and receiving of messages, it is more difficult when dealing with other lines of code. For example, a belief may be adopted to record the fact that a bid has been accepted in an auction, which may be used at a later time when deciding whether to accept bids. In this case, the adoption of a belief does not directly relate to the sending and receiving of messages within a conversation, but is still closely related to the protocol implementation. A more complex example may involve the adoption of beliefs that relate to multiple protocols, in which case assigning this to one particular protocol may be impossible. An example of this would be if an agent adopts a belief about a quantity of stock that it owns. This may be relevant to a conversation where it wishes to sell this stock (a \texttt{broker.sell} conversation) and also in deciding whether to respond to a ``sell'' recommendation from the Guru agent (the \texttt{guru.subscribe} protocol). For these reasons, ``lines per protocol'' is used as the metric, as this gives a general indication of the amount of work that is required to be done per protocol implemented.

\begin{table}[!hbt]
\centering
\caption{Objective measures of programmer effort for Fudan students.}
\label{tab:evaluation:fudan:speed}
\begin{tabular}{|l|c|c|}
\hline
 & Protocols Implemented & Lines per Protocol  \\
\hline
 ACRE & 5.43 & 18.35 \\
 Non-ACRE & 5.85 & 27.06 \\
\hline
\end{tabular}
\end{table}

The results of this objective measurement are laid out in Table~\ref{tab:evaluation:fudan:speed}. In each case, the metric is the average per participant in the appropriate group. For the ``protocols implemented'' metric, the difference between the groups is not statistically significant using an unpaired \textit{t}-test for $p=0.05$. The difference in ``lines per protocol'' is, however, statistically significant using the same test.

From this table, it can be seen that there is a marginal difference between the average number of protocols implemented within the three-hour time period. As this difference is not significant, it indicates that the speed of developement (in terms of protocols as units of work) is comparable whether ACRE is being used or not. This suggests that ACRE does not impose a steep learning curve above that of learning how to use more traditional methods of message and conversation handling.

An interesting finding is that ACRE resulted in significantly fewer lines of code being written per protocol implemented. Although this is a somewhat crude metric, it does suggest that the automated conversation handling of ACRE does reduce the overall amount of code that a developer is required to write in order to implement protocols.

\subsection{Subjective Analysis} \label{sec:evaluation:fudan:subjective}

The previous Section examines the solutions that were received in terms of quantitative analysis. The metrics chosen attempt to measure the speed at which participants were able to develop implementations of protocols, along with the amount of effort involved in the creation of each implementation. However, none of these metrics capture the quality of the code.

Capturing code quality using quantitative metrics is a very difficult task. For this reason, the solutions submitted were examined manually so as to subjectively analyse each. Particular attention was paid to how complete the non-ACRE implementations were. During this analysis, a number of common issues were identified in the non-ACRE code. In some cases, these issues would allow the agent to be exploited by another malicious agent. In other cases, the code submitted was very closely tied to the scenario that was presented, and would require many changes in order to be reused within a larger MAS or in an extended scenario. In all cases, the issues that arose could not arise in ACRE-enabled agents as a result of how ACRE is designed.

The issues identified were typically in the AF-AgentSpeak rules that handle communication. Typically, an AF-AgentSpeak agent will contain rules that cause some actions to be taken in specified circumstances. An AF-AgentSpeak rule has a \textit{triggering event} and a \textit{context} that are used to identify these situations where the rule should fire. The triggering event is some event that has occurred (i.e. a change in the belief base) whereas the context is a set of beliefs that should be present in order for the rule to be relevant. When writing non-ACRE code in AF-AgentSpeak, the event triggering the rule is typically the receipt of an incoming message and the context consists of beliefs about the state of the conversation (checking that the sender of the message is as expected, ensuring that the message is received in the correct sequence when compared with necessary preceding messages, etc.). An example of this can be seen in Figure~\ref{fig:evaluation:non-ACRE-code}. In the non-ACRE code that was received, these triggering events and contexts were frequently written in a suboptimal way.

%TODO: is `local' a valid address?
\begin{figure}[!htb]
   \footnotesize
	\begin{Verbatim}[fontsize=\footnotesize]
+initialized : true <-
    +openingAccount(banker),
    .send(request,agentID(banker,addresses("local")),openAccount);
 
+message(inform,agentID(?sender,?addr), openedAccount(?id,?amt)) :
  openingAccount(?sender) <-
    +bankAccount;
 	\end{Verbatim}
 	\caption{Sample non-ACRE code.}
 	\label{fig:evaluation:non-ACRE-code}
 \end{figure}

The specific common issues that were identified during the course of this subjective analysis are as follows. Words in parentheses are short descriptions that will be used to refer to each issue in the ensuing analysis:

\begin{itemize}
   \item \textit{No Checking of Message Senders (Sender):} An AF-AgentSpeak belief about the receipt of a FIPA message includes the details of the sender of the message. In many cases, this information was ignored, meaning that a message triggering a rule could have been sent by any agent. Instead, only the content and/or performative of the message were checked in the triggering event of rules.
   \item \textit{No Checking of Conversation Progress (Progress):} According to the scenario description, every message is sent according to the rules set out in the supplied protocols. As such, they are sent and received in a clearly defined sequence (e.g. advice will not be sent by the Guru agent unless the Player agent has subscribed to its updates). Many participants did not attempt to check the context of messages to ensure that the required earlier messages had indeed been previously exchanged. Instead, each message was treated individually.
   \item \textit{Hard-coded name checking (Name):} In cases where the sender of a message was checked, it was frequently the case that the specific name of the sender agent was hard-coded in the AF-AgentSpeak code.
   \item \textit{Checking addresses (Address):} The addresses of core agents was also hard-coded in some cases.
\end{itemize}

The issues identified will not prevent a Player agent from successfully participating in the trading scenario as currently defined. However, the presence of such issues means that additional effort is required to adapt the solution for a more open agent system, for example if multiple Player agents were present in the same MAS and interacting with the same core agents. In many cases, heavy modification to existing code would also be required if the scenario were to be extended with the addition of more protocols and/or core agents as the code is too closely tied to the specifics of the scenario presented.

The \textit{Sender} issue arises whenever a Player agent fails to check the identity of the sender of a message that it has received. In certain situations this will have unintended adverse consequences. A typical example of this issue arising is when a Player agent reacts to a recommendation from the Guru agent to buy a particular stock. Most implementations react to these recommendations by buying a quantity of the specified stock. In a more open MAS, a rival malicious Player agent could send recommendations to other agents to cause them to buy stock that is not expected to perform well, contrary to the advice from the real Guru agent. In the absence of a check for the expected sender, it would be required only that the correct performative and content be present in the message.

The failure to verify the state of a conversation also makes a Player agent susceptible to exploitation. An example of where this may occur is in the \texttt{broker.buy} protocol shown in Figure~\ref{fig:evaluation:broker-buy}, which is used for buying stock from the Stockbroker. The protocol insists that the Stockbroker makes a proposal to the Player agent, which must be accepted before the purchase proceeds. If, however, the Player has no concept of a conversation, it may be convinced that it has bought some quantity of a stock (via an \texttt{inform} message of the type that leads the conversation to the ``done'' state), without ever having participated in the negotiations and without having initiated the transaction. In isolation, this issue would allow a malicious Stockbroker agent to sell stock to a Player that it does not want. Worse, if combined with a failure to verify the sender of messages, another Player agent could potentially convince the agent to buy stock in which it has no interest.

As an aside, it is interesting to note that when participants did check the progress of conversations, they did so using the same state names that are included in the ACRE FSM diagrams. This suggests that even for developers who cannot or choose not to use ACRE, the availability of formalised protocols is a useful tool for visualising and reasoning about conversations for the developers themselves.

The two other common issues identified are of the type that cause difficulty if the scenario is altered in any way, for example if the conversations are to be held with different agents and/or multiple agent platforms. For agents that did check that the sender of a message was as expected, it was frequently the case that the name of the expected sender was hard-coded into the Player agent code. This limits the code to being able to deal with only a single agent with a specific name. Changing the name of a core agent would require all the rules relating to its conversations to be rewritten. More seriously, the rules cannot be reused in the situation where a second agent capable of engaging in a particular conversation is introduced (e.g. the addition of a second Stockbroker that deals with a different stock portfolio). Existing code would require rewriting to deal with this type of situation. In some cases, the addresses of core agents were hard-coded also, restricting the solution to a single agent platform and a single message transport mechanism.

The Conversation Manager of ACRE checks that the sender of a message is a participant in the relevant conversation. It also ensures that the message is received in the appropriate sequence by consulting the appropriate protocol FSM. For this reason, the common issues identified in the non-ACRE code cannot arise in code that uses ACRE. In AF-AgentSpeak code using ACRE, the triggering event of a rule is typically that a conversation has advanced, with the context being used to check other details about the conversation. The participants of a conversation need only be named when the conversation is initiated.

For a pre-existing conversation, a \texttt{conversationAdvanced} event can only be triggered by a message that has been sent by an agent that is an existing participant in the conversation, has the correct performative according to the protocol and has content that matches the next expected message. Thus there is no requirement on the developer of a Player agent to implement these checks manually. If any of these criteria are not met, an \texttt{unmatchedMessage} event is raised and the conversation does not advance. Thus the types of problem identified above cannot occur in an ACRE-enabled agent.

An ACRE agent cannot be fooled by another Player agent sending messages that should be sent by a core agent, as it is not a participant to a conversation\footnote{ACRE does not protect against messages that are sent by an agent other than the agent named in the \texttt{sender} field of the message. This form of secure communication, if required, is considered to be a task for the underlying Message Transport Service.}. Similarly, out-of-sequence messages cannot be used to manipulate a Player agent either. For example, a Stockbroker agent cannot inform a Player agent of a successful purchase unless the Player has previously agreed to the purchase.

The use of ACRE thus protects an agent against these issues. This leads to more reusable code as an extension to the scenario or a more open MAS would require far less effort to adapt the existing ACRE code to the new circumstances. Thus ACRE can be seen to prevent certain styles of coding that inadvertently restrict the extensibility and reusability of communication-handling code.

\subsubsection{Prevalence of Issues}

For each of the issues identified during the subjective analysis of the code, it was necessary to record how prevalent it was amongst participants. To facilitate meaningful analysis, each implementation was assigned one of the three following classifications relating to each of the four issues identified:
\begin{itemize}
	\item \textit{Not susceptible:} The issue in question was not present for any rule in the implementation.
	\item \textit{Totally susceptible:} The issue was present in the implementation for every rule where it could possibly be present.
	\item \textit{Somewhat susceptible:} The issue in question was present in at least one relevant rule, but not all rules where it could have occurred.
\end{itemize}

For Player agents that were totally susceptible to the \textit{Sender} issue, they could not possibly be affected by the \textit{Name} or \textit{Address} issues as they never attempted to check the details of the sender of the message. Therefore, in the following analysis the figures shown for these latter two issues are a percentage of those agents where it was possible for them to arise (i.e. those Player agents that were either somewhat susceptible or not susceptible to the \textit{Sender} issue).

\begin{figure}[!htb]
   \includegraphics[width=\textwidth]{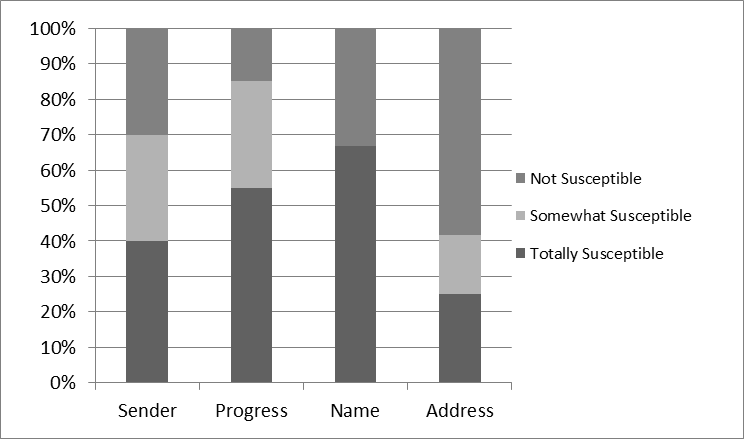}
   \caption{Issues present in Fudan students' non-ACRE code.}
   \label{fig:evaluation:fudan-issues}
\end{figure}

The prevalence of the issues identified in non-ACRE code is shown in Table~\ref{tab:evaluation:fudan-issues} and illustrated graphically in Figure~\ref{fig:evaluation:fudan-issues}. In the Table, the figures in parentheses are the absolute number of subjects to which the percentage relates. For the \textit{Sender} and \textit{Progress} issues, these percentages are of the 20 non-ACRE solutions considered. The \textit{Name} and \textit{Address} issues are out of the 12 solutions that are not totally susceptible to the \textit{Sender} issue.

\begin{table}[!htb]
	\centering
	\caption{Issues present in Fudan students' non-ACRE code.}
	\label{tab:evaluation:fudan-issues}
	\setlength{\extrarowheight}{5pt}
	\begin{tabular}{|l|l|l|l|l|}
\hline

& \textbf{Sender} & \textbf{Progress} & \textbf{Name} & \textbf{Address} \\
\hline
	Totally Susceptible 	& 	40\% (8) & 	55\% (11) & 	67\% 	(8) & 	25\% (3)\\
	Somewhat Susceptible 	& 	30\% (6) & 	30\% (6) & 	0\% (0)	& 	17\% (2) \\
	Not Susceptible 		& 	30\% (6) & 	15\% (3) & 	33\% 	(4) & 	58\% (7) \\
	\hline
	\end{tabular}
\end{table}

From these, it can be seen that the common issues identified during the subjective analysis were widespread amongst the non-ACRE group. The only issue that was found in less than half of the relevant agents was the hard-coding of addresses.

Over two thirds of agents were susceptible to reacting to messages sent by the wrong agent at least some of the time. This issue occurred all the time in 40\% of agents, who never checked the identity of message senders. Of the agents that did check the message sender, two thirds hard-coded the agent names directly in their code, which would require extensive rewriting of rules if the scenario were to be extended or if the code was to be applied in another situation.

Only three participants (15\%) wrote code that always verified that messages were received in the order specified by the underlying protocol. Further analysis found that all three of these solutions were somewhat susceptible to the \textit{Sender} issue. This means that no submissions were received that were immune to all the issues identfied.

The identification of these issues, along with their widespread presence amongst the solutions received is a strong argument in favour of using a conversation-handling technology such as ACRE. The automated conversation checking and exception handling of ACRE automatically guards against these issues and so encourages more robust and reusable code. As can be seen from the objective metrics, this is done without adding to the programming effort that is required of developers. Left to handle conversations themselves, developers can fall into bad habits that adversely affect the overall quality of their code.

\section{Postgraduate Experiment}

Following the evaluation experiment conducted with undergraduate students, the same scenario was also presented to a different group of students, to investigate whether the patterns seen in the first group are more generally applicable. The second group of students were part of a part-time Masters-level Agent Oriented Software Engineering module held in University College Dublin, Ireland. Unlike the undergraduate students from the first evaluation, these students are experienced software developers in industry, though none had previous experience with AOP prior to taking this module.

Classes were conducted every day for five days, with students being assigned practical work each afternoon to become familiar with AOP. Communication handling and ACRE were introduced on the fourth day and the evaluation was conducted on the fifth day. This meant that these participants had less preparation for the evaluation than those in the first experiment.

As with the initial experiment, the evaluation was conducted in a laboratory setting over a three-hour period. This again ensured that measures of programmer effort relate to the same time period for each student. Unlike that experiment, however, the postgraduate students were permitted to submit an updated version of their solution two weeks after the original experiment. While these second submissions are not useful for objective measures (as the time spent on solutions by each student is no longer constant), it is interesting to subjectively compare the solutions submitted during the initial experiment with those resubmitted by the same student later.

Students were again split into two groups by random assignment. In a class of 19 students, 10 submitted ACRE-based solutions while the remaining 9 students created non-ACRE agents.

\subsection{Objective Analysis}

The same objective metrics as for the first evaluation were also applied to this experiment. The results of this are set out in Table~\ref{tab:evaluation:ucd:speed}. As with the Fudan students, the ACRE group on average implemented marginally fewer protocols in the time available. Overall, the number of protocols developed was lower than the Fudan students for both the ACRE and non-ACRE groups. This may have been as a result of the shorter time available to these students in which to become familiar with ACRE and with AOP in general. With regard to the number of lines of code per protocol, it can be seen that the ACRE participants wrote a very similar amount of code to their Fudan counterparts. A dramatic difference can be seen in the non-ACRE code, however. The UCD students wrote substantially fewer lines of code than those in Fudan. For the UCD experiment, the average number of lines per protocol in a non-ACRE solution was less than that of an ACRE-enabled agent.

\begin{table}[!hbt]
\centering
\caption{Objective measures of programmer effort for UCD students.}
\label{tab:evaluation:ucd:speed}
\begin{tabular}{|l|c|c|}
\hline
 & Protocols Implemented & Lines per Protocol  \\
\hline
 ACRE & 4.4 & 18.93 \\
 Non-ACRE & 4.67 & 14.87 \\
\hline
\end{tabular}
\end{table}

For both metrics, however, the difference in the measurements was found not to be statistically significant using an unpaired \textit{t}-test for $p=0.05$. This adds weight to the argument that ACRE does not add to the amount of programmer effort that is required to create a communication-heavy agent program.

\subsection{Subjective Analysis}

In order to be consistent with the earlier experiment in Fudan, the UCD non-ACRE students' code was analysed to investigate whether they were susceptible to the same common issues as the undergraduate participants. Since both groups worked in the same conditions, comparisons can be drawn between the two groups. With the postgraduate students, further comparisons can be made between the solutions that were submitted during the fixed-time lab session and the improved solutions that were submitted after the students had done further work on their code.

\subsubsection{Prevalence of Issues}

\begin{figure}[!htb]
   \includegraphics[width=\textwidth]{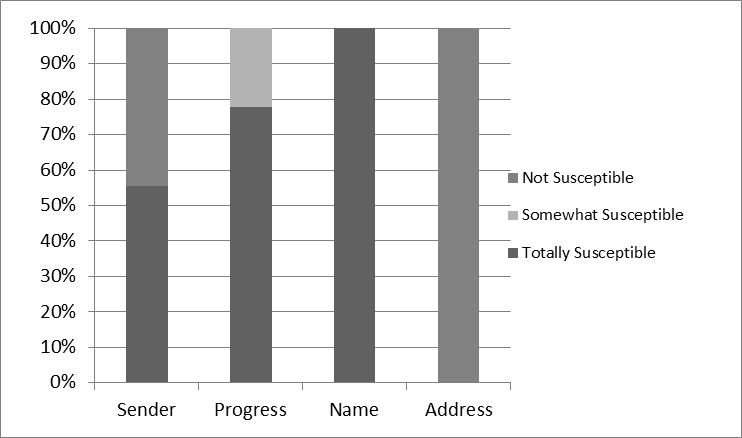}
   \caption{Issues present in UCD students' non-ACRE code in fixed time period.}
   \label{fig:evaluation:ucd-friday-issues}
\end{figure}

The results of the subjective analysis of the postgraduate students' non-ACRE code as it was after the fixed-length laboratory session can be seen in Table~\ref{tab:evaluation:ucd-friday-issues} and Figure~\ref{fig:evaluation:ucd-friday-issues}.

The percentages in Table~\ref{tab:evaluation:ucd-friday-issues} are of the total of 9 non-ACRE students for the \textit{Sender} and \textit{Progress} issues. The percentages \textit{Name} and \textit{Address} issues are calculated from the students that were not totally susceptible to the \textit{Sender} issue, as was done for the Fudan experiment analysis. In this case, that amounts to 4 participants.

\begin{table}[!htb]
	\centering
	\caption{Issues present in UCD students' non-ACRE code in fixed time period.}
	\label{tab:evaluation:ucd-friday-issues}
	\setlength{\extrarowheight}{5pt}
	\begin{tabular}{|l|l|l|l|l|}
\hline

& \textbf{Sender} & \textbf{Progress} & \textbf{Name} & \textbf{Address} \\
\hline
	Totally Susceptible 	& 	56\% (5) & 78\% (7) & 100\% (4)	& 	0\% (0)\\
	Somewhat Susceptible 	& 	0\% (0)	 & 22\% (2) & 0\% (0)	& 	0\% (0) \\
	Not Susceptible 		& 	44\% (4) & 0\% (0)	& 0\% (0) 	& 	100\% (4) \\
	\hline
	\end{tabular}
\end{table}

A significant difference between these results and those for Fudan is that no UCD student hard-coded addresses when checking message senders. Of interest, however, is that several students hard-coded addresses for some outgoing messages. This was not a feature of the Fudan submissions.

As with the Fudan submissions, no solution was submitted that was immune from all of the common issues. Over half of participants failed to ever check the sender of incoming messages. When this check was performed, it was always in the form of a hard-coded agent name. For the \textit{Progress} issue, almost a quarter of participants made some attempt to check that messages were being received in the intended sequence, however none of the solutions were successful in doing this every time it was appropriate.

\begin{table}[!htb]
	\centering
	\caption{Issues present in UCD students' non-ACRE code without time restriction.}
	\label{tab:evaluation:ucd-final-issues}
	\setlength{\extrarowheight}{5pt}
	\begin{tabular}{|l|l|l|l|l|}
\hline

& \textbf{Sender} & \textbf{Progress} & \textbf{Name} & \textbf{Address} \\
\hline
	Totally Susceptible 	& 	56\% (5) & 	56\% (5) & 	75\% (3)	& 	0\% (0)\\
	Somewhat Susceptible 	& 	11\% (1) & 	44\% (4) & 	25\% (1)	& 	0\% (0) \\
	Not Susceptible 		& 	33\% (3) & 	0\% (0) & 	0\% (0) 	& 	100\% (4) \\
	\hline
	\end{tabular}
\end{table}

\begin{figure}[!htb]
   \includegraphics[width=\textwidth]{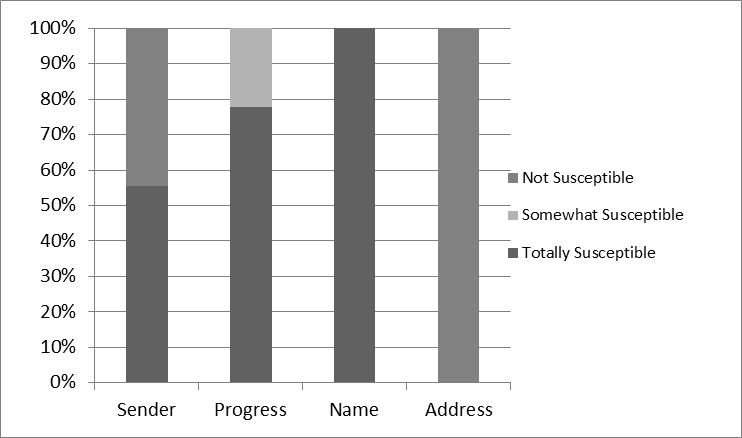}
   \caption{Issues present in UCD students' non-ACRE code without time restriction.}
   \label{fig:evaluation:ucd-final-issues}
\end{figure}

An extension to this analysis is to compare the time-constrained solutions that were initially submitted with the final submissions made a few weeks later. Although it is difficult to draw reliable conclusions from such a relatively small sample size, it is interesting to examine the differences between the two sets of results. These final submissions were analysed in the same way, with the results being presented in Table~\ref{tab:evaluation:ucd-final-issues} and Figure~\ref{fig:evaluation:ucd-final-issues}. The calculations here are done in the same way as for the previous experiments.

The observations taken here can be directly compared with the previous UCD experiment as the solutions were implemented by the same students. With regard to the \textit{Sender} issue, it is interesting to note that one student who had previously not been susceptible to the issue introduced a bug into the code after the initial submission. This caused one submission to be moved to the ``somewhat susceptible'' category.

For the \textit{Progress} issue, an improvement was observed in the later submissions. In one case, a protocol that had been implemented after the original submission included checks to ensure the conversation progressed correctly. In the other case, a check was introduced in exactly one place, where a belief that triggered the purchase or sale of stock was required to be still present at the end.

The other change was in the \textit{Name} issue, where one student who had previously hard-coded all names implemented a protocol where this was not done. The protocol in question was the \texttt{bidder.sell} protocol, which requires the Player agent to interact with three Bidder agents. It is because the same conversation can be conducted with more than a single core agent that this check was left more generic. Thus it does not appear to reflect any particular change in approach from the participant.

Thus an interesting observation about these later submissions is that code written after the initial experiment did not show any consistent improvement in terms of susceptibility to common issues. Although it must be noted that the sample size is small, this suggests that the problems noted with the code submitted after the initial experiment is unlikely to be solely attributable to the fact that participants were subject to time constraints.

% Notes on UCD
% - note that nobody hardcoded addresses
% - however, some did hardcode outgoing addresses
% - reduction in hard-coded name checking before final submission was because they got more protocols done and didn't hard-code bidders names for the auction to sell items
% - improvement in checking progress: for one, it was that the student got little done in the original time period. A check was included in one of the protocols implemented later. For the other, it isn't obviously checking conversation progress per se, just that a belief that triggers a purchase/sale must be present when the purchase/sale completes. Other than this there's no checking.
% - UCD students had a shorter time to prepare (only learning agents for a week, and ACRE for a day beforehand). This might explain the fewer students that are not susceptible at all to these issues.
% - protocols only partially implemented are not included (this is defined as any protocol where player action is required to complete the task for which the protocol is designed, such as approving the purchase or sale of a stock).

\section{Wider Applicability of Issues}

Having identified a number of problems that may arise in code written without the use of ACRE conversation handling, it is important to consider whether other features of the programming environment may have caused this to happen. The findings of this experiment would not be convincing if the issues were found to be applicable only to the Agent Factory framework and/or its implementation of AF-AgentSpeak.

For this reason, it is desirable to demonstrate that a system such as ACRE has wider benefits than to one single framework or configuration. To this end, an alternative framework that also lacks built-in conversation and protocol management was sought. In particular, best-practice communication code was desirable, so that any issues identified could not be attributed to poor programming practice or a lack of familiarity with the full capabilities of the platform or language.

Jason is a MAS development platform that uses an extended version of AgentSpeak(L) as its AOP language~\cite{Bordini2007}. It supports communication between agents via KQML-like messages. As with the pre-existing version of Agent Factory used for evaluation, it also lacks any built-in support for conversation handling or the definition of interaction protocols. There are two principal reasons for choosing Jason for this analysis. Firstly, it is a popular platform that is well-known and well-established within the AOP community. Secondly, the developers of Jason have released a book that includes sample code for performing a variety of tasks, including inter-agent communication~\cite{Bordini2007}. As this code is written by the developers of the framework themselves, it can be assumed that this represents recommended best-practice in the area of communicating between agents in Jason.

\begin{figure}[!ht]
\begin{Verbatim}[numbers=left,fontsize=\footnotesize,frame=single]
@lc1[atomic]
+!contract(CNPId) : cnp_state(CNPId,propose) <-
   -+cnp_state(CNPId,contract);
   .findall(offer(O,A),propose(CNPId,O)[source(A)],L);
   .print("Offers are ",L);
   L \== []; // constraint the plan execution to at least one offer
   .min(L,offer(WOf,WAg)); // sort offers, the first is the best
   .print("Winner is ",WAg," with ",WOf);
   !announce_result(CNPId,L,WAg);
   -+cnp_state(Id,finished).
\end{Verbatim}
	\caption{Sample Jason rule forming part of an implementation for  Contract Net protocol.}
	\label{fig:evaluation:jason-contract-net}
\end{figure}

Figure~\ref{fig:evaluation:jason-contract-net} is an extract from an implementation of a contract net interaction protocol for Jason~\cite[p. 134]{Bordini2007}. This implementation is provided by the developers of the Jason framework to show how a conversation might be implemented for that platform using AgentSpeak. The extract chosen shows a plan that makes up part of the program of the agent that initiates and coordinates the contract net. This particular section is intended to be invoked whenever all bids have been received, so as to identify the winner (line 7) and create an intention to announce this result to all the participants in the contract net (line 9).

The MAS within which this agent is intended to run is a fixed set of agents that each has a particular purpose. Specifically, all other agents in the system are intended to be participants in the contract net (three submit bids when called upon to do so, one refuses to participate and one is designed not to respond). As such, the code does not cater for proposals being received from agents that were not party to the original call for proposals. This can be seen in line 4 of the extract, which creates a list of offers that have been received based on any proposal received from any source, regardless of whether they had been invited to participate or not. This is an example of the \textit{Sender} issue that was identified in the ACRE evaluation. If this code were to be re-used for a more open MAS, the code would require modification to ensure that only agents that are expected to do so will submit proposals. Related to this is a method named \texttt{SocAcc} (meaning ``socially acceptable'') that is provided by Jason to ensure that some types of messages will not be processed if they were sent by inappropriate agents. Although this can be used to prevent non-participating agents from sending proposals, once an agent has been given permission to send proposals it cannot be prevented from participating in any contract net.

The extract also illustrates that a Jason agent can also be susceptible to the \textit{Progress} issue. Here, a belief named \texttt{cnp\_state} is used to track the state of the conversation. The code in question is written by experts and as such this belief is present in many plans so the state of the contract net is known at all times. However, as our evaluation has shown, less experienced programmers are more prone to omitting this type of check. Even experienced programmers who do not use agents can make this type of error when switching from their preferred programming paradigm, as the UCD evaluation illustrated.

One further issue arises in the choice of a name to uniquely identify a conversation. In the extract shown, this is referred to as the \texttt{CNPId} variable. In the Jason program, the value of this identifier is manually specified in the initial intention that originally triggers the start of the contract net (not shown). ACRE automatically generates identifiers for conversations, meaning that an agent programmer is not required to be concerned with this aspect of conversation handling.

\section{Summary}

To gauge the effectiveness and usefulness of ACRE, two user evaluations were conducted. One of these was conducted by final year undergraduate students in Fudan University, Shanghai, China. The other was done with postgraduate students in University College Dublin, Ireland. The latter group were professional software developers in industry, albeit none had prior experience of developing Agent-Oriented software.

In preparation for these evaluations, a scenario was developed whereby a participant was required to write an agent to perform a particular task. To achieve this task, it was necessary to engage in communication with a number of core agents, which were supplied. The scenario was designed to focus on communication but without requiring complex reasoning to be successful.

Each class of participants was divided into two groups, both of which used the AF-AgentSpeak AOP language on the Agent Factory framework. In each case, one group was required to develop their solution using ACRE and the other was required to do without. So that meaningful quantitative analysis could be conducted, participants were given a fixed time period within which to create their solutions. Two quantitative metrics were used: the number of protocols implemented was used as a measure of progress while the number of lines of code written per protocol was used to measure the programmer effort. For the first metric, no statistically significant difference was found between the ACRE and non-ACRE group for either class. In the latter metric, the ACRE students in Fudan wrote significantly fewer lines of code per protocol implemented, though the difference was not significant in UCD. These measurements support the argument that ACRE does not cause developers to spent more time or effort on implementing communication-heavy MASs than a system developed without the use of ACRE.

Following this objective evaluation, subjective analysis was conducted on the submitted code to identify common problems and issues in non-ACRE code of the type that cannot occur in ACRE code. These issues had two key consequences. Firstly, a susceptible agent could be persuaded by a malicious agent to perform tasks it did not intend to perform. Secondly, many solutions were too closely tied to the specific scenario presented and an extension to the scenario in terms of additional agents or platforms would require re-writing of existing rules to handle the extended scenario. This would have a negative impact on code reuse. Although participants were not explicitly required to write solutions that were optimal with regard to extensibility and reusability, these findings still represent strong evidence that these problems will commonly arise unless due care is taken to avoid them. Between the two classes, no non-ACRE programmer's code was completely immune to the issues identified.

Following the evaluation, a parallel was drawn with the Jason agent framework, for which code is available that was written by experts (the developers of the system) to implement a similar type of communication-focused agent. The analysis done on this code indicated that the issues identified in the evaluation experiments above are not specific to Agent Factory and can be found elsewhere if automated conversation-handling capabilities are not present, such as in Jason. Although capable and  diligent programmers may still write robust and reusable code, the presence of a system such as ACRE aids developers in avoiding the pitfalls that have been identified during the evaluation process.
\chapter{Conclusions and Further Work} \label{chap:conclusions}

\section{Introduction}
This final Chapter concludes the thesis by firstly discussing possible avenues for further work to extend that presented in the preceding Chapters. This is done in Section~\ref{sec:conclusions:further-work}. Following this, Section~\ref{sec:conclusions:conclusions} presents some concluding remarks, with particular focus on how the objectives of the thesis have been met.

\section{Further Work} \label{sec:conclusions:further-work}

There are many ways in which the work presented in this thesis may be built upon. This Section briefly discusses a number of such avenues, though this is by no means an exhaustive list.

\begin{description}
   \item \textbf{Improved Inheritance:} The present implementation of inheritance within protocols is relatively simple, allowing only for single inheritance where one protocol is, in effect, copied directly into another so that its states can be referred to within the importing protocol. Allowing for multiple inheritance in this way increases the possibilities of duplicate state names. For this to be a useful feature, the referencing of states both by the state name and its containing protocol name would be permissible. An additional extension allowing two states in different imported protocols to be marked as equivalent may also be useful. An further extension would be to allow the same protocol to be imported twice by defining an alias for each instance within the larger protocol.

   \item \textbf{Code Generation:} Currently, ACRE provides the capabilities necessary for agent developers to implement conversation-based communications. However, all the code is still written by the developers themselves. From a given protocol definition, an interesting extension would be to automatically generate skeleton AOP code that a developer would extend to fully implement the domain-specific aspects of the protocol. A further extension becomes possible as inheritance increasingly becomes a feature of AOP languages. Libraries of standard protocol implementations could be made available, containing hooks for domain-specific code to be integrated into the conversation. An example of this would be in a protocol that contains a request for action, followed by a reply to state that either the action has been done or that the recipient of the request has refused to comply. A hook into this conversation could result in the receipt of the request generating a goal to decide whether or not to perform the requested action and respond accordingly.

   \item \textbf{Compatibility with other Conversation Models:} At present, ACRE is not directly compatible with other models of conversations that differ from ACRE's Finite State Machine (FSM) representation. However, an interesting avenue of research would be to investigate the compatibility of FSMs with other representations. For instance, recent work on the integration of global session types (as discussed in Section~\ref{sec:communication:global-session-types}) has shown promise. Translating ACRE FSMs into global session types and vice-versa would improve interoperability between diverse platforms.

   \item \textbf{Improved Debugging:} The definition of ACRE protocols as centralised entities has advantages from the point of view of run-time debugging and monitoring. A tool similar to the Sniffer Agent shipped with JADE that is capable of intercepting and reading messages within a MAS could be equipped with ACRE capabilities in order to provide a centralised view of all the interaction occurring within the MAS.
   
   \item \textbf{Integration with Other Frameworks:} To date, a concrete implementation of ACRE within the Agent Factory framework has been created. However, as ACRE's abstract architecture has been designed to be as framework-independent as possible, further integrations with other systems are also possible.

\end{description}

\section{Conclusions} \label{sec:conclusions:conclusions}

This thesis has presented a comprehensive discussion of the Agent Conversation Reasoning Engine (ACRE). The objectives of this work were originally outlined in Section~\ref{sec:introduction:core-contributions}. This Section discusses how these have been achieved.

A holistic view of conversation management has been presented. This has been achieved in a number of stages:
\begin{itemize}
   \item The formal operational semantics that were contained in Chapter~\ref{chap:semantics} mean that ACRE is not tied to one particular implementation language. Any developer, following these semantics, can create a consistent implementation in any language of their choosing.
   \item Chapter~\ref{chap:acre-architecture} proposed an abstract architecture that has been designed with minimal assumptions so as to be integrable with many MAS frameworks.
   \item A reference implementation was created using the Java programming language that integrates with the Agent Factory MAS framework. Java was chosen as it is the most popular base language upon which to build MAS toolkits and frameworks. As such, much of the work done in creating this reference implementation can be directly reused in integrations with other frameworks. This work was presented in Chapter~\ref{chap:agent-factory}.
   \item To aid in the adoption of the type of conversation management provided by ACRE, tool support has also been provided. This includes graphical programs to allow for the creation and editing of agent protocols, the management of protocols in repositories and, for Agent Factory, a plugin to the framework's native debugger to allow a developer to view the interaction between agents in a structured and informative manner.
\end{itemize}

The evaluation of ACRE was presented in Chapter~\ref{chap:evaluation}. This indicated that the adoption of a system like ACRE does not add to the workload required to create a communication-centric MAS. In addition, subjective analysis of the code that participants in the evaluation created demonstrated a number of common pitfalls that can adversely affect the security, predictability and reliability of a solution created without the use of ACRE's conversation management. This evaluation was conducted using a testbed that was heavily focussed on communication between the agents in the system. Participants undertaking the same programming tasks using different implementation technologies is a common approach in the software engineering community. This work applies this to the agents domain.

The ability of agents to share common definitions of protocols is facilitated through a standard XML format, which is discussed in Chapter~\ref{chap:acre-architecture} and supplied in full in Appendix~\ref{app:schemas}. Along with the provision of graphical tools to create and manage these descriptions, this facilitates the adoption of this type of shared protocol independently of any particular AOP language or framework. Even in cases where conversation management is not a feature of the AOP language used, such protocol definitions can still be utilised as a form of Application Programming Interface to aid communication between the developers themselves.

Finally, the last requirement of the core contributions outlined in Section~\ref{sec:introduction:core-contributions} was that certain restrictions were followed in the design of the system so as to aid greater compatibility:
\begin{itemize}
   \item The definition of protocols was centralised (as defined in the XML documents) so that a single view of the progress of conversations can be maintained. This is in contrast to some previous approaches (discussed in Chapter~\ref{chap:communication}) that maintained separate definitions for the various participants in an interaction.
   \item In designing the generic architecture discussed in Chapter~\ref{chap:acre-architecture}, very few assumptions were made about the capabilities of frameworks within which ACRE could be integrated. The assumptions that are made are as follows:
   \begin{itemize}
      \item Some mechanism exists whereby an ACRE implementation can access the incoming messages of an agent;
      \item The framework contains some message-sending service that ACRE can use to send messages; and
      \item Agents can use some action to interact with ACRE in the sending of messages.
   \end{itemize}
    In terms of language requirements, ACRE does not rely on any particular implementation of communication semantics to be present. These semantics can be used in a complimentary manner to the conversation management if they are present.
   \item Although ACRE makes use of the conversation management aspects of FIPA-ACL (discussed in Chapter~\ref{chap:fipa}), it is not dependent on these being used. Thus ACRE can still be used by agents that communicate with others that do not feature conversation management, as the matching of messages to conversations is performed on the fly, according to the definitions of the underlying protocols being used.
\end{itemize}

\bibliography{PhDThesis}
\appendix
\part{Appendices}
\chapter{Agent UML Diagrams for FIPA Interaction Protocols}
\label{app:fipa-protocols}

This Appendix contains the Agent UML diagrams for the FIPA Interaction Protocols listed in Chapter~\ref{chap:fipa}. These are ordered in such a way to be consistent with the order in which they are discussed in Chapter~\ref{chap:fipa}.

\begin{figure}[!hbtp]
   \includegraphics{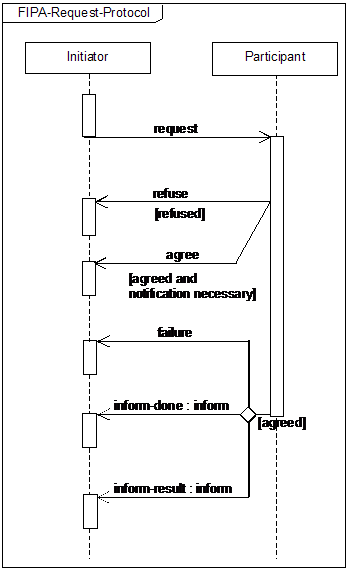}
   \caption{FIPA Request Interaction Protocol (SC00026).}
   \label{fig:fipa-request}
\end{figure}

\begin{figure}[!hbtp]
   \includegraphics{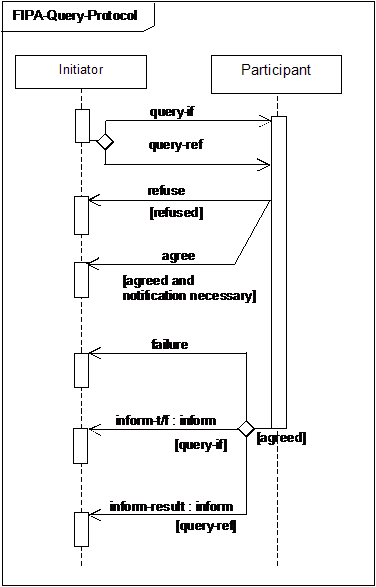}
   \caption{FIPA Query Interaction Protocol (SC00027).}
   \label{fig:fipa-query}
\end{figure}

\begin{figure}[!htp]
   \includegraphics{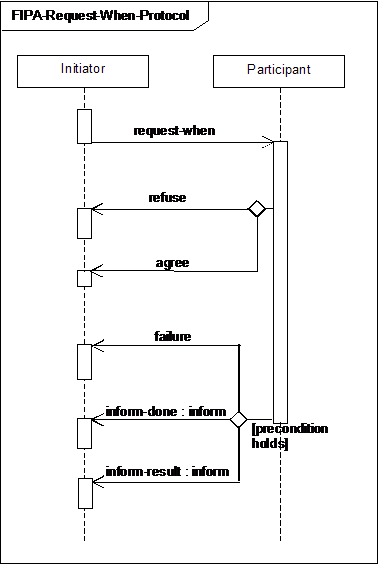}
   \caption{FIPA Request When Interaction Protocol (SC00028).}
   \label{fig:fipa-request-when}
\end{figure}

\begin{figure}[!htp]
   \includegraphics[width=410px]{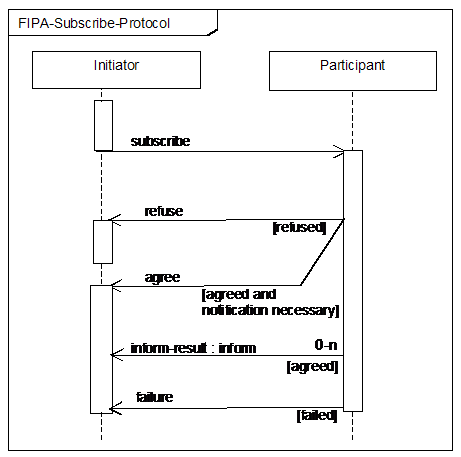}
   \caption{FIPA Subscribe Interaction Protocol (SC00035).}
   \label{fig:fipa-subscribe}
\end{figure}

\begin{figure}[!htp]
   \includegraphics{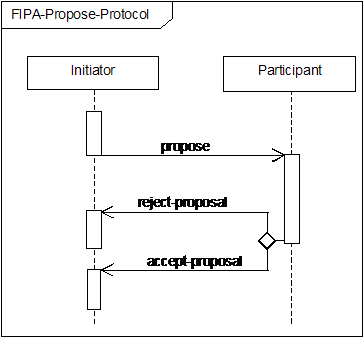}
   \caption{FIPA Propose Interaction Protocol (SC00036).}
   \label{fig:fipa-propose}
\end{figure}

\begin{figure}[!htp]
   \includegraphics{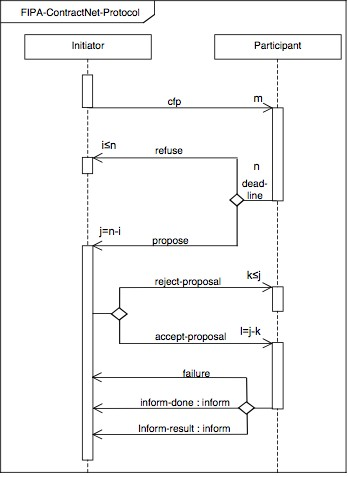}
   \caption{FIPA Contract Net Interaction Protocol (SC00029).}
   \label{fig:fipa-contract-net}
\end{figure}

\begin{figure}[!htp]
   \includegraphics{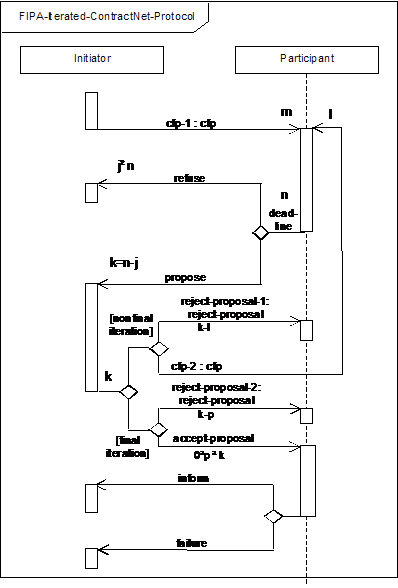}
   \caption{FIPA Iterated Contract Net Interaction Protocol (SC00030).}
   \label{fig:fipa-iterated-contract-net}
\end{figure}

\begin{figure}[!htp]
   \includegraphics[width=410px]{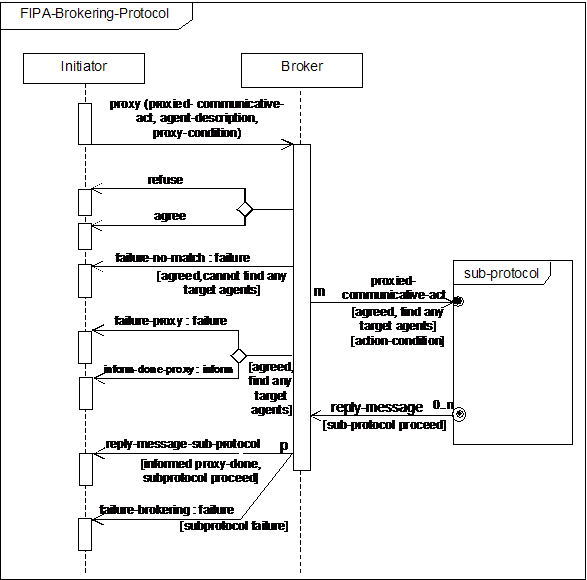}
   \caption{FIPA Brokering Interaction Protocol (SC00033).}
   \label{fig:fipa-brokering}
\end{figure}

\begin{figure}[!htp]
   \includegraphics[width=410px]{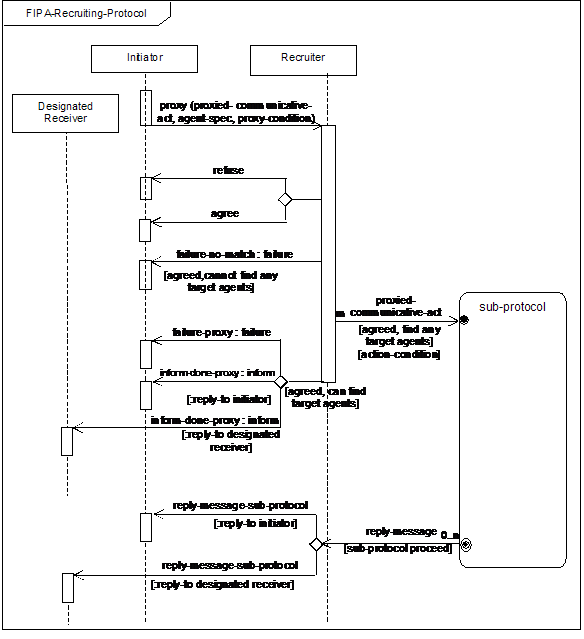}
   \caption{FIPA Recruiting Interaction Protocol (SC00034).}
   \label{fig:fipa-recruiting}
\end{figure}

\begin{figure}[!htp]
   \includegraphics{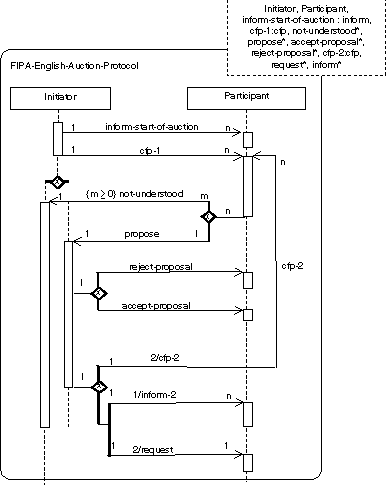}
   \caption{FIPA English Auction Interaction Protocol (XC00031).}
   \label{fig:fipa-english-auction}
\end{figure}

\begin{figure}[!htp]
   \includegraphics{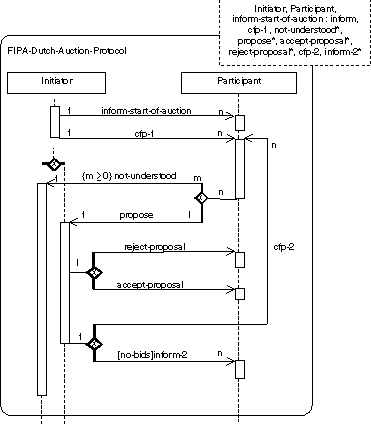}
   \caption{FIPA Dutch Auction Interaction Protocol (XC00032).}
   \label{fig:fipa-dutch-auction}
\end{figure}

\begin{figure}[!ht]
   \includegraphics{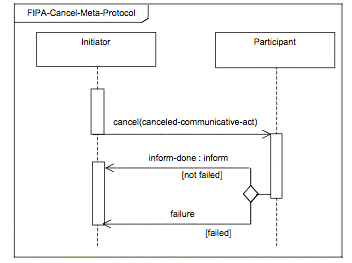}
   \caption{FIPA Cancel Meta Protocol.}
   \label{fig:fipa-cancel}
\end{figure}
\chapter{Schemas for ACRE Repositories}
\label{app:schemas}

\begin{Verbatim}[frame=single,fontsize=\footnotesize]
<xs:schema 
  xmlns:xs="http://www.w3.org/2001/XMLSchema"
  xmlns:acre="http://acre.lill.is/XMLSchema"
  xmlns="http://acre.lill.is" 
  targetNamespace="http://acre.lill.is" 
  elementFormDefault="qualified">
  <xs:annotation>
    <xs:documentation xml:lang="en">
      Schema for ACRE Protocol Repository
      Author: Dave Lillis [ dave /at/ lill /dot/ is ]
    </xs:documentation>
  </xs:annotation>
  <xs:element name="repository">
    <xs:complexType>
      <xs:sequence>
        <xs:element name="base" type="xs:anyURI"/>
        <xs:element name="namespaces" type="namespaces-type"/>
      </xs:sequence>
    </xs:complexType>
  </xs:element>
  <xs:complexType name="namespaces-type">
    <xs:sequence>
      <xs:element name="namespace" type="namespace-type" 
        minOccurs="0" maxOccurs="unbounded"/>
   </xs:sequence>
 </xs:complexType>
 <xs:complexType name="namespace-type">
   <xs:sequence>
     <xs:element name="protocol" type="protocol-type" 
       maxOccurs="unbounded"/>
   </xs:sequence>
   <xs:attribute name="name" type="namespace-name-type"
     use="required"/>
  </xs:complexType>
  <xs:complexType name="protocol-type">
    <xs:attribute name="name" type="name-type" use="required"/>
    <xs:attribute name="version" type="version-type"
      use="required"/>
  </xs:complexType>

<!-- definition of ACRE namespace type -->
  <xs:simpleType name="namespace-name-type">
    <xs:restriction base="xs:string">
      <xs:pattern value=
       "[a-z\d]([a-z\d-]*[a-z\d])?(\.[a-z\d]([a-z-\d]*[a-z\d])?)*"/>
    </xs:restriction>
  </xs:simpleType>

<!-- definition of ACRE name type -->
  <xs:simpleType name="name-type">
    <xs:restriction base="xs:string">
      <xs:pattern value="[a-z\d]([a-z\d-]*[a-z\d])?"/>
    </xs:restriction>
  </xs:simpleType>

<!-- definition of ACRE version type -->
  <xs:simpleType name="version-type">
    <xs:restriction base="xs:string">
      <xs:pattern value="\d+\.\d+"/>
    </xs:restriction>
  </xs:simpleType>
</xs:schema>
\end{Verbatim}
\begin{figure}[!hb]
\caption{XML Schema Document for repository.xml files.}
\label{fig:repository-xsd}
\end{figure}

\begin{Verbatim}[frame=single,fontsize=\footnotesize]
<xs:schema
  xmlns:xs="http://www.w3.org/2001/XMLSchema"
  xmlns:acre="http://acre.lill.is/XMLSchema"
  targetNamespace="http://acre.lill.is"
  xmlns="http://acre.lill.is"
  elementFormDefault="qualified" >

  <xs:annotation>
    <xs:documentation xml:lang="en">
      Schema for ACRE Protocol Definitions
      Author: Dave Lillis [ dave /at/ lill /dot/ is ]
    </xs:documentation>
  </xs:annotation>

  <!-- overall definition of 'protocol' tag -->
  <xs:element name="protocol">
    <xs:complexType>
      <xs:sequence>
        <xs:element name="namespace" type="namespace-type" />
        <xs:element name="name" type="name-type" />
        <xs:element name="version" type="version-type"/>
        <xs:element name="description" type="xs:string" 
          minOccurs="0"/>
        <xs:element name="import" type="import-type"
          minOccurs="0" maxOccurs="unbounded"/>
        <xs:element name="states" type="states-type"
          minOccurs="0"/>
        <xs:element name="transitions" type="transitions-type"
          minOccurs="0"/>
      </xs:sequence>
    </xs:complexType>
  </xs:element>

  <!-- 'states' tag contains a sequence of 'state' tags -->
  <xs:complexType name="states-type">
    <xs:sequence>
      <xs:element name="state" type="state-type"
        maxOccurs="unbounded" minOccurs="0"/>  
    </xs:sequence> 
  </xs:complexType>

  <!-- 'state' tags must have a 'name' attribute -->
  <xs:complexType name="state-type">
    <xs:attribute name="name" type="xs:string" use="required" />
  </xs:complexType>

  <!-- 'transitions' tag has a sequence of 'transition' tags -->
  <xs:complexType name="transitions-type">
    <xs:sequence>
      <xs:element name="transition" type="transition-type"
        maxOccurs="unbounded" />
    </xs:sequence> 
  </xs:complexType>

  <!-- 'import' must include the namespace, name
    and version of the imported protocol -->

  <xs:complexType name="import-type">
    <xs:sequence>
      <xs:element name="namespace" type="namespace-type" />
      <xs:element name="name" type="name-type" />
      <xs:element name="version" type="version-type" />
      </xs:sequence>
  </xs:complexType>

  <!-- 'transition' tag has 6 attributes: 3 are mandatory -->
  <xs:complexType name="transition-type">
    <xs:attribute name="performative" type="xs:string"
      use="required" />
    <xs:attribute name="from-state" type="xs:string"
      use="required" />
    <xs:attribute name="to-state" type="xs:string"
      use="required" />
    <xs:attribute name="sender" type="variable-type"
      default="?" />
    <xs:attribute name="receiver" type="variable-type"
      default="?" />
    <xs:attribute name="content" type="xs:string" 
      default="?"/>
  </xs:complexType>

  <!-- definition of ACRE namespace type -->
  <xs:simpleType name="namespace-type">
    <xs:restriction base="xs:string">
      <xs:pattern value=
       "[a-z\d]([a-z\d-]*[a-z\d])?(\.[a-z\d]([a-z\d-]*[a-z\d])?)*"/>
    </xs:restriction>
  </xs:simpleType>
  
  <!-- definition of ACRE name type -->
  <xs:simpleType name="name-type">
    <xs:restriction base="xs:string">
      <xs:pattern value="[a-z\d]([a-z\d-]*[a-z\d])?"/>
    </xs:restriction>
  </xs:simpleType>
  
  <!-- definition of ACRE version type -->
  <xs:simpleType name="version-type">
    <xs:restriction base="xs:string">
      <xs:pattern value="\d+\.\d+"/>
    </xs:restriction>
  </xs:simpleType>
  
  <!-- definition of ACRE variable type -->
  <xs:simpleType name="variable-type">
    <xs:restriction base="xs:string">
      <xs:pattern value="\?{1,2}?\w*"/>
    </xs:restriction>
  </xs:simpleType>
</xs:schema>
\end{Verbatim}
\begin{figure}[!hb]
\caption{XML Schema Document for protocol declarations.}
\label{fig:protocol.xsd}
\end{figure}

\chapter{Interaction Protocols for Trading Evaluation Scenario}
\label{chap:appendix-3}

This appendix presents the interaction protocols used in the trading evaluation scenario set out in Chapter~\ref{chap:evaluation}, along with the instructions given to the participants. The protocols are organised under the heading of the agent they relate to. The description of each agent and their associated protocols that was given to evaluation participants is also included. Changes that have been made to the descriptions provided in the original scenario are for typesetting and formatting reasons. One short section detailing how participants should submit their code has been omitted from this appendix.

Each protocol is based on one or more of the standard FIPA Interaction Protocols discussed in Chapter~\ref{chap:fipa}. The system agents that were provided to participants were discoverable via Agent Factory's Directory Facilitator Service (DFS). The ``DFS Name'' provided for each agent is the name by which the DFS identifies it.

\section{Agent Trading Game} \label{sec:appendix3:agent_trading_game}

The Agent Trading Game is designed to help you gain experience both with interacting with other agents and implementing strategic planning within your agent programming language of choice.

You are required to create an agent that buys and sells items with the aim of earning as much money as possible. To do this, your agent must interact with a number of other agents using a variety of well-defined protocols. The protocols that the agents should use have been created in advance and are expressed as ACRE protocols.

As the game runs, a graphical user interface (GUI) will show how the stocks are performing, as well as all the transactions you engage in (buying or selling items). You can also see how much money you have in the bank and what items you own.

The game is controlled by a clock (you can see the clock ticking in the top right corner of the GUI). Each clock tick represents one day. Your aim is to earn as much money as possible at the end of one year (i.e. 365 days). At this point, the game will add your bank balance to the value of the items you own at that time to find your total score for the game.

The agents you need to interact with are as follows (more in-depth descriptions of how they work are provided later):

\begin{itemize}
   \item \textbf{The Banker Agent:} The role of the Banker agent is to keep track of how much money you have. The first thing you will have to do is contact the Banker to open a bank account. At any time, the Banker can tell you how much money you have in this account.
   \item \textbf{The StockBroker Agent:} The StockBroker agent is in charge of buying and selling stocks and shares. You can earn money by buying stocks at low prices and selling them later at a profit. To help you with this, the StockBroker can give you a list of all the stocks and their prices, as well as buying or selling stocks for you.
   \item \textbf{The Guru Agent:} The best way to find out which stocks are the right ones to buy is by asking an expert. The Guru agent can tell you which stocks are likely to go up in value and are worth investing in. By subscribing to the Guru's information, you can make better decisions about what stocks to buy from the StockBroker.
   \item \textbf{The Auctioneer Agent:} Stocks and shares are only one way of making money. The Auctioneer Agent, however, allows you to buy and sell property, which can make you far more money than on the stock market. From time to time, the Auctioneer will advertise properties for sale that you can bid for. If you are successful, the value of your property will rise very quickly, guaranteed! However, at the start you won't be able to afford to take part in these auctions, so you'll have to make some money on the stock market first.
   \item \textbf{The Bidder Agents:} The Bidder agents are the people you can sell your properties to after buying them from the Auctioneer. These compete against each other, so you should be able to get the best deal by engaging them in an auction.
\end{itemize}

\subsection{Protocols}

All of the protocols used in this scenario can be viewed in detail using the ACRE Repository Manager. To view and explore these protocols, you should open the remote repository located at: \url{http://www.lill.is/acre/trading}.

The protocols are grouped into namespaces: one for each of the system agents you need to interact with (e.g. all of the Banker Agent’s protocols are stored under the "\texttt{trading.bank}" namespace).

In most cases, the protocols presented here are based on a FIPA standard interaction protocol. The full list of FIPA interaction protocols is located at the following URL: (\url{http://www.fipa.org/repository/ips.html}). Frequently, however, the protocols here do not follow the FIPA protocols exactly. In some cases, elements of the FIPA protocol are optional, and additionally the FIPA standard protocols do not specify any restrictions on the content of messages, just the performatives they contain.

\subsection{Suggested Strategy}

The best way to tackle this game is to take the following steps in order:
\begin{enumerate}
   \item Open a bank account: you won't be able to buy or sell anything until you have done this.
   \item Find out what stocks the StockBroker has available: You can't buy stocks until you know their names! Also, you won't be able to afford to participate in any auctions with the amount of money you begin with.
   \item Buy some stocks: Stocks increase in value, so you can make money by buying stocks and selling them later.
   \item Ask the Guru for advice: The Guru knows which stocks will rise in value the quickest. Once you're able to buy stocks, you can concentrate on buying the best stocks.
   \item Sell your stocks again: If you want to participate in any auctions, you'll need cash. You get cash by selling stocks for more than you paid for them.
   \item Subscribe to the Auctioneer to find out about upcoming auctions: Buying items in auctions is the best way to make lots of money quickly. The market price of items doubles within a couple of days of you buying them.
   \item Buy items in auction.
   \item Sell items you've bought: After you buy items in auctions, their market value increases very quickly during the following two days. After this happens, the value will not change again in the future. Selling items you've already bought means that you can use the money to bid on more auctions, or re-invest in the stock market.
\end{enumerate}

\clearpage

\section[The Banker Agent]{The Banker Agent\\\small(DFS Name: ``banker'')}

The Banker agent is responsible for storing your money for you. Whenever you buy items from the Stockbroker or Auctioneer, the money comes from your bank account. Any money you gain from selling items goes to your bank account too. There are two different types of conversation you can have with your bank. Firstly, you can open a bank account and secondly you can check to see how much money you have. When you buy and sell items, the other agents will contact your bank directly to arrange for the money to be transferred.

\subsection{Protocol: trading.bank.open}
\begin{description}
	\item[Purpose:] Open a bank account
	\item[Based on:] FIPA Request Interaction Protocol \\ (\url{http://www.fipa.org/specs/fipa00026})
	\item[Variables:]~
	\begin{description}
		\item[?id:] The identifier used as the name of the bank account.
		\item[?amount:] The amount of money in the account when it has been opened.
	\end{description}
\end{description}

\textbf{Description:} This protocol allows you to request that the Banker agent open an account for your agent. In certain situations, the Banker may refuse (if, for example, your agent already has a bank account).

If the creation of the account succeeds, the Banker agent will reply with an \texttt{inform} message that includes the identifier by which your account should be referred (the \texttt{?id} variable) and the opening balance in the account (this is shown in the \texttt{?amount} variable).

\begin{figure}[!ht]
	\caption{Finite State Machine representing the ``trading.bank.open'' protocol.}
	\includegraphics[width=\textwidth]{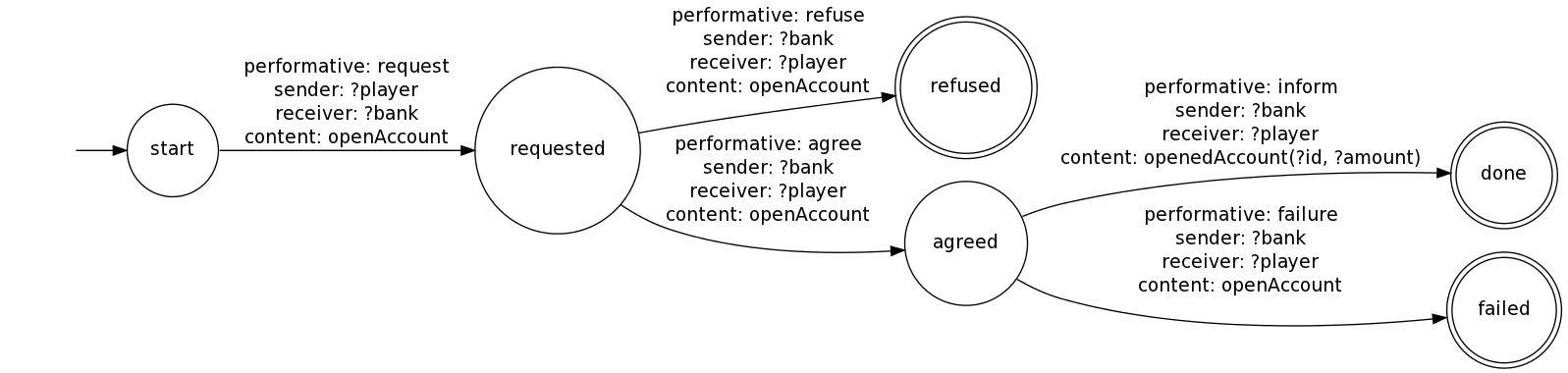}
\end{figure}

\subsection{Protocol: trading.bank.enquiry}
\begin{description}
	\item[Purpose:] Query the Banker agent to find out how much money is in a bank account.
	\item[Based on:] FIPA Query Interaction Protocol \\ (\url{http://www.fipa.org/specs/fipa00027})
	\item[Variables:]~
	\begin{description}
		\item[?id:] The identifier of the bank account in question (this name would have been supplied by the Banker agent when you opened your account).
		\item[?amount:] The amount of money in the account.
	\end{description}
\end{description}

\textbf{Description:} The amount of money available in your bank account will change as you buy and sell stocks and other items. This protocol is provided so that you can ask the Banker about your account balance at any time. The Banker may refuse if you ask about an account that you don't own, or the request may fail if you have not previously opened an account.

\begin{figure}[!ht]
	\caption{Finite State Machine representing the ``trading.bank.enquiry'' protocol.}
	\includegraphics[width=\textwidth]{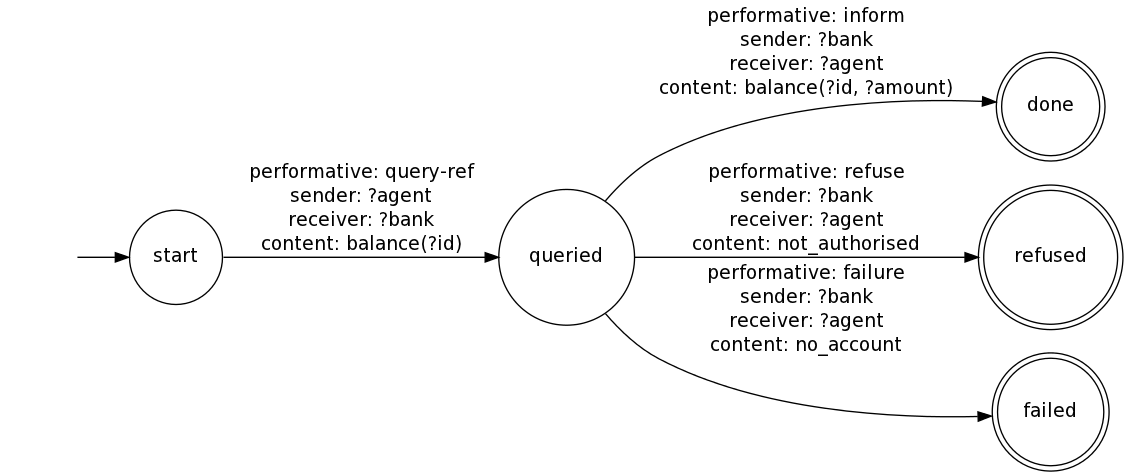}
\end{figure}

\clearpage

\section[The StockBroker Agent]{The StockBroker Agent\\\small(DFS Name: ``stockbroker'')}
The StockBroker Agent allows you to buy or sell stocks. You should aim to buy stocks at a low price and later sell them at a profit. When you first interact with the StockBroker, you will need to check what stocks it has listings for. Following this, you can check the price of a stock, buy and sell stocks, and check what stocks you have in your portfolio.

\subsection{Protocol: trading.broker.listing}
\begin{description}
	\item[Purpose:]: Find out what stocks it is possible to buy.
	\item[Based on:] FIPA Query Interaction Protocol \\ (\url{http://www.fipa.org/specs/fipa00027})
	\item[Variables:]~
	\begin{description}
		\item[?stocklist:] A list of the names of the stocks that are available.
	\end{description}
\end{description}

\textbf{Description:} With this protocol, you are asking the StockBroker which stocks you may buy and sell to/from it. This is done using a \texttt{query-ref} message, in which you ask for a listing.

If the StockBroker doesn't refuse, it will send a message containing the listing as a list. As an example, the content of this message will look something like the following (for three sample stocks):

\texttt{listing(["Nile Ltd.", "Shannon Inc.",} \\ \hspace*{1cm} \texttt{"Mississippi Corp."]);}

\begin{figure}[!ht]
	\caption{Finite State Machine representing the ``trading.broker.listing' protocol.}
	\includegraphics[width=\textwidth]{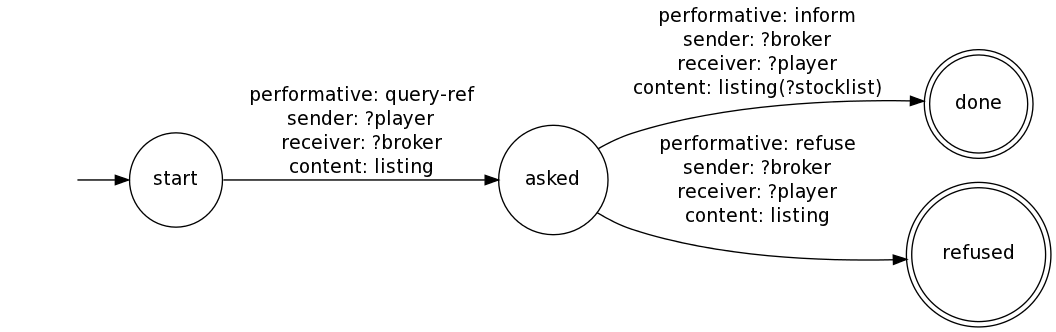}
\end{figure}

\subsection{Protocol: trading.broker.price}
\begin{description}
	\item[Purpose:] out the current price of a particular stock.
	\item[Based on:] FIPA Query Interaction Protocol \\ (\url{http://www.fipa.org/specs/fipa00027})
	\item[Variables:]~
	\begin{description}
		\item[?stock:] The name of the stock in question.
		\item[?price:] The current price of the stock.
	\end{description}
\end{description}

\textbf{Description:} This protocol allows you to ask the Broker about the price of a particular stock. This is done by sending a \texttt{query-ref} message that includes the name of the stock about which you are enquiring (in the \texttt{?stock} parameter). The Broker will normally respond with an \texttt{inform} message that tells of the price of the stock that was requested. In some situations (e.g. if the stock's name is unknown), the request may fail.

\begin{figure}[!ht]
	\caption{Finite State Machine representing the ``trading.broker.price'' protocol.}
	\includegraphics[width=\textwidth]{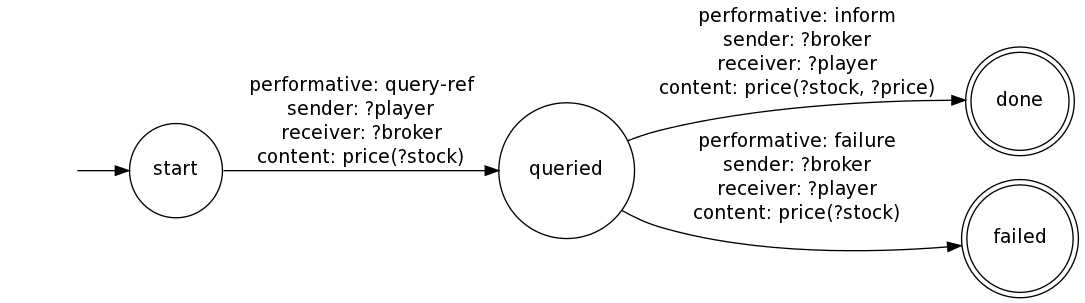}
\end{figure}

\subsection{Protocol: trading.broker.portfolio}
\begin{description}
	\item[Purpose:] Find out details of stocks you currently own, including the quantity of the stock, the price it was bought at and its current value.
	\item[Based on:] FIPA Query Interaction Protocol \\ (\url{http://www.fipa.org/specs/fipa00027})
	\item[Variables:]~
	\begin{description}
		\item[?portfolio:] A list of the details of the stocks you own.
	\end{description}
\end{description}

\textbf{Description:} This is a simple query protocol that allows you to find out what stocks you own. The StockBroker may refuse if you have not previously bought any stocks. If you do own stocks, it will reply with an inform message that includes your portfolio in the \texttt{?portfolio} parameter.

This parameter is a list of lists (i.e. a 2-dimensional list). Each individual list is made up of two values: the first value is the name of a stock, the second value is the quantity of that stock you own. For example, if you owned 400 units of ABC Ltd. stock, 300 units of XYZ Inc. stock and 250 units of REM LLC. stock, the content of this inform message would be:

\texttt{portfolio([["ABC Ltd.",400],["XYZ Inc.",300], \\ \hspace*{1cm}
   ["Rem LLC.",250]])}

\begin{figure}[!ht]
	\caption{Finite State Machine representing the ``trading.broker.portfolio'' protocol.}
	\includegraphics[width=\textwidth]{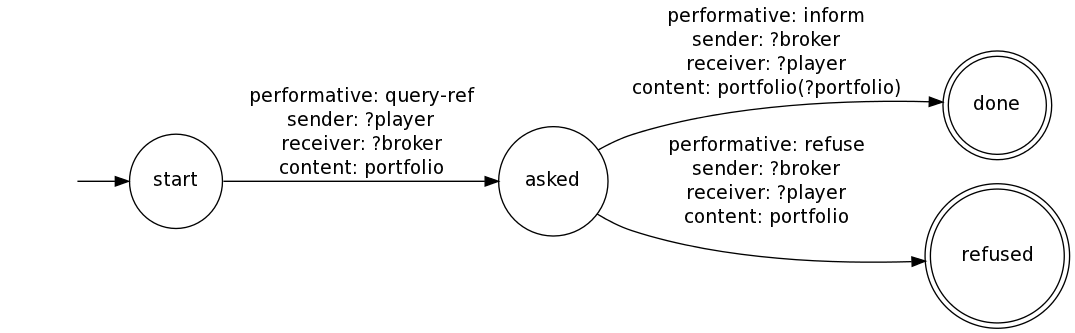}
\end{figure}

\subsection{Protocol: trading.broker.buy}
\begin{description}
	\item[Purpose:] Buy some quantity of an available stock.
	\item[Based on:] FIPA Request Interaction Protocol \\ (\url{http://www.fipa.org/specs/fipa00026})
	\item[Variables:]~
	\begin{description}
		\item[?stock:] The name of the stock that is to be bought.
		\item[?value:] The total value of the stock that is to be bought.
		\item[?qty:] The quantity of the stock to be bought.
	\end{description}
\end{description}

\textbf{Description:} The ``buy'' protocol allows an agent to buy stock from the Broker. The initial request to buy stocks can be expressed in either of two ways. In each case, the \texttt{?item} parameter is used to indicate the stock to be bought, however the requesting agent has the option of specifying either a quantity of stock that is desired or else the value of the stock it wants to buy.

In either case, the broker will typically respond with a proposal to sell a particular quantity of the stock at a particular total value. In some cases, the total value will not be exactly the same as what was initially requested. For example, if the price of a stock is 3, and you have requested to buy a value of 100, the broker will respond with a proposal to sell a quantity of 33 units, leading to an overall value of 99, rather than 100.

The player may now either accept this proposal or reject it. If the proposal is accepted, the broker will respond to indicate whether the transaction was successful or not (this may happen if the player does not have enough money in its account to complete the purchase).

\begin{figure}[!ht]
	\caption{Finite State Machine representing the ``trading.broker.buy'' protocol.}
	\includegraphics[width=\textwidth]{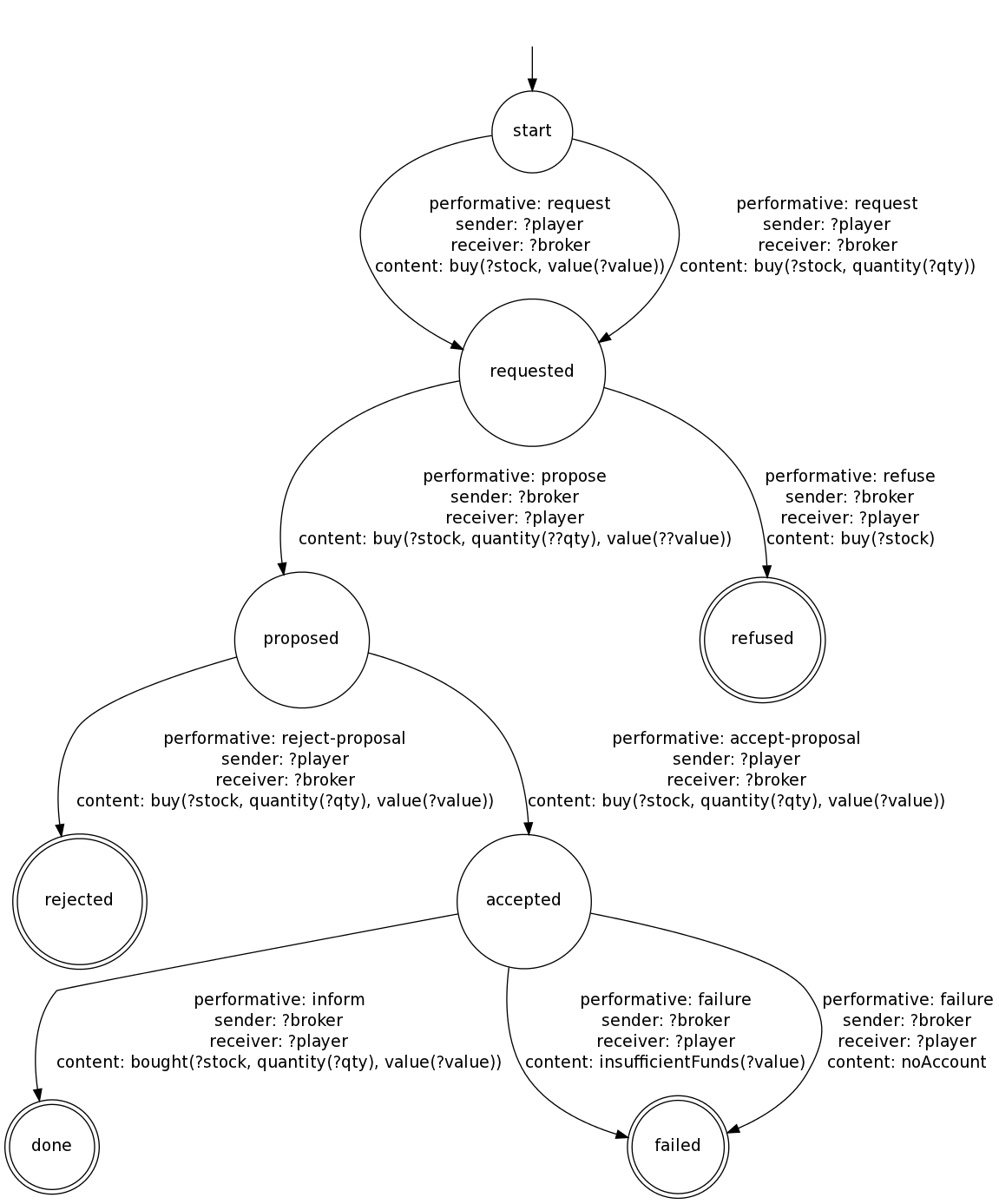}
\end{figure}

\subsection{Protocol: trading.broker.sell}
\begin{description}
	\item[Purpose:] some quantity of stock that you have previously bought.
	\item[Based on:] FIPA Request Interaction Protocol \\ (\url{http://www.fipa.org/specs/fipa00026})
	\item[Variables:]~
	\begin{description}
		\item[?stock:] The name of the stock that is being sold.
		\item[?value:] The quantity of stock being sold.
		\item[?qty:] The value of the stock being sold, at current market prices.
	\end{description}
\end{description}

\textbf{Description:} The ``sell'' protocol is almost identical to the ``buy''" protocol described above. In this case, you wish to sell a quantity of stock that it has previously bought. The initial request to sell can be expressed as a quantity of stock that is to be sold, or a value of stock that is to be sold.

Failure of this protocol may come about if the player does not own enough stock for the sale to go through.

\begin{figure}[!ht]
	\caption{Finite State Machine representing the ``trading.broker.sell'' protocol.}
	\includegraphics[width=\textwidth]{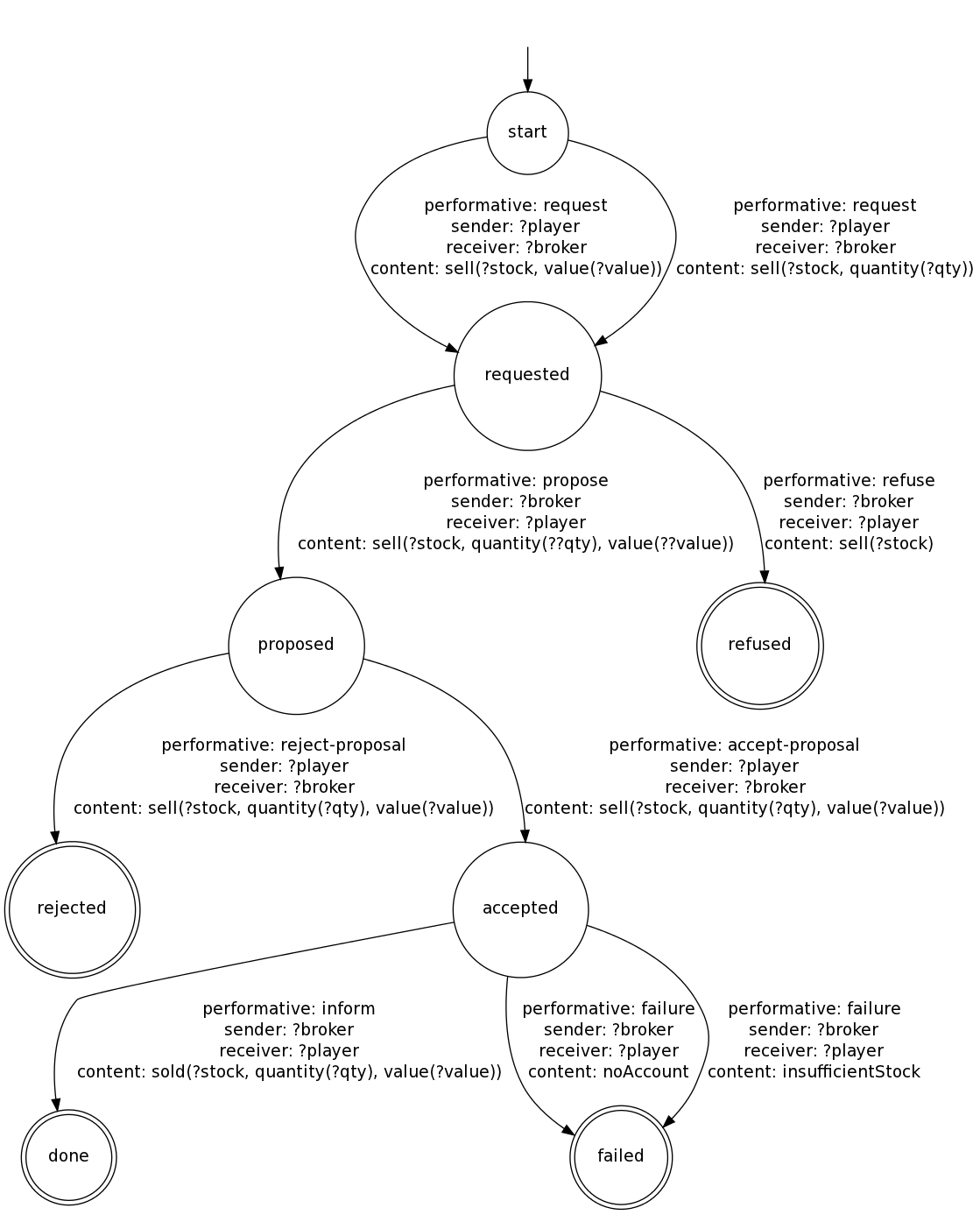}
\end{figure}

\clearpage

\section[The Guru Agent]{The Guru Agent\\\small(DFS Name: ``guru'')}

The Guru Agent knows what is likely to happen in the stock market, so it can tell you which stocks are good to buy (or sell) during the game. The guru only knows one protocol, which allows you to subscribe to its stock tips.

\subsection{Protocol: trading.guru.subscribe}
\begin{description}
	\item[Purpose:] Ask the Guru agent to provide information on future stock movements.
	\item[Based on:] FIPA Subscribe Interaction Protocol \\ (\url{http://www.fipa.org/specs/fipa00035})
	\item[Variables:]~
	\begin{description}
		\item[?stock:] The name of the stock about which the guru has information.
		\item[?rise:] The \% increase in the price of the stock that is expected.
		\item[?days:] The number of days it is expected for the full stock rise to occur.
	\end{description}
\end{description}

\textbf{Description:} When advice comes from the Guru agent, it can either be advice to buy a particular stock, or to sell it. Advice to sell a stock indicates that its value is not expected to rise significantly in the near term. If a player already has bought this stock, it should consider selling it and investing the money in other stocks. Players that do not own this stock should not buy it at present.

Advice to buy is more complex, however. The guru will also tell you by how much it is to rise (the \texttt{\texttt{??rise}} parameter contains this information as a percentage) and how long this rise will take (in the \texttt\texttt{{??days}} parameter). For example, if a stock called ``Blue'' is to double in price over the next week, this would be indicated by the message \texttt{buy(Blue,100,7)}.

The Guru will continue sending advice for as long as you're subscribed. If you want to stop receiving the Guru's advice, you will have to cancel the conversation.

\begin{figure}[!ht]
	\caption{Finite State Machine representing the ``trading.guru.subscribe'' protocol.}
	\includegraphics[width=\textwidth]{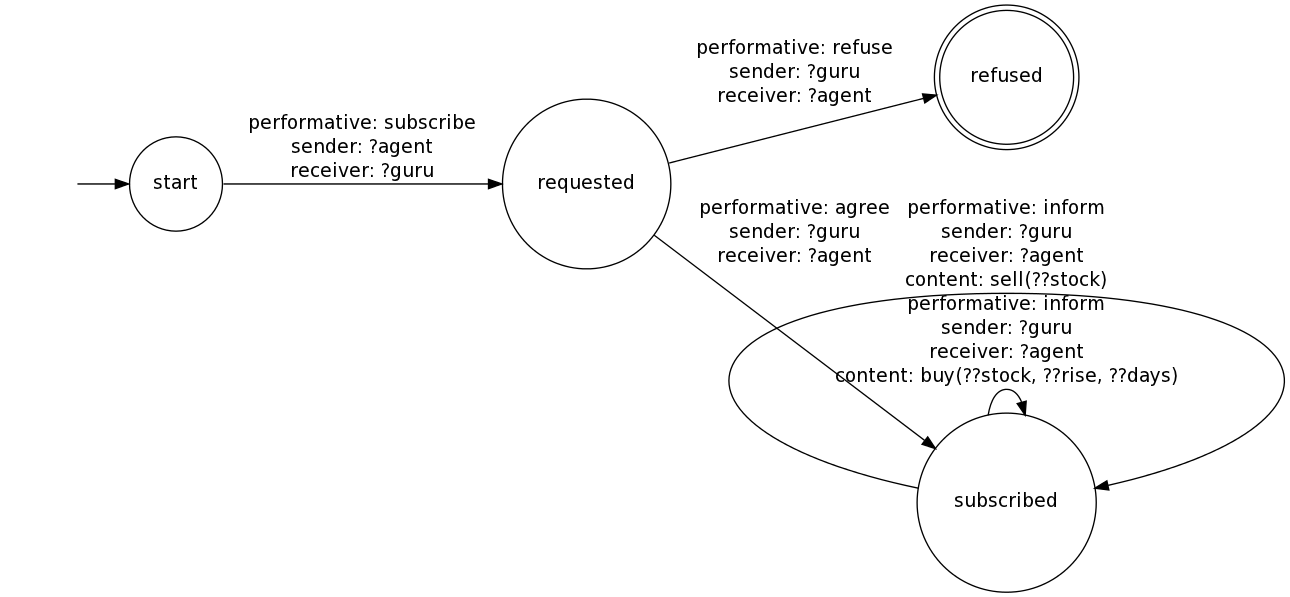}
\end{figure}

\clearpage

\section[The Auctioneer Agent]{The Auctioneer Agent\\\small(DFS Name: ``auctioneer'')}

The Auctioneer Agent allows you to buy high-value items that will quickly rise in value. From time-to-time, it will inform subscribed agents of an item that is available for auction. Winning an auction (by offering the most money for the item on sale) means that you now own the item and can subsequently sell it for a profit (see the Bidder agents section below for details of selling items).

The Auctioneer will only run one auction at a time, so it waits for one auction to finish before announcing the beginning of the next auction.

\subsection{Protocol: trading.auctioneer.subscribe}
\begin{description}
	\item[Purpose:] Subscribe to any auctions that the Auctioneer wishes to hold and bid on those in which we're interested.
	\item[Based on:] FIPA Subscribe Interaction Protocol \\ (\url{http://www.fipa.org/specs/fipa00035}) \\ and FIPA English Auction Interaction Protocol \\ (\url{http://www.fipa.org/specs/fipa00031})
	\item[Variables:]~
	\begin{description}
		\item[?item:] The identifier of the item that is being auctioned.
		\item[?amt:] The amount of money the auctioneer is seeking, or the bidder is offering.
	\end{description}
\end{description}

\textbf{Description:} This is a complex protocol with two parts: the first is the process of subscribing to the Auctioneer agent's auctions. The second is the auction itself, where the Auctioneer asks for bids to be made for the item in question.

Subscribing is done by sending a \texttt{subscribe} message to the Auctioneer. If the Auctioneer agrees, you are now subscribed to announcements about auctions (this is the \textit{subscribed} state).

From here, the Auctioneer will begin an auction by issuing a call for proposals (via a \texttt{cfp} message) that asks for bids for a particular item to be made. If you do not wish to take part in the auction, you may refuse, which returns the conversation to the \textit{subscribed} state in which you await the next auction to be advertised.

Submitting a bid is done using a \texttt{propose} message, which brings the conversation to the \textit{proposed} state. From here, the Auctioneer may either accept or reject the bid. If it rejects the bid (usually because another agent made the same bid earlier), it will then issue a new call for proposals at a higher price. This results in the conversation returning to the \textit{called} state, where you again have the option to bid or to refuse.

If the offer is accepted, then the Auctioneer will attempt to complete the transaction. This may fail if, for example, you don't have enough money to cover the bid you made. A successful transaction results in the conversation returning to the \textit{subscribed} state again, where you will hear about new auctions as they arise.

\begin{figure}[!ht]
	\caption{Finite State Machine representing the ``trading.auctioneer.subscribe'' protocol.}
	\includegraphics[width=\textwidth]{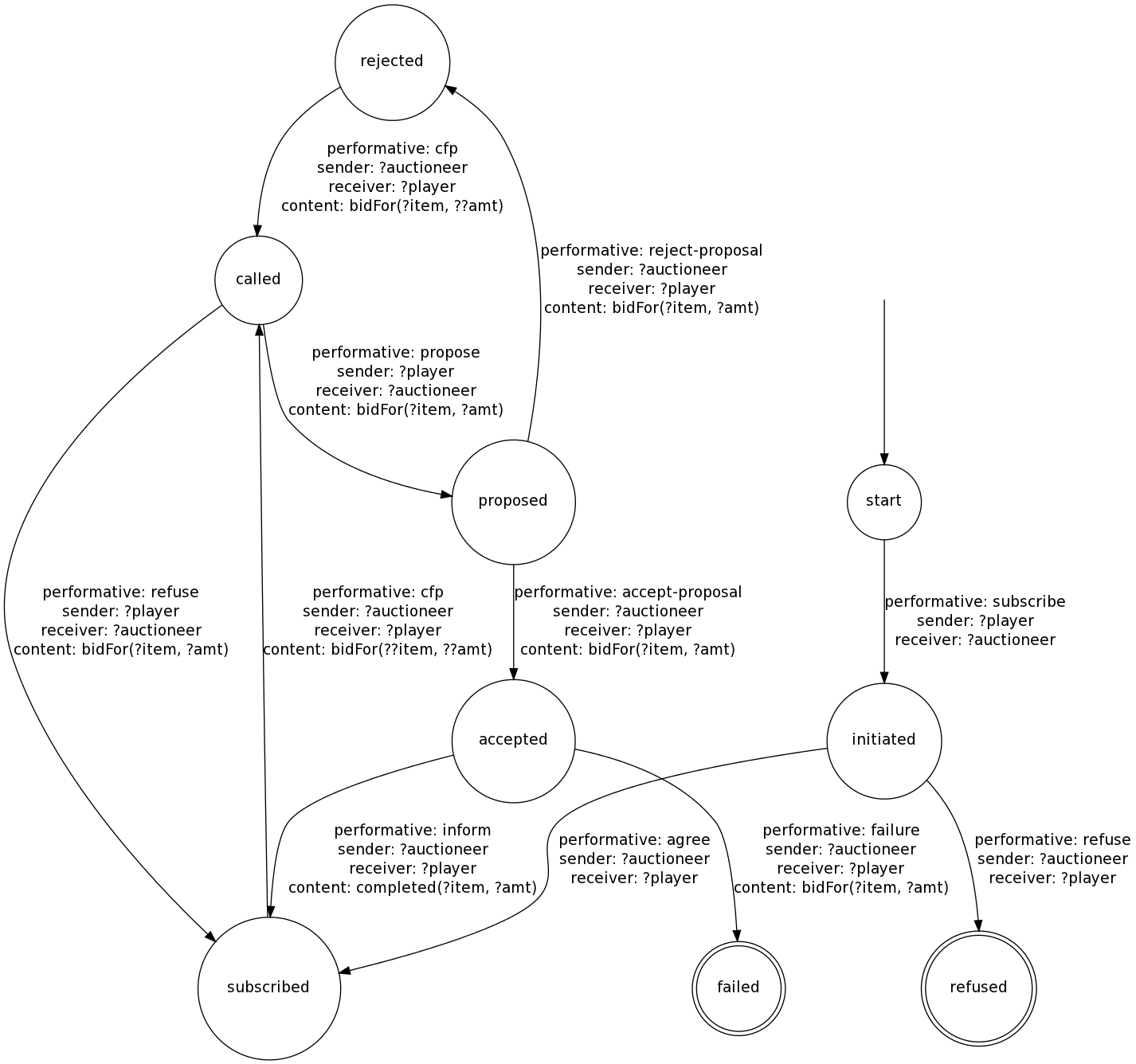}
\end{figure}

\clearpage

\section[The Bidder Agents]{The Bidder Agents\\\small(DFS Names: ``bidder1'', ``bidder2'', ``bidder3'')}

The Bidder agents are usually willing to buy items that you have bought after an auction with the Auctioneer agent. To get the highest price for the item you are selling, you should engage all bidders in an auction and choose the highest bid.

\subsection{Protocol: trading.bidder.sell}
\begin{description}
	\item[Purpose:] : Sell an item that was previously bought from the Auctioneer.
	\item[Based on:] FIPA Contract Net Interaction Protocol \\ (\url{http://www.fipa.org/specs/fipa00029}))
	\item[Variables:]~
	\begin{description}
		\item[?item:] The identifier of the item being sold.
		\item[?amt:] The amount of money being bid.
	\end{description}
\end{description}

\textbf{Description:} Unlike the protocols that the other agents can engage in, this protocol is intended to be used in a group, where a number of different conversations are occurring at the same time. This works like a one-shot auction whereby each bidder is asked to bid on an item and each submits one bid (if it wishes: it is permitted to refuse). You, acting as the auctioneer in this case, should accept the highest bid and reject the others.

The protocol begins with the Bidder being asked to submit a bid for an item. Each bidder may refuse to participate in the auction or it may make a bid using a \texttt{propose} message. Once bids have been received, the player has the option of accepting or rejecting the proposal. If a proposal is accepted, the bidder will then inform the player that the transaction has been completed, or alternatively that the transaction has failed (e.g. if the player is attempting to sell an item it does not own).

\begin{figure}[!ht]
	\caption{Finite State Machine representing the ``trading.bidder.sell'' protocol.}
	\includegraphics[width=\textwidth]{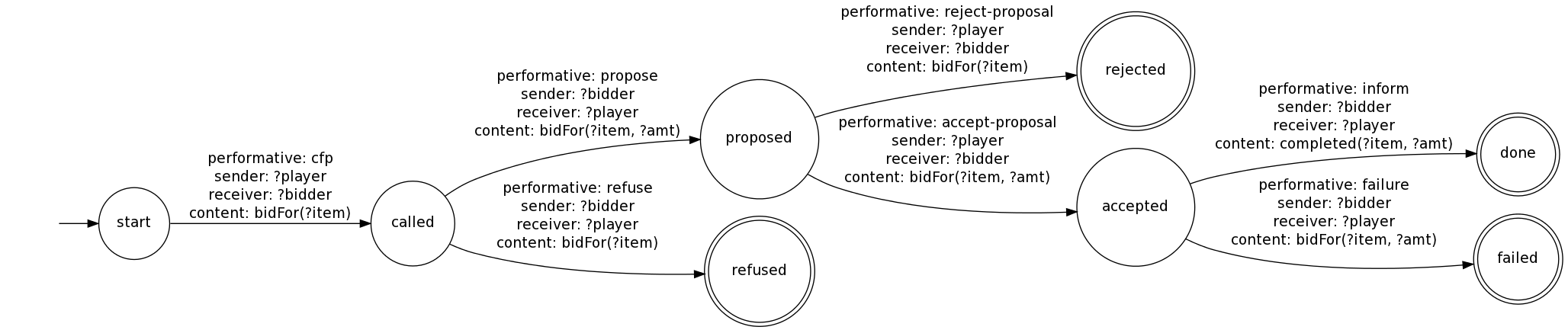}
\end{figure}
\singlespacing
\end{document}